# The Kepler Giant Planet Search. I. A Decade of Kepler Planet-host Radial Velocities from W. M. Keck Observatory

Lauren M. Weiss[1], Howard Isaacson[2,3], Andrew W. Howard[4], Benjamin J. Fulton[5,6], Erik A. Petigura[7], Daniel Fabrycky[8], Daniel Jontof-Hutter[9], Jason H. Steffen[10,11], Hilke E. Schlichting[12], Jason T. Wright[13,14,15], Corey Beard[16,25], Casey L. Brinkman[17], Ashley Chontos[18,19,26], Steven Giacalone[20], Michelle L. Hill[21], Molly R. Kosiarek[22,27], Mason G. MacDougall[7], Teo Močnik[23], Alex S. Polanski[24], Emma V. Turtelboom[3], Dakotah Tyler[7], and Judah Van Zandt[7]

[1] Department of Physics and Astronomy, University of Notre Dame, Notre Dame, IN 46556, USA; lweiss4@nd.edu
[2] Centre for Astrophysics, University of Southern Queensland, Toowoomba, QLD, Australia
[3] Department of Astronomy, 501 Campbell Hall, University of California, Berkeley, Berkeley, CA 94720, USA
[4] Department of Astronomy, California Institute of Technology, Pasadena, CA 91125, USA
[5] Cahill Center for Astronomy & Astrophysics, California Institute of Technology, Pasadena, CA 91125, USA
[6] IPAC-NASA Exoplanet Science Institute, Pasadena, CA 91125, USA
[7] Department of Physics and Astronomy, University of California, Los Angeles, CA 90095, USA
[8] Dept. of Astronomy & Astrophysics, University of Chicago, 5640 S. Ellis Avenue, Chicago, IL 60637, USA
[9] Dept. of Physics, University of the Pacific, Stockton, CA 95211, USA
[10] Department of Physics and Astronomy, University of Nevada, Las Vegas, 4505 South Maryland Parkway, P.O. Box 454002, Las Vegas, NV 89154, USA
[11] Nevada Center for Astrophysics, 4505 South Maryland Parkway, P.O. Box 454002, Las Vegas, NV 89154, USA
[12] Department of Earth, Planetary, and Space Sciences, The University of California, Los Angeles, 595 Charles E. Young Drive East, Los Angeles, CA 90095, USA
[13] Department of Astronomy & Astrophysics, 525 Davey Laboratory, The Pennsylvania State University, University Park, PA, 16802, USA
[14] Center for Exoplanets and Habitable Worlds, 525 Davey Laboratory, The Pennsylvania State University, University Park, PA, 16802, USA
[15] Penn State Extraterrestrial Intelligence Center, 525 Davey Laboratory, The Pennsylvania State University, University Park, PA, 16802, USA
[16] Department of Physics & Astronomy, The University of California, Irvine, Irvine, CA 92697, USA
[17] Institute for Astronomy, University of Hawai'i, 2680 Woodlawn Drive, Honolulu, HI 96822 USA
[18] Department of Astrophysical Sciences, Princeton University, 4 Ivy Lane, Princeton, NJ 08540, USA
[19] Institute for Astronomy, University of Hawai'i, 2680 Woodlawn Drive, Honolulu, HI 96822, USA
[20] Department of Astronomy, University of California, Berkeley, Berkeley, CA 94720, USA
[21] Department of Earth and Planetary Sciences, University of California, Riverside, 900 University Avenue, Riverside, CA 92521, USA
[22] Department of Astronomy and Astrophysics, University of California, Santa Cruz, Santa Cruz, CA 95064, USA
[23] Gemini Observatory/NSF's NOIRLab, 670 N. A'ohoku Place, Hilo, HI 96720, USA
[24] Department of Physics and Astronomy, University of Kansas, Lawrence, KS 66045, USA
Received 2023 June 27; revised 2023 September 25; accepted 2023 September 26; published 2023 December 27

## Abstract

Despite the importance of Jupiter and Saturn to Earth's formation and habitability, there has not yet been a comprehensive observational study of how giant exoplanets correlate with the architectural properties of close-in, sub-Neptune-sized exoplanets. This is largely because transit surveys are particularly insensitive to planets at orbital separations $\gtrsim 1$ au, and so their census of Jupiter-like planets is incomplete, inhibiting our study of the relationship between Jupiter-like planets and the small planets that do transit. To investigate the relationship between close-in, small and distant, giant planets, we conducted the Kepler Giant Planet Survey (KGPS). Using the W. M. Keck Observatory High Resolution Echelle Spectrometer, we spent over a decade collecting 2844 radial velocities (RVs; 2167 of which are presented here for the first time) of 63 Sunlike stars that host 157 transiting planets. We had no prior knowledge of which systems would contain giant planets beyond 1 au, making this survey unbiased with respect to previously detected Jovians. We announce RV-detected companions to 20 stars from our sample. These include 13 Jovians ($0.3\,M_J < M \sin i < 13\,M_J$, 1 au $< a <$ 10 au), eight nontransiting sub-Saturns, and three stellar-mass companions. We also present updated masses and densities of 84 transiting planets. The KGPS project leverages one of the longest-running and most data-rich collections of RVs of the NASA Kepler systems yet, and it will provide a basis for addressing whether giant planets help or hinder the growth of sub-Neptune-sized and terrestrial planets. Future KGPS papers will examine the relationship between small, transiting planets and their long-period companions.

---

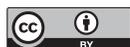

*Unified Astronomy Thesaurus concepts:* Exoplanets (498); Exoplanet catalogs (488); Exoplanet systems (484); Radial velocity (1332); Transits (1711); Orbital elements (1177); Exoplanet detection methods (489); Mini Neptunes (1063); Super Earths (1655); Extrasolar rocky planets (511); Binary stars (154)

*Supporting material:* machine-readable tables

## 1. Introduction

The formation of the solar system and of many exoplanet systems is a major unsolved problem in astrophysics and planetary science. Jupiter and Saturn are thought to have influenced the architecture of the solar system by delaying and stunting the formation of terrestrial planets (Walsh et al. 2011) and shepherding water-bearing comets toward Earth (Meech & Raymond 2020). Do exoplanet systems with small inner planets also have exterior giant planets? If that is the case, how do those planets' properties correlate with inner planetary system properties?

The NASA Kepler mission has discovered thousands of transiting planet candidates between the sizes of Mars and Neptune within 1 au of their host stars (Borucki et al. 2010; Batalha et al. 2013; Mullally et al. 2015; Thompson et al. 2018), the majority of which are bona fide, confirmed planets (Rowe et al. 2014; Morton et al. 2016). The Kepler sample of small planets provides a unique opportunity to investigate whether giant planets help or hinder the growth of small worlds in other planetary systems. However, Kepler's short (4 yr) mission duration impaired its sensitivity to Jovian analogs (planets of $0.3 M_J$–$13 M_J$ with orbits at 1–10 au). Furthermore, planets at the orbital distance of Jupiter (5 au) have only a 0.1% chance of transiting their star—a prerequisite for detection of their transits with Kepler. Thus, the occurrence of Jovian siblings to the small Kepler planets cannot readily be determined from the Kepler data alone.

Dynamical measurements of the Kepler planet-hosting stars provide an excellent means for detecting Jovian companions to small planets. Radial velocity (RV) measurements have led to the identification of Jovian and substellar companions in 16 Kepler systems, including precise characterization of their orbital periods, eccentricities, and minimum masses in six systems (e.g., Marcy et al. 2014; Gettel et al. 2016; Otor et al. 2016; Mills et al. 2019b; Weiss et al. 2020; Zhang et al. 2021). Although the ESA Gaia mission has great promise for finding Jovians around nearby stars, the majority of Kepler stars are too distant (∼1000 pc) to produce sufficient single-measurement astrometric precision to characterize planetary orbits from Jovian-mass companions at 1–10 au.

Because Jovian planets could have orbital periods substantially longer than the Kepler mission, detecting Jovians via RVs requires long-term monitoring. (Note that Jupiter, at 5 au, has a period of 12 yr.) Since 2009, we have conducted the Kepler Giant Planet Search (KGPS). In this survey, we have been measuring RVs of Kepler stars hosting small ($< 4 R_\oplus$) planets to discover and characterize giant planet companions to these sub-Neptunes. KGPS includes several major objectives, which are addressed in a series of papers. This paper (Paper I) includes our sample selection criteria, a presentation of the RVs collected, a new algorithm to find giant planets in systems with known small transiting planets, and orbital and $M \sin i$ properties of the nontransiting planets in our survey, seven of which are new discoveries. In Paper II, we will use a subsample of KGPS that is carefully curated to avoid biases to measure the occurrence of Jovian planets in systems with sub-Neptunes and compare this Jovian frequency to the frequency of Jovians in the solar neighborhood. The KGPS sample will provide future opportunities to look for correlations between the physical and dynamical properties of the small, inner planets and the giant, outer planets (Papers III and up; see Table 1), as well as many other papers from the community.

## 2. Sample Selection

This paper and others in the KGPS series include RV follow-up of Kepler planetary systems with the following properties, based on their initial characterization in Marcy et al. (2014) (and with revised properties from Petigura et al. 2017):

1. A single, Sunlike host star, defined as:
   (a) a Sunlike effective temperature (late F, G, or K spectral type, 4300 K $< T_{\text{eff}} <$ 6300 K);
   (b) slowly rotating ($v \sin i <$ 10.1 m s$^{-1}$);
   (c) a main-sequence or only moderately evolved star ($\log g > 4.0$);
   (d) low magnetic activity ($\log R'_{\text{HK}} < -4.8$);
   (e) no companions within 2″ of comparable brightness in V band ($\Delta V < 5$);
   (f) bright in V band ($V < 13.6$).[28]
2. Having a sufficient RV baseline to detect Jupiter analogs, defined as:
   (a) the earliest RV from 2018 or earlier;
   (b) at least 10 RVs.
3. Having at least one small ($R_p < 4 R_\oplus$)[29] transiting planet with $P < 100$ days detected in the Kepler prime mission.
4. Having no knowledge of the presence or absence of giant planets beyond 1 au.[30]

The majority of the planetary systems in our sample have confirmed or validated small planets with Kepler names, although there are two exceptions (KOI-2720 and KOI-3083), and so we generally refer to the Kepler Object of Interest (KOI) designation of each star. Some of the stars in this paper were also selected as part of an RV follow-up survey of Kepler transit planet host stars with three or more transiting planets (as of early 2015) with effective temperatures ranging from 4700 to 6150 K, low rotation speeds ($v \sin i <$ 10 m s$^{-1}$), and Kepler magnitudes brighter than 13.2. Those targets were KOI-94, KOI-282, KOI-316, KOI-623, KOI-1781, KOI-1909, KOI-2169, KOI-2732, and KOI-3083.

Some of the stars in this paper were also selected as part of an RV follow-up survey of multiplanet systems with interesting dynamical architectures and/or significant transit timing variations (TTVs). Those targets are KOI-84, KOI-85, KOI-

---

[28] Gaia magnitudes were not available when targets were first selected
[29] Based on additional information from Gaia, all of the planetary systems that were initially selected for this survey still satisfy the criterion that at least one planet has $R_p < 4 R_\oplus$.
[30] Transits would be one way to identify candidate Jovian planets, although Jupiter-sized transiting planets can have ultralow densities. Jovian companions can also be inferred dynamically (e.g., Kepler-56), where RVs were obtained because the planets were known to be misaligned from the star's rotation axis by a presumed Jovian perturber.

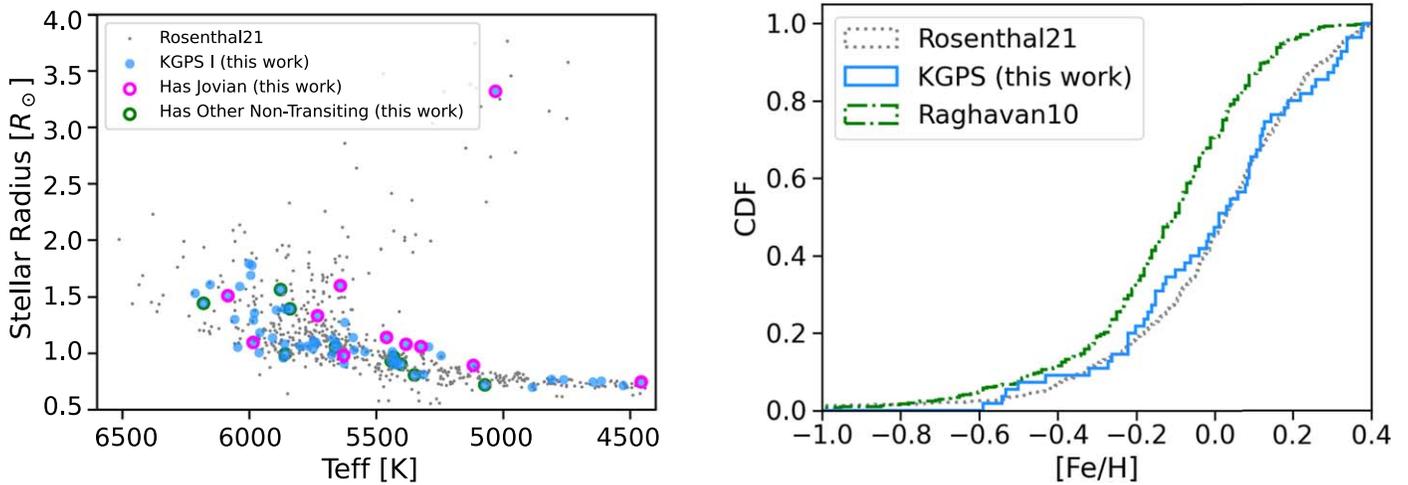

**Figure 1.** Left: stellar radius vs. effective temperature for stars in the KGPS I sample (blue) compared to the CLS (Rosenthal et al. 2021; gray). Stars for which we detected Jovian companions are ringed in magenta; other stars with nontransiting companions are ringed in green. The outlier in the giant branch is KOI-1241 (Kepler-56). Right: the continuous distribution function (CDF) of stellar metallicities from three different surveys: Raghavan et al. (2010), Rosenthal et al. (2021), and KGPS.

**Table 1**
Kepler Giant Planet Search (KGPS) Paper Series

| No. | Title |
| --- | --- |
| KGPS I | A Decade of Kepler Planet-host Radial Velocities from W. M. Keck Observatory |
| KGPS II | The Occurrence of Jovian Companions to Small Transiting Planets |
| KGPS III+ | Additional Patterns in the Architectures of Systems with Small and Jovian Planets |

103, KOI-244, KOI-245, KOI-246, KOI-260, KOI-262, KOI-274, KOI-275, KOI-277, and KOI-1930. The systems KOI-85, KOI-244, and KOI-246 were analyzed in Mills et al. (2019b). We present additional RVs of those systems and an updated analysis here.

Some of the stars in this paper have recently been observed with moderate-cadence RVs as part of a program to measure precise planet masses in multiplanet systems. Those targets are KOI-41, KOI-111, and KOI-316. Here we report the discoveries and upper limits on the masses of nontransiting companions in those systems. Detailed studies of the physical properties of the transiting planets in those systems are in preparation. We have removed RVs that were obtained for Rossiter–McLaughlin (RM) observations, since those RVs do not represent the center-of-mass motion of the host star and are not suitable for detecting additional companions. Those stars were KOI-94 and KOI-244, both of which have their RM observations and analysis described in Albrecht et al. (2013).

Some of the stars in this paper are currently (since the 2022B observing semester) monitored at moderate to high cadence on the High Resolution Echelle Spectrometer (HIRES) based on student-led observing proposals. RVs from those student-led proposals are not included in this paper but will be reported in student-led papers soon.

In this paper, we also report the long-term HIRES RVs of several stars that fall outside of the sample criteria described. Nonetheless, we have long RV baselines that will likely be useful to the community. These systems include several systems that are famous for their high multiplicity, small planets, and habitable-zone planets but are fainter than our magnitude cutoff. KOI-157 (Kepler-11, $V > 13.6$) was the first six-transiting-planet system and one of the Kepler mission's earliest discoveries (Lissauer et al. 2011). KOI-351 (Kepler-90, $V > 13.6$) is a system with eight transiting planets (Cabrera et al. 2014; Schmitt et al. 2014; Shallue & Vanderburg 2018), making it the highest-multiplicity exoplanet system known. Its outermost transiting planet, KOI-351 h, is nearly the size of Jupiter and has an orbital period of 331 days. KOI-377 (Kepler-9, $V > 13.6$) was the first system with multiple transiting planets discovered by Kepler, as well as the first example of TTVs discovered by Kepler (Holman et al. 2010). KOI-701 (Kepler-62, $V > 13.6$) is a Sunlike star with five transiting planets, including one with $R_p < 1.5 R_\oplus$ that is arguably the most "Earthlike" exoplanet currently known based on its size, orbital distance, and host star properties (Borucki et al. 2013). KOI-1241 (Kepler-56) is an evolved star that was originally suspected (and later confirmed) to harbor a cold giant planet that is responsible for misaligning its transiting planets, and so it does not meet the strictly "giant-blind" aspect of our sample selection (Huber et al. 2013; Otor et al. 2016). KOI-1781, which was originally selected as part of the KGPS survey, is moderately active (which we only determined after obtaining multiple spectra).

Our full catalog is generally RV-quiet, main-sequence stars (Figure 1, left). We did not place any cuts on the metallicity of stars in our sample. This is noteworthy because the occurrence of giant planets in RV surveys correlates significantly with host star metallicity (Fischer & Valenti 2005; Petigura et al. 2018). To assess the possible effect of the stellar metallicity of our sample on the results, we compared the distribution of iron abundances of our host stars to two samples from the literature (Figure 1, right). Rosenthal et al. (2021; CLS-I) is a 30 yr, blind RV survey of nearby Sunlike stars that were selected in a manner similar to the KGPS selection criteria (e.g., for their viability for precision RV follow-up), whereas Raghavan et al. (2010) is a volume-limited survey of stars selected to determine

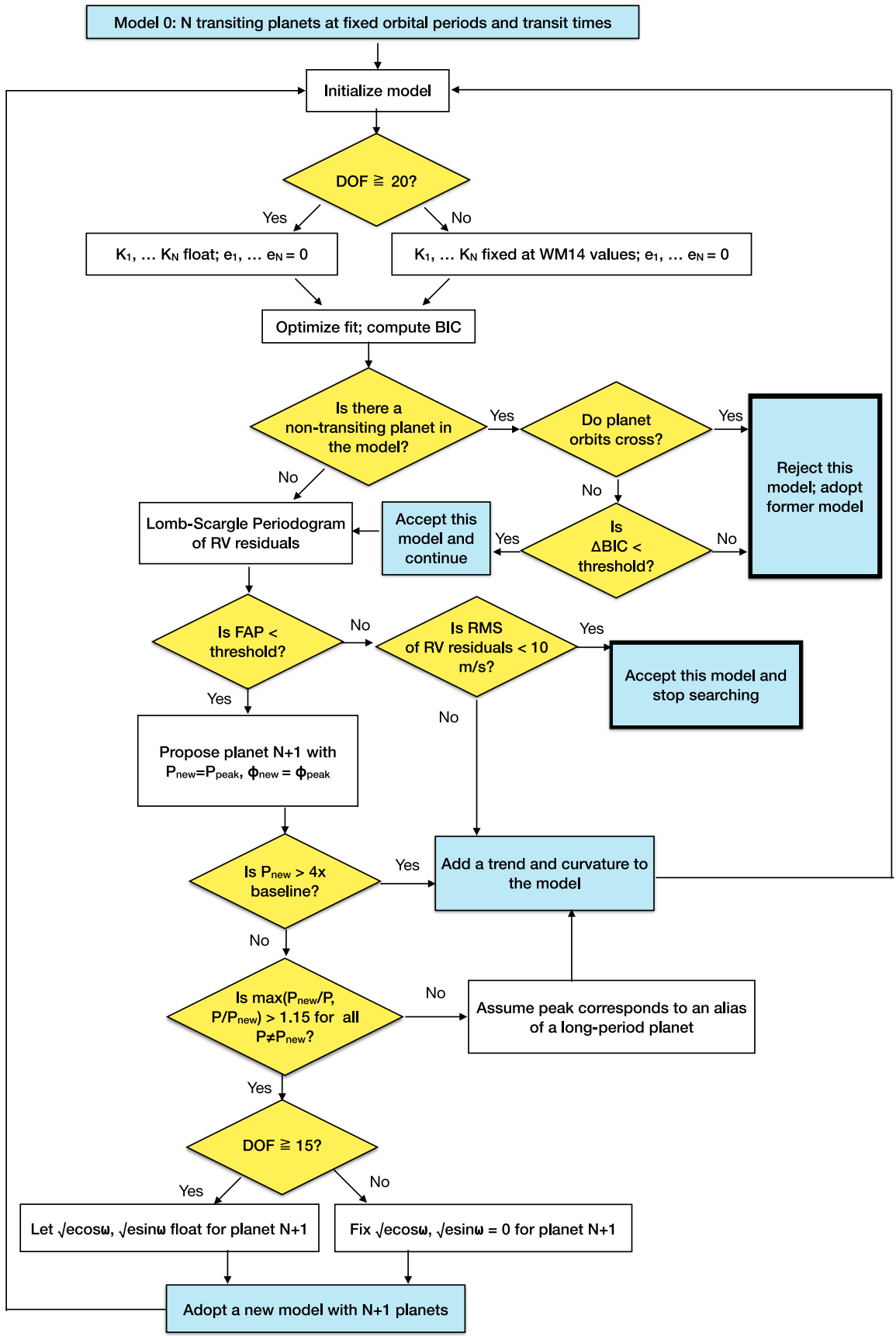

**Figure 2.** KGPS algorithm. This algorithm was applied to 63 systems to search for nontransiting giant planets while accounting for the known (transiting) planets.

**Table 2**
Radial Velocities from Keck-HIRES of 63 Kepler Planet-hosting Stars

| Star | BJD −2450000 | RV (m s$^{-1}$) | RV Error (m s$^{-1}$) | S-value | log $R'_{HK}$ |
|---|---|---|---|---|---|
| hip94931 | 6109.920119 | 14.01 | 1.12 | 0.176 | −5.013 |
| hip94931 | 6110.830622 | 17.28 | 1.26 | 0.178 | −5.008 |
| hip94931 | 6111.896948 | 13.29 | 1.15 | 0.177 | −5.011 |
| hip94931 | 6112.869244 | 14.57 | 1.18 | 0.180 | −5.001 |
| hip94931 | 6113.809189 | 20.16 | 1.22 | 0.175 | −5.016 |
| hip94931 | 6114.831727 | 19.32 | 1.22 | 0.177 | −5.010 |
| hip94931 | 6115.853397 | 17.51 | 1.05 | 0.175 | −5.018 |
| hip94931 | 6133.929013 | 18.50 | 1.22 | 0.176 | −5.012 |
| hip94931 | 6134.967381 | 19.87 | 1.25 | 0.177 | −5.012 |
| hip94931 | 6138.030813 | 14.89 | 1.27 | 0.178 | −5.008 |

**Note.** Columns are star name, barycentric Julian date, RV, RV error, Mount Wilson S-value, and stellar activity index log $R'_{HK}$. The first few rows of the table are shown for content.

(This table is available in its entirety in machine-readable form.)

stellar multiplicity. The metallicity distribution of the host stars studied in KGPS is statistically indistinguishable from that of the CLS-I sample ($p > 0.1$ via a Kolmogorov–Smirnov test), whereas both CLS-I and KGPS contain stars that are more metal-rich on average than the typical star in the local neighborhood ($p = 0.004$). The high metallicities of the CLS-I and KGPS samples might be related to their focus on Sunlike stars, which have higher metallicity than is typical for the solar neighborhood (Nordström et al. 2004).

We accessed the orbital ephemerides of all planetary systems from the NASA Exoplanet Archive, searching both the composite planetary systems table and the cumulative Kepler candidate table to ensure that no transiting planetary candidates in these systems that are not yet "confirmed" were missed (NASA Exoplanet Archive 2022b, 2022a).[31] We considered all confirmed planets, confirmed stellar companions, and planet candidates in our analysis (see the Appendix). We did not consider objects with false-positive dispositions.

### 3. Observations

The discovery of planets via dynamical effects requires some component of orbit determination, which is only tractable with multiple epochs of data. RVs sample the line-of-sight velocity component of the primary star in a manner that has been successfully used for discovering giant planets and determining their orbits over the past three decades (Mayor & Queloz 1995; Howard et al. 2010; Rosenthal et al. 2021). When modeling RV data, at least three free parameters are required to determine a closed orbit of a single, circular planet (orbital period, orbital phase, and RV semi-amplitude). Additional free parameters allow for eccentric orbits and/or additional planets in the model. We obtained at least 10 RVs covering a baseline of at least 3 yr for each star in our sample, which were sufficient to detect orbital motion of the star from a Jovian companion (see Section 5.2). For stars with RVs that clearly indicated the presence of at least one Jupiter-mass planet, we collected far more than 10 RVs in order to improve our sampling of the orbit of the planet, including probing its eccentricity and giant planet multiplicity. In total, we collected 2844 RVs of 63 targets.

---

[31] Accessed on 2022 February 17.

RVs were obtained at the W. M. Keck Observatory using the HIRES instrument. We used the typical setup and observational strategy of the California Planet Search team. This setup has been described extensively in prior literature (e.g., Howard et al. 2010), but we will recapitulate the main points here. The CPS observing setup ensures long-term stability of the wavelength solution of HIRES by (i) positioning the emission lines from a thorium-argon calibration lamp to ensure that they fall within 0.5 pixels of a nominal configuration every night and (ii) inserting a cell of gas-phase (50°C) molecular iodine into the light path of the instrument to imprint an I2 absorption spectrum onto each stellar spectrum for which we wish to measure an RV. The I2-imprinted spectra were forward modeled with a laboratory spectrum of I2 multiplied by a deconvolved stellar spectral template of the target star, which is then convolved with the local point-spread function (PSF) of HIRES. The deconvolved stellar spectral template was taken without iodine immediately before and after observations of rapidly rotating B-type stars through iodine, to sample the immediate, local PSF of the Keck telescope plus HIRES instrument. This strategy produces RVs with an rms scatter of $\sim$2.5 m s$^{-1}$ over a timescale of three decades (Howard & Fulton 2016; Weiss et al. 2016; Rosenthal et al. 2021).

Some modifications of the CPS observing setup, which was first developed for bright stars ($V < 10.5$ Butler et al. 1996), are necessary to attain accurate RVs of the fainter Kepler stars ($V \sim 12.5$). As described in Marcy et al. (2014), sky subtraction becomes important at these magnitudes, particularly in conditions with a full Moon and/or highly reflective cirrus. For stars of $V > 10.5$, we used the C2 decker, which subtends $14''\!.0 \times 0''\!.86$, rather than the traditionally used B5 decker ($3''\!.5 \times 0''\!.86$), to measure the contribution of sky photons at the ends of the long slit, enabling robust sky subtraction. Only a few stars in our program (KOI-3158, KOI-244, KOI-245, and KOI-246) were sufficiently bright for B5 observations.

Integration times varied among the targets based on the program goals but were generally constrained by our need to obtain sufficient signal-to-noise ratios (S/Ns) for precise RVs, balanced by a short enough integration time to avoid noise from cosmic rays, barycentric correction errors, and lost opportunities elsewhere on the sky. Using the HIRES exposure meter, we monitored the S/N of the incoming exposure, requiring a minimum of 30,000 counts (arbitrary units that result in S/N per pixel >70). The maximum exposure time was 45 minutes. Stars with $V < 9$ were observed with two consecutive exposures to average over $p$-mode oscillations, but the only stars for which this applied were HIP 94931 (KOI-3158 = Kepler-444) and KOI-69 (Kepler-93). We did not observe any targets in weather conditions that did not permit our minimum S/N in 45 minutes.

Only one star (KOI-3158) had a bright companion within $2''\!.0$; for this target, we used the image rotator to ensure that the companion was out of the slit, minimizing contamination from the companion. Other stars that had companions within $2''\!.0$ were at least 5 mag brighter than their companions in the V band, and so contamination from the companion was comparable to the Poisson noise for S/N = 100.

Twenty-two of the Kepler planet-hosting stars in our sample were described in Marcy et al. (2014), which presented 677 HIRES RVs. Here we present 2167 RVs, augmenting the total HIRES RV catalog of Kepler's planet-hosting stars to 2844 RVs (Table 2). Furthermore, we have updated the 677

Table 3
Results of KGPS Automated Planet and Trend Search

| KOI | Kepler No. | nRVs | Baseline (yr) | Last Obs. | New Comp. Num. | Period (days) | $K$ (m s$^{-1}$) | rms (m s$^{-1}$) | FAP | $\Delta$BIC | Flag |
|---|---|---|---|---|---|---|---|---|---|---|---|
| 41 | 100 | 134 | 13.4 | 2022-11-14 | ... | ... | ... | 4.5 | ... | ... | |
| 41 | 100 | 134 | 13.4 | 2022-11-14 | 1 | 60.9 | 3.8 | 3.8 | 0.00002 | −45.9 | |
| 69 | 93 | 184 | 13.1 | 2022-09-06 | ... | ... | ... | 29.8 | ... | ... | |
| 69 | 93 | 184 | 13.1 | 2022-09-06 | trend | ... | ... | 2.7 | 0.0 | −837.2 | |
| 70 | 20 | 143 | 9.1 | 2018-09-30 | ... | ... | ... | 6.2 | ... | ... | |
| 70 | 20 | 143 | 9.1 | 2018-09-30 | 1 | 35.0 | 4.3 | 5.4 | 0.00493 | −29.5 | |
| 72 | 10 | 226 | 10.0 | 2019-08-20 | ... | ... | ... | 4.0 | ... | ... | |
| 72 | 10 | 226 | 10.0 | 2019-08-20 | 1 | 25.5 | 1.7 | 3.8 | 0.03815 | −18.9 | |
| 82 | 19 | 73 | 9.3 | 2019-08-29 | ... | ... | ... | 3.9 | ... | ... | |
| 84 | 19 | 124 | 7.9 | 2017-07-29 | ... | ... | ... | 5.6 | ... | ... | |
| 84 | 19 | 124 | 7.9 | 2017-07-29 | 1 | 28.5 | 3.8 | 4.9 | 0.02454 | −24.0 | |
| 85 | 65 | 79 | 11.1 | 2022-06-15 | ... | ... | ... | 13.7 | ... | ... | |
| 85 | 65 | 79 | 11.1 | 2022-06-15 | 1 | 257.9 | 20.7 | 7.4 | 0.0 | −80.2 | |
| 94 | 89 | 72 | 7.3 | 2017-04-10 | ... | ... | ... | 9.0 | ... | ... | 1,2 |
| 103 | ... | 21 | 12.2 | 2021-10-29 | ... | ... | ... | 8.0 | ... | ... | |
| 104 | 94 | 39 | 12.0 | 2022-06-14 | ... | ... | ... | 111.4 | ... | ... | |
| 104 | 94 | 39 | 12.0 | 2022-06-14 | 1 | 811.7 | 231.4 | 8.4 | 0.00079 | −1205.7 | |
| 108 | 103 | 28 | 12.2 | 2021-10-29 | ... | ... | ... | 6.6 | ... | ... | |
| 111 | 104 | 44 | 11.3 | 2021-10-26 | ... | ... | ... | 4.5 | ... | ... | |
| 116 | 106 | 48 | 9.8 | 2019-08-18 | ... | ... | ... | 6.0 | ... | ... | |
| 116 | 106 | 48 | 9.8 | 2019-08-18 | 1 | 90.6 | 7.0 | 3.5 | 0.00174 | −44.6 | |
| 122 | 95 | 36 | 7.9 | 2017-07-29 | ... | ... | ... | 4.9 | ... | ... | |
| 123 | 109 | 38 | 10.9 | 2020-08-10 | ... | ... | ... | 6.3 | ... | ... | |
| 137 | 18 | 25 | 8.8 | 2018-06-29 | ... | ... | ... | 5.2 | ... | ... | |
| 142 | 88 | 58 | 9.0 | 2022-06-19 | ... | ... | ... | 44.1 | ... | ... | |
| 142 | 88 | 58 | 9.0 | 2022-06-19 | 1 | 22.2 | 56.2 | 27.8 | 0.00043 | −144.8 | |
| 142 | 88 | 58 | 9.0 | 2022-06-19 | 2 | 1370.3 | 54.9 | 10.2 | 0.00175 | −114.7 | 1 |
| 148 | 48 | 59 | 12.8 | 2022-06-14 | ... | ... | ... | 32.6 | ... | ... | |
| 148 | 48 | 59 | 12.8 | 2022-06-14 | 1 | 998.1 | 45.5 | 9.1 | 0.0 | −173.2 | 3 |
| 153 | 113 | 42 | 12.1 | 2022-06-07 | ... | ... | ... | 8.1 | ... | ... | |
| 157 | 11 | 31 | 11.6 | 2022-04-20 | ... | ... | ... | 7.8 | ... | ... | |
| 244 | 25 | 69 | 11.2 | 2021-09-27 | ... | ... | ... | 8.6 | ... | ... | |
| 244 | 25 | 69 | 11.2 | 2021-09-27 | 1 | 91.5 | 9.0 | 5.9 | 0.00009 | −45.0 | |
| 245 | 37 | 218 | 12.4 | 2022-09-06 | ... | ... | ... | 3.6 | ... | ... | |
| 245 | 37 | 218 | 12.4 | 2022-09-06 | 1 | 458.0 | 2.0 | 3.2 | 0.00001 | −20.0 | |
| 245 | 37 | 218 | 12.4 | 2022-09-06 | 2 | 15.0 | 1.7 | 2.9 | 0.00197 | −22.2 | |
| 246 | 68 | 96 | 12.2 | 2022-07-15 | ... | ... | ... | 14.8 | ... | ... | |
| 246 | 68 | 96 | 12.2 | 2022-07-15 | 1 | 635.7 | 18.2 | 3.8 | 0.0 | −244.3 | |
| 246 | 68 | 96 | 12.2 | 2022-07-15 | 2 | 362.3 | 5.1 | 2.9 | 0.00003 | −55.9 | 4 |
| 260 | 126 | 35 | 5.3 | 2019-11-11 | ... | ... | ... | 10.1 | ... | ... | |
| 260 | 126 | 35 | 5.3 | 2019-11-11 | trend | ... | ... | 9.8 | 0.98921 | 5.2 | |
| 261 | 96 | 55 | 10.9 | 2021-06-19 | ... | ... | ... | 6.7 | ... | ... | |
| 262 | 50 | 39 | 9.8 | 2022-04-02 | ... | ... | ... | 14.0 | ... | ... | |
| 262 | 50 | 39 | 9.8 | 2022-04-02 | trend | ... | ... | 13.9 | 0.99485 | 6.7 | |
| 265 | 507 | 49 | 10.8 | 2022-06-11 | ... | ... | ... | 4.6 | ... | ... | |
| 265 | 507 | 49 | 10.8 | 2022-06-11 | 1 | 93.9 | 4.5 | 3.3 | 0.03644 | −27.6 | |
| 273 | 454 | 116 | 12.1 | 2022-09-12 | ... | ... | ... | 63.8 | ... | ... | |
| 273 | 454 | 116 | 12.1 | 2022-09-12 | 1 | 525.4 | 113.5 | 12.4 | 0.0 | −1318.0 | |

Table 3
(Continued)

| KOI | Kepler No. | nRVs | Baseline (yr) | Last Obs. | New Comp. Num. | Period (days) | $K$ (m s$^{-1}$) | rms (m s$^{-1}$) | FAP | $\Delta$BIC | Flag |
|---|---|---|---|---|---|---|---|---|---|---|---|
| 273 | 454 | 116 | 12.1 | 2022-09-12 | 2 | 4447.5 | 24.3 | 5.1 | 0.0 | −76.9 | |
| 274 | 128 | 17 | 7.9 | 2018-06-25 | ⋯ | ⋯ | ⋯ | 8.1 | ⋯ | ⋯ | |
| 275 | 129 | 35 | 7.8 | 2022-06-15 | ⋯ | ⋯ | ⋯ | 65.4 | ⋯ | ⋯ | |
| 275 | 129 | 35 | 7.8 | 2022-06-15 | 1 | 1901.3 | 89.3 | 4.5 | 0.00017 | −424.7 | |
| 277 | 36 | 25 | 9.2 | 2021-10-11 | ⋯ | ⋯ | ⋯ | 7.5 | ⋯ | ⋯ | |
| 281 | 510 | 11 | 10.1 | 2021-09-15 | ⋯ | ⋯ | ⋯ | 4.1 | ⋯ | ⋯ | |
| 282 | 130 | 10 | 6.9 | 2021-07-27 | ⋯ | ⋯ | ⋯ | 5.9 | ⋯ | ⋯ | |
| 283 | 131 | 46 | 6.9 | 2017-08-05 | ⋯ | ⋯ | ⋯ | 4.8 | ⋯ | ⋯ | |
| 285 | 92 | 23 | 9.9 | 2021-06-24 | ⋯ | ⋯ | ⋯ | 5.8 | ⋯ | ⋯ | |
| 292 | 97 | 31 | 8.0 | 2018-07-09 | ⋯ | ⋯ | ⋯ | 5.8 | ⋯ | ⋯ | |
| 295 | ⋯ | 14 | 10.7 | 2022-05-16 | ⋯ | ⋯ | ⋯ | 17.6 | ⋯ | ⋯ | |
| 295 | ⋯ | 14 | 10.7 | 2022-05-16 | trend | ⋯ | ⋯ | 11.7 | 0.99948 | −12.6 | |
| 299 | 98 | 42 | 7.9 | 2018-06-23 | ⋯ | ⋯ | ⋯ | 7.1 | ⋯ | ⋯ | |
| 305 | 99 | 45 | 7.0 | 2017-08-06 | ⋯ | ⋯ | ⋯ | 5.3 | ⋯ | ⋯ | |
| 316 | 139 | 38 | 11.8 | 2022-06-10 | ⋯ | ⋯ | ⋯ | 9.3 | ⋯ | ⋯ | |
| 321 | 406 | 58 | 11.9 | 2022-05-13 | ⋯ | ⋯ | ⋯ | 4.6 | ⋯ | ⋯ | |
| 351 | 90 | 34 | 11.2 | 2022-06-14 | ⋯ | ⋯ | ⋯ | 9.2 | ⋯ | ⋯ | 1 |
| 365 | 538 | 28 | 7.0 | 2017-07-09 | ⋯ | ⋯ | ⋯ | 3.1 | ⋯ | ⋯ | |
| 370 | 145 | 11 | 6.8 | 2021-07-09 | ⋯ | ⋯ | ⋯ | 13.9 | ⋯ | ⋯ | |
| 370 | 145 | 11 | 6.8 | 2021-07-09 | trend | ⋯ | ⋯ | 12.9 | 1.0 | 0.8 | |
| 377 | 9 | 49 | 11.7 | 2022-02-22 | ⋯ | ⋯ | ⋯ | 11.4 | ⋯ | ⋯ | |
| 377 | 9 | 49 | 11.7 | 2022-02-22 | trend | ⋯ | ⋯ | 10.0 | 0.98636 | −7.0 | |
| 623 | 197 | 11 | 7.0 | 2021-09-12 | ⋯ | ⋯ | ⋯ | 9.0 | ⋯ | ⋯ | |
| 701 | 62 | 17 | 4.8 | 2017-05-13 | ⋯ | ⋯ | ⋯ | 5.3 | ⋯ | ⋯ | |
| 719 | 220 | 10 | 9.8 | 2021-07-09 | ⋯ | ⋯ | ⋯ | 5.4 | ⋯ | ⋯ | |
| 1241 | 56 | 47 | 10.1 | 2022-06-18 | ⋯ | ⋯ | ⋯ | 49.6 | ⋯ | ⋯ | |
| 1241 | 56 | 47 | 10.1 | 2022-06-18 | 1 | 971.5 | 97.6 | 7.9 | 0.00019 | −295.6 | |
| 1241 | 56 | 47 | 10.1 | 2022-06-18 | trend | ⋯ | ⋯ | 48.0 | 0.00019 | −10.4 | 1 |
| 1442 | 407 | 98 | 11.0 | 2022-05-10 | ⋯ | ⋯ | ⋯ | 123.6 | ⋯ | ⋯ | |
| 1442 | 407 | 98 | 11.0 | 2022-05-10 | 1 | 2068.6 | 172.4 | 5.6 | 0.0 | −3854.9 | |
| 1612 | ⋯ | 56 | 11.1 | 2022-06-21 | ⋯ | ⋯ | ⋯ | 6.5 | ⋯ | ⋯ | |
| 1612 | ⋯ | 56 | 11.1 | 2022-06-21 | trend | ⋯ | ⋯ | 4.6 | 0.01541 | −30.0 | |
| 1692 | 314 | 14 | 10.8 | 2022-06-11 | ⋯ | ⋯ | ⋯ | 13.5 | ⋯ | ⋯ | |
| 1692 | 314 | 14 | 10.8 | 2022-06-11 | trend | ⋯ | ⋯ | 13.1 | 1.0 | 1.8 | |
| 1781 | 411 | 14 | 9.7 | 2022-02-21 | ⋯ | ⋯ | ⋯ | 22.6 | ⋯ | ⋯ | |
| 1781 | 411 | 14 | 9.7 | 2022-02-21 | trend | ⋯ | ⋯ | 20.0 | 1.0 | −1.2 | |
| 1909 | 334 | 11 | 6.1 | 2021-09-19 | ⋯ | ⋯ | ⋯ | 8.8 | ⋯ | ⋯ | |
| 1925 | 409 | 97 | 10.7 | 2022-11-14 | ⋯ | ⋯ | ⋯ | 4.1 | ⋯ | ⋯ | |
| 1925 | 409 | 97 | 10.7 | 2022-11-14 | 1 | 344.9 | 4.1 | 3.2 | 0.00037 | −36.3 | |
| 1930 | 338 | 11 | 7.0 | 2021-09-21 | ⋯ | ⋯ | ⋯ | 8.8 | ⋯ | ⋯ | |
| 2169 | 1130 | 34 | 10.2 | 2022-09-13 | ⋯ | ⋯ | ⋯ | 1649.3 | ⋯ | ⋯ | |
| 2169 | 1130 | 34 | 10.2 | 2022-09-13 | 1 | 6226.4 | 2244.6 | 56.7 | 0.00133 | $-2.2 \times 10^5$ | 3 |
| 2687 | 1869 | 26 | 9.3 | 2021-06-23 | ⋯ | ⋯ | ⋯ | 7.3 | ⋯ | ⋯ | |
| 2720 | ⋯ | 22 | 6.3 | 2018-08-06 | ⋯ | ⋯ | ⋯ | 4.4 | ⋯ | ⋯ | |
| 2732 | 403 | 11 | 6.0 | 2021-08-15 | ⋯ | ⋯ | ⋯ | 9.2 | ⋯ | ⋯ | |
| 3083 | ⋯ | 11 | 6.2 | 2021-10-29 | ⋯ | ⋯ | ⋯ | 6.4 | ⋯ | ⋯ | |
| 3158 | 444 | 207 | 9.8 | 2022-04-11 | ⋯ | ⋯ | ⋯ | 19.5 | ⋯ | ⋯ | |

**Table 3**
(Continued)

| KOI | Kepler No. | nRVs | Baseline (yr) | Last Obs. | New Comp. Num. | Period (days) | $K$ (m s$^{-1}$) | rms (m s$^{-1}$) | FAP | $\Delta$BIC | Flag |
|---|---|---|---|---|---|---|---|---|---|---|---|
| 3158 | 444 | 207 | 9.8 | 2022-04-11 | trend | … | … | 2.6 | 0.0 | −820.5 | |
| 3158 | 444 | 207 | 9.8 | 2022-04-11 | 1 | 72.4 | 1.5 | 2.4 | 0.00169 | −27.1 | 5 |
| 3179 | 1911 | 11 | 7.9 | 2021-07-09 | … | … | … | 9.9 | … | … | |

**Note.** Columns are KOI, Kepler number (e.g., "10" corresponds to Kepler-10), number of RVs, the RV baseline, date of last observation, proposed companion number added to the model (if any), period and RV semi-amplitude of the proposed companion, rms of the RV residuals (before MCMC fitting), FAP of the proposed companion, $\Delta$BIC of the model with the proposed companion compared to the previous model (without the proposed companion), and a flag. Target names appear in the Keck Observatory Archive (KOA) as KNNNNN, with K preceding the five-digit Kepler number (left-filled with zeros). Flags indicate the following: 1 = the KOA name is preceded by a C (e.g., CK00094); 2 = the KGPS automatic pipeline did not include eccentricity for the massive planet, but our final analysis did; 3 = the KGPS automatic pipeline did not detect a companion that was apparent by eye; 4 = the KGPS automatic pipeline detected what is likely an alias of the true orbital period; 5 = the KOA name is the Hipparcos name (e.g., Kepler-444 is desginated as HIP 94931).

(This table is available in machine-readable form.)

Table 4
Stellar Properties and RV Data Summary

| KOI | Kepler No. | Gaia mag | $M_\star$ ($M_\odot$) | [Fe/H] | NTP | NSS | NGP | NSC | NRV | $\Delta T$ (yr) | RV rms (m s$^{-1}$) | $\sigma_{jit}$ (m s$^{-1}$) | $dv/dt$ (S/N) | <$M \sin i$ at 10 au ($M_J$, 99.7% conf.) |
|---|---|---|---|---|---|---|---|---|---|---|---|---|---|---|
| 41 | 100 | 11.1 | 1.12 | 0.12 | 3 | 1 | 0 | 0 | 134 | 13.4 | 3.3 | 4.8 | 2.7 | 0.22 |
| 69 | 93 | 9.9 | 0.89 | −0.20 | 1 | 0 | 0 | 0 | 184 | 13.1 | 2.6 | 4.9 | 112.3 | 7.41 |
| 70 | 20 | 12.5 | 0.93 | 0.04 | 5 | 1 | 0 | 0 | 143 | 9.1 | 5.4 | 2.7 | 2.0 | 1.43 |
| 72 | 10 | 10.9 | 0.90 | −0.17 | 2 | 1 | 0 | 0 | 226 | 10.0 | 4.0 | 4.8 | 0.1 | 0.29 |
| 82 | 102 | 11.5 | 0.80 | 0.10 | 5 | 0 | 0 | 0 | 73 | 9.3 | 3.9 | 3.2 | 1.6 | 0.75 |
| 84 | 19 | 11.9 | 0.88 | −0.12 | 1 | 1 | 0 | 0 | 124 | 7.9 | 5.0 | 2.7 | 3.2 | 1.39 |
| 85 | 65 | 11.0 | 1.24 | 0.09 | 3 | 0 | 1 | 0 | 79 | 11.1 | 7.4 | 5.0 | 1.5 | 1.51 |
| 94 | 89 | 12.2 | 1.20 | 0.04 | 4 | 0 | 0 | 0 | 72 | 12.7 | 8.9 | 4.8 | 1.8 | 1.44 |
| 103 | ... | 12.6 | 0.93 | −0.07 | 1 | 0 | 0 | 0 | 21 | 12.2 | 8.0 | 5.2 | 0.3 | 1.23 |
| 104 | 94 | 12.9 | 0.82 | 0.22 | 1 | 0 | 1 | 0 | 39 | 12.0 | 5.4 | 3.4 | 5.4 | 2.89 |
| 108 | 103 | 12.2 | 1.15 | 0.14 | 2 | 0 | 0 | 0 | 28 | 12.2 | 6.6 | 4.7 | 0.4 | 0.93 |
| 111 | 104 | 12.6 | 0.84 | −0.53 | 3 | 0 | 0 | 0 | 44 | 11.3 | 4.5 | 4.5 | 0.1 | 0.33 |
| 116 | 106 | 12.8 | 0.97 | −0.10 | 4 | 1 | 0 | 0 | 48 | 9.8 | 3.5 | 4.6 | 1.2 | 0.69 |
| 122 | 95 | 12.3 | 1.12 | 0.33 | 1 | 0 | 0 | 0 | 36 | 7.9 | 4.9 | 4.8 | 1.1 | 1.07 |
| 123 | 109 | 12.3 | 1.03 | −0.06 | 2 | 0 | 0 | 0 | 38 | 10.9 | 6.3 | 4.5 | 2.3 | 0.67 |
| 137 | 18 | 13.5 | 0.97 | 0.32 | 3 | 0 | 0 | 0 | 25 | 8.8 | 5.2 | 2.7 | 0.2 | 1.02 |
| 142 | 88 | 13.1 | 0.98 | 0.29 | 1 | 0 | 2 | 0 | 58 | 9.0 | 8.3 | 2.7 | 2.0 | 1.96 |
| 148 | 48 | 13.1 | 0.91 | 0.26 | 3 | 0 | 2 | 0 | 59 | 12.8 | 4.6 | 2.9 | 6.3 | 1.97 |
| 153 | 113 | 13.5 | 0.77 | 0.12 | 2 | 0 | 0 | 0 | 42 | 12.1 | 8.1 | 1.8 | 0.8 | 1.23 |
| 157 | 11 | 13.7 | 1.00 | 0.05 | 6 | 0 | 0 | 0 | 31 | 11.6 | 7.8 | 4.7 | 1.2 | 1.58 |
| 244 | 25 | 10.6 | 1.14 | −0.15 | 2 | 1 | 0 | 0 | 69 | 11.2 | 5.9 | 4.6 | 0.1 | 0.55 |
| 245 | 37 | 9.5 | 0.80 | −0.43 | 3 | 0 | 0 | 0 | 218 | 12.4 | 3.2 | 2.7 | 1.4 | 0.26 |
| 246 | 68 | 9.9 | 1.06 | 0.11 | 2 | 0 | 2 | 0 | 96 | 12.2 | 2.9 | 4.5 | 0.3 | 0.21 |
| 260 | 126 | 10.5 | 1.08 | −0.32 | 3 | 0 | 0 | 0 | 35 | 5.3 | 9.8 | 4.7 | 1.0 | 12.03 |
| 261 | 96 | 10.3 | 1.01 | 0.01 | 1 | 0 | 0 | 0 | 55 | 10.9 | 6.7 | 5.2 | 3.0 | 1.08 |
| 262 | 50 | 10.4 | 1.10 | −0.16 | 2 | 0 | 0 | 0 | 39 | 9.8 | 13.9 | 4.9 | 0.8 | 7.43 |
| 265 | 507 | 11.9 | 1.17 | 0.12 | 1 | 1 | 0 | 0 | 49 | 10.8 | 3.3 | 3.0 | 0.1 | 0.51 |
| 273 | 454 | 11.4 | 1.08 | 0.34 | 1 | 0 | 2 | 0 | 116 | 12.1 | 5.7 | 2.8 | 8.7 | 2.51 |
| 274 | 128 | 11.3 | 1.12 | −0.13 | 2 | 0 | 0 | 0 | 17 | 7.9 | 8.1 | 3.4 | 0.3 | 1.50 |
| 275 | 129 | 11.6 | 1.24 | 0.25 | 2 | 0 | 1 | 0 | 35 | 7.8 | 4.3 | 3.2 | 2.1 | 4.09 |
| 277 | 36 | 12.1 | 0.97 | −0.22 | 2 | 0 | 0 | 0 | 25 | 9.2 | 7.5 | 2.9 | 0.1 | 0.99 |
| 281 | 510 | 11.9 | 0.86 | −0.50 | 1 | 0 | 0 | 0 | 11 | 10.1 | 4.1 | 2.7 | 0.4 | 0.66 |
| 282 | 130 | 11.8 | 0.90 | −0.26 | 3 | 0 | 0 | 0 | 10 | 6.9 | 5.9 | 4.3 | 0.1 | 1.91 |
| 283 | 131 | 11.5 | 1.06 | 0.18 | 2 | 0 | 0 | 0 | 46 | 6.9 | 4.8 | 5.0 | 1.8 | 1.23 |
| 285 | 92 | 11.7 | 1.25 | 0.19 | 3 | 0 | 0 | 0 | 23 | 9.9 | 5.8 | 3.2 | 0.1 | 0.94 |
| 292 | 97 | 12.8 | 0.91 | −0.22 | 1 | 0 | 0 | 0 | 31 | 8.0 | 5.8 | 4.6 | 1.3 | 1.24 |
| 295 | ... | 12.2 | 0.97 | −0.22 | 2 | 0 | 0 | 0 | 14 | 10.7 | 11.7 | 4.5 | 4.7 | 4.11 |
| 299 | 98 | 12.9 | 0.98 | 0.09 | 1 | 0 | 0 | 0 | 42 | 7.9 | 7.1 | 2.7 | 0.5 | 1.19 |
| 305 | 99 | 13.0 | 0.81 | 0.13 | 1 | 0 | 0 | 0 | 45 | 7.0 | 5.3 | 3.4 | 0.5 | 1.18 |
| 316 | 139 | 12.7 | 1.08 | 0.38 | 3 | 0 | 1 | 0 | 38 | 11.8 | 4.0 | 2.7 | 0.4 | 1.34 |
| 321 | 406 | 12.5 | 1.06 | 0.30 | 2 | 0 | 0 | 0 | 58 | 11.9 | 4.6 | 2.7 | 0.3 | 0.54 |
| 351 | 90 | 13.7 | 1.11 | 0.14 | 8 | 0 | 1 | 0 | 34 | 11.2 | 9.9 | 4.5 | 0.5 | 1.80 |
| 365 | 538 | 11.2 | 0.87 | −0.18 | 2 | 0 | 0 | 0 | 28 | 7.0 | 3.1 | 2.8 | 0.3 | 0.83 |
| 370 | 145 | 11.9 | 1.20 | 0.01 | 2 | 0 | 0 | 0 | 11 | 6.8 | 12.9 | 4.7 | 1.2 | 6.33 |
| 377 | 9 | 13.8 | 1.03 | 0.00 | 3 | 0 | 0 | 0 | 49 | 11.7 | 10.0 | 5.0 | 0.4 | 1.89 |
| 623 | 197 | 11.8 | 0.85 | −0.59 | 4 | 0 | 0 | 0 | 11 | 7.0 | 9.0 | 4.4 | 0.1 | 3.05 |
| 701 | 62 | 13.7 | 0.68 | −0.37 | 5 | 0 | 0 | 0 | 17 | 4.8 | 5.3 | 4.6 | 0.6 | 2.56 |
| 719 | 220 | 12.9 | 0.73 | 0.08 | 4 | 0 | 0 | 0 | 10 | 9.8 | 5.4 | 1.8 | 0.1 | 1.36 |
| 1241 | 56 | 12.4 | 1.46 | 0.42 | 2 | 0 | 2 | 0 | 47 | 10.1 | 5.7 | 3.5 | 2.0 | 6.98 |
| 1442 | 407 | 12.5 | 1.08 | 0.38 | 1 | 0 | 1 | 0 | 98 | 11.0 | 4.9 | 2.7 | 8.4 | 4.27 |
| 1612 | ... | 8.7 | 1.02 | −0.27 | 1 | 0 | 0 | 0 | 56 | 11.1 | 4.6 | 4.4 | 1.3 | 1.42 |
| 1692 | 314 | 12.5 | 1.00 | 0.34 | 2 | 0 | 0 | 0 | 14 | 10.8 | 13.1 | 2.9 | 0.8 | 3.18 |
| 1781 | 411 | 12.2 | 0.83 | 0.18 | 3 | 0 | 0 | 0 | 14 | 9.7 | 20.0 | 3.9 | 2.2 | 4.32 |
| 1909 | 334 | 12.7 | 1.01 | 0.03 | 3 | 0 | 0 | 0 | 11 | 6.1 | 8.8 | 4.6 | 0.7 | 3.27 |
| 1925 | 409 | 9.4 | 0.93 | 0.09 | 1 | 1 | 0 | 0 | 97 | 10.7 | 3.2 | 2.7 | 0.6 | 0.39 |
| 1930 | 338 | 12.1 | 1.06 | −0.02 | 4 | 0 | 0 | 0 | 11 | 7.0 | 8.8 | 4.9 | 0.4 | 3.37 |
| 2169 | 1130 | 12.4 | 0.94 | 0.06 | 4 | 0 | 0 | 1 | 34 | 10.2 | 6.5 | 2.8 | 0.5 | 19.14 |
| 2687 | 1869 | 10.1 | 1.01 | −0.01 | 2 | 0 | 0 | 0 | 26 | 9.3 | 7.3 | 2.8 | 2.1 | 1.90 |
| 2720 | ... | 10.3 | 1.00 | −0.15 | 1 | 0 | 0 | 0 | 22 | 6.3 | 4.4 | 4.6 | 1.3 | 1.43 |
| 2732 | 403 | 12.8 | 1.21 | 0.00 | 4 | 0 | 0 | 0 | 11 | 6.0 | 9.2 | 4.8 | 0.3 | 3.74 |
| 3083 | ... | 12.8 | 1.16 | 0.29 | 3 | 0 | 0 | 0 | 11 | 6.2 | 6.4 | 2.7 | 0.2 | 2.23 |
| 3158 | 444 | 8.6 | 0.73 | −0.54 | 5 | 0 | 0 | 1 | 207 | 9.8 | 2.3 | 2.8 | 35.7 | 5.67 |
| 3179 | 1911 | 12.1 | 0.99 | −0.04 | 1 | 0 | 0 | 0 | 11 | 7.9 | 9.9 | 5.0 | 0.1 | 3.17 |

**Note.** Columns are KOI, Kepler number (e.g., "10" corresponds to Kepler-10), Gaia magnitude, stellar mass and metallicity, the number of transiting planets, number of RV-detected sub-Saturns, number of RV-detected giant planets, number of RV-detected stellar companions, number of RVs, RV baseline, rms of the RV residuals to the best-fit model, the expected stellar jitter computed in Section 4.2, the measured RV trend, and the $3\sigma$ upper limit on $M \sin i$ at 10 au based on the trend. KOI-142, KOI-351, and KOI-1241 each have one giant planet interior to 1 au that counts toward the total number of giant planets in this table. KOI-142 c was detected based on TTVs, whereas KOI-351 h and KOI-1241 c are transiting planets with $M \sin i > 0.3 M_J$ that are interior to 1 au. Additional columns associated with this table are available in the machine-readable version.

(This table is available in its entirety in machine-readable form.)

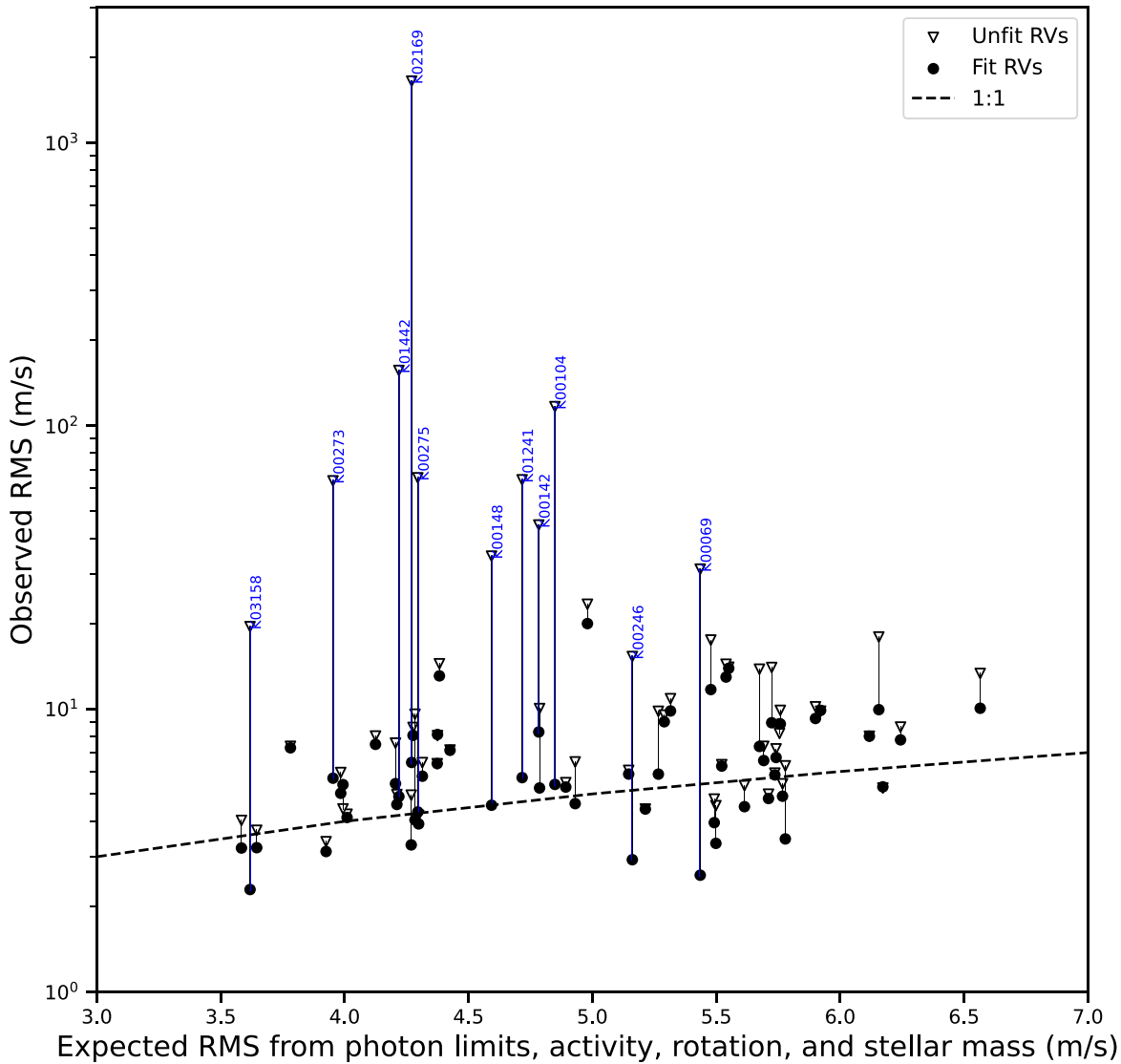

**Figure 3.** Observed rms scatter of the RVs vs. expected rms from jitter. The jitter is computed based on the average of the Mount Wilson $S_{HK}$ index, the stellar mass, the stellar rotation, and photon limits. The vertical axis for triangles shows the rms of the unfit (raw) RVs, and the vertical axis for circles shows the rms of the residuals to the best-fit model that includes transiting planets and any detected nontransiting planets and companions. The RVs for some systems had significant reductions in their rms (factor of 5 or more) when the model included a massive companion (blue).

previously published RVs in a manner that is self-consistent with the newly determined RVs. The 2844 RVs of our 63 targets are available in a machine-readable version of Table 2.

### 4. Orbit Fitting

Multiplanet systems pose a variety of challenges to accurate orbital fitting, especially if the orbital periods are not known a priori. It is useful to sample the shortest-period planets at least as often as every half-orbital period to disambiguate the true signals from the aliases (Dawson & Fabrycky 2010); meanwhile, long-term monitoring is also necessary to detect the longest-period planets, such as those identified in Blunt et al. (2019) and Rosenthal et al. (2021).

#### 4.1. The KGPS Algorithm

When fitting systems with transiting planets, (i) the orbital period and phase of the transiting planets are known, and (ii) the assumption of long-term orbital stability implies that additional planets, particularly massive giant planets, do not have orbits so close to the transiting planets that they would destabilize the system. These key advantages of searching for new planets in a system with known transiting planets, as compared to a blind planet search, inspired us to develop a custom algorithm for identifying nontransiting planets in the Kepler systems, which we call the "KGPS" algorithm (Figure 2). The key steps are as follows:

1. Search for additional published RVs with RV precision of $<5\,\mathrm{m\,s^{-1}}$. Other instruments included HARPS-N (Cosentino et al. 2012) and SOPHIE (Perruchot et al. 2011).
2. Model and optimize a linear combination of orbits for the $N$ transiting planets. We assumed that the transiting planets were in circular orbits. This assumption is motivated by previous studies showing that the stars with multiple transiting planets typically have $e < 0.1$ (Mills et al. 2019a; He et al. 2020; Yee et al. 2021). Further, introducing eccentricity and argument of periastron as free parameters for planets with low RV semi-amplitudes can absorb noise and/or signals from

Table 5
KGPS Planet Physical Properties

| Name[b] | KOI | Per. (days) | $a$ (au) | $R_p$ ($R_\oplus$) | $K$[a] (m s$^{-1}$) | $M \sin i$[a] ($M_\oplus$) | $\rho_p$[a] (g cm$^{-3}$) | Flag |
|---|---|---|---|---|---|---|---|---|
| Kepler-100 b | 41.02 | 6.89 | 0.073 | 1.34 ± 0.13 | 1.7 ± 0.4 | 5.5 ± 1.3 | 13 ± 4 | |
| Kepler-100 c | 41.01 | 12.8 | 0.11 | 2.34 ± 0.08 | 1.0 ± 0.4 | 4.0 ± 1.7 | 1.7 ± 0.7 | |
| Kepler-100 d | 41.03 | 35.3 | 0.22 | 1.54 ± 0.23 | 0.32 ± 0.33 | 1.8 ± 1.8 | 2.7 ± 2.9 | |
| Kepler-100 e | 41.10 | 60.89 ± 0.05 | 0.31 | ⋯ | 3.6 ± 0.5 | 23.8 ± 3.2 | ⋯ | |
| Kepler-93 b | 69.01 | 4.73 | 0.053 | 1.63 ± 0.06 | 1.50 ± 0.24 | 3.6 ± 0.6 | 4.6 ± 0.8 | |
| Kepler-93 B | 69.10 | > 20,000 | > 15 | ⋯ | > 89 | ⋯ | ⋯ | |
| Kepler-20 b | 70.02 | 3.7 | 0.046 | 2.01 ± 0.18 | 4.1 ± 0.6 | 9.5 ± 1.5 | 6.5 ± 1.4 | |
| Kepler-20 e | 70.04 | 6.1 | 0.064 | 0.78 ± 0.10 | ⋯ | 0.56 | 6.5 | |
| Kepler-20 c | 70.01 | 10.9 | 0.094 | 2.88 ± 0.13 | 4.1 ± 0.7 | 13.6 ± 2.3 | 3.1 ± 0.6 | |
| Kepler-20 f | 70.05 | 19.6 | 0.14 | 0.94 ± 0.04 | ⋯ | 1.2 | 7.8 | |
| Kepler-20 g | 70.10 | 34.95 ± 0.04 | 0.21 | ⋯ | 4.3 ± 0.7 | 21.0 ± 3.4 | ⋯ | |
| Kepler-20 d | 70.03 | 77.6 | 0.35 | 2.49 ± 0.07 | 2.0 ± 0.7 | 13 ± 4 | 4.6 ± 1.5 | |
| Kepler-10 b | 72.01 | 0.837 | 0.017 | 1.49 ± 0.04 | 2.72 ± 0.32 | 3.7 ± 0.4 | 6.1 ± 0.8 | |
| Kepler-10 d | 72.10 | 25.6 ± 2.9 | 0.16 | ⋯ | 1.4 ± 0.4 | 5.9 ± 1.7 | ⋯ | A |
| Kepler-10 c | 72.02 | 45.3 | 0.24 | 2.34 ± 0.06 | 2.4 ± 0.5 | 12.1 ± 2.6 | 5.2 ± 1.1 | |
| Kepler-102 b | 82.05 | 5.29 | 0.055 | 0.471 ± 0.024 | ⋯ | 0.074 | 3.9 | |
| Kepler-102 c | 82.04 | 7.07 | 0.067 | 0.554 ± 0.026 | ⋯ | 0.14 | 4.6 | |
| Kepler-102 d | 82.02 | 10.3 | 0.086 | 1.34 ± 0.09 | 1.5 ± 0.7 | 4.5 ± 1.9 | 10 ± 5 | |
| Kepler-102 e | 82.01 | 16.1 | 0.12 | 2.41 ± 0.14 | 1.4 ± 0.6 | 4.7 ± 2.1 | 1.9 ± 0.8 | |
| Kepler-102 f | 82.03 | 27.5 | 0.17 | 0.753 ± 0.033 | ⋯ | 0.48 | 6.2 | |
| Kepler-19 b | 84.01 | 9.29 | 0.084 | 2.30 ± 0.06 | 2.4 ± 0.7 | 7.4 ± 2.1 | 3.3 ± 1.0 | |
| Kepler-19 c | 84.10 | 28.52 ± 0.08 | 0.18 | ⋯ | 3.7 ± 0.8 | 16.6 ± 3.5 | ⋯ | |
| Kepler-65 b | 85.02 | 2.15 | 0.035 | 1.52 ± 0.09 | 1.0 ± 1.0 | 2.5 ± 2.4 | 4 ± 4 | |
| Kepler-65 c | 85.01 | 5.86 | 0.069 | 2.58 ± 0.06 | 1.5 ± 1.2 | 5 ± 4 | 1.6 ± 1.3 | |
| Kepler-65 d | 85.03 | 8.13 | 0.085 | 1.78 ± 0.11 | 1.2 ± 1.1 | 4 ± 4 | 4 ± 4 | |
| Kepler-65 e | 85.10 | 257.9 ± 0.8 | 0.85 | ⋯ | 19.7 ± 2.0 | 217 ± 22 | ⋯ | |
| KOI-94 b | 94.04 | 3.74 | 0.05 | 1.64 ± 0.12 | 1.3 ± 1.2 | 3.6 ± 3.2 | 5 ± 4 | |
| KOI-94 c | 94.02 | 10.4 | 0.099 | 3.86 ± 0.10 | 1.9 ± 1.4 | 7 ± 6 | 0.7 ± 0.5 | |
| KOI-94 d | 94.01 | 22.3 | 0.16 | 10.31 ± 0.24 | 15.7 ± 1.6 | 77 ± 8 | 0.39 ± 0.04 | |
| KOI-94 e | 94.03 | 54.3 | 0.3 | 6.12 ± 0.14 | 2.1 ± 1.3 | 14 ± 9 | 0.33 ± 0.21 | |
| K00103.01 | 103.01 | 14.9 | 0.12 | 2.59 ± 0.07 | 1.0 ± 1.6 | 3 ± 6 | 1.1 ± 1.8 | |
| Kepler-94 b | 104.01 | 2.51 | 0.034 | 3.04 ± 0.12 | 5.9 ± 1.4 | 11.0 ± 2.6 | 2.1 ± 0.5 | |
| Kepler-94 c | 104.10 | 816.4 ± 0.7 | 1.6 | ⋯ | 236.0 ± 3.1 | (2.82 ± 0.04) × 10$^3$ | ⋯ | |
| Kepler-103 b | 108.01 | 16.0 | 0.13 | 3.26 ± 0.08 | 3.2 ± 1.6 | 14 ± 7 | 2.2 ± 1.1 | |
| Kepler-103 c | 108.02 | 180.0 | 0.65 | 5.61 ± 0.20 | 3.2 ± 1.9 | 30 ± 19 | 0.9 ± 0.6 | |
| Kepler-104 b | 111.01 | 11.4 | 0.093 | 2.38 ± 0.06 | 3.2 ± 0.9 | 9.9 ± 2.8 | 4.1 ± 1.2 | |
| Kepler-104 c | 111.02 | 23.7 | 0.15 | 2.36 ± 0.08 | 1.8 ± 1.0 | 7 ± 4 | 3.0 ± 1.6 | |
| Kepler-104 d | 111.03 | 51.8 | 0.25 | 2.63 ± 0.14 | 1.1 ± 0.9 | 6 ± 5 | 1.7 ± 1.4 | |
| Kepler-106 b | 116.03 | 6.16 | 0.065 | 0.86 ± 0.04 | ⋯ | 0.82 | 7.1 | |
| Kepler-106 c | 116.01 | 13.6 | 0.11 | 2.39 ± 0.07 | 3.7 ± 0.7 | 13.3 ± 2.4 | 5.4 ± 1.0 | |
| Kepler-106 d | 116.04 | 24.0 | 0.16 | 1.01 ± 0.06 | ⋯ | 1.5 | 8.3 | |
| Kepler-106 e | 116.02 | 43.8 | 0.24 | 2.62 ± 0.17 | 0.1 ± 0.5 | 6.6 ± 2.4 | 2.0 ± 0.8 | |
| Kepler-106 f | 116.10 | 90.64 ± 0.24 | 0.39 | ⋯ | 6.8 ± 0.8 | 46 ± 6 | ⋯ | A |
| Kepler-95 b | 122.01 | 11.5 | 0.1 | 3.12 ± 0.09 | 2.8 ± 0.9 | 10.2 ± 3.4 | 1.9 ± 0.6 | |
| Kepler-109 b | 123.01 | 6.48 | 0.069 | 2.33 ± 0.07 | 1.5 ± 1.3 | 5 ± 4 | 2.0 ± 1.7 | |
| Kepler-109 c | 123.02 | 21.2 | 0.15 | 2.54 ± 0.07 | 0.4 ± 1.3 | 2 ± 6 | 0.6 ± 2.0 | |
| Kepler-18 b | 137.03 | 3.5 | 0.045 | 1.81 ± 0.21 | 5.4 ± 1.7 | 13 ± 4 | 12 ± 4 | |
| Kepler-18 c | 137.01 | 7.64 | 0.075 | 4.27 ± 0.11 | 6.5 ± 1.6 | 20 ± 5 | 1.40 ± 0.35 | |
| Kepler-18 d | 137.02 | 14.9 | 0.12 | 5.11 ± 0.13 | 6.3 ± 1.7 | 24 ± 6 | 0.99 ± 0.27 | |
| KOI-142 b | 142.01 | 10.9 | 0.096 | 3.80 ± 0.20 | 3.3 ± 1.5 | 11 ± 5 | 1.1 ± 0.5 | |
| KOI-142 c | 142.10 | 22.2672 ± 0.0006 | 0.15 | ⋯ | 47.7 ± 1.7 | 208 ± 7 | ⋯ | |
| KOI-142 d | 142.11 | 1425 ± 14 | 2.5 | ⋯ | 63.8 ± 3.3 | (1.00 ± 0.05) × 10$^3$ | ⋯ | |
| Kepler-48 b | 148.01 | 4.78 | 0.054 | 1.85 ± 0.09 | 2.7 ± 1.2 | 6.8 ± 2.9 | 5.9 ± 2.5 | |
| Kepler-48 c | 148.02 | 9.67 | 0.086 | 2.56 ± 0.07 | 3.5 ± 1.0 | 11.1 ± 3.2 | 3.6 ± 1.1 | |
| Kepler-48 d | 148.03 | 42.9 | 0.23 | 1.98 ± 0.07 | 1.7 ± 1.1 | 9 ± 6 | 6 ± 4 | |
| Kepler-48 e | 148.10 | 998 ± 4 | 1.9 | ⋯ | 46.6 ± 1.4 | 687 ± 21 | ⋯ | |
| Kepler-48 f | 148.11 | (5.2 ± 0.4) × 10$^3$ | 5.7 | ⋯ | 12 ± 4 | (2.9 ± 0.9) × 10$^2$ | ⋯ | |
| Kepler-113 b | 153.02 | 4.75 | 0.051 | 2.00 ± 0.07 | 3.3 ± 1.5 | 7.5 ± 3.4 | 5.2 ± 2.3 | |
| Kepler-113 c | 153.01 | 8.93 | 0.078 | 2.66 ± 0.17 | 0.4 ± 1.2 | 1.1 ± 3.3 | 0.3 ± 1.0 | |
| Kepler-11 b | 157.06 | 10.3 | 0.092 | 1.92 ± 0.06 | 2.2 ± 2.2 | 7 ± 7 | 6 ± 6 | |
| Kepler-11 c | 157.01 | 13.0 | 0.11 | 3.05 ± 0.08 | 3.1 ± 2.1 | 11 ± 8 | 2.2 ± 1.5 | |
| Kepler-11 d | 157.02 | 22.7 | 0.16 | 3.38 ± 0.10 | 0.0 ± 1.4 | 8 ± 6 | 1.2 ± 0.9 | |

Table 5
(Continued)

| Name[b] | KOI | Per. (days) | $a$ (au) | $R_p$ ($R_\oplus$) | $K$[a] (m s$^{-1}$) | $M \sin i$[a] ($M_\oplus$) | $\rho_p$[a] (g cm$^{-3}$) | Flag |
|---|---|---|---|---|---|---|---|---|
| Kepler-11 e | 157.03 | 32.0 | 0.2 | 4.04 ± 0.11 | 1.6 ± 2.0 | 8 ± 10 | 0.7 ± 0.8 | |
| Kepler-11 f | 157.04 | 46.7 | 0.25 | 2.85 ± 0.21 | 3.3 ± 2.2 | 18 ± 12 | 4.4 ± 3.0 | |
| Kepler-11 g | 157.05 | 118.0 | 0.47 | 3.59 ± 0.10 | 0.0 ± 1.2 | 9 ± 9 | 1.1 ± 1.1 | |
| Kepler-25 b | 244.02 | 6.24 | 0.069 | 2.70 ± 0.06 | 3.4 ± 1.0 | 10.8 ± 3.1 | 3.0 ± 0.9 | |
| Kepler-25 c | 244.01 | 12.7 | 0.11 | 4.63 ± 0.10 | 6.0 ± 1.0 | 24 ± 4 | 1.34 ± 0.23 | |
| Kepler-25 d | 244.10 | 91.61 ± 0.24 | 0.42 | ⋯ | 8.9 ± 1.1 | 69 ± 9 | ⋯ | |
| Kepler-37 b | 245.03 | 13.4 | 0.1 | 0.276 ± 0.028 | ⋯ | 0.0087 | 2.3 | |
| Kepler-37 c | 245.02 | 21.3 | 0.14 | 0.718 ± 0.024 | ⋯ | 0.4 | 5.9 | |
| Kepler-37 d | 245.01 | 39.8 | 0.21 | 1.91 ± 0.05 | 0.20 ± 0.29 | 4.9 ± 1.3 | 3.9 ± 1.1 | |
| Kepler-37 e | 245.04 | 51.2 | 0.25 | ⋯ | 1.63 ± 0.34 | 8.1 ± 1.7 | ⋯ | A |
| Kepler-37 f | 245.10 | 2001.0 ± 3.0 | 2.9 | ⋯ | 2.4 ± 0.6 | 40 ± 10 | ⋯ | A |
| Kepler-68 b | 246.01 | 5.4 | 0.061 | 2.31 ± 0.05 | 3.0 ± 0.5 | 8.4 ± 1.5 | 3.8 ± 0.7 | |
| Kepler-68 c | 246.02 | 9.61 | 0.09 | 0.927 ± 0.034 | 0.8 ± 0.6 | 2.7 ± 1.9 | 18 ± 13 | |
| Kepler-68 d | 246.10 | 632.3 ± 2.3 | 1.5 | ⋯ | 17.4 ± 0.7 | 236 ± 10 | ⋯ | |
| Kepler-68 e | 246.11 | (4.46 ± 0.10) × 10$^3$ | 5.4 | ⋯ | 6.5 ± 1.8 | (1.6 ± 0.5) × 10$^2$ | ⋯ | |
| Kepler-126 b | 260.01 | 10.5 | 0.097 | 1.58 ± 0.06 | 0.1 ± 1.9 | 4 ± 7 | 6 ± 10 | |
| Kepler-126 c | 260.03 | 21.9 | 0.16 | 1.61 ± 0.11 | 0.3 ± 1.5 | 1 ± 7 | 2 ± 9 | |
| Kepler-126 d | 260.02 | 100.0 | 0.44 | 2.54 ± 0.06 | 7.0 ± 3.0 | 55 ± 23 | 18 ± 8 | |
| Kepler-96 b | 261.01 | 16.2 | 0.13 | 2.37 ± 0.06 | 3.2 ± 1.2 | 13 ± 5 | 5.3 ± 1.9 | |
| Kepler-50 b | 262.01 | 7.81 | 0.081 | 1.54 ± 0.05 | 0.0 ± 2.6 | 4 ± 9 | 6 ± 13 | |
| Kepler-50 c | 262.02 | 9.38 | 0.091 | 1.82 ± 0.18 | 2.8 ± 2.8 | 10 ± 10 | 9 ± 9 | |
| Kepler-507 b | 265.01 | 3.57 | 0.048 | 1.28 ± 0.04 | 2.2 ± 0.6 | 5.7 ± 1.5 | 15 ± 4 | |
| Kepler-507 c | 265.10 | 93.7 ± 0.4 | 0.42 | ⋯ | 4.3 ± 0.7 | 34 ± 5 | ⋯ | A |
| Kepler-454 b | 273.01 | 10.6 | 0.096 | 1.84 ± 0.06 | 1.8 ± 0.6 | 6.4 ± 2.0 | 5.6 ± 1.8 | |
| Kepler-454 c | 273.10 | 524.61 ± 0.29 | 1.3 | ⋯ | 112.1 ± 0.8 | 1470 ± 10 | ⋯ | |
| Kepler-454 d | 273.11 | (4.8 ± 1.4) × 10$^3$ | 5.7 | ⋯ | 33 ± 7 | (9.0 ± 2.0) × 10$^2$ | ⋯ | |
| Kepler-128 b | 274.01 | 15.1 | 0.13 | 1.43 ± 0.06 | ⋯ | 6.2 | 12.0 | |
| Kepler-128 c | 274.02 | 22.8 | 0.17 | 1.42 ± 0.17 | ⋯ | 6.1 | 12.0 | |
| Kepler-129 b | 275.01 | 15.8 | 0.13 | 2.29 ± 0.07 | 2.2 ± 1.4 | 10 ± 6 | 4.6 ± 3.0 | |
| Kepler-129 c | 275.02 | 82.2 | 0.4 | 2.40 ± 0.08 | 3.3 ± 1.2 | 26 ± 10 | 10 ± 4 | |
| Kepler-129 d | 275.10 | (2.1 ± 0.8) × 10$^3$ | 3.4 | ⋯ | 92 ± 33 | (2.1 ± 0.8) × 10$^3$ | ⋯ | |
| Kepler-36 b | 277.02 | 13.8 | 0.11 | 1.49 ± 0.06 | 0.3 ± 1.0 | 1 ± 4 | 2 ± 6 | |
| Kepler-36 c | 277.01 | 16.2 | 0.13 | 3.96 ± 0.14 | 4.0 ± 2.2 | 16 ± 9 | 1.4 ± 0.8 | |
| Kepler-510 b | 281.01 | 19.6 | 0.14 | 2.62 ± 0.17 | ⋯ | 6.6 | 2.0 | |
| Kepler-130 b | 282.02 | 8.46 | 0.079 | 1.04 ± 0.04 | ⋯ | 1.7 | 8.6 | |
| Kepler-130 c | 282.01 | 27.5 | 0.17 | 2.88 ± 0.07 | ⋯ | 7.2 | 1.7 | |
| Kepler-130 d | 282.03 | 87.5 | 0.38 | 1.38 ± 0.08 | ⋯ | 5.4 | 11.0 | |
| Kepler-131 b | 283.01 | 16.1 | 0.13 | 2.06 ± 0.05 | 1.7 ± 0.9 | 7 ± 4 | 4.3 ± 2.3 | |
| Kepler-131 c | 283.02 | 25.5 | 0.17 | 0.806 ± 0.035 | ⋯ | 0.63 | 6.6 | |
| Kepler-92 b | 285.01 | 13.7 | 0.12 | 3.70 ± 0.12 | 4.4 ± 1.9 | 19 ± 8 | 2.1 ± 0.9 | |
| Kepler-92 c | 285.02 | 26.7 | 0.19 | 2.45 ± 0.07 | 2.3 ± 1.9 | 13 ± 10 | 5 ± 4 | |
| Kepler-92 d | 285.03 | 49.4 | 0.29 | 2.06 ± 0.07 | 1.7 ± 1.7 | 12 ± 11 | 7 ± 7 | |
| Kepler-97 b | 292.01 | 2.59 | 0.036 | 1.62 ± 0.09 | 1.7 ± 1.3 | 3.4 ± 2.5 | 4.4 ± 3.3 | |
| Kepler-134 b | 295.01 | 5.32 | 0.059 | 1.63 ± 0.04 | ⋯ | 4.2 | 5.4 | |
| Kepler-134 c | 295.02 | 10.1 | 0.09 | 1.03 ± 0.04 | ⋯ | 1.7 | 8.5 | |
| ⋯ | 295.10 | > 16,000 | > 12 | ⋯ | > 31 | ⋯ | ⋯ | |
| Kepler-98 b | 299.01 | 1.54 | 0.026 | 1.87 ± 0.13 | 1.1 ± 1.2 | 2.1 ± 2.1 | 1.7 ± 1.8 | |
| Kepler-99 b | 305.01 | 4.6 | 0.051 | 1.81 ± 0.14 | 1.7 ± 0.9 | 3.8 ± 2.1 | 3.5 ± 2.0 | |
| Kepler-139 d | 316.03 | 7.31 | 0.076 | 1.70 ± 0.06 | 1.5 ± 0.6 | 4.7 ± 2.0 | 5.3 ± 2.3 | |
| Kepler-139 b | 316.01 | 15.8 | 0.13 | 2.38 ± 0.07 | 1.3 ± 0.5 | 5.3 ± 2.1 | 2.2 ± 0.9 | |
| Kepler-139 c | 316.02 | 157.0 | 0.58 | 2.45 ± 0.08 | 2.6 ± 0.5 | 23 ± 5 | 8.7 ± 1.8 | |
| Kepler-139 e | 316.10 | (2.05 ± 0.07) × 10$^3$ | 3.2 | ⋯ | 20.4 ± 2.0 | (4.3 ± 0.4) × 10$^2$ | ⋯ | |
| Kepler-406 b | 321.01 | 2.43 | 0.036 | 1.45 ± 0.04 | 2.4 ± 0.8 | 5.3 ± 1.8 | 9.6 ± 3.2 | |
| Kepler-406 c | 321.02 | 4.62 | 0.055 | 0.83 ± 0.04 | ⋯ | 0.72 | 6.9 | |
| KOI-351 b | 351.06 | 7.01 | 0.074 | 1.30 ± 0.07 | ⋯ | 4.3 | 11.0 | |
| KOI-351 c | 351.05 | 8.72 | 0.086 | 1.45 ± 0.07 | ⋯ | 6.5 | 12.0 | |
| Kepler-90 i | ⋯ | 14.4 | 0.12 | 1.32 ± 0.21 | ⋯ | 4.5 | 11.0 | |
| KOI-351 d | 351.03 | 59.7 | 0.31 | 2.87 ± 0.09 | ⋯ | 7.2 | 1.7 | |
| KOI-351 e | 351.04 | 91.9 | 0.41 | 2.61 ± 0.09 | ⋯ | 6.6 | 2.0 | |
| KOI-351 f | 351.07 | 125.0 | 0.51 | 2.77 ± 0.12 | ⋯ | 6.9 | 1.8 | |
| KOI-351 g | 351.02 | 211.0 | 0.72 | 7.72 ± 0.21 | 2.3 ± 2.5 | 23 ± 25 | 0.27 ± 0.29 | |
| KOI-351 h | 351.01 | 332.0 | 0.97 | 11.25 ± 0.31 | 20.3 ± 2.9 | 236 ± 33 | 0.91 ± 0.14 | |

Table 5
(Continued)

| Name[b] | KOI | Per. (days) | a (au) | $R_p$ ($R_\oplus$) | $K$[a] (m s$^{-1}$) | $M \sin i$[a] ($M_\oplus$) | $\rho_p$[a] (g cm$^{-3}$) | Flag |
|---|---|---|---|---|---|---|---|---|
| Kepler-538 b | 365.01 | 81.7 | 0.35 | $2.16 \pm 0.05$ | $1.7 \pm 0.8$ | $10 \pm 5$ | $5.7 \pm 2.7$ | |
| Kepler-538 c | 365.02 | 118.0 | 0.45 | $0.63 \pm 0.06$ | ⋯ | 0.24 | 5.2 | |
| Kepler-145 b | 370.02 | 23.0 | 0.17 | $2.22 \pm 0.08$ | ⋯ | 5.7 | 2.8 | |
| Kepler-145 c | 370.01 | 42.9 | 0.26 | $3.88 \pm 0.22$ | ⋯ | 9.5 | 0.9 | |
| Kepler-9 d | 377.03 | 1.59 | 0.027 | $1.51 \pm 0.05$ | $4.5 \pm 2.3$ | $8 \pm 4$ | $14 \pm 7$ | |
| Kepler-9 b | 377.01 | 19.3 | 0.14 | $8.09 \pm 0.21$ | $6.1 \pm 2.3$ | $26 \pm 10$ | $0.27 \pm 0.10$ | |
| Kepler-9 c | 377.02 | 38.9 | 0.23 | $8.11 \pm 0.22$ | $5.1 \pm 2.7$ | $27 \pm 14$ | $0.28 \pm 0.15$ | |
| Kepler-197 b | 623.03 | 5.6 | 0.059 | $1.112 \pm 0.035$ | ⋯ | 2.3 | 9.2 | |
| Kepler-197 c | 623.01 | 10.3 | 0.088 | $1.31 \pm 0.04$ | ⋯ | 4.4 | 11.0 | |
| Kepler-197 d | 623.02 | 15.7 | 0.12 | $1.19 \pm 0.04$ | ⋯ | 3.0 | 9.8 | |
| Kepler-197 e | 623.04 | 25.2 | 0.16 | $0.93 \pm 0.11$ | ⋯ | 1.1 | 7.6 | |
| Kepler-62 b | 701.02 | 5.71 | 0.055 | $1.44 \pm 0.05$ | ⋯ | 6.4 | 12.0 | |
| Kepler-62 c | 701.05 | 12.4 | 0.093 | $0.64 \pm 0.04$ | ⋯ | 0.24 | 5.2 | |
| Kepler-62 d | 701.01 | 18.2 | 0.12 | $2.12 \pm 0.07$ | ⋯ | 5.4 | 3.1 | |
| Kepler-62 e | 701.03 | 122.0 | 0.43 | $1.87 \pm 0.07$ | ⋯ | 4.8 | 4.0 | |
| Kepler-62 f | 701.04 | 267.0 | 0.72 | $1.54 \pm 0.08$ | ⋯ | 4.0 | 6.1 | |
| Kepler-220 b | 719.04 | 4.16 | 0.046 | $0.834 \pm 0.035$ | ⋯ | 0.72 | 6.9 | |
| Kepler-220 c | 719.01 | 9.03 | 0.076 | $1.63 \pm 0.08$ | ⋯ | 4.2 | 5.4 | |
| Kepler-220 d | 719.02 | 28.1 | 0.16 | $1.03 \pm 0.11$ | ⋯ | 1.7 | 8.5 | |
| Kepler-220 e | 719.03 | 45.9 | 0.23 | $1.32 \pm 0.05$ | ⋯ | 4.6 | 11.0 | |
| Kepler-56 b | 1241.02 | 10.5 | 0.11 | $4.77 \pm 0.21$ | $7.5 \pm 1.5$ | $36 \pm 7$ | $1.9 \pm 0.4$ | |
| Kepler-56 c | 1241.01 | 21.4 | 0.18 | $10.9 \pm 0.6$ | $38.5 \pm 1.4$ | $236 \pm 9$ | $1.01 \pm 0.11$ | |
| Kepler-56 d | 1241.10 | $994 \pm 5$ | 2.3 | ⋯ | $95.0 \pm 2.2$ | $(2.04 \pm 0.05) \times 10^3$ | ⋯ | |
| ⋯ | 1241.11 | > 14686 | > 10 | ⋯ | > 120 | ⋯ | ⋯ | |
| Kepler-407 b | 1442.01 | 0.669 | 0.015 | $1.16 \pm 0.04$ | $1.2 \pm 0.6$ | $1.8 \pm 0.9$ | $6.2 \pm 3.2$ | |
| Kepler-407 c | 1442.10 | 2096 ± | 3.3 | ⋯ | $172.2 \pm 0.8$ | $3523 \pm 17$ | ⋯ | |
| Kepler-408 b | 1612.01 | 2.47 | 0.036 | $0.710 \pm 0.028$ | ⋯ | 0.38 | 5.9 | |
| ⋯ | 1612.10 | > 16214 | > 12 | ⋯ | > 15 | ⋯ | ⋯ | |
| Kepler-314 b | 1692.02 | 2.46 | 0.036 | $0.78 \pm 0.04$ | ⋯ | 0.55 | 6.4 | |
| Kepler-314 c | 1692.01 | 5.96 | 0.064 | $2.77 \pm 0.15$ | ⋯ | 6.9 | 1.8 | |
| Kepler-411 b | 1781.02 | 3.01 | 0.038 | $1.87 \pm 0.06$ | ⋯ | 4.8 | 4.1 | |
| Kepler-411 c | 1781.01 | 7.83 | 0.073 | $3.24 \pm 0.11$ | ⋯ | 8.0 | 1.3 | |
| Kepler-411 d | 1781.03 | 58.0 | 0.28 | $3.38 \pm 0.12$ | ⋯ | 8.3 | 1.2 | |
| Kepler-334 b | 1909.02 | 5.47 | 0.061 | $1.03 \pm 0.04$ | ⋯ | 1.7 | 8.5 | |
| Kepler-334 c | 1909.01 | 12.8 | 0.11 | $1.37 \pm 0.04$ | ⋯ | 5.3 | 11.0 | |
| Kepler-334 d | 1909.03 | 25.1 | 0.17 | $1.39 \pm 0.05$ | ⋯ | 5.6 | 11.0 | |
| Kepler-409 b | 1925.01 | 69.0 | 0.32 | $0.98 \pm 0.04$ | $0.06 \pm 0.18$ | $1.4 \pm 1.1$ | $8 \pm 6$ | |
| Kepler-409 c | 1925.10 | $(3.0 \pm 0.9) \times 10^3$ | 4.0 | ⋯ | $3.5 \pm 1.3$ | $77 \pm 28$ | ⋯ | |
| Kepler-338 e | 1930.04 | 9.34 | 0.09 | $1.77 \pm 0.27$ | ⋯ | 4.6 | 4.6 | |
| Kepler-338 b | 1930.01 | 13.7 | 0.12 | $2.49 \pm 0.08$ | ⋯ | 6.3 | 2.2 | |
| Kepler-338 c | 1930.02 | 24.3 | 0.17 | $2.39 \pm 0.08$ | ⋯ | 6.1 | 2.4 | |
| Kepler-338 d | 1930.03 | 44.4 | 0.25 | $2.83 \pm 0.31$ | ⋯ | 7.1 | 1.7 | |
| Kepler-1130 e | 2169.04 | 2.19 | 0.032 | $0.388 \pm 0.029$ | ⋯ | 0.034 | 3.2 | |
| Kepler-1130 c | 2169.02 | 3.27 | 0.042 | $0.644 \pm 0.030$ | ⋯ | 0.26 | 5.3 | |
| Kepler-1130 d | 2169.03 | 4.27 | 0.051 | $0.639 \pm 0.029$ | ⋯ | 0.25 | 5.3 | |
| Kepler-1130 b | 2169.01 | 5.45 | 0.06 | $0.85 \pm 0.10$ | ⋯ | 0.77 | 7.0 | |
| Kepler-1130 B | 2169.10 | $(1.61 \pm 0.04) \times 10^4$ | 12.0 | ⋯ | $2195.9 \pm 2.4$ | $(7.047 \pm 0.007) \times 10^4$ | ⋯ | |
| K02687.01 | 2687.01 | 1.72 | 0.028 | $0.730 \pm 0.022$ | ⋯ | 0.42 | 6.0 | |
| K02687.02 | 2687.02 | 8.17 | 0.08 | $0.961 \pm 0.028$ | ⋯ | 1.3 | 7.9 | |
| K02720.01 | 2720.01 | 6.57 | 0.069 | $1.17 \pm 0.04$ | $0.8 \pm 1.2$ | $2.3 \pm 3.5$ | $8 \pm 12$ | |
| Kepler-403 b | 2732.01 | 7.03 | 0.078 | $1.77 \pm 0.19$ | ⋯ | 4.6 | 4.5 | |
| Kepler-403 d | 2732.02 | 13.6 | 0.12 | $1.76 \pm 0.07$ | ⋯ | 4.6 | 4.6 | |
| K02732.04 | 2732.04 | 24.6 | 0.18 | $1.40 \pm 0.07$ | ⋯ | 5.7 | 12.0 | |
| Kepler-403 c | 2732.03 | 54.3 | 0.31 | $2.32 \pm 0.32$ | ⋯ | 5.9 | 2.6 | |
| K03083.02 | 3083.02 | 6.23 | 0.069 | $0.71 \pm 0.05$ | ⋯ | 0.37 | 5.8 | |
| K03083.03 | 3083.03 | 8.29 | 0.084 | $0.65 \pm 0.05$ | ⋯ | 0.27 | 5.4 | |
| K03083.01 | 3083.01 | 10.2 | 0.096 | $0.82 \pm 0.05$ | ⋯ | 0.67 | 6.7 | |
| Kepler-444 b | 3158.01 | 3.6 | 0.04 | $0.348 \pm 0.030$ | ⋯ | 0.022 | 2.9 | |
| Kepler-444 c | 3158.02 | 4.55 | 0.047 | $0.44 \pm 0.04$ | ⋯ | 0.056 | 3.6 | |
| Kepler-444 d | 3158.03 | 6.19 | 0.058 | $0.45 \pm 0.04$ | ⋯ | 0.063 | 3.7 | |
| Kepler-444 e | 3158.04 | 7.74 | 0.067 | $0.48 \pm 0.04$ | ⋯ | 0.076 | 3.9 | |

Table 5
(Continued)

| Name[b] | KOI | Per. (days) | $a$ (au) | $R_p$ ($R_\oplus$) | $K$[a] (m s$^{-1}$) | $M \sin i$[a] ($M_\oplus$) | $\rho_p$[a] (g cm$^{-3}$) | Flag |
|---|---|---|---|---|---|---|---|---|
| Kepler-444 f | 3158.05 | 9.74 | 0.078 | $0.57 \pm 0.05$ | ⋯ | 0.16 | 4.7 | |
| Kepler-444 BC | 3158.10 | $(1.19 \pm 0.10) \times 10^5$ | $52.2 \pm 3.0$ | ⋯ | ⋯ | $(2.00 \pm 0.07) \times 10^5$ | ⋯ | |
| K03179.01 | 3179.01 | 5.99 | 0.065 | $0.84 \pm 0.08$ | ⋯ | 0.75 | 6.9 | |

**Note.**

[a] Values for $K$, $M \sin i$, and $e$ are fixed for planets with no reported errors on these values. Fixed planet masses are based on a mass–radius relation (Weiss & Marcy 2014), and fixed eccentricities are circular.

[b] Nontransiting companions are given KOI indices beginning with KNNNN.10 for consistency with Marcy et al. (2014). Flags: A–planet candidate due to uncertain orbital period, low significance, or confusion with stellar activity; see the Appendix for details. For names, companions with planetary masses are given lowercase letters, companions with stellar masses are given uppercase letters, and companions with ambiguous masses are left nameless. Additional columns associated with this table are available in the machine-readable version.

(This table is available in its entirety in machine-readable form.)

Table 6
KGPS Nontransiting Companion Orbital Properties

| Name | KOI | Per. (days) | $a$ (au) | $K$ (m s$^{-1}$) | $M \sin i$ ($M_J$) | $e$ | FAP | $\Delta$BIC | Notes |
|---|---|---|---|---|---|---|---|---|---|
| Kepler-100 e | 41.10 | $60.89 \pm 0.04$ | 0.31 | $3.7 \pm 0.5$ | $0.076 \pm 0.010$ | $0.03 \pm 0.13$ | 0.00002 | −45.9 | |
| Kepler-93 B | K00069 B | >20,000 | >15 | >89 | >15.5 | ⋯ | 0.00000 | −837.2 | A |
| Kepler-20 g | 70.10 | $34.95 \pm 0.04$ | 0.21 | $4.3 \pm 0.7$ | $0.066 \pm 0.011$ | 0.0 | 0.00493 | −29.5 | |
| Kepler-19 c | 84.10 | $28.52 \pm 0.08$ | 0.18 | $3.7 \pm 0.8$ | $0.052 \pm 0.011$ | 0.0 | 0.02454 | −24.0 | |
| Kepler-65 e | 85.10 | $257.2 \pm 0.8$ | 0.85 | $20.4 \pm 2.0$ | $0.70 \pm 0.07$ | $0.306 \pm 0.018$ | 0.00000 | −80.2 | |
| Kepler-94 c | 104.10 | $816.4 \pm 0.7$ | 1.6 | $236.0 \pm 3.1$ | $8.89 \pm 0.12$ | $0.34593 \pm 0.00024$ | 0.00079 | −1205.7 | |
| Kepler-106 f | 116.10 | $90.64 \pm 0.24$ | 0.39 | $6.8 \pm 0.8$ | $0.146 \pm 0.018$ | 0.0 | 0.00174 | −44.6 | |
| KOI-142 c | 142.10 | $22.2672 \pm 0.0006$ | 0.15 | $47.7 \pm 1.7$ | $0.654 \pm 0.024$ | 0.0 | 0.00043 | −144.8 | B |
| KOI-142 d | 142.11 | $1425 \pm 14$ | 2.5 | $63.8 \pm 3.3$ | $3.15 \pm 0.17$ | $0.430 \pm 0.029$ | 0.00175 | −114.7 | |
| Kepler-48 e | 148.10 | $998 \pm 4$ | 1.9 | $46.6 \pm 1.4$ | $2.16 \pm 0.07$ | $0.003 \pm 0.028$ | 0.00000 | −173.2 | |
| Kepler-48 f | 148.11 | $(5.2 \pm 0.4) \times 10^3$ | 5.7 | $12 \pm 4$ | $0.93 \pm 0.29$ | $0.01 \pm 0.15$ | ⋯ | ⋯ | C |
| Kepler-25 d | 244.10 | $91.61 \pm 0.24$ | 0.42 | $8.9 \pm 1.1$ | $0.216 \pm 0.027$ | 0.0 | 0.00009 | −45.0 | D |
| Kepler-68 d | 246.10 | $632.3 \pm 2.3$ | 1.5 | $17.4 \pm 0.7$ | $0.742 \pm 0.031$ | $0.190 \pm 0.018$ | 0.00000 | −244.3 | |
| Kepler-68 e | 246.11 | $(4.46 \pm 0.10) \times 10^3$ | 5.4 | $6.5 \pm 1.8$ | $0.52 \pm 0.14$ | $0.29 \pm 0.13$ | 0.00003 | −55.9 | E |
| Kepler-507 c | 265.10 | $93.7 \pm 0.4$ | 0.42 | $4.3 \pm 0.7$ | $0.106 \pm 0.017$ | 0.0 | 0.03644 | −27.6 | |
| Kepler-454 c | 273.10 | $524.61 \pm 0.29$ | 1.3 | $112.1 \pm 0.8$ | $4.626 \pm 0.033$ | $0.001 \pm 0.006$ | 0.00000 | −1318.0 | F |
| Kepler-454 d | 273.11 | $(4.8 \pm 1.4) \times 10^3$ | 5.7 | $33 \pm 7$ | $2.8 \pm 0.6$ | $0.148 \pm 0.027$ | 0.00000 | −76.9 | |
| Kepler-129 d | 275.10 | $(2.1 \pm 0.8) \times 10^3$ | 3.4 | $92 \pm 33$ | $6.6 \pm 2.4$ | $0.09 \pm 0.06$ | 0.00017 | −424.7 | G |
| ⋯ | 295.10 | >15619 | >12 | >31 | >3.9 | ⋯ | 0.99948 | −12.6 | A |
| Kepler-139 e | 316.10 | $(2.05 \pm 0.07) \times 10^3$ | 3.2 | $20.4 \pm 2.0$ | $1.34 \pm 0.13$ | $0.01 \pm 0.05$ | ⋯ | ⋯ | C |
| Kepler-56 d | 1241.10 | $994 \pm 5$ | 2.3 | $95.0 \pm 2.2$ | $6.43 \pm 0.15$ | $0.2048 \pm 0.0028$ | 0.00019 | −295.6 | |
| ⋯ | 1241.11 | >15,000 | >10 | >120 | >1.5 | ⋯ | 0.00019 | −10.4 | A |
| Kepler-407 c | 1442.10 | $2096 \pm 5$ | 3.3 | $172.2 \pm 0.8$ | $11.09 \pm 0.05$ | $0.2039 \pm 0.0004$ | 0.00000 | −3854.9 | |
| ⋯ | 1612.10 | >16,000 | >12 | >15 | >0.7 | ⋯ | 0.01541 | −30.0 | A |
| Kepler-409 c | 1925.10 | $(3.0 \pm 0.9) \times 10^3$ | 4.0 | $3.5 \pm 1.3$ | $0.24 \pm 0.09$ | $0.01 \pm 0.33$ | 0.00037 | −36.3 | |
| Kepler-1130 B | 2169.10 | $(1.61 \pm 0.04) \times 10^4$ | 12.0 | $2195.9 \pm 2.4$ | $221.71 \pm 0.21$ | $0.677409 \pm 0.000031$ | 0.00133 | $-2.3 \times 10^5$ | H |
| Kepler-444 BC | 3158.10 | $(1.19 \pm 0.10) \times 10^5$ | $52.2 \pm 3.0$ | >45 | $629 \pm 21$ | $0.55 \pm 0.05$ | 0.00000 | −820.5 | I |

**Note.** Companions indicated in bold are newly discovered in the KGPS data; other companions were previously announced in the literature, and we update their orbits and masses based on our new RVs. Flags: A—trend. B—TTV detection preceded RV discovery; the eccentricity of KOI-142 c is 0.05 based on TTVs. C—automated routine fails to find planet (long-period structure is evident, but no peak is strongly preferred). D—period ambiguous because of aliasing (91 or 122 days). E—peak power at $P \approx 360$ days; 2000 day orbit preferred. F—Kepler-454 d is contemporaneously announced in Bonomo et al. (2023), but their RVs were not available at the time of our analysis. G—the eccentricity of the companion is poorly constrained because of time sampling, with either a circular or substantially eccentric model providing comparably good fits. H—automated routine places companion at $P = 6000$ days because of the window function; a much longer period is preferred. I—known stellar companion, ephemeris and mass from Zhang et al. (2023).

(This table is available in machine-readable form.)

nontransiting planets. Additional free parameters were the RV zero-point $\gamma$ and jitter ($\sigma_j$, which is added to the measured errors in quadrature) for the HIRES RVs, as well as $\gamma$ and $\sigma_j$ terms for any additional instruments. (Note that the only cases where two or more observatories were used had many more than 10 RVs.) For data sets with at least 20 degrees of freedom (dof) and for planets larger than 1.15 $R_\oplus$, we also allowed the $N$ RV semi-amplitudes of the planets ($K_1,...,K_N$) to vary, allowing a determination of the masses of the small planets. In cases with fewer than 20 dof or for planets smaller than 1.15 $R_\oplus$, we used the Weiss & Marcy (2014) mass–radius relation to estimate the most likely mass of each transiting planet to determine fixed values of $K_1,...,K_N$. This assumed mass–radius relationship did not produce any results that were inconsistent with our observations. We used the software package RadVel to develop and optimize all of our models (Fulton et al. 2018). Note that for fixed orbital periods, phases, and a circular orbit, this is a linear problem (in parameters $K \cos \omega$, $K \sin \omega$, $\gamma$, and $\sigma_j$), so there is a single global best-fit solution (Wright & Howard 2012).

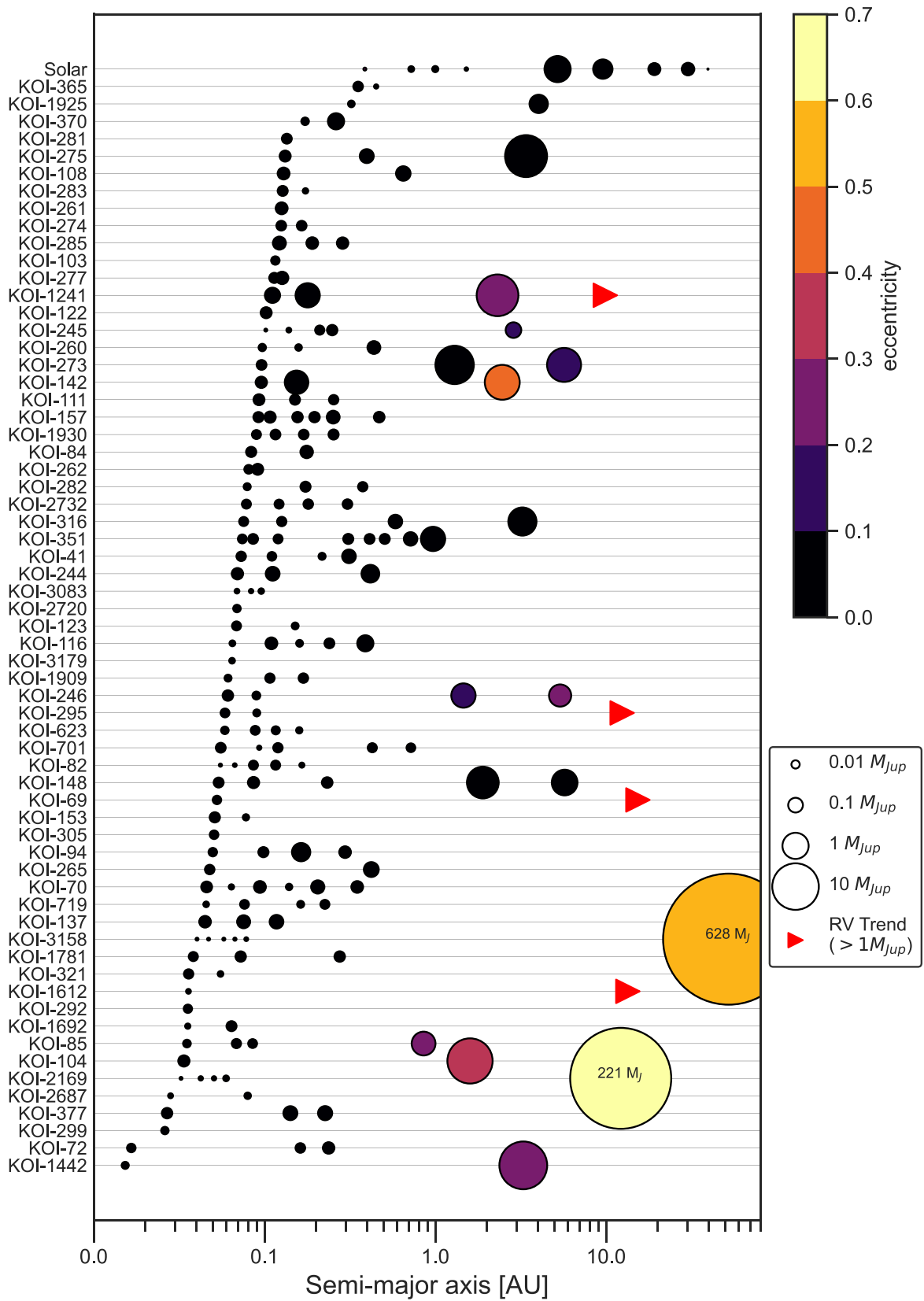

**Figure 4.** Architectures of the KGPS I planetary systems. Systems are ranked by the semimajor axis of the innermost known (transiting) planet. Point sizes scale with the square root of planet masses, $M \sin i$ values, or mass estimates from the Weiss & Marcy (2014) mass–radius relationship. Colors correspond to eccentricities (if measured); low-mass planets are assumed to have circular orbits.

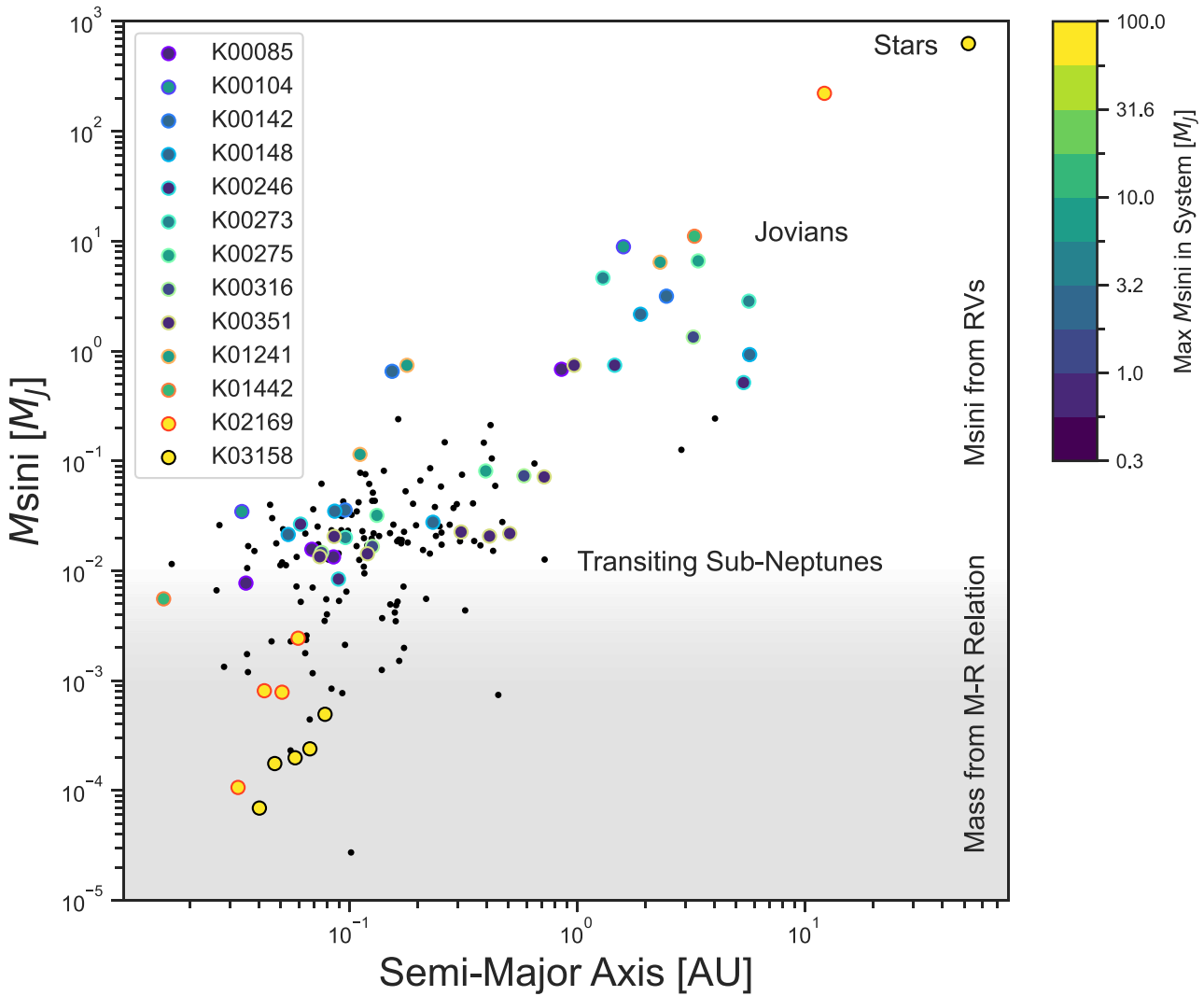

**Figure 5.** The masses (or $M \sin i$ values or mass estimates) vs. semimajor axes of the KGPS I transiting planets and companions. For systems with a giant planet ($M \sin i > 0.3 M_J$), the circle fill color corresponds to the maximum detected companion mass in the system (capped at 100 Jupiter masses), and the edge color encircling each point corresponds to the planetary system. Black circles correspond to transiting planets in our survey for which giant planets were not detected. Planet masses or $M \sin i$ values are reported where they are measured, but mass estimates are used for $m \lesssim 3 \times 10^{-2} M_J$ (see text for details). Companions that have eccentric, stellar-mass companions at large separations (KOI-2169 and KOI-3158; yellow) have planets that are smaller and closer to their stars than the typical transiting planets from the KGPS sample, whereas transiting planets that have giant planet companions (nonyellow colors) have comparable sizes and orbits to the transiting planets without detected giant planet companions.

3. After subtracting the best N-planet fit, take a "fast periodogram" of the residual RVs, using the algorithm fasper developed in Press & Rybicki (1989). Assess whether there is a significant peak by estimating the false-alarm probability (FAP) given the peak height and the number of effectively independent frequencies considered: $\text{FAP} \approx 1 - [1 - \exp(-P_{\max})]^{2N_\nu/f_{\text{over}}}$, where $P_{\max}$ is the maximum power and $N_\nu/f_{\text{over}}$ is the total number of frequencies sampled divided by the typical rate of oversampling the average Nyquist frequency (Press & Rybicki 1989). For all targets, we oversampled the Nyquist frequency by $f_{\text{over}} = 100$.

4. If there is a significant peak (FAP < 0.05), assess whether the period of the peak is an appropriate guess for the orbit of an additional planet. This involves checking that the period is (1) within the window function (<4× the RV baseline), (2) longer than 10.0 days (signals near peaks in the window function of the RVs, such as 1 day, are often aliases of long-period structure visible by eye in the RVs), and (3) distinct from the periods of other planets in the model (with a minimum period ratio of 1.15). If any of these conditions are not met, instead propose including a linear RV trend in the model, effectively adding one free parameter ($dv/dt$) to represent an additional high-mass, long-period body. In cases where at least one condition (1–3) is not met and with at least 15 dof, we also included RV curvature ($d^2v/dt^2$).

5. If the new proposed period passes the above criteria, fit a new model with all of the previous planets, plus a planet at the new period. We modeled each new planet orbit with two to four free parameters: the RV semi-amplitude $K_{N+1}$ and the time nearest conjunction, $t_c$ (even if the planet does not transit). Note that the orbital period was fixed during this step. We used the phase of the Lomb–

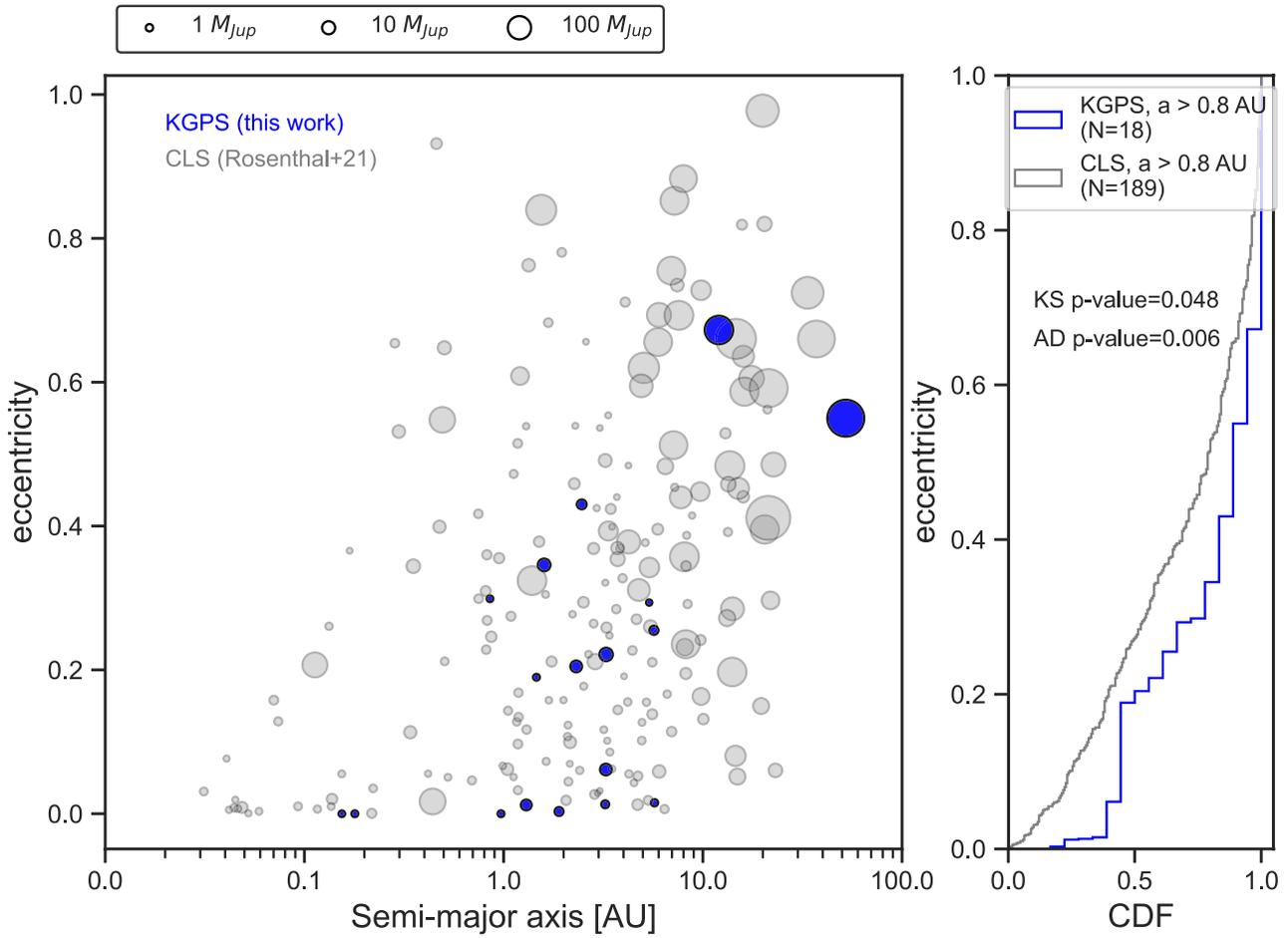

**Figure 6.** Left: the eccentricity vs. semimajor axis of companions with $M \sin i > 0.3 M_J$ in KGPS (blue) and CLS (Rosenthal et al. 2021, gray). The eccentricities of the KGPS transiting planets are assumed to be zero. The point size scales with the square root of the companion $M \sin i$ (see legend). Right: the cumulative distribution function of the KGPS and CLS companions for $a > 1$ au, with Kolmogorov–Smirnov and Anderson–Darling p-values shown.

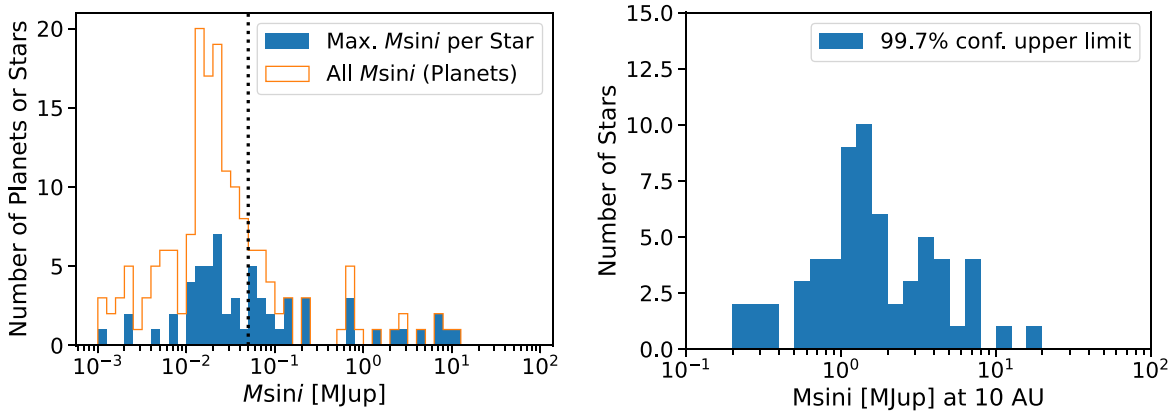

**Figure 7.** Left: the $M \sin i$ values (measured or estimated from the Weiss & Marcy 2014 mass–radius relation) of all the transiting and nontransiting planets in KGPS I (orange histogram), and the maximum $M \sin i$ value per star (blue histogram). Right: upper limits on $M \sin i$ for Kepler stars with RVs. The upper limits are based on models in which a linear trend ($dv/dt$) is adopted.

Scargle periodogram at each period to guess a time of conjunction. In cases with at least 15 dof, we also included two eccentricity parameters, modeled using the basis of $\sqrt{e} \sin \omega$ and $\sqrt{e} \cos \omega$, which effectively assert a uniform prior for $e \in [0, 1)$ and $\omega \in [0, 2\pi)$.

6. Compute and compare the Bayesian information criterion (BIC, with difference $\Delta$BIC) of the $N + 1$-planet and $N$-planet models (Kass & Raftery 1995). In our formulation, $\Delta$BIC = BIC($N + 1$-planet model) – BIC ($N$-planet model), with negative values indicating that an N+1-planet model is preferred. While some authors consider $\Delta$BIC of −10 to be sufficient to distinguish the models, we adopt a more conservative $\Delta$BIC threshold of −20. This is particularly important since we effectively tried thousands of different orbital periods before selecting the best one to model via the Lomb–Scargle periodogram. The choice of −20 is arbitrary but based on a qualitative calibration based on trial and error; a substantially lower value leads to several dubious planetary signals, whereas a substantially higher value misses signals that are apparent by eye. The choice of ΔBIC of −20 is also advantageous, since it matches that adopted by Rosenthal et al. (2021) for claiming a planet detection. For trends, we only required ΔBIC of −10, since we only searched for one trend per system (as compared to thousands of orbital periods).

7. While the $N + 1$-planet model is preferred, increment $N$ and repeat steps 3–6.
8. When the search has converged on a preferred model, use a Markov Chain Monte Carlo (MCMC) algorithm to explore uncertainties in the free parameters. We used RadVel, which employs the affine-invariant sampler of Goodman & Weare (2010), as implemented for Python in emcee (Foreman-Mackey et al. 2013). Our priors were as follows: (a) the RV semi-amplitudes are positive ($K_1,\ldots,K_N > 0$), (b) eccentricities are bounded between 0 and 1, and (c) the RV jitter is bounded between 0 and 20 m s$^{-1}$.

The KGPS algorithm includes steps to account for the number of RVs and the dof to reduce model complexity and speed up model selection compared to purely blind algorithms (e.g., rvsearch; Rosenthal et al. 2021). The KGPS algorithm uses tools from the astronomy-specific Python packages astropy and radvel (Astropy Collaboration et al. 2013, 2018; Fulton et al. 2018; Astropy Collaboration et al. 2022). The output of the KGPS search algorithm for each system is shown in Table 3.

### 4.2. Empirical Estimates of Stellar Jitter

One nuisance parameter in our model is the stellar jitter, $\sigma_j$. This parameter is added in quadrature to the intrinsic RV errors in a manner that accounts for stellar-induced RV variability. Often, the stellar jitter is measured empirically for a given star based on the rms of the RV residuals. However, in cases where we have only 10 RVs per star and are looking to detect nontransiting planets, a larger-than-expected rms of the RV residuals could be due to nontransiting planets rather than stellar variability.

To disentangle whether the scatter in the RVs is likely caused by a nontransiting planet or by stellar physics, we make empirical estimates of the expected stellar jitter for each star in our sample. The empirical estimates are based on measured RV scatter for large samples of stars with many observations and known stellar properties. Our stellar jitter estimate incorporates Mount Wilson calcium activity index ($S_{HK}$), measured as an excess above an empirically determined activity baseline ($\Delta S = S_{HK} - S_{BL}$) as a function of photometric color ($B - V$)

$$\sigma_{j,S} = 2.3 + 17.4\, \Delta S \text{ m s}^{-1} \quad (0.4 < B - V < 0.7)$$
$$\sigma_{j,S} = 2.1 + 4.7\, \Delta S \text{ m s}^{-1} \quad (0.7 < B - V < 1.0)$$
$$\sigma_{j,S} = 1.6 - 0.003\, \Delta S \text{ m s}^{-1} \quad (1.0 < B - V < 1.3)$$
$$\sigma_{j,S} = 2.1 + 2.7\, \Delta S \text{ m s}^{-1} \quad (1.3 < B - V < 1.6)$$

(Isaacson & Fischer 2010), the line-of-sight projected stellar rotational velocity $v \sin i$

$$\sigma_{j,\text{rot}} = 0.876 + 0.140 \, \frac{v \sin i}{\text{km s}^{-1}} + 0.009 \left(\frac{v \sin i}{\text{km s}^{-1}}\right)^2 \text{ m s}^{-1} \tag{1}$$

(Chontos et al. 2022), and the stellar mass $M_\star$ (Luhn et al. 2020)

$$\sigma_{j,M} = 2.5 \text{ m s}^{-1} \quad (M_\star < 1.2)$$
$$\sigma_{j,M} = 3.5 \text{ m s}^{-1} \quad (1.2 < M_\star < 1.3)$$
$$\sigma_{j,M} = 4.0 \text{ m s}^{-1} \quad (1.3 < M_\star < 1.5)$$
$$\sigma_{j,M} = 5.0 \text{ m s}^{-1} \quad (1.5 < M_\star < 1.7)$$
$$\sigma_{j,M} = 5.5 \text{ m s}^{-1} \quad (M_\star > 1.7).$$

We computed the expected jitter from each of these relations and chose the mean value as the expected intrinsic stellar jitter for each star in our sample. In addition to the physical jitter, we defined "photometric" jitter by empirically fitting a lower envelope to the observed rms of the RVs as a function of the stellar flux relative to a $V = 12.0$ star ($F/V_{12}$) for our sample:

$$\sigma_{j,\text{phot}} = \frac{0.81}{F/V_{12}} + 2.2 \text{ m s}^{-1}. \tag{2}$$

We combined the expected intrinsic stellar jitter and the photometric jitter for a total expected value of

$$\text{rms}_{\text{expected}} = \sqrt{\text{Max}(\sigma_{j,S}, \sigma_{j,\text{rot}}, \sigma_{j,M})^2 + \sigma_{j,\text{phot}}^2}. \tag{3}$$

The expected jitter values and observed rms of the RV residuals are given in Table 4. We plotted the observed rms as a function of expected rms in Figure 3. The rms of the RVs for each system is shown twice: once before any planets are modeled (open triangles), and once after fitting the small transiting planets and any nontransiting companions as prescribed in Figure 2 (filled circles). Systems for which our model improved the rms of the RVs by at least a factor of 5 are highlighted in blue. These are the systems in which we have detected giant planets.

More than half of the points in Figure 3 are above the "expected" curve. One possibility is that our method for estimating the expected RV scatter by averaging the three jitter estimates from the literature is overly optimistic. For example, in cases where one factor contributes strongly to the RV scatter (such as stellar activity), the rms of the residuals would be better described as the maximum than the mean of the three values. However, when we adopt the maximum of the three jitter estimates, the majority of predictions overestimate the total rms because taking the maximum estimated jitter is a conservative approach. Another possibility is that there might still be planets of various masses that we have not yet detected in our sample that contribute significantly to the rms of the residual RVs.

A careful study of RV variability across a large sample of stars that incorporates multiple jitter predictors (stellar activity, mass, rotation, and photon limits) and identifies as many low-mass planets as possible would enable a more precise estimate of the jitter of each star, but such an effort is beyond the scope of this paper.

### 4.3. Refined Orbital Estimates

The KGPS algorithm yielded several systems for which the best-fit model included one or more nontransiting, eccentric companions. The KGPS algorithm has limitations in its ability to identify the most likely orbital periods of long-period companions, particularly in cases where the companion is eccentric and has a longer period than the RV baseline. For every system in which one or more nontransiting companions were detected, we examined whether a superior fit could be found by eye to the best-fit model from the KGPS algorithm. We then used the best fit (as determined either by eye or by the KGPS algorithm) to seed a new MCMC analysis in which the orbital periods and eccentricities of the nontransiting companion(s) were allowed to vary, as well as any transiting companions suspected of having significant RV amplitudes. The MCMC chains had 48 walkers each and ran until the chains were "well mixed," based on having a maximum Gelman–Rubin statistic of $\leqslant 1.03$, a minimum chain length to autocorrelation length of 75, a maximum relative change in autocorrelation time of $\leqslant .01$, and at least 1000 independent draws. In cases where the periodogram yielded peaks of similar heights but at different orbital periods, we tested models with the companion at each of the possible periods.

In most cases, the MCMC algorithm yielded nearly identical results to the KGPS algorithm. A few notable exceptions are KOI-148, KOI-246, and KOI-316, in which an additional long-period planet was detected by eye. In KOI-148 and KOI-246 these additional companions are based on low-amplitude, moderately eccentric signals with periods comparable to the RV baseline, whereas in KOI-316 a high-amplitude signal is missed by the KGPS algorithm because the signal has very uneven time sampling, leading to a low-significance peak in the periodogram.

Other substantial updates were in KOI-94 and KOI-2169. In KOI-94, the RV data yield a significant improvement in the orbital period of one of the transiting planets (KOI-94 d), a warm Jupiter-sized planet with an RV semi-amplitude of $\sim 20\,\mathrm{m\,s^{-1}}$. The system is known to have planet–planet interactions (based on observed TTVs; Masuda et al. 2013), and the baseline from our RV survey (12 yr) is substantially longer than the baseline of Kepler photometry (4 yr), and so it is not surprising that we are detecting a modest deviation from the best-fit linear ephemeris for this planet. In KOI-2169, the orbit of the stellar-mass companion is very eccentric and has a much longer period than the RV baseline, which resulted in a poor fit in the KGPS algorithm. We used the publicly available algorithm `rvsearch` to explore a fine grid in the period–eccentricity solution space for this companion and identify a better fit (Rosenthal et al. 2021).

## 5. Results

The best-fit models for individual planetary systems are summarized in Table 3 and the Appendix. Among the 63 planetary systems here, we detected 27 nontransiting companions, 7 of which are first announced here, 13 of which are Jovian ($M\sin i > 0.3 M_J$, 1 au $< a <$ 10 au), and 8 of which are sub-Saturnian ($M\sin i < 0.3 M_J$). We also detected five significant RV trends, three of which are demonstrably from stellar or brown dwarf (rather than planetary-mass) companions. In cases where the planet or stellar companions were already known, we provide updated orbital parameters and $M\sin i$ values. An overview of each star's observing program is presented in Table 4. The resulting planetary system architectures and physical properties are presented in Table 5, and the orbital properties of the nontransiting planets are presented in Table 6.

### 5.1. Preliminary Architectural Patterns

The system architectures are presented in Figure 4. Each row corresponds to a planetary system, and the systems are ranked by the semimajor axis of the innermost known planet (typically, a transiting planet). Each body (planetary or stellar) orbiting the primary star is indicated with a circle, and the circle size is scaled to the square root of the body's mass or $M\sin i$, if known. In cases where the RVs were insufficient to meaningfully constrain the planet mass, a mass–radius relationship was assumed (Weiss & Marcy 2014). Each circle is colored by the object's eccentricity, if known (the transiting planets are assumed to have zero eccentricity).

A few notable patterns emerge in Figure 4. The two most massive objects in our survey, Kepler-444 BC (represented by a single circle with the combined mass of Kepler-444 B and Kepler-444 C) and Kepler-1130 B, are also the most eccentric objects in our survey, and they accompany systems of high-multiplicity, small planets: the Kepler-444 planets are Mars sized, and the Kepler-1130 planets are Earth sized. In these systems, the periastron passage distances of the stellar-mass companion(s) with respect to the primary are 20 and 4 au, respectively, well inside the typical physical extent of a forming protoplanetary disk of $\sim$100 au (Weiss et al. 2022). It is likely that these stellar-mass companions truncated the protoplanetary disks in which the small planets formed. Such disk truncation would prevent the flow of dust and rocky material at the truncation radius, reducing the rocky material available in the innermost $\sim$1 au of the disk from which the planets formed. Discovering more systems with architectures analogous to Kepler-444 and Kepler-1130 is critical for building a sample from which to study the effect of eccentric S-type binary stars on the formation and architectures of planetary systems.

In contrast, the Jovian-mass planets we detected are not associated with particularly small or particularly close-in planets. Figure 5 shows the semimajor axes and masses of the bodies of Figure 4, including the transiting planets and their RV-detected companions. Systems that contain a Jovian-mass or larger body are shown in color, with the color corresponding to the mass of the most massive companion detected in the system. Two of these massive companions are transiting, KOI-94 d (Kepler-89 d) and KOI-351 h (Kepler-90 h), whereas the other companions were all detected via RVs (and in the case of Kepler-444, also high spatial resolution imaging) and either do not transit or have not yet been detected in transits. The KGPS survey spans seven orders of magnitude in companion mass and four orders of magnitude in orbital separation.

Figure 6 shows how the KGPS I companions compare to field star companions from the California Legacy Survey (CLS; Rosenthal et al. 2021) in their eccentricity and semimajor axis distributions, for companions with $M\sin i > 0.3 M_J$. These surveys have different selection functions, and so any comparison between them should be examined with caution. An important distinction at short orbital periods is that the KGPS planets at $a < 0.8$ au are generally detected via transits,[32] whereas the CLS planets are all detected via RVs. There might also be differences at long periods because the CLS planets were detected in a 30 yr baseline, whereas the KGPS planets were detected in a 15 yr baseline, although the longest-period object is actually in the KGPS sample. We compare these samples for $M \sin i > 0.3 M_J$, for which the RVs of both surveys were sufficient to determine precise eccentricities, based on the results of Table 6. KGPS companions (blue) have lower eccentricities than the CLS companions (gray). The difference in the distributions of KGPS and CLS eccentricities is statistically significant based on the Anderson–Darling tests ($p = 0.006$) and marginally significant based on the Kolmogorov–Smirnov test ($p = 0.048$). One possible reason for an astrophysical difference in their eccentricity distributions is that the KGPS companions all have at least one surviving planet with $a < 1$ au in addition to any planets with $a > 1$ au, whereas the CLS planets are not required to have inner companions, and so they can have a higher envelope on eccentricity as a function of semimajor axis. Future work exploring the relationship between eccentricity, planet multiplicity, and orbital stability will clarify whether the discrepancy in eccentricity between KGPS and CLS is (1) statistically significant and (2) dynamical in origin.

---

[32] The two planets with $M\sin i > 0.3 M_J$ and $a < 0.8$ au from KGPS are KOI-142 c, which was detected in TTVs, and Kepler-56 c, which was detected in transits.

### 5.2. Upper Limits on Companion Masses

It is desirable to have a constraint on the $M \sin i$ upper limit for companions in each system based on the RVs. For each system, we used the best-fit model output from KGPS but allowed $dv/dt$ to vary to test the hypothesis that the RVs have some long-term trend consistent with the presence of a long-period companion on a circular orbit ($P \gtrsim 4\times$ the RV baseline, typically $\gtrsim 10$ au). The absence of a significant trend or significant peaks in the Lomb–Scargle periodogram via KGPS generally rules out the presence of any companions with $M \sin i \gtrsim 1 M_J$ within 10 au with $3\sigma$ confidence, although the exact upper limits depend on the RV baseline, the number of RVs, and rms. The $3\sigma$ upper limits on $M \sin i$ for all the Kepler systems are shown in Figure 7. All of the cases where the upper limit on $M \sin i$ is larger than $10 M_J$ at 10 au correspond to systems with detected long-period planets. There is typically sufficient freedom in the fit to the orbital parameters of the long-period nontransiting planets, especially massive ones, that a large value of $dv/dt$ cannot be ruled out. These same $dv/dt$ upper limits typically correspond to $M \sin i < 1 M_J$ at 5 au, which is interior to the nominal lowest period of $4\times$ the RV baseline for most systems, but in combination with a nondetection in the Lomb–Scargle periodogram, the lack of a significant trend rules out the majority of possible Jovian companions interior to 5 au. Detailed injection recovery to determine each star's sensitivity to planets as a function of mass and orbital period will be addressed in Paper II of this series.

### 6. Discussion

Intriguingly, the presence of Jupiter-class planets has an unknown effect on the formation of temperate terrestrial planets in the Earth and Venus regime. Giant planets form early during planet formation while the protoplanetary disk still has substantial gases, typically within 10 Myr (Pollack et al. 1996). A nascent giant planet sculpts the distribution of gases and solids in the protoplanetary disk, potentially creating gaps and/or ring structures. These radial features form sites where the agglomeration of solids—which is essential for forming a next generation of small planets—is either enhanced or suppressed. Accordingly, giant planets are predicted to either hinder or help the formation of small planets, depending on whether pairwise planetesimal collisions (e.g., Levison & Agnor 2003; Walsh et al. 2011) or pebble accretion (e.g., Chen et al. 2020; Schlecker et al. 2021) is the dominant mode of small planet formation.

Several studies have attempted to determine the occurrence of Jupiter-like planets in systems with small planets using heterogeneously constructed samples (Zhu & Wu 2018; Bryan et al. 2019). These studies include some systems with giant planets detected in RV surveys (but relatively poor sensitivity to small, inner planets); some systems with small, transiting planets (but an unknown occurrence of giant planets); and some systems in which both small and giants planets are detectable. One underlying assumption of Zhu & Wu (2018), which applies Bayes's theorem, is that the occurrence of a planet of a given mass at a given location is independent of the masses and locations of other planets in the system. Such an assumption is valid for planets sufficiently far apart but does not hold for planets with closely packed orbits because dynamical stability becomes important (Rosenthal et al. 2022). Thus, a homogeneous survey that does not depend on Bayes's theorem is desirable.

The RV data presented here belong to one of the largest surveys of planet-hosting stars detected by the Kepler primary mission, in terms of the number of systems observed, the number of RVs, and the total RV baseline. The KGPS sample, in conjunction with the contemporaneous study by Bonomo et al. (2023), will be a powerful tool for analyzing the architectures of cold Jupiters in systems that have small, transiting planets. These data sets are complementary, in that the HIRES observations often span longer time intervals, which means that they provide good coverage for longer periods, while the HARPS-N visits to targets are often obtained at high cadence over timescales similar to the stellar activity cycles, which makes them more amenable to Gaussian process analysis. These studies bear similarity to the Cumming et al. (2008) result that provided early insight in the frequency of massive planets around Sunlike stars, based on one decade of Keck-HIRES RV data. Recent studies by Rosenthal et al. (2021, 2022) have extended the RV baselines from 10 to 30 yr and increased the total number of stars, yielding ever more distant giant planets and an improved characterization of giant planet occurrence in field stars. While these legacy RV studies have established the occurrence of Jupiter-like planets for Sunlike field stars, they are largely insensitive to the Kepler-like planets ($\sim 10 M_\oplus$, $P \sim 30$ days) that we know are also abundant around Sunlike stars (Batalha et al. 2013; Fressin et al. 2013; Petigura et al. 2013; Bryson et al. 2021, and references therein).

The current work presents a similar search for giant planets to the RV legacy surveys, but every star hosts a transiting planet or planets with known radii, and in some cases masses. While knowing the frequency of cool and cold giant planets is valuable in itself, the presence of those planets around systems of small transiting planets in the super-Earth and mini-Neptune regime can be used to test the predictions of various theories of planet formation. The presence or absence of Jupiter-mass planets in the systems of compact multiplanet systems (Jontof-Hutter et al. 2018), especially those in and near orbital resonance, will be invaluable to the theorists exploring how these systems arose, changed, and settled into their final states. For several stars, the first approximately three RVs collected were insufficient for detecting a giant planet (e.g., KOI-316) or accurately characterizing orbital properties (e.g., KOI-2169). These cases provide nice examples of why stopping RV follow-up after obtaining just a few RVs can lead to a gross mischaracterization of the system architecture. Long-term RV monitoring with many visits (at least 10) seems necessary to accurately map the architectures of long-period companions.

While the RV precision needed to detect Jupiter-mass planets has existed for more than 30 yr (Campbell et al. 1988; Marcy & Butler 1992), a significant portion of that time had to pass in order to measure their orbital periods. Designing and maintaining a long-term RV monitoring program is challenging, particularly in an era in which telescope time-allocation committees expect short-term results.[33]

This work reflects the richness of a data set that has had the proper time to age. This work presents a marked improvement on the early glimpse of the giant planet companions to Kepler planets revealed in Marcy et al. (2014), in terms of the number of giant planets and the characterization of their orbits, as well as a more complete view of the systems of small transiting planets that do not have giant companions. Future papers in the KGPS series will explore the relationship between small and giant planets revealed through this unique data set. Patterns in the relationship between giant planets and small (including terrestrial) planets might reveal how Jupiter affected our own small planets and whether this formation channel is common or rare.

## 7. Conclusion

This is the first paper in a series called the Kepler Giant Planet Search. In this paper, we presented RVs and best-fit Keplerian orbital solutions based on a decade of observations of the NASA Kepler field with the W. M. Keck Observatory HIRES instrument. These constitute 2844 RVs in total, collected for 63 stars, the majority of which (56) constitute a homogeneous, magnitude-limited survey of RV-quiet, Sunlike stars. Our RVs improved the mass determinations for 84 transiting planets and revealed 27 companions that are not known to transit, 7 of which are announced here for the first time. Future papers will address architectural patterns observed in these planetary systems, our detection efficiency, and the joint occurrence of giant planets and small transiting planets. More extensive long-term surveys are needed to address the issue of the occurrence rate for giant planets in outer orbits around close-in systems of smaller planets identified by the Kepler, K2, and TESS missions.

## Acknowledgments

In the preparation of this manuscript, the following people contributed project leadership, project design, proposal writing, and major contributions to telescope observations (>400 spectra each): Lauren M. Weiss (2013–2023), Howard T. Isaacson (2013–2023), and Geoffrey W. Marcy (2013–2015). L.M.W. developed the KGPS algorithm, conducted the analysis, generated all figures, and wrote the manuscript, with H.T.I. contributing substantially to project management, sample selection, and manuscript preparation. The following people made major contributions to database maintenance, characterization of stellar properties, data reduction, observing logistics, and telescope observations: Andrew W. Howard, B.J. Fulton, and Erik A. Petigura. The following people wrote telescope proposals that secured the observations presented in this paper and also provided feedback on the manuscript: Eric Agol, Daniel Fabrycky, Eric B. Ford, Daniel Jontof-Hutter, Daniel Huber, Miki Nakajima, James Owen, Alan Reyes, Leslie A. Rogers, Jason Rowe, Jason Steffen, Hilke E. Schlichting, and Jason T. Wright. The following people collected at least 10 nights of observations at Keck Observatory: H.T.I. (851 observations over 152 nights), L.M.W. (456 observations over 122 nights), E.A.P. (174 observations over 66 nights), B.J.F. (254 observations over 59 nights), Evan Sinukoff (260 observations over 57 nights), A.W.H. (329 observations over 56 nights), Lea Hirsch (254 observations over 53 nights), Ashley Chontos (103 observations over 48 nights), G.W.M. (456 observations over 45 nights), Steven Giacalone (113 observations over 45 nights), Joseph M. Akana Murphy (76 observations over 42 nights), Judah van Zandt (83 observations over 42 nights), Corey Beard (88 observations over 37 nights), Malena Rice (81 observations over 37 nights), Molly Kosiarek (83 observations over 35 nights), Sarah Blunt (75 observations over 34 nights), Fei Dai (59 observations over 30 nights), Jack Lubin (61 observations over 29 nights), Paul Dalba (74 observations over 28 nights), Ryan Rubenzahl (67 observations over 27 nights), Alex Polanski (41 observations over 26 nights), Casey Brinkman (53 observations over 23 nights), Aida Behmard (50 observations over 23 nights), Teo Mocnik (58 observations over 20 nights), Michelle Hill (39 observations over 19 nights), Lee Rosenthal (42 observations over 18 nights), Dakotah Tyler (37 observations over 16 nights), Emma Turtelboom (41 observations over 16 nights), Daria Pidhorodetska (25 observations over 13 nights), Mason MacDougall (25 observations over 13 nights), Ian Crossfield (29 observations over 12 nights), Samuel K. Grunblatt (39 observations over 12 nights), Sean M. Mills (30 observations over 11 nights), Rae J. Holcomb (30 observations over 11 nights), and Samuel Yee (16 observations over 10 nights). The following people contributed fewer than 10 nights of observations at Keck Observatory: Andrew Mayo, Ben Montet, Ji Wang, Marta Bryan, Isabel Angelo, Debra Fischer, Rebecca Jensen-Clem, Stephen Kane, Nicholas Saunders, Emma Louden, Jingwen Zhang, Jon Swift, John Brewer, David Shaw, Tabby Boyajian, Gaspar Bakos, Aaron Householder, Jonathan Giguere, Gregory Gilbert, Luke Handley, George Zhou, Dan Bayliss, Luke Bouma, Shannon Dulz, Phil Muirhead, Chas Beichman, Grant Regen, Sarah Lange, and Joel Hartman. J.T.W. and M.H. acknowledge the astronomers who declined authorship on this paper for reasons described in https://www.science.org/content/article/after-outcry-disgraced-sexual-harasser-removed-astronomy-manuscript.

This work would not have been possible without the generosity of various time-allocation committees that provided support in the form of Keck-HIRES observing time over a decade. We acknowledge support in the form of observational resources at W. M. Keck Observatory from the following institutions: NASA, the University of Hawai'i, the University of California, California Institute of Technology, and the University of Notre Dame. We acknowledge support from the NASA-Keck Key Strategic Mission Support program (grant No. 80NSSC19K1475), NASA Exoplanet Research Program (grant No. 80NSSC23K0269), and NASA JPL RSAs 1537000, 1607073, and 1633061.

---

[33] Even the so-called long-term, multisemester programs offered at major observatories are typically capped at 2 yr, or one-fifth the length of our observing effort.

This work was supported by a NASA-Keck PI Data Award, administered by the NASA Exoplanet Science Institute. Data presented herein were obtained at the W. M. Keck Observatory from telescope time allocated to the National Aeronautics and Space Administration through the agency's scientific partnership with the California Institute of Technology and the University of California. The Observatory was made possible by the generous financial support of the W. M. Keck Foundation. The Center for Exoplanets and Habitable Worlds and the Penn State Extraterrestrial Intelligence Center are supported by Penn State and its Eberly College of Science.

This data set made use of the NASA Exoplanet Science Institute at IPAC, which is operated by the California Institute of Technology under contract with the National Aeronautics and Space Administration. This research has made use of NASA's Astrophysics Data System Bibliographic Services. This research has made use of the SIMBAD database, operated at CDS, Strasbourg, France.

The authors wish to recognize and acknowledge the very significant cultural role and reverence that the summit of Maunakea has always had within the indigenous Hawaiian community. We are most fortunate to have the opportunity to conduct observations from this mountain.

*Facilities:* Kepler, Keck:I, TNG, OHP:1.93m.

*Software:* astropy (Astropy Collaboration 2013, 2018, 2022), RadVel, rvsearch, kgps.

## Appendix
## KGPS I Stars, Planets, and RVs

### A.1. HIP 94931 (Kepler-444)

Kepler-444 (KOI-3158, HIP 94931) is a bright ($V = 8.9$), nearby ($d = 35.7$ pc) K dwarf (component A) hosting five Mars-sized transiting planets within 0.1 au. The planetary system, including asteroseismology of the host star, was first characterized in detail by Campante et al. (2015). The system has a wide-separation M dwarf binary companion (components B and C) characterized by Dupuy et al. (2016), which has a nearly edge-on orbit and high eccentricity with periastron passage of just a few astronomical units. Mills & Fabrycky (2017) conducted a detailed analysis of the TTVs of the planets and measured mass upper limits for two of the planets comparable to that of Mars.

The RVs of Kepler-444 (Figure 8) exhibit a linear trend that is broadly consistent with the orbit described in Dupuy et al. (2016). An updated study that uses Gaia data and dynamical analysis confirms the eccentricity of the BC pair with respect to the primary found previously (0.8; Stalport et al. 2022). However, a separate analysis by Zhang et al. (2023) that includes direct adaptive optics of the BC pair, Gaia data, and new HIRES RVs of the systemic BC component finds an eccentricity of BC orbit of 0.55, which changes the periastron passage of the BC component from 5 to 8 au. The slight increase in periastron passage suggests a less severely truncated, more massive disk in which the planets formed than what was originally proposed in Dupuy et al. (2016), resolving the apparent conflict between the estimated primordial solid mass in the disk and the total planet mass.

The periodogram of our best-fit residual to the trend includes a marginally significant peak at 72 days. The KGPS automated pipeline preferred including a planet at this period (based on $\Delta \mathrm{BIC} = -27$), but the Keplerian associated with this signal had an unphysically high eccentricity (of 0.6, our imposed upper limit) that would result in orbit crossing. We do not believe that this signal is associated with a planet; it is probably related to some complexity in the long-term RV signal of the Kepler-444 ABC system, and it might even be related to the orbit of the BC pair. Further analysis is needed to test the significance of the 72 day signal and identify its origin.

### A.2. KOI-41 (Kepler-100)

Kepler-100 (KOI-41) is a $V = 11.2$ Sunlike star at $d = 307$ pc. Transiting planets Kepler-100 b, c, and d, which have orbital periods of 6.89, 12.82, and 35.33 days and radii of 1.3, 2.2, and 1.6 Earth radii, respectively, were confirmed with RVs in Marcy et al. (2014). The mass measurements at the time were of order $2\sigma$ significance and/or mass upper limits. A detailed photodynamical analysis, which reproduced the transit depth variations of the planets as well as their TTVs, resulted in more precise masses: $M_b = 5.1 \pm 1.7 \, M_\oplus$, $M_c = 14.6 \pm 2.8 \, M_\oplus$, and $M_d = 1.1 \pm 0.5 \, M_\oplus$ (Judkovsky et al. 2022). The architecture of this system cannot be explained by photoevaporation alone because a low-mass, gas-enveloped planet exists interior to a highly irradiated rocky planet, and so other physical mechanisms (including atmospheric mass loss) were essential for sculpting the planet compositions (Owen & Campos Estrada 2020).

This star was observed for multiple programs at Keck Observatory. It was originally selected as a bright Kepler planet-hosting star for follow-up in Marcy et al. (2014). After several years of low-cadence observations, this target was selected for moderate-cadence analysis to improve the mass measurements of the transiting planets. Our analysis of the full RV time series spanning 2009–2022 (Figure 9) yields $M_b = 5.5 \pm 1.3 \, M_\oplus$, $M_c = 3.8 \pm 1.7 \, M_\oplus$, and $M_d = 1.2 \pm 1.4 \, M_\oplus$, in agreement with the TTV solution. In addition, we discover a nontransiting planet, Kepler-100 e, at $P_e = 60.88 \pm 0.04$ days and with minimum mass $M_e \sin i_e = 24.8 \pm 3.5 \, M_\oplus$. Fixing the eccentricity of Kepler-100 e at zero results in $M_b = 4.8 \pm 1.3 \, M_\oplus$, $M_c = 4.3 \pm 1.7 \, M_\oplus$, $M_d = 2.4 \pm 1.8 \, M_\oplus$, and $M_e = 25 \pm 3 \, M_\oplus$, yielding an overall upper limit on the mass of planet d: $M_d < 7.8 \, M_\oplus$ ($3\sigma$ conf.). A joint analysis of the TTV and RVs would yield even better masses and orbital constraints of the four known planets. In our 12 yr RV baseline, we find no significant trend in the RV residuals (rms = $3.3 \, \mathrm{m \, s^{-1}}$), consistent with $M \sin i < 0.06 M_J$ at 5 au or $M \sin i < 0.23 M_J$ at 10 au ($3\sigma$ conf.), after assuming a four-planet model (including nontransiting planet e). A run of high-cadence RVs collected in 2022 as part of a student project are not presented here.

### A.3. KOI-69 (Kepler-93)

Kepler-93 (KOI-69) has one transiting planet that is notable for having one of the most precise radius measurements of a super-Earth yet reported: $R_b = 1.48 \pm 0.02 \, R_\oplus$, which has a radius uncertainty of 120 km (Ballard et al. 2014). The precision of the planet radius measurement is based on asteroseismology of the host star (for a precise stellar radius) and transits observed by the NASA Spitzer space telescope (for a precise transit depth). RVs from Keck/HIRES yielded a planetary mass of $M_b = 3.8 \pm 1.5\, M_\oplus$ and also a long-term trend (Marcy et al. 2014). Additional RVs from TNG/HARPS-N improved the mass characterization of the planet, $M_b = 4.0 \pm 0.7\, M_\oplus$, a value consistent with an Earthlike rocky composition for the planet (Dressing et al. 2015).

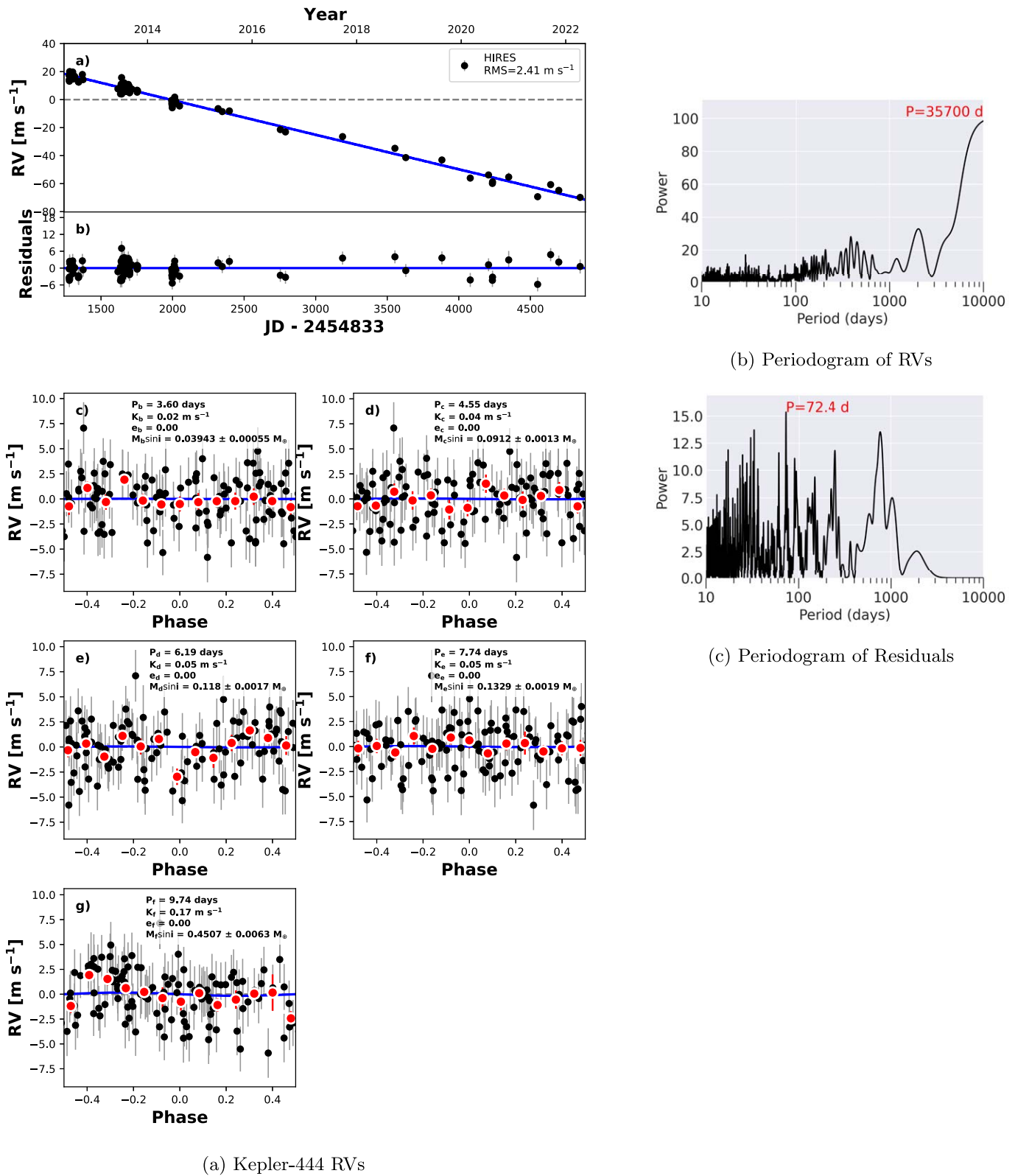

(a) Kepler-444 RVs

(b) Periodogram of RVs

(c) Periodogram of Residuals

**Figure 8.** HIP 94931; KOI-3158 (Kepler-444). Left: RVs from Keck-HIRES. Panels include (top to bottom) the full RV time series, RV residuals, and the RVs phase-folded to the orbits of the individual planets (after subtracting the contributions from the other known planets). The phase-folded plots include weighted mean RVs binned by orbital phase (red circles). Right: periodograms of the HIRES RVs, before (top) and after (bottom) subtracting the best-fit model that includes the transiting planets and one nontransiting stellar companion modeled as an RV trend.

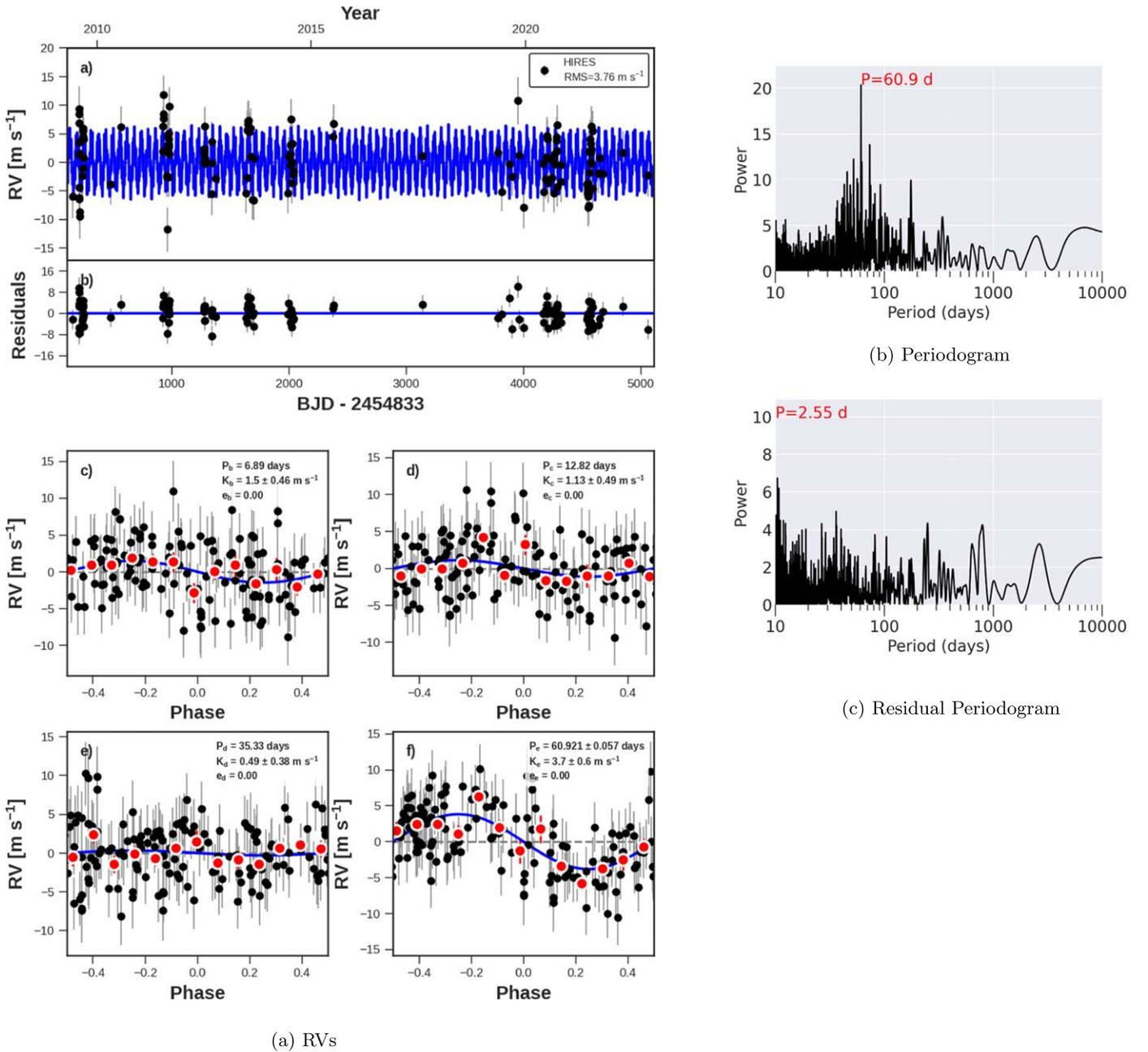

(a) RVs

(b) Periodogram

(c) Residual Periodogram

**Figure 9.** Same as Figure 8, but for KOI-41 (Kepler-100). The best-fit model (blue line) includes a nontransiting planet, Kepler-100 e, at 60 days (orbit details inset in phase-folded panel). The periodograms in the right column are of the RVs (top) and the RVs after the best-fit model has been subtracted (bottom).

With 64 new RVs from Keck/HIRES collected since Marcy et al. (2014), we have extended the RV baseline of Kepler-93 to 12 yr (Figure 10). The RV trend is $dv/dt = 0.0373 \pm 0.0003 \, \mathrm{m\,s^{-1}\,day^{-1}}$. For a companion at 15 au (which corresponds to 4 times the RV baseline), this trend yields a minimum mass of $M \sin i > 15 M_J$, meaning that the companion is stellar or a brown dwarf, rather than planetary. The additional RVs also improve the mass precision for the transiting planet to a $6\sigma$ detection: $M_b = 3.6 \pm 0.6 \, M_\oplus$.

### A.4. KOI-70 (Kepler-20)

Kepler-20 (KOI-70) has five transiting planets that are notable for their unusual size ordering, with alternating super-Earth-sized and Mars-sized planets. Buchhave et al. (2016) characterized this system using RVs from HARPS-N and HIRES, finding a nontransiting planet with a period of 34.9 days in between the 19.6 and 77.6 day transiting planets. We have collected one new HIRES RV since 2016 that is consistent with the solution from Buchhave et al. (2016). While the single new RV data point does not substantially improve the mass measurement of the transiting planets, it does further constrain the nondetection of an RV trend, yielding an upper limit of $M \sin i < 0.35 M_J$ at 5 au or $M \sin i < 1.4 M_J$ at 10 au ($3\sigma$ conf., Figure 11).

### A.5. KOI-72 (Kepler-10)

Kepler-10 (KOI-72) contains the first rocky planet discovered by the Kepler mission with $P_b < 1$ day and a sub-Neptune-sized planet at $P_c = 45.45$ days (Batalha et al. 2011). RVs were collected on both Keck-HIRES and TNG-HARPS-N, and while both sets of data are consistent with a rocky composition for planet b, the data were discrepant regarding the mass and composition of planet c, with HIRES data favoring a low mass $M_c = 7\,M_\oplus$ and HARPS-N data favoring a high mass $M_c = 17\,M_\oplus$ (Dumusque et al. 2014; Weiss et al. 2016). The discrepancy might be caused by additional nontransiting planets in the system sampled at different phases by the two instruments, an idea that is supported by the detection of TTVs of planet c (Weiss et al. 2016). Modeling the RVs with Gaussian processes appears to somewhat resolve the discrepancy, although the host star is an old thick-disk star that is not expected to be magnetically active (Rajpaul et al. 2017).

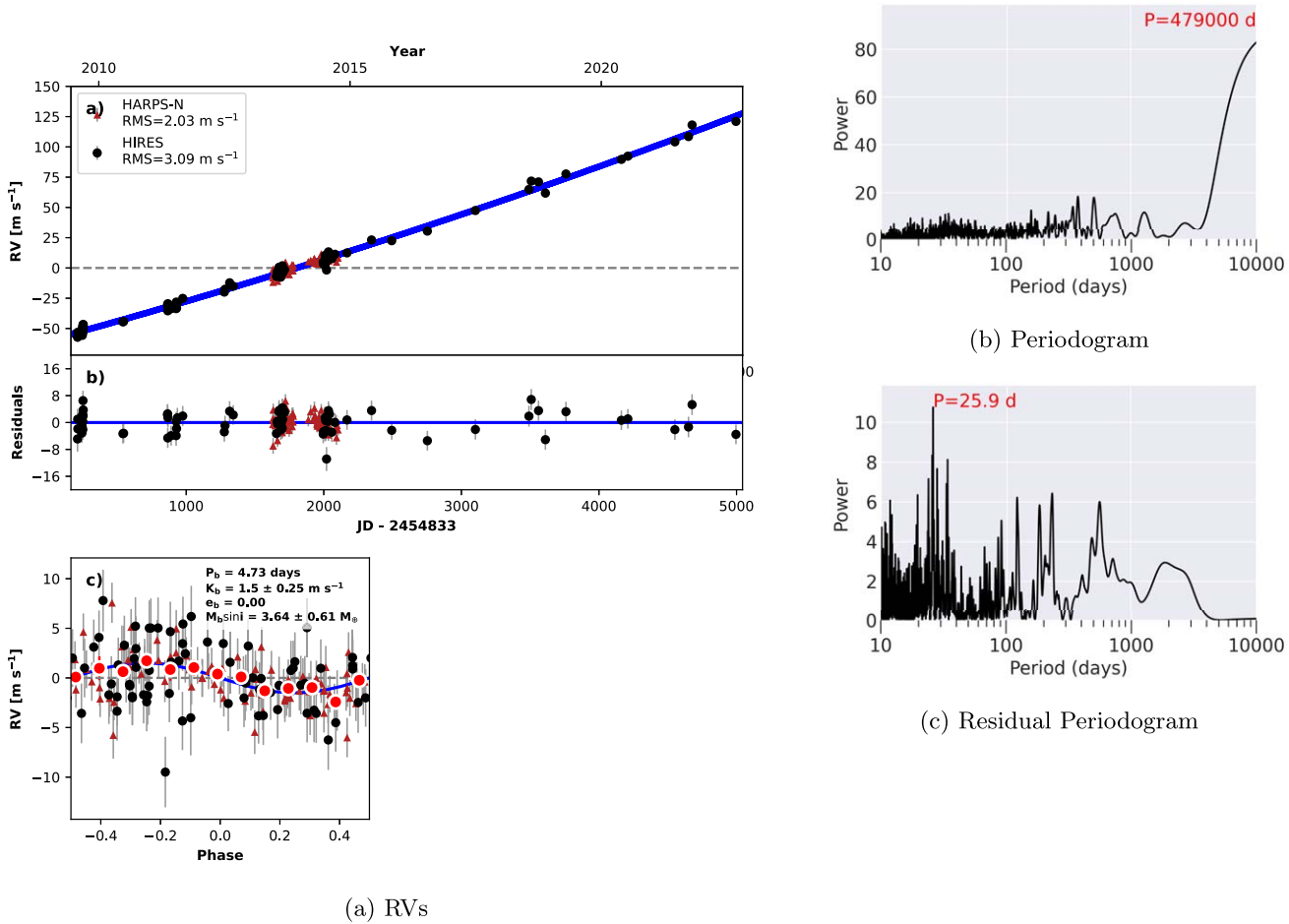

(a) RVs

(b) Periodogram

(c) Residual Periodogram

**Figure 10.** Same as Figure 8, but for KOI-69 (Kepler-93). Left: the RVs were taken at Keck-HIRES (black circles) and TNG-HARPS-N (maroon triangles). The best-fit model includes a stellar companion modeled as an RV trend.

We have collected 14 new HIRES RVs of Kepler-10 since Weiss et al. (2016), producing an RV baseline of 12 yr (Figure 12). The residuals to the best two-planet fit have rms = 4.6 m s$^{-1}$ (HIRES) and 3.9 m s$^{-1}$ (HARPS-N) and exhibit no RV trend, yielding an upper limit of $M\sin i < 0.11 M_J$ at 5 au or $M\sin i < 0.46 M_J$ at 10 au (3$\sigma$ conf.) for additional companions. Note that there is a third low-mass planet candidate ($m_p \sim 1\,M_\oplus$) in the system, as indicated in Weiss et al. (2016), but the period is ambiguous. Because the third planet candidate is low-mass and has $P < 200$ days, our choice to exclude it from our model here does not substantially affect the upper limits on Jovian planets at long periods. However, we do detect significant structure in the periodogram of the RV residuals, with a prominent peak at $P = 25$ days (FAP = 0.04, $\Delta$BIC = $-18.9$). This period was previously modeled as the stellar rotation period in a fit that used a Gaussian process method to remove stellar noise (Rajpaul et al. 2017), although a rotation period of 25 days is uncharacteristically fast for a 10 Gyr old thick-disk star. We consider this signal as a third planet candidate in the system, to be confirmed with more RVs. A careful analysis that jointly fits the HIRES and HARPS-N RVs as well as the TTVs might yield an improved characterization and/or confirmation of the planet candidate at 25 days.

### A.6. KOI-82 (Kepler-102)

KOI-82 (Kepler-102) is a system with five transiting planets, three of which are smaller than Earth. The system was confirmed in Marcy et al. (2014) with HIRES RVs establishing mass upper limits for the planets. The star is active (with a visible stellar rotation signal in the Kepler photometry, and Mount Wilson $S_{HK} = 4.48$), which contributes to substantial RV scatter. A joint analysis of HIRES and HARPS-N RVS that included a Gaussian process trained on the rotational modulation in the Kepler light curve found $M_d = 2.5 \pm 1.4\,M_\oplus$, $M_e = 4.7 \pm 1.7\,M_\oplus$ (Brinkman et al. 2022). Here we employ a simple circular fit to planets d and e based on the HIRES data alone, with masses for the other (smaller) planets based on an empirical mass–radius relationship (Weiss & Marcy 2014). The HIRES RVs extend from 2010

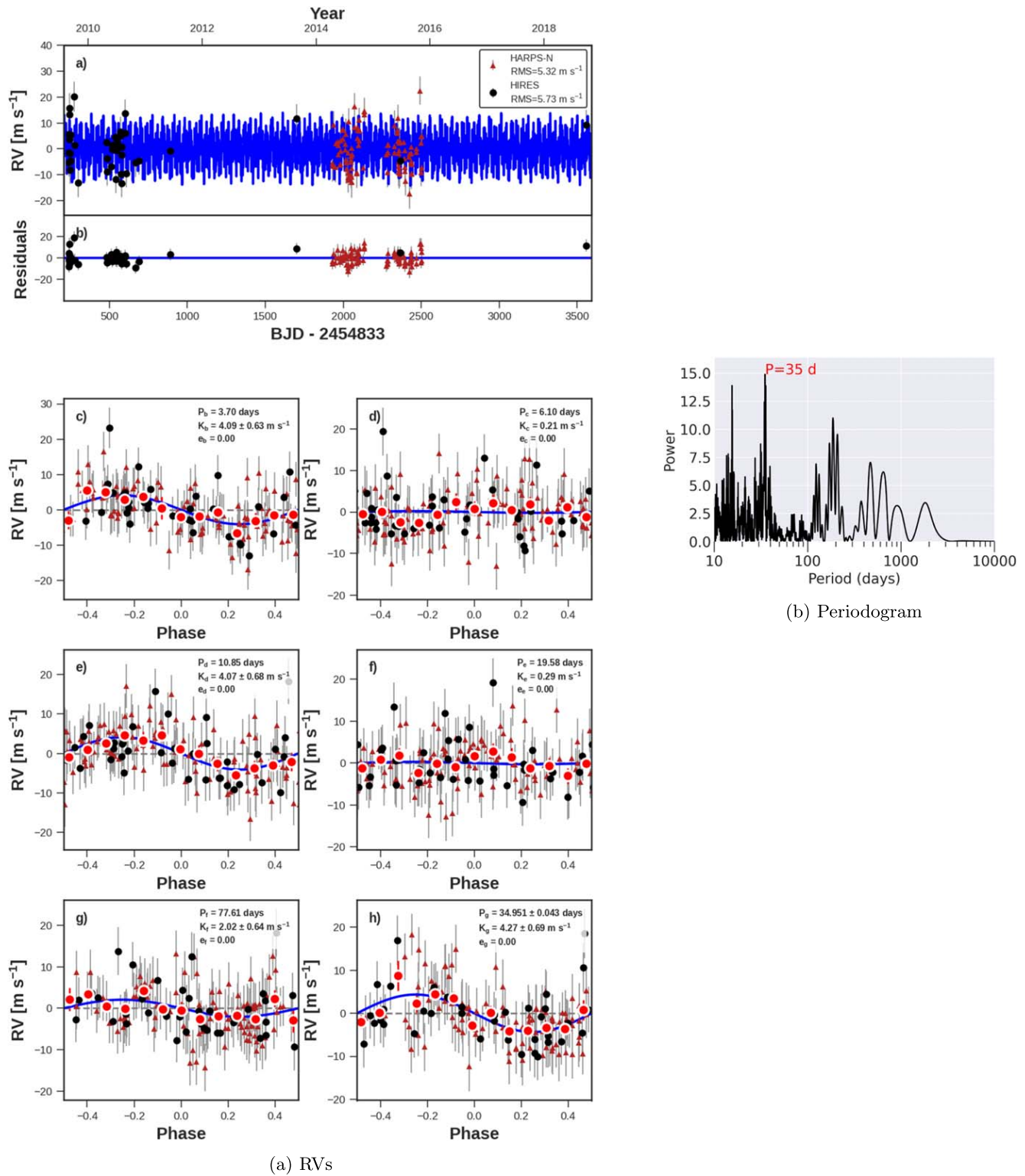

(a) RVs

(b) Periodogram

**Figure 11.** Same as Figure 8, but for KOI-70 (Kepler-20). Left: the RVs were taken at Keck-HIRES (black circles) and TNG-HARPS-N (maroon triangles). The best-fit model includes a nontransiting planet, Kepler-20 g, at 35 days (orbital parameters inset in phase-folded panel).

April to 2021 June, comprising a baseline of 11 yr. The residuals to our best fit have rms = 3.9 m s$^{-1}$ and no RV trend, yielding a 3$\sigma$ upper limit of $M \sin i < 0.2 M_J$ at 5 au or $M \sin i < 0.8 M_J$ at 10 au for additional companions (Figure 13). Note that the Lomb–Scargle periodogram of the RVs has a marginal peak at 28 days, which is consistent with the stellar rotation identified in the Kepler light curve (Brinkman et al. 2022).

### A.7. KOI-84 (Kepler-19)

Kepler-19 (KOI-84) was first confirmed and characterized in Ballard et al. (2011). The system has one transiting planet (b) with $P_b = 9.3$ days and $R_b = 2.2\,R_\oplus$. Significant TTVs of Kepler-19 b indicated the presence of a nontransiting perturber (Ballard et al. 2011). Precise RVs were collected with HARPS-N (101) in order to conduct a joint TTV-RV analysis (Malavolta et al. 2017), which resulted in a mass measurement of $M_b = 8.4 \pm 1.6\,M_\oplus$ and the detection of two nontransiting planets: Kepler-19 c ($P_c = 28.7$ days, $M_c = 13.1 \pm 2.7\,M_\oplus$) and Kepler-19 d ($P_d = 63$ days, $M_d = 20.3 \pm 3.4\,M_\oplus$). The mass and radius of Kepler-19 b are consistent with a rocky interior overlaid with a substantial volatile envelope, and the masses of the nontransiting planets are also consistent with volatile-enveloped planets (Weiss & Marcy 2014; Rogers 2015).

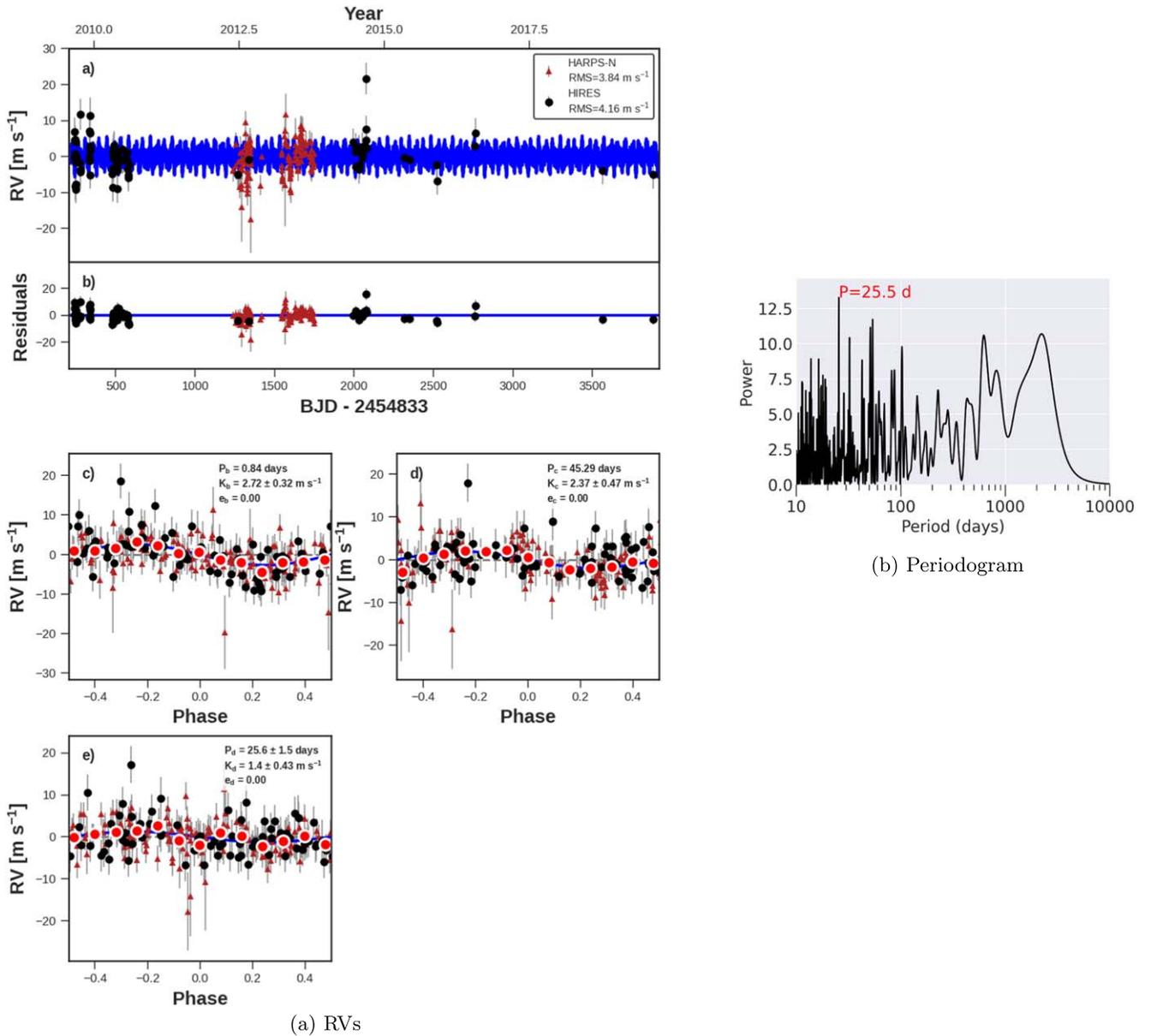

(a) RVs

(b) Periodogram

**Figure 12.** Same as Figure 8, but for KOI-72 (Kepler-10). Left: the RVs were taken at Keck-HIRES (black circles) and TNG-HARPS-N (maroon triangles). A planet candidate is marginally detected at 25 days ($\Delta$BIC = −18).

We collected 24 HIRES RVs between 2009 and 2017, resulting in a baseline of 8 yr (Figure 14). We jointly analyze the HIRES and HARS-N RVs. We recover the planet at 28.54 days (Kepler-19 c, FAP = 0.02, $\Delta$BIC = −24), although we do not detect the planet at 63 days (Kepler-19 d). In the periodogram of the residuals, the highest peak is near 38 days and has FAP = 0.15. We obtain masses of $M_b = 7.5 \pm 1.9\,M_\oplus$ and $M_c = 17.3 \pm 3.0\,M_\oplus$. A joint analysis of all the RVs and TTVs is needed to reconfirm the planet at 63 days. The residual rms of the RVs is 4.7 m s$^{-1}$ with no significant trend, yielding an upper limit of $M\sin i < 0.3 M_J$ at 5 au or $M\sin i < 1.8 M_J$ at 10 au (3$\sigma$ conf.).

### A.8. KOI-85 (Kepler-65)

KOI-85 (Kepler-65) has three transiting planets originally validated in Chaplin et al. (2013), all with $P < 10$ days and $R_p < 4\,R_\oplus$. A nontransiting planet was identified in 7 yr of Keck-HIRES RVs (Mills et al. 2019b). The compact configuration of the three inner planets generates TTVs, and a photodynamical analysis of the Kepler photometry and HIRES RVs yielded precise masses for the four known planets: $M_b = 2.4^{+2.4}_{-1.6}\,M_\oplus$, $M_c = 5.4 \pm 1.7\,M_\oplus$, $M_d = 4.14 \pm 0.80$, and $M_e = 260^{+200}_{-50}\,M_\oplus$ (planet e is nontransiting and decoupled from the inner planets, and so its inclination is not well constrained). We have collected 76 HIRES RVs between 2011 May and 2022 June (including two new RVs since Mills et al. 2019b), providing a baseline of 11 yr (Figure 15). After fitting a four-planet model (including planet e), the residual RVs have an rms = 7.0 m s$^{-1}$ and no apparent trend, yielding a 3$\sigma$ upper limit of $M \sin i < 0.30 M_J$ at 5 au or $M \sin i < 1.2 M_J$ at 10 au.

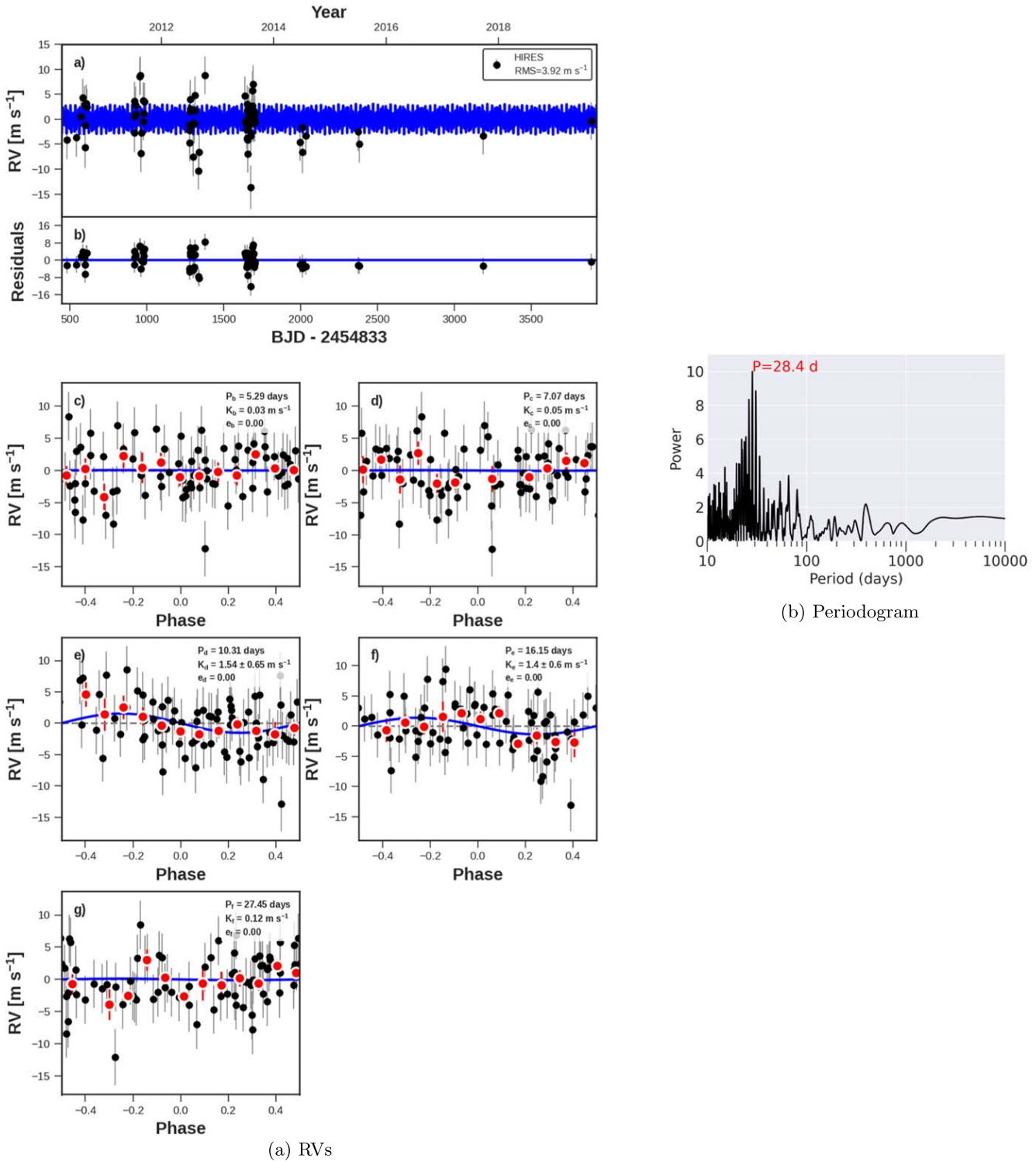

Figure 13. Same as Figure 8, but for KOI-82 (Kepler-102). Left: the RVs were taken at Keck-HIRES (black circles) and TNG-HARPS-N (maroon triangles). No nontransiting companions are detected.

*A.9. KOI-94 (Kepler-89)*

KOI-94 (Kepler-89) is a $V = 12.2$ late F-type star ($T_{eff} = 6000$ K) with $R_\star = 1.20 M_\odot$. It has four transiting planets, including a warm Jupiter (KOI-94 d, $P_d = 22.34$ days, $R_d = 11.3 \pm 1.1 R_\oplus$) surrounded by three sub-Neptunes ($P_b = 3.73$ days, $P_c = 10.4$ days, $P_d = 54.3$ days). Hirano et al. (2012) characterized the photometry of the system and detected a planet–planet occultation, thereby inferring that the line-of-sight projected mutual inclination between KOI-94.01 and KOI-94.03 is $-1°.15 \pm 0°.55$. The system was confirmed with RVs in Weiss et al. (2013), in which the mass of KOI-94 d was measured ($M_d = 106 \pm 11 \, M_\oplus$) and upper limits were set for the smaller transiting planets. A contemporaneous analysis of TTVs in this system yielded a substantially different mass for the warm Jupiter (Masuda et al. 2013), although an erroneously low stellar mass was used in the TTV analysis (1.0 vs. 1.27 $M_\odot$), contributing to the discrepancy.

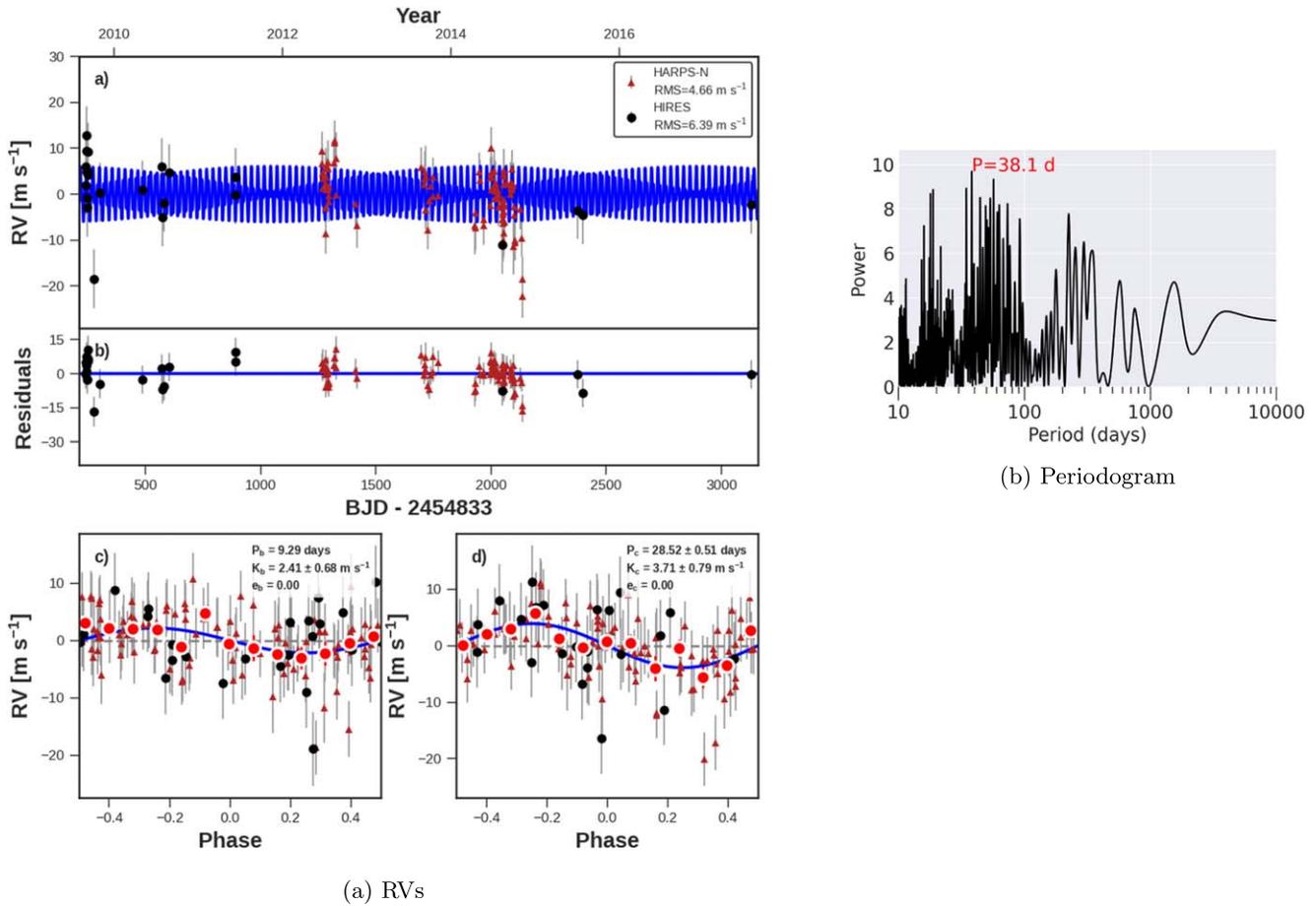

**Figure 14.** Same as Figure 8, but for KOI-84 (Kepler-19). Left: the RVs were taken at Keck-HIRES (black circles) and TNG-HARPS-N (maroon triangles). The best-fit model includes a nontransiting planet, Kepler-19 c, at 28.54 days (orbital parameters inset in phase-folded panel).

Several transits of the Kepler-89 planets (two of planet d, four of planet c, one of planet e, and none of planet b) were identified in TESS photometry, yielding an updated solution to the best fit to the TTVs (Jontof-Hutter et al. 2022). A four-planet model fit the observed TTVs poorly (reduced $\chi^2 = 2.48$), and Jontof-Hutter et al. (2022) proposed two possible five-planet models that better reproduced the TTVs, although still with reduced $\chi^2 > 1.8$. In model A the fifth planet candidate was at 118.0 days and had 7.0 $M_\oplus$, whereas in model B the fifth planet was at 39.99 days and had 0.71 $M_\oplus$.

KOI-94 was selected for follow-up as part of a program to survey stars with at least three transiting planets from 2015 onward. We have continued monitoring KOI-94 with HIRES and have collected 44 new RVs since 2013 and 97 RVs total on 73 unique nights (Figure 16). Note that RVs from 2012 August 10, which were part of an RM observation (Albrecht et al. 2013), have been removed from our analysis. Because the planets have significant TTVs, we tested a model in which we allowed the period of the warm Jupiter (which is strongly detected in the RVs) to vary, but with a prior informed from the TTV analysis of Jontof-Hutter et al. (2022): $P_d = 22.3425 \pm 0.0004$. For this model, the best-fit mass from the RVs is $M_d = 72 \pm 9 \, M_\oplus$, which significantly reduces the tension between the RV solution and the TTV solution. We tentatively detected a moderate eccentricity of the warm Jupiter of $e_d = 0.2 \pm 0.1$, although (1) this eccentricity might be from mixing of near-resonant signals and (2) the introduction of a fifth planet to the model might reduce the eccentricity. We also tested a model with zero eccentricity for the warm Jupiter (which might be more realistic for a planet in a near-resonant chain) and found $M_d = 77 \pm 8 \, M_\oplus$. We prefer the zero-eccentricity model (Figure 16), although both models had very similar results for the masses of the transiting planets. The residual RVs have rms = 8.9 m s$^{-1}$, no significant peak in the Lomb–Scargle periodogram (right panel of Figure 16), and no apparent trend, yielding an upper limit of $M \sin i < 0.90 M_J$ at 5 au ($M \sin i < 3.6 M_J$ at 10 au) for additional companions. These upper limits do not exclude the putative TTV planet candidates identified in Jontof-Hutter et al. (2022). The star's moderate rotation ($v \sin i = 7.5$ km s$^{-1}$) produces relatively large RV errors, which will make an RV detection of the putative fifth planet challenging. Additional RVs collected in 2022 as part of a student project are not presented here and are consistent with the nondetection of long-period giant companions.

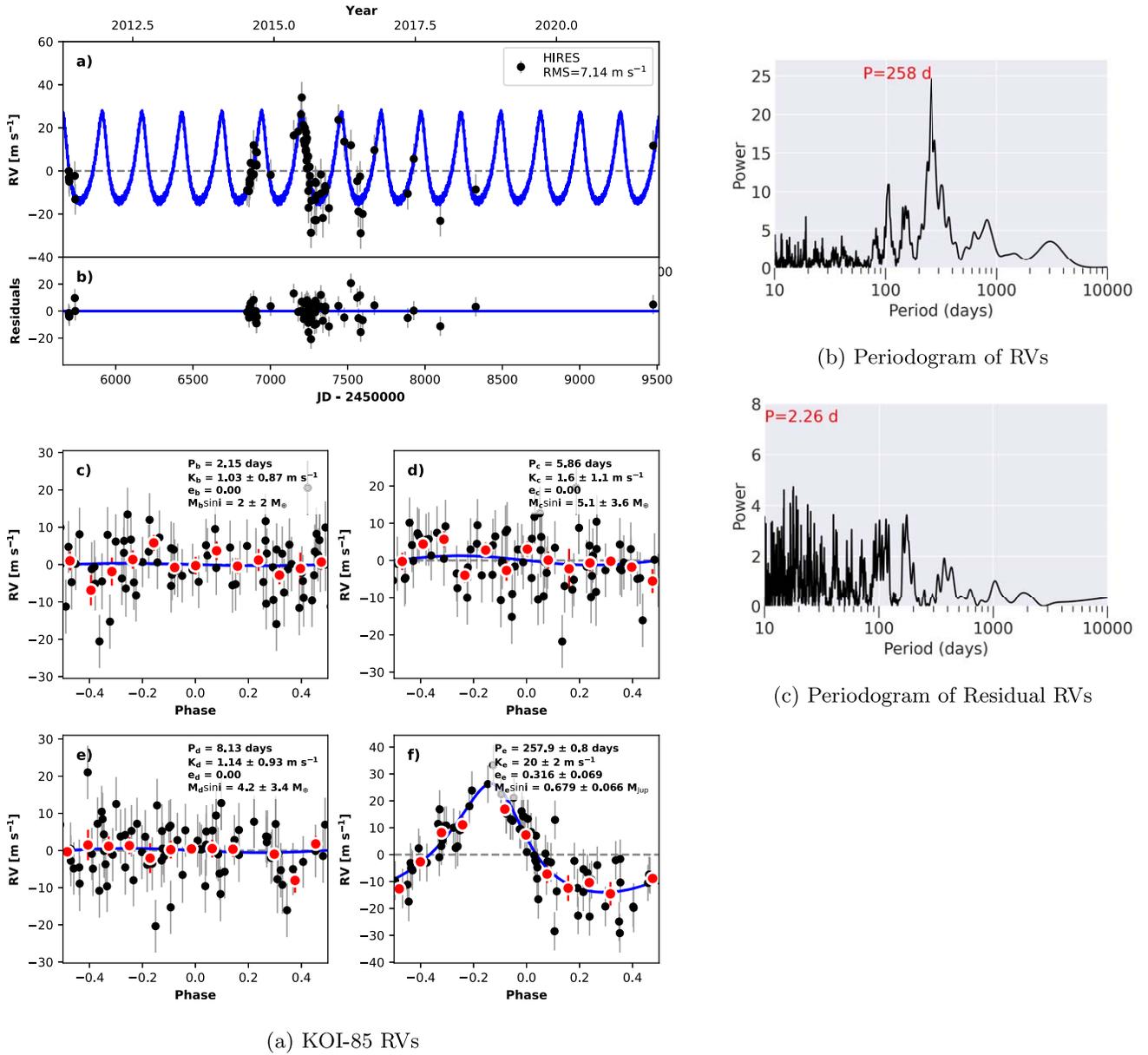

(a) KOI-85 RVs

**Figure 15.** Same as Figure 8, but for KOI-85 (Kepler-65). Left: the best-fit model (blue line) includes a nontransiting planet, Kepler-65 e (orbit details inset in phase-folded panel). The periodograms in the right column are of the RVs (top) and the RVs after the best-fit model has been subtracted (bottom).

### A.10. KOI-103 (Kepler-1710)

Kepler-1710 (KOI-103) is a $V = 12.6$ Sunlike star. It has one transiting planet with $P_b = 14.91$ days and $R_b = 3.20\,R_\oplus$, which was recently reconfirmed with ExoMiner, a machine-learning validation tool (Valizadegan et al. 2022). We have collected 21 HIRES RVs with over a decade of baseline (2009–2021; Figure 17). After fitting for the transiting planet, the residual RVs have an rms of 8.3 m s$^{-1}$ and no trend, yielding a $3\sigma$ upper limit of $M\sin i < 0.3 M_J$ at 5 au ($M\sin i < 1.1 M_J$ at 10 au) for any additional companions. The RV scatter in this system is unexplained in terms of the expected amplitude of the transiting planet and the stellar activity (log $R'_{HK} = -4.77$). There are no stellar companions within $2''$ of the primary identified with either imaging or spectroscopy that could be causing increased RV scatter. Perhaps there are unidentified nontransiting planets that are confounding a mass measurement of the transiting planet.

### A.11. KOI-104 (Kepler-94)

Kepler-94 (KOI-104) is a $V = 12.9$ late K-type or early M-type star with discrepant various temperatures reported in the literature ($T_{\rm eff} = 4200$–4750 K; Muirhead et al. 2012; Yee et al. 2017; Fulton & Petigura 2018; Berger et al. 2018; Brewer & Fischer 2018). The system has one short-period transiting sub-Neptune ($P_b = 2.51$ days, $R_b = 3.5\,R_\oplus$). The transiting planet was confirmed with Keck-HIRES RVs in Marcy et al. (2014) ($M_b = 10.8 \pm 1.4\,M_\oplus$), which also revealed a long-period companion with $P_c = 820 \pm 3$ days and $M\sin i_c = 9.7 M_J \pm 0.6 M_J$. We have collected 10 new RVs since Marcy et al. (2014), which yield refined properties for the long-period object: $P_c = 816.3 \pm 0.7$ days, $M\sin i_c = 8.9 M_J \pm 0.2 M_J$, and $e_c = 0.35 \pm 0.0086$ (Figure 18). Because $M\sin i$ corresponds to a minimum mass, it is still unclear whether the nontransiting object is below or above the deuterium-burning limit of $13 M_J$. The HIRES RVs extend from 2010 to 2022, composing a 12 yr baseline. The residuals to our best two-planet fit (including the nontransiting companion) have rms = 5.4 m s$^{-1}$ and no apparent RV trend, consistent with a $3\sigma$ upper limit of $M \sin i < 0.30 M_J$ at 5 au ($M \sin i < 1.2 M_J$ at 10 au) on any additional companions.

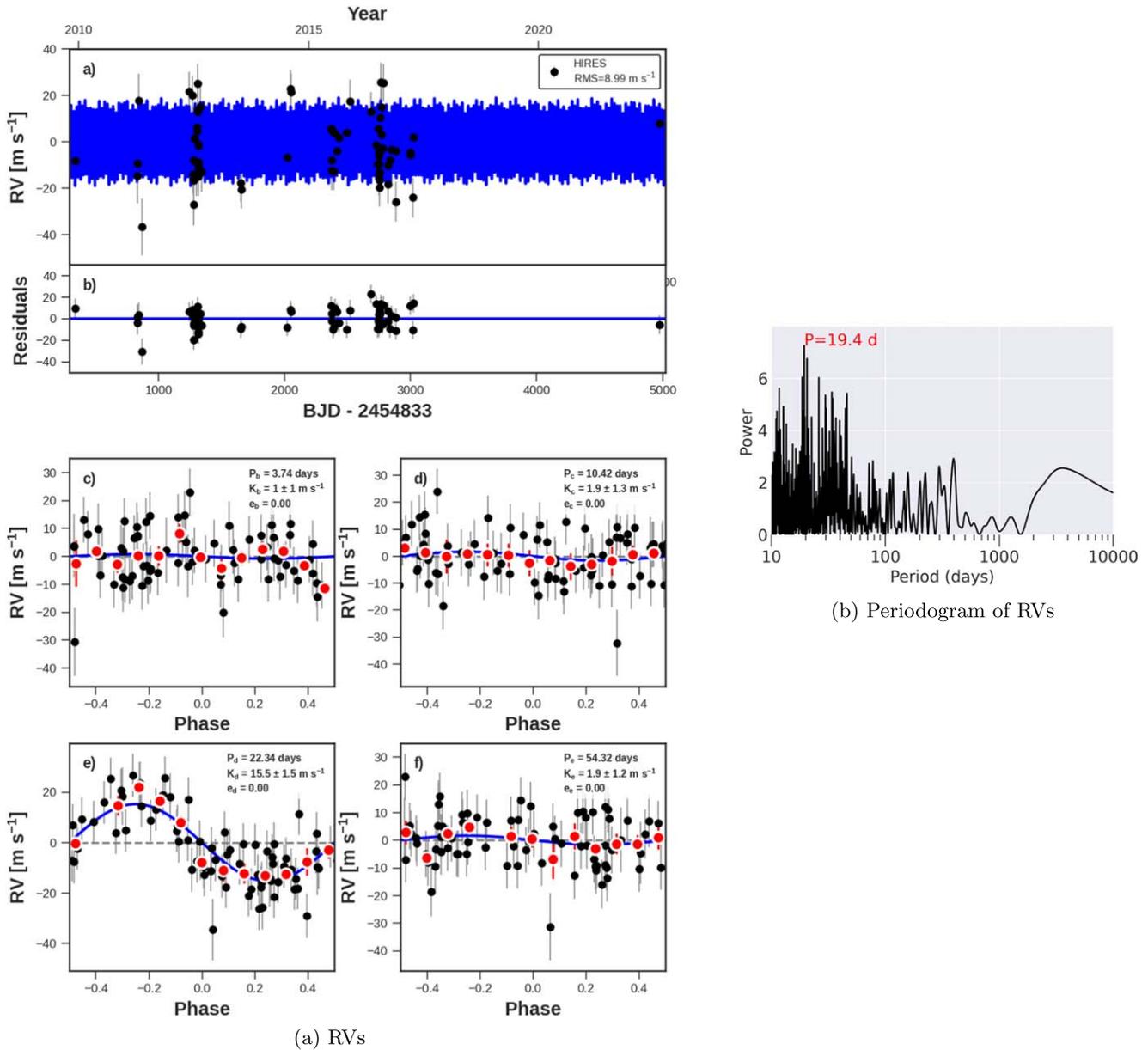

(a) RVs

(b) Periodogram of RVs

**Figure 16.** Same as Figure 8, but for KOI-94 (Kepler-89). No nontransiting companions are detected.

Brewer & Fischer (2018) reported unusual abundances for this star. The star is apparently rich in nitrogen ([N/H] = 0.46, [C/H] = 0.19, [O/H] = 0.10) and very rich in sodium ([Na/H] = 0.70). The star is also rich in iron and nickel ([Fe/H]=0.38, [Ni/H] = 0.41), as well as magnesium and aluminum ([Mg/H] = 0.30, [Al/H] = 0.51). The uncertainties are reported as less than 0.1 dex for all abundances, although the star is a late K/early M type, for which determining abundances can be challenging. It is noteworthy that this star has apparently high metallicities for metals and also has a super-Jovian companion (which is rare for late K-type and early M-type stars). Perhaps the high metal abundances permitted the formation of the planet through core accretion, as suggested based on the correlation between giant planet occurrence and host star metallicity (Fischer & Valenti 2005). A more detailed examination of the host star properties and abundances and their possible relationship to the planet properties is warranted but is beyond the scope of this paper.

### A.12. KOI-108 (Kepler-103)

Kepler-103 (KOI-108) is a Sunlike star with two transiting planets: a sub-Neptune ($R_b = 3.37 R_\oplus$) with $P_b = 16.0$ days, and a sub-Saturn ($R_c = 5.14 R_\oplus$) with $P_c = 180$ days. The planets were confirmed with RVs from Keck-HIRES, with reported best-fit masses of $9.7 \pm 8.6$ and $36 \pm 25$ for planets b and c, respectively (Marcy et al. 2014). Although the planet masses were not detected with high confidence, these values correspond to upper limits on the planet masses that demonstrate that they are planetary (rather than stellar). We have collected six HIRES RVs since 2013, which have not substantially improved the planet mass estimates (Figure 19).

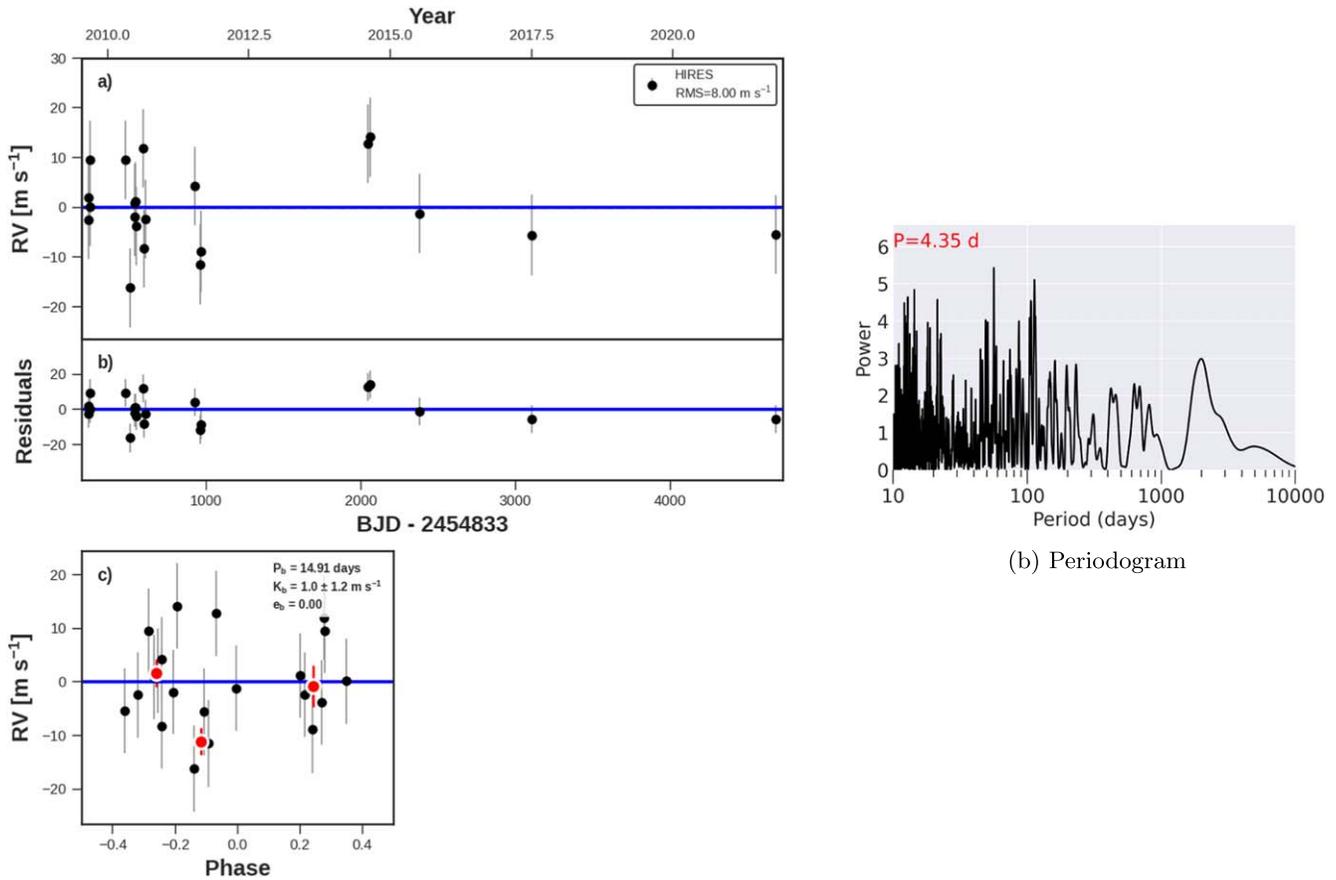

**Figure 17.** Same as Figure 8, but for KOI-103 (Kepler-1710). No nontransiting companions are detected.

The rms of the RV residuals is 6.8 m s$^{-1}$. The 12 yr RV baseline, which extends from 2009 to 2022, has no trend, placing a 3$\sigma$ upper limit of $M \sin i < 0.24 M_J$ at 5 au ($M \sin i < 1.0 M_J$ at 10 au). TTVs have been identified by Gajdoš et al. (2019), but the planet masses were not measured based on the TTVs.

### A.13. KOI-116 (Kepler-106)

Kepler-106 (KOI-116) is a Sunlike star with four small transiting planets, with masses and/or mass upper limits published in Marcy et al. (2014). A periodogram of the RV residuals (rms = 6.0 m s$^{-1}$) reveals peaks near $P = 90$, 180, and 365 days (Figure 20). These peaks could be driven by the seasonal window function, combined with the seasonal downward trend of the RVs in 2012, 2013, and 2014. The star is not particularly active and does not have S-value variations that correspond to these seasonal RV slopes. The RV slopes are likely caused by some combination of a (real) giant planet with an undetermined period and seasonal aliasing. Another possibility is that planets at 90, 180, and/or 365 days would continue the near-resonant chain of the inner four transiting planets. If the planet candidate at 90 days is real (FAP = 0.003, $\Delta$BIC = −43), it produces an RV semi-amplitude of 7.5 m s$^{-1}$ and has $M \sin i = 50 M_\oplus$. Continued monitoring at a variety of cadences, including RVs early and late in the Kepler season, will help resolve what additional planet(s) exist(s). TTVs have been identified by Gajdoš et al. (2019), but the planet masses were not measured. Although the mass, period, and veracity of planets with $P < 1$ yr are uncertain, the RVs have no significant long-term trend, yielding a 3$\sigma$ upper limit of $M \sin i < 0.25 M_J$ at 5 au ($M \sin i < 1.0 M_J$ at 10 au). This mass upper limit does not change with the addition of a fifth planet at 90 days.

### A.14. KOI-122 (Kepler-95)

Kepler-95 (KOI-122) has one transiting planet at $P_b = 11.5$ days and $R_b = 3.4 R_\oplus$. The planet was confirmed with Keck-HIRES RVs in Marcy et al. (2014), which yielded a mass of $M_b = 13.0 \pm 2.9 M_\oplus$. The star is Sunlike, but with high metallicity ([Fe/H] = 0.3). We have collected three new RVs since 2013, which have not substantially improved the mass of the transiting planet (Figure 21). The RV baseline extends from 2009 to 2017 with no trend, yielding a 3$\sigma$ upper limit of $M \sin i < 0.30 M_J$ at 5 au ($M \sin i < 1.2 M_J$ at 10 au) for additional planets in the system.

### A.15. KOI-123 (Kepler-109)

Kepler-109 (KOI-123) is a Sunlike star with two transiting planets at $P_b = 6.48$ and $P_c = 21.2$ days, both of which are smaller than Neptune ($R_b = 2.37 R_\oplus$, $R_c = 2.52 R_\oplus$). The planets were confirmed with Keck-HIRES RVs in Marcy et al. (2014), which established mass upper limits for the planets. Since 2013, we have collected 24 new RVs, resulting in 38 total RVs collected between 2009 and 2021 (an 11 yr baseline). Our best two-planet fit yields mass upper limits of $M_b = 5.0 \pm 4.4 M_\oplus$ and $M_c = 5.5 \pm 3.6 M_\oplus$, consistent with both planets being low-density sub-Neptunes with extended gas envelopes (Figure 22). The RV residuals have rms = 6.4 m s$^{-1}$ and no apparent trend, yielding a $3\sigma$ upper limit of $M \sin i < 0.16 M_J$ at 5 au ($M \sin i < 0.65 M_J$ at 10 au) for additional planets in the system.

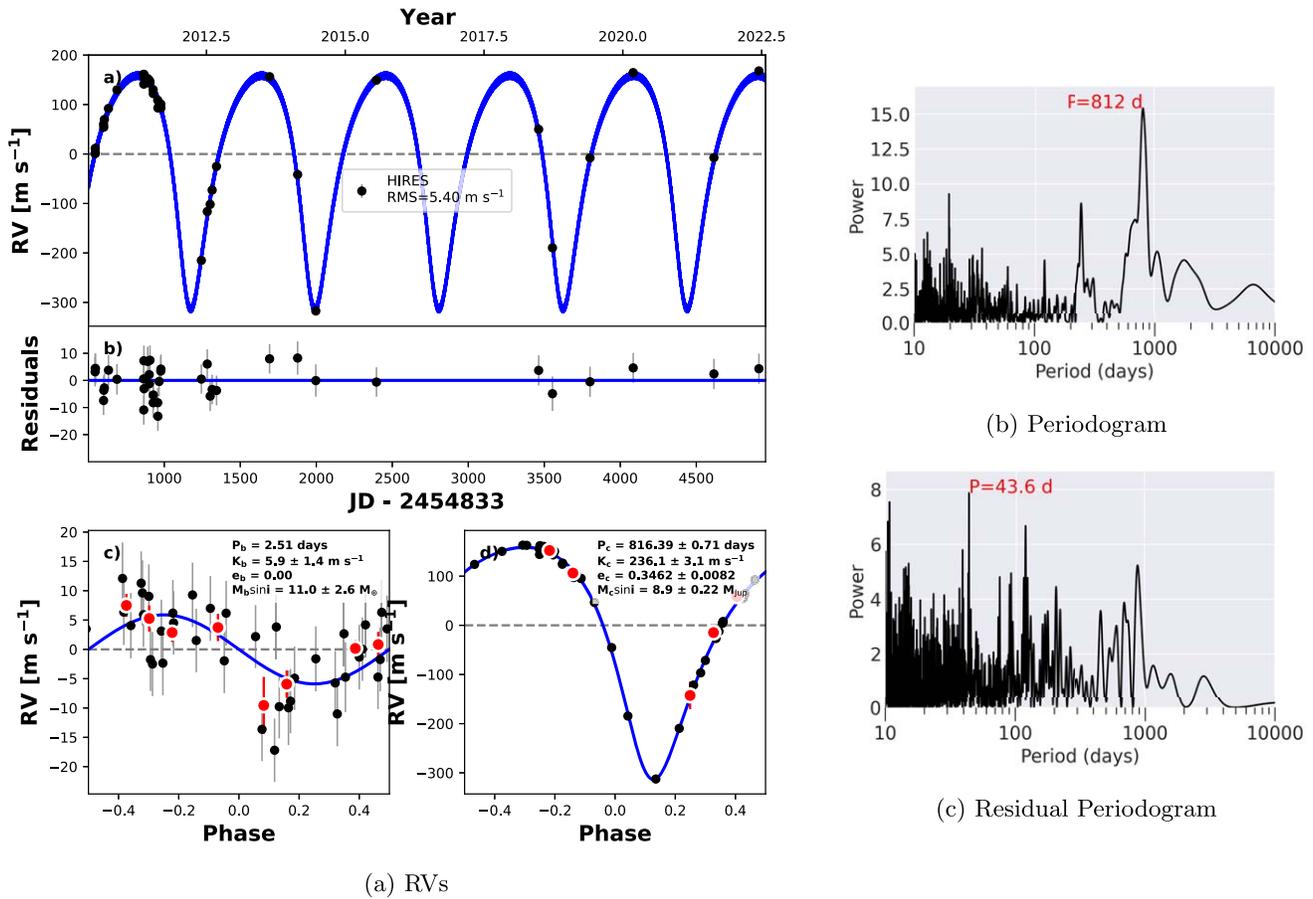

**Figure 18.** Same as Figure 8, but for KOI-104 (Kepler-94). Left: the best-fit model includes a nontransiting planet, Kepler-94 c (orbital parameters inset in phase-folded panel). Right: the periodogram of the RVs after removing the signals from the transiting planet (top) and both planets (bottom).

### A.16. KOI-137 (Kepler-18)

Kepler-18 (KOI-137) is a Sunlike star with three transiting planets interior to 15 days ($P_b = 3.50$ days, $P_c = 7.64$ days, $P_d = 14.86$ days) that were first announced and confirmed in Cochran et al. (2011). The innermost planet, Kepler-18 b, is notably smaller than the other two planets ($R_b = 2.0 R_\oplus$, $R_c = 5.49 R_\oplus$, $R_d = 6.98 R_\oplus$). The planets are near (but not in) a 4:2:1 Laplace resonant chain, and planets c and d have significant TTVs as the result of their dynamical interactions. Cochran et al. (2011) also collected RVs with Keck-HIRES to determine a Doppler mass for the planets, in comparison to (and in combination with) the TTV-determined masses. Their best-fit masses using TTVs and RVs were $M_b = 6.9 \pm 3.4 M_\oplus$, $M_c = 17.3 \pm 1.9 M_\oplus$, and $M_d = 16.4 \pm 1.4 M_\oplus$.

The solutions in Cochran et al. (2011) were based on Kepler photometry from quarters 0 to 6, whereas a more recent analysis by Hadden & Lithwick (2017) was based on the full Kepler time series (quarters 0–17). One challenge of TTVs for near-resonant systems is that the determination of mass and eccentricity can be degenerate (Lithwick et al. 2012). In Kepler-18, the mass of planet c is sensitive to whether a high-mass (and low-eccentricity) prior was adopted, with the solution for the default prior yielding $M_c = 12.9 \pm 6.0 M_\oplus$ and the high-mas prior yielding $M_c = 21.6 \pm 3.6 M_\oplus$, whereas the mass of planet d did not sensitively depend on the prior ($M_d = 16.2 \pm 1.4 M_\oplus$; Hadden & Lithwick 2017). Planet b has flat TTVs, and the reanalysis in Hadden & Lithwick (2017) did not update the mass of this planet.

Since 2011, we have collected 11 new RVs of Kepler-18. Our analysis of the RVs alone (without incorporating TTVs) yields $M_b = 12.9 \pm 4.1 M_\oplus$, $M_c = 19.5 \pm 5.1 M_\oplus$, and $M_d = 24.0 \pm 6.5 M_\oplus$, consistent with gas-enveloped compositions for all three planets (Figure 23). There is no apparent RV trend over the 9 yr baseline (2009–2018), yielding a $3\sigma$ upper limit of $M \sin i < 0.40 M_J$ at 5 au ($M \sin i < 1.6 M_J$ at 10 au). This system would benefit from a joint analysis of the new RVs and TTVs to improve the masses and eccentricities of the planets.

### A.17. KOI-142 (Kepler-88)

KOI-142 (Kepler-88) is a system with one transiting planet ($P_b = 10.95$ days, $R_b = 3.43, R_\oplus$) that has TTVs with semi-amplitude of half a day (0.05 of the orbital period). The system was first noted in Steffen et al. (2013) and was colloquially dubbed the "King of TTVs" on account of the large TTVs of planet b. Nesvorný et al. (2013) conducted an in-depth dynamical analysis of the system, finding a unique solution for the mass and period of a giant planet perturbing the transiting planet from just outside the 2:1 mean motion resonance. The giant planet was confirmed with RVs collected at SOPHIE (Barros et al. 2014). This system is unique in that it is the only Kepler system we followed up for which TTVs clearly indicated an outer perturber. However, the period of the perturber was well-known (22 days). The TTVs gave no prior knowledge about whether this system would harbor a giant planet beyond 1 au.

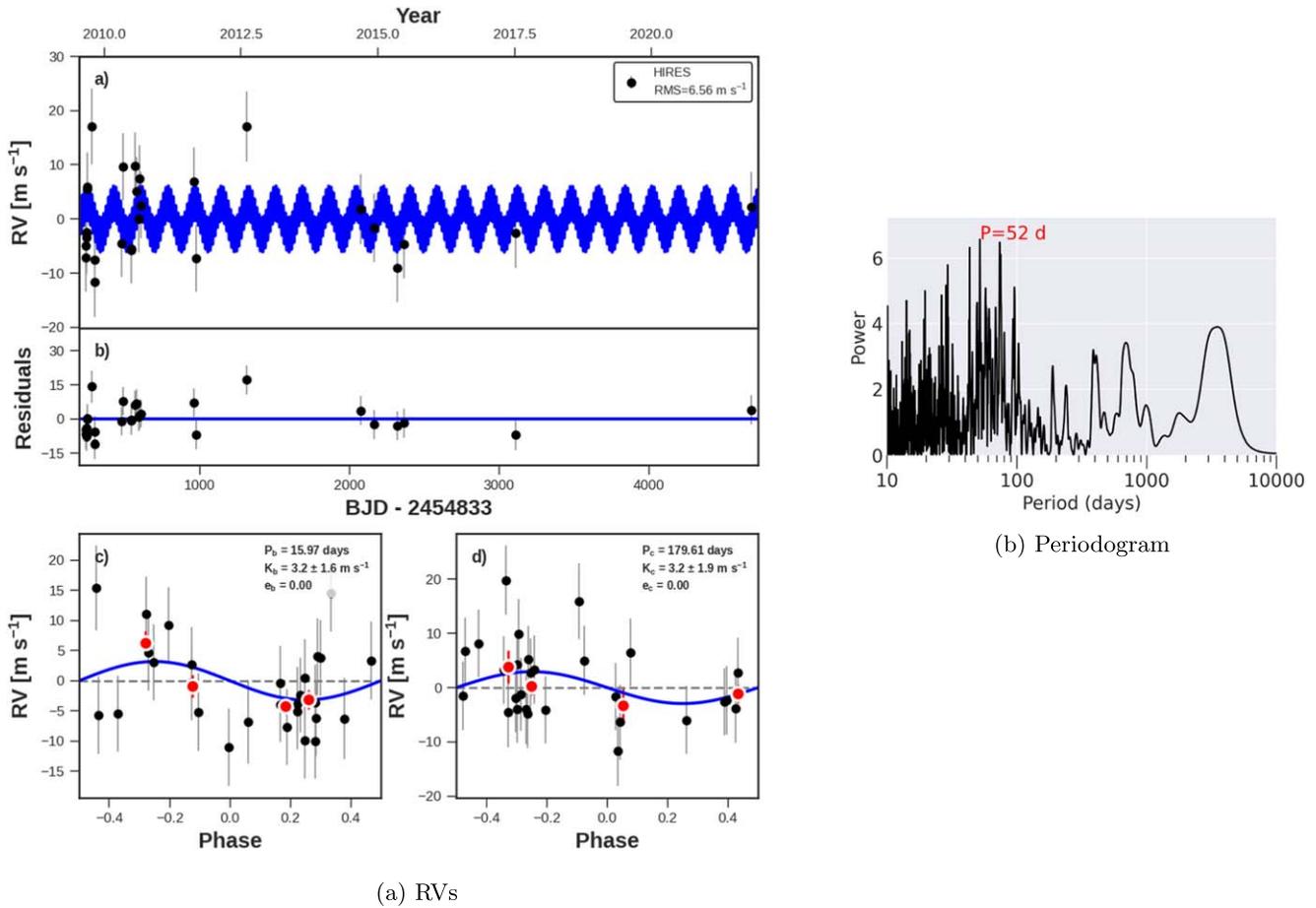

(a) RVs

(b) Periodogram

**Figure 19.** Same as Figure 8, but for KOI-108 (Kepler-103). No nontransiting companions are detected.

A total of 8 yr of Keck-HIRES RVs were also collected and analyzed in a joint RV and photodynamical model in Weiss et al. (2020), which confirmed and refined the mass of the giant planet in resonance and also revealed a distant Jovian-mass companion at 2.6 au. Although the host star is Sunlike, it has high metallicity ([Fe/H] = 0.3). We have collected two additional Keck-HIRES RVs since 2020, which are consistent with the solution in Weiss et al. (2020; see Figure 24). Because our baseline is comparable to the longest detected orbit in the system, any putative trend in the residuals to our best three-planet fit is not well constrained, and so our upper limit on additional companions is only $M \sin i < 0.67 M_J$ at 5 au ($M \sin i < 2.7 M_J$ at 10 au). Continued low-cadence monitoring will improve the ephemeris determination of the distant Jovian planet and might yield additional long-period planets.

### A.18. KOI-148 (Kepler-48)

KOI-148 (Kepler-48) has three transiting planets with dynamical interactions that produce TTVs, which confirmed their planetary nature (Steffen et al. 2012). Marcy et al. (2014) presented 28 RVs of the system collected with HIRES, yielding masses for the transiting planets and revealing the existence of one nontransiting giant planet (Marcy et al. 2014). There are four stellar companions within 6″ of the primary summarized in Marcy et al. (2014). For the RV measurements, care was taken to align the field to avoid secondary light into the slit. The primary star is Sunlike.

We have obtained 28 new RVs with HIRES since Marcy et al. (2014), resulting in 56 RVs total, which we used to improve the characterization of the masses and orbits of the planets (Figure 25). We detected a new nontransiting planet, Kepler-48 f, based on our long-term monitoring, with high confidence (FAP = $2 \times 10^{-8}$, $\Delta$BIC = $-530$). This planet's orbital period ($P_f = 5400$ days) is distinct from the peak periodicity of the Mount Wilson S-values ($\sim$2500 days). In order of increasing orbital distance, the planets have periods $P_b = 4.78$ days, $P_c = 9.67$ days, $P_d = 42.9$ days, $P_e = 1001$ days, and $P_f = 5205$ days and minimum masses of 6.1 $M_\oplus$, 11.6 $M_\oplus$, 8 $M_\oplus$, 2.1 $M_J$, and 1.3 $M_J$. We assume that the three transiting planets are circular, and we find upper limits on the eccentricities of the nontransiting planets of $e_e < 0.09$ and $e_f < 0.36$ (3$\sigma$ confidence).

Because our baseline is comparable to the longest detected orbit in the system, any putative trend in the residuals to our best five-planet fit is not well constrained, and so our upper limit on additional companions is only $M \sin i < 0.50 M_J$ at 5 au ($M \sin i < 2.0 M_J$ at 10 au). Future RVs will improve the characterization of the period, eccentricity, and minimum mass of the long-period companion and will also place better constraints on the presence of additional companions.

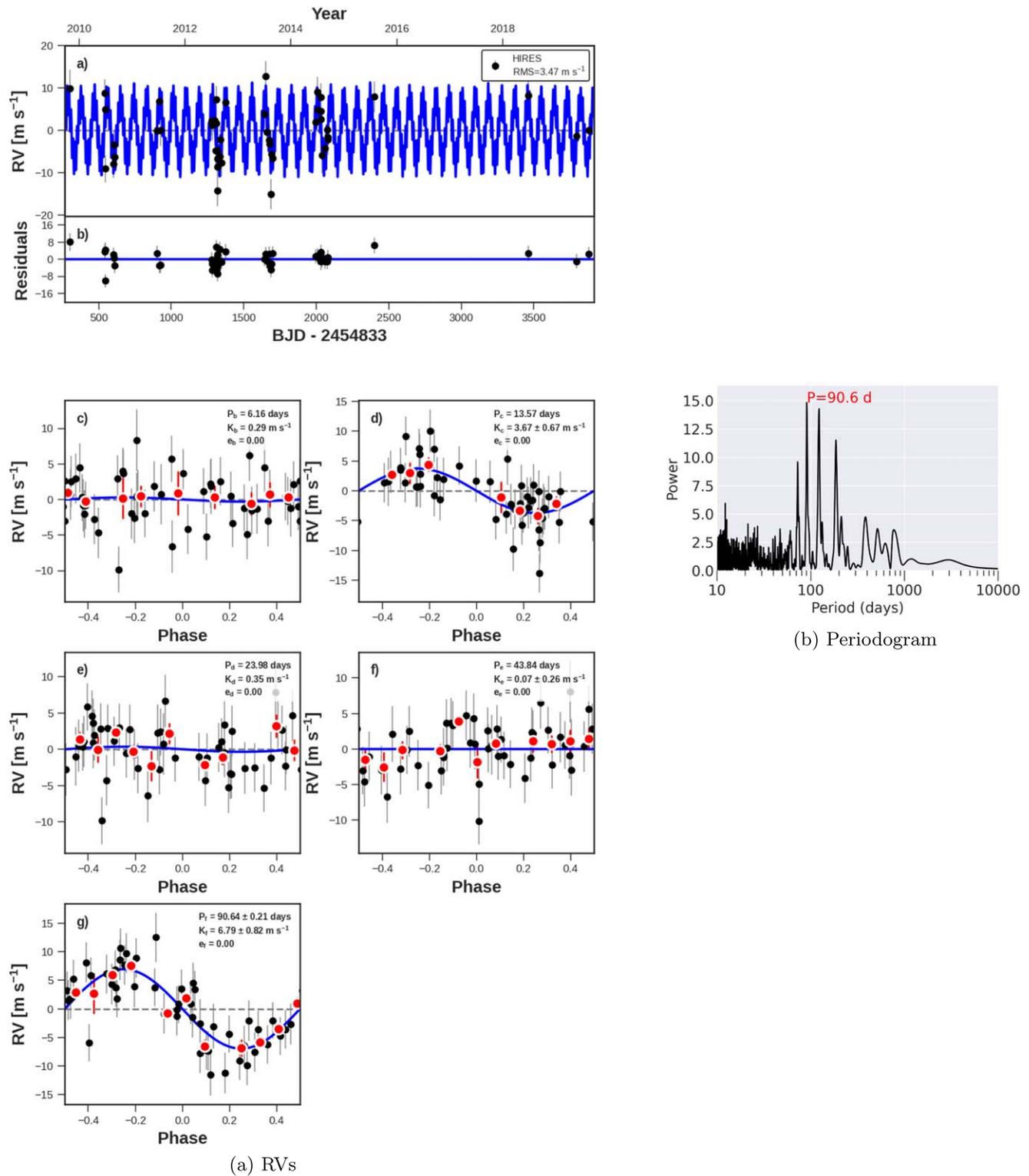

(a) RVs

(b) Periodogram

**Figure 20.** Same as Figure 8, but for KOI-116 (Kepler-106). The best-fit model includes a nontransiting planet, although the period of that planet is uncertain owing to aliasing (right).

### A.19. KOI-153 (Kepler-113)

Kepler-113 (KOI-153) is a $V = 13.7$ K-type star ($T_{\rm eff} = 4791$ K; Fulton & Petigura 2018). The system has two transiting sub-Neptunes ($R_b = 1.82 \pm 0.6\, R_\oplus$, $R_c = 2.18 \pm 0.06\, R_\oplus$) with periods of $P_b = 4.75$ days and $P_c = 8.93$ days. The planets were confirmed with Keck-HIRES RVs in Marcy et al. (2014), which yielded a mass upper limit for the outer planet and a 2.7$\sigma$ measurement for the inner planet of $11.7 \pm 4.2\, M_\oplus$. With 18 new HIRES RVs since 2013 (42 RVs total), we have refined the planet masses, finding $M_b = 7.0 \pm 3.3\, M_\oplus$ and $M_c = 2.5 \pm 2.5\, M_\oplus$ (Figure 26). Planet b, which is near the distinction between planets with rocky surfaces and planets with volatile envelopes (Weiss & Marcy 2014; Rogers 2015; Fulton et al. 2017), is consistent with having a predominantly rocky (by volume) composition, with a possible thin gas envelope, whereas planet c is consistent with having a larger gas envelope more typical of sub-Neptunes. The residual RVs (rms = 8.2 m s$^{-1}$) have structure at 55 and >500 days, although the false-positive probability (0.16) is too high to consider these signals as planet candidates. The possibility of an additional planet in the system was also noted in Marcy et al. (2014), but at periods of 1.065, 16, 0.984, and 0.515 days, none of which have become stronger signals with more data. Additional RVs will help determine whether one or both of the new candidate periods are associated with bona fide planets. There are no strong peaks in the periodogram of the RV residuals. There is no apparent trend in the RV residuals time series to our two-planet fit, yielding a 3$\sigma$ upper limit of $M \sin i < 0.31 M_J$ at 5 au ($M \sin i < 1.3 M_J$ at 10 au).

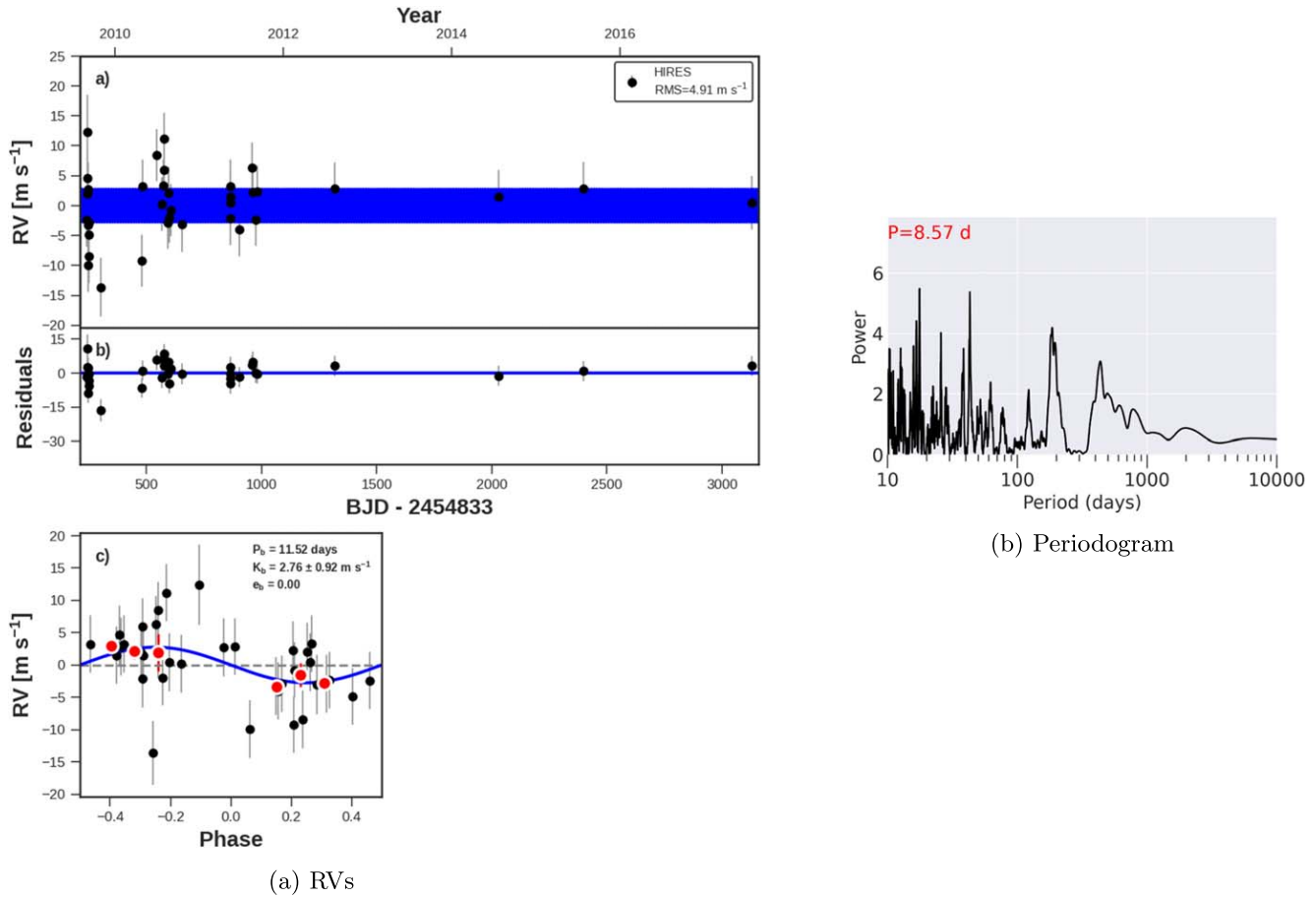

(a) RVs

(b) Periodogram

**Figure 21.** Same as Figure 8, but for KOI-122 (Kepler-95). No nontransiting companions are detected.

### A.20. KOI-157 (Kepler-11)

Kepler-11 (KOI-157) is a faint Kepler star ($V = 14$) with six transiting planets that has been extensively studied as an exemplary multiplanet system with significant TTVs (Lissauer et al. 2011). The planets range in size from 1.8 to 4.2 $R_\oplus$ and have ultralow densities, based on their TTVs (Lissauer et al. 2013), and the host star is a solar twin (Bedell et al. 2017). A subset of the RVs presented here were used in a joint RV and TTV analysis, the results of which were that (i) the RVs alone ruled out rocky compositions for the planets, (ii) the combined RV-TTV analysis was consistent with low densities for the planets, and (iii) the RVs reduced the mass upper limit of planet g, which is dynamically decoupled from the inner five planets and is therefore not well characterized by their TTVs (Weiss 2016). Between 2010 and 2022, we collected 31 RVs (Figure 27). The RVs provide mass upper limits for the transiting planets, as described in Weiss (2016). The residual RVs to our best six-planet fit have rms = 7.3 m s$^{-1}$ and no apparent trend, yielding an upper limit of $M \sin i < 0.6 M_J$ at 5 au ($M \sin i < 1.6 M_J$ at 10 au, 3$\sigma$ conf.)

### A.21. KOI-244 (Kepler-25)

Kepler-25 (KOI-244) is a Sunlike star with two transiting planets near the 2:1 mean motion resonance. Dynamical interactions between the transiting planets produce TTVs, which confirm the planetary nature of the system (Steffen et al. 2012). Marcy et al. (2014) collected 62 RVs of this system with Keck-HIRES. The RVs also revealed a nontransiting planet at $P_d = 91$ or 122 days. Hadden & Lithwick (2014) performed N-body forward modeling to fit the TTVs of the transiting planets and found upper limits on the planet masses. A photodynamical analysis that jointly fit the Kepler photometry and the HIRES RVs (including five new RVs) revealed a reduction in the planet mass–eccentricity degeneracy when using photometry alone (Mills et al. 2019b). The results are consistent with a low-eccentricity resonant state but do not indicate that the planets are definitively librating within the resonance. Two complete transits were observed with Keck-HIRES Doppler spectroscopy to measure the RM effect, finding the orbital plane of the planets and the axis of stellar rotation to be well aligned (Albrecht et al. 2013). The detection of the planets transiting in NASA TESS photometry was leveraged to update their orbital ephemerides, improving the precision of the linear ephemerides by an order of magnitude (Battley et al. 2021). Jontof-Hutter et al. (2022) combine legacy Kepler photometry with new TESS photometry to determine that the TTV solution does not change with the inclusion of TESS data for this system.

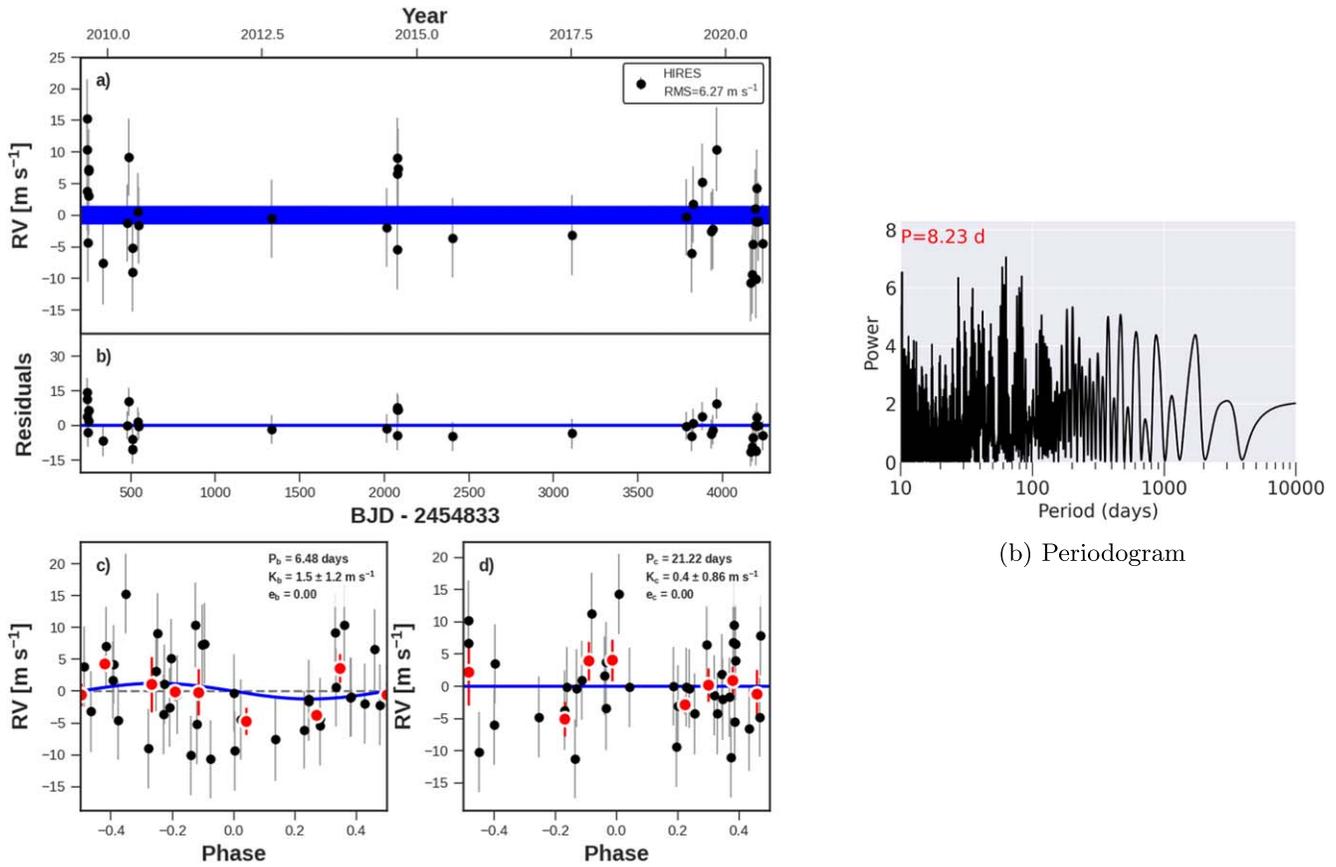

(a) RVs

(b) Periodogram

**Figure 22.** Same as Figure 8, but for KOI-123 (Kepler-109). No nontransiting companions are detected.

We have collected two new RVs since 2019, bringing the total number of non-RM RVs to 69 (Figure 28). The new RVs and updated ephemerides have not substantially changed the characterization of any of the known planets. However, the nondetection of an RV trend in the residuals to a three-planet model (including planet d at either 91 or 122 days) over the baseline of 2011–2022 yields a $3\sigma$ upper limit on additional long-period companions of $M \sin i < 0.70 M_J$ at 10 au.

### A.22. KOI-245 (Kepler-37)

Kepler-37 (KOI-245) is a $V = 9.7$ G-type ($T_{\rm eff} = 5300$ K) star. The system has three transiting sub-Neptunes ($P_b = 13.4$ days, $P_c = 21.3$ days, $P_d = 39.8$ days), one of which is smaller than Mercury ($R_b = 0.30 R_\oplus$), that were identified and confirmed in Barclay et al. (2013). The star has solar-like oscillations, which enabled a precise determination of the physical stellar properties through the combination of astero-oseismology and spectroscopy. Marcy et al. (2014) used Keck-HIRES RVs to place upper limits on the masses of transiting planets b, c, and d. A fourth transiting candidate at $P = 51$ days was identified in the photometry but was suspected to be due to noise or an instrumental false positive (Barclay et al. 2013). Nonetheless, the fourth candidate has been considered in various TTV analyses (Hadden & Lithwick 2014; Holczer et al. 2016). Most recently, Rajpaul et al. (2021) further examined Kepler photometry data, concluding that Kepler-37 e, with a 51 day period, should be stripped of its confirmed/validated status, citing the extremely weak photometric evidence. Assuming a circular orbit near 51 days, we find marginal evidence for an RV signal at $P_e = 50.25 \pm 0.15$ days, with $K_e = 1.6 \pm 0.3$ m s$^{-1}$ and $M_e = 8.4 \pm 1.7 M_\oplus$. We consider models both with and without a transiting planet candidate at this period.

Since the publication of Marcy et al. (2014), we have acquired 48 new RVs on Keck-HIRES, resulting in 81 RVs from HIRES and 218 RVs total (including literature RVs from HARPS-N; Figure 29). We applied models with either three or four small planets (to account for the candidate at 51 days) to the combined HIRES and HARPS-N RVs. In both cases, there is significant structure in the residual RVs, with peaks near 460 and 2000 days. When either the 460 day peak or 2000 day peak is fit, the other peak disappears, and so one of these peaks is an alias.

We initially adopted the signal at 2000 days to test as a new planet candidate in the system. The amplitude of this signal is low ($K = 3.3$ m s$^{-1}$), and the fit is quite eccentric ($e = 0.5$). The effective temperature of this star (6265 K) places it very near the Kraft break, and the slightly evolved stellar surface gravity (log $g = 4.31$) means that this star is unlikely to be heavily influenced by chromospheric activity. However, there appears to be modest correlation between the Mount Wilson $S$-values and RVs, with the $S$-value periodogram peaking at approximately 2000 days (Figure 29). Further RV, stellar activity, and photometric monitoring, as well as a joint fit to the RVs and TTVs, is needed to ascertain whether the long-period signal is planetary in nature. For the purposes of this work, we do not consider the signal a detected planet since its provenance is questionable. Whether or not the long-period signal is modeled with a Keplerian and removed, the RV residuals yield upper limits of $M \sin i < 0.22 M_J$ at 10 au with $3\sigma$ conf.

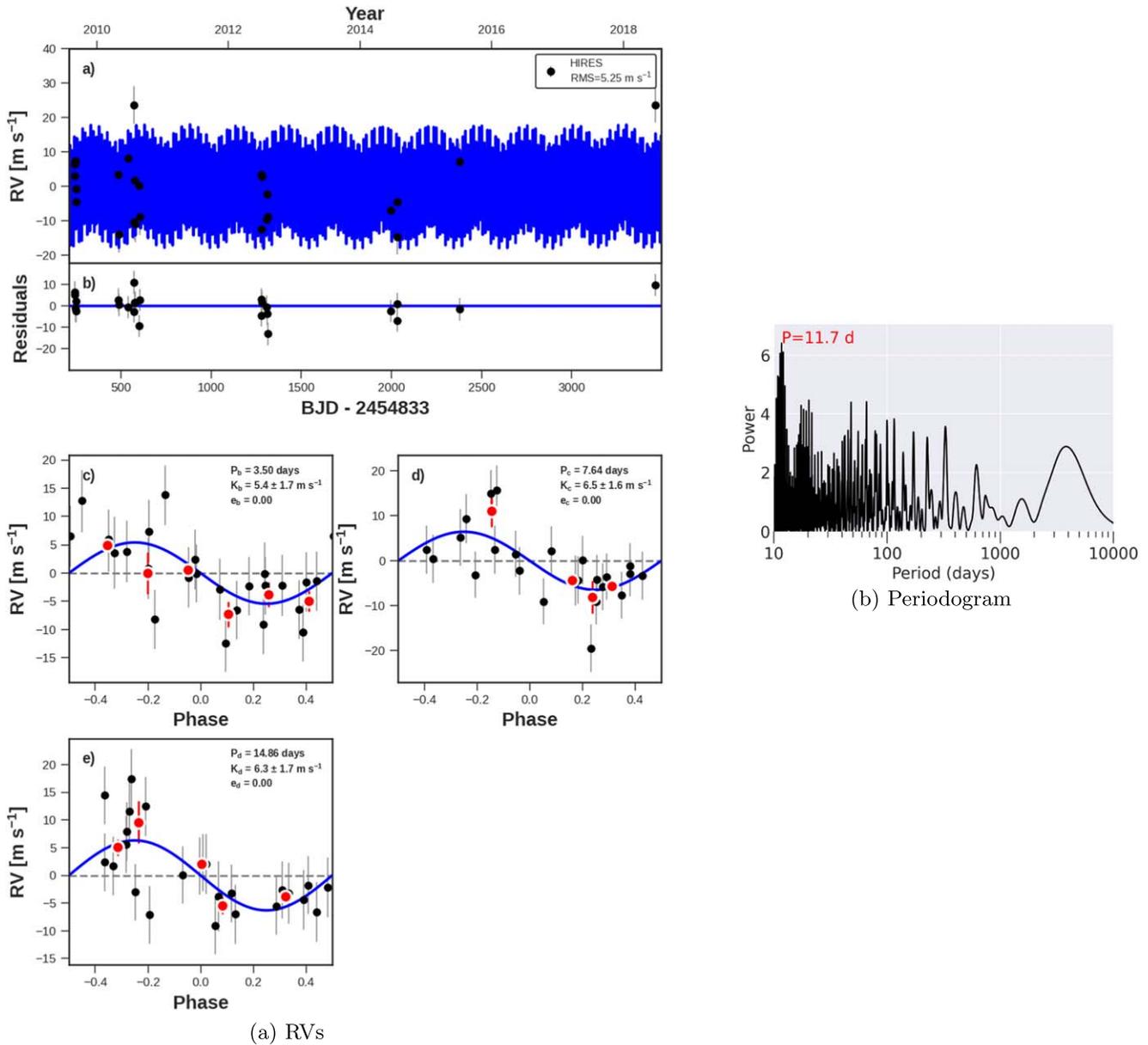

(a) RVs

(b) Periodogram

Figure 23. Same as Figure 8, but for KOI-137 (Kepler-18). No nontransiting companions are detected.

### A.23. KOI-246 (Kepler-68)

Kepler-68 (KOI-246; Figure 30) is a system with a sub-Neptune-sized and an Earth-sized transiting planet. The transiting planets were validated via photometric techniques, including odd–even tests and color diagnostics, and dynamically confirmed using RVs in Gilliland et al. (2013). The stellar mass and radius were determined via asteroseismology. The RVs also revealed a nontransiting planet, Kepler-68 d, with a best-fit period of 581 days. Additional RVs were obtained and published in Marcy et al. (2014), resulting in an update of the period of the nontransiting planet to 625 days and modestly refined transiting planet masses. Another update in Mills et al. (2019b) again revised the period of the nontransiting planet to 634 days and improved the masses of the transiting planets via dynamical analysis.

With several additional seasons of RVs, we confirm the properties of planet d: $P_d = 633$ days, $M \sin i_d = 0.765 M_J \pm 0.045 M_J$, $e_d = 0.19 \pm 0.05$ (Figure 30). We also announce a nontransiting planet, Kepler-68 e. The peak of our Lomb–Scargle periodogram is at 362 days, which is likely a yearly alias of the (presumed) long period of planet e. The Mount Wilson $S$-values also have a peak at 362 days, which is likely due to the window function. We obtained a new RV template observation and reran our barycentric correction analysis for this target, but the peak at 362 days persisted even after our new reduction and RV determinations. The best-fit model includes planets at both 362 and 633 days with moderate eccentricities that would result in orbit crossing, and so this configuration is unlikely. Our interpretation is that the peak at 362 days is the alias of a planet, Kepler-68 e, at a period of $P_e \approx 4400$ days with $M \sin i_e \approx 0.4 M_J$ and $e_e \approx 0.4$. Adopting these properties in our four-planet model removes the 362 day peak from the Lomb–Scargle periodogram of the residuals. The mass upper limit of an additional (third) giant planet is not well constrained because of the uncertainty in the orbit of planet e, yielding a 3σ upper limit of $M \sin i < 0.11 M_J$ at 5 au ($0.43 M_J$ at 10 au). If we ignore the second giant planet and instead consider the residuals to a three-planet model, the upper limits on an additional companion increase by 50%.

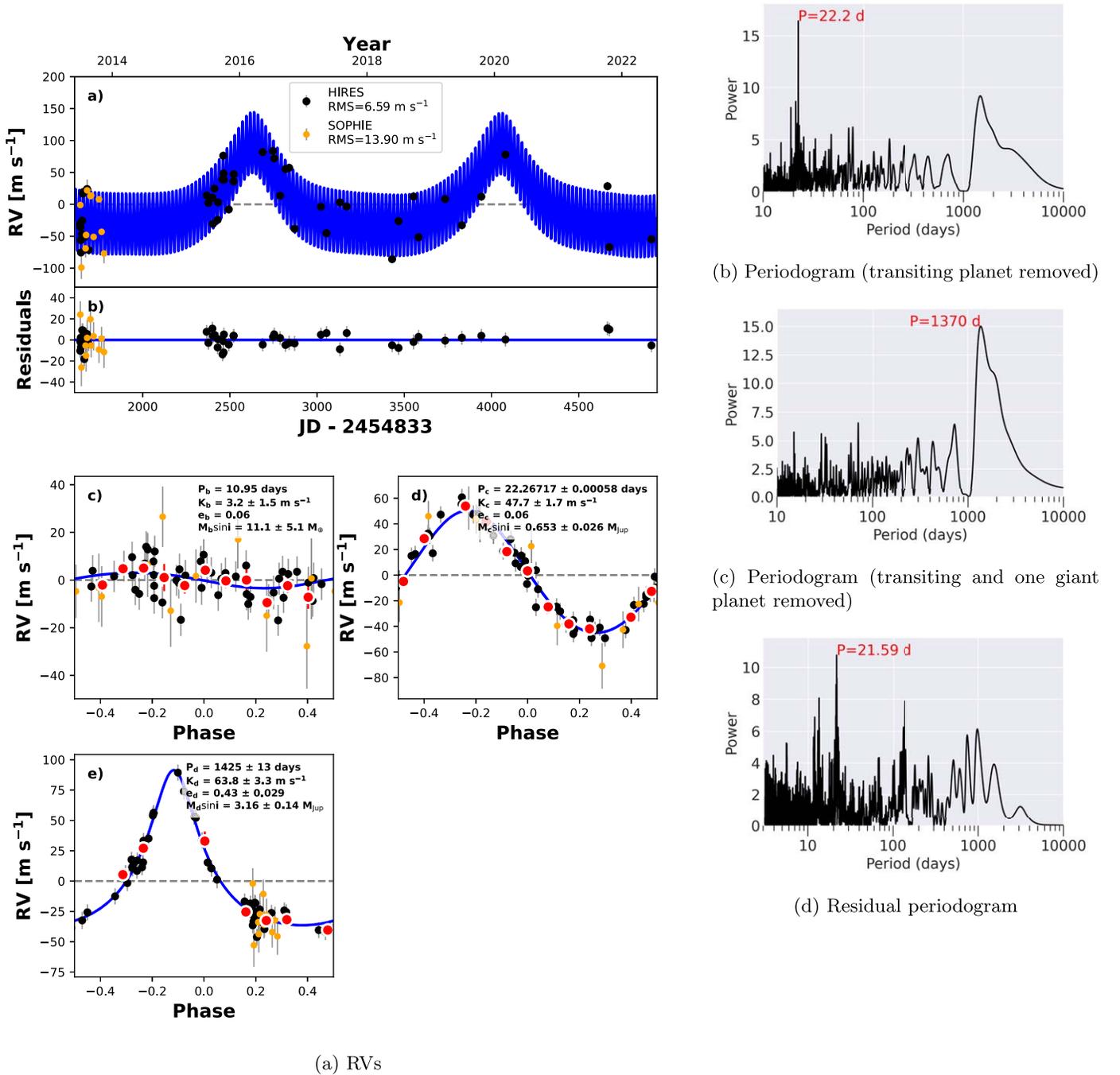

(a) RVs

**Figure 24.** Same as Figure 8, but for KOI-142. Left: RVs are from HIRES (black circles) and SOPHIE (yellow circles). The best-fit model has two nontransiting giant planets (details inset in RV panels, left).

### A.24. KOI-260 (Kepler-126)

Kepler-126 (KOI-260) is a $V = 10.5$ star near the Kraft break ($T_{\rm eff} = 6200$ K). The system has three transiting planets at $P_b = 10.5$, $P_c = 21.9$, and $P_d = 100$ days that were statistically validated (Lissauer et al. 2014; Rowe et al. 2014). All three planets have significant TTVs that have been used to set upper limits on their eccentricities (Van Eylen & Albrecht 2015). Planet d appears to be dynamically decoupled from the inner two planets, and so its TTVs are likely caused by at least one planet that has not been detected yet.

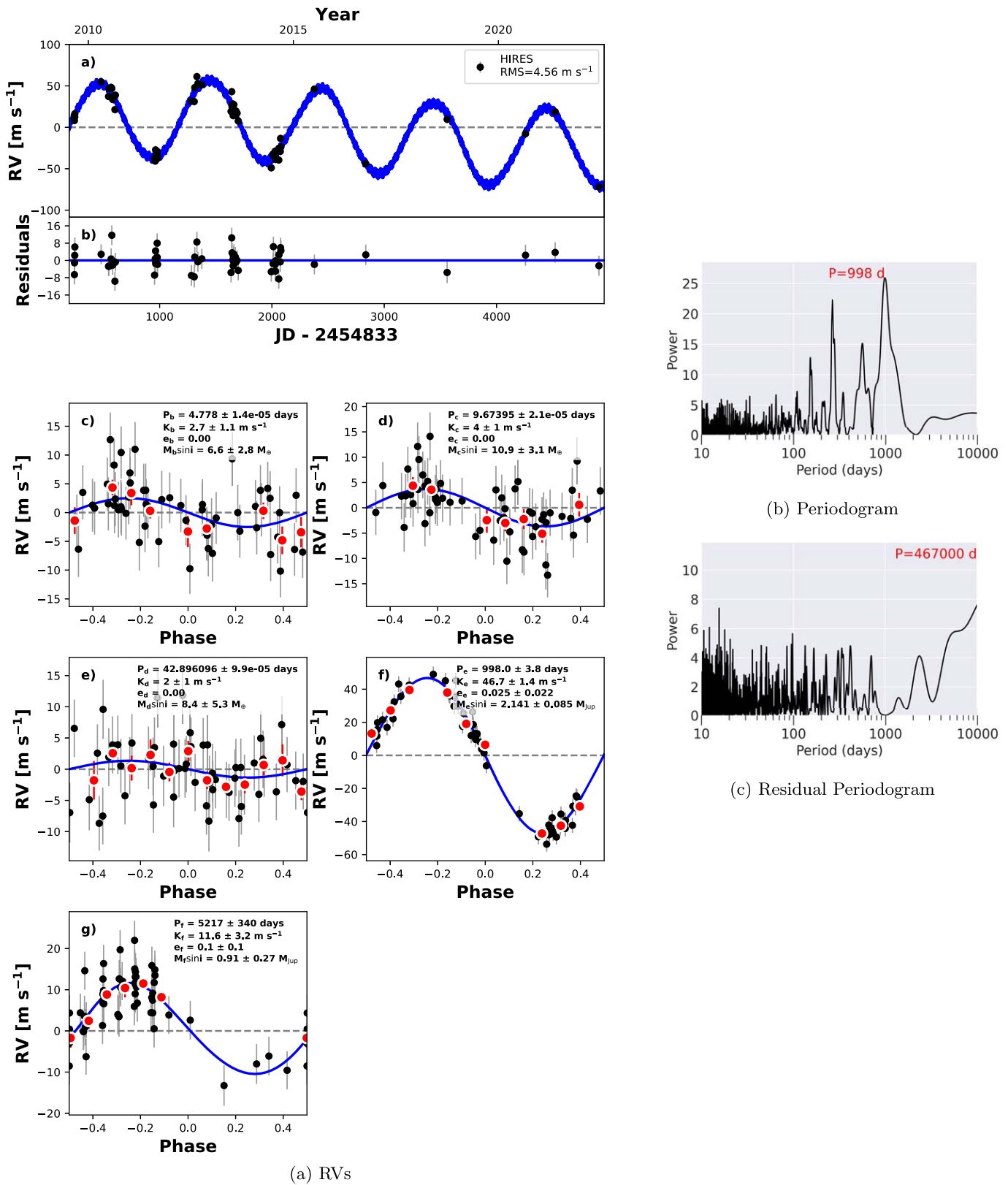

**Figure 25.** Same as Figure 8, but for KOI-148 (Kepler-48). The longest-period companion presented here was not detected by the KGPS algorithm but was identified in the residuals by eye.

We present 35 RVs from Keck-HIRES, which were collected between 2014 July 23 and 2019 November 11 (Figure 31). The time series began in 2014 for a KOI follow-up program that examined planets in multiplanet systems with period ratios near mean motion resonance. The residual to our best three-planet fit has rms = 9.9 m s$^{-1}$, which is typical for a star with $T_{\rm eff}$ = 6150 K and $v \sin i$ = 8.7 km s$^{-1}$ (Isaacson & Fischer 2010). The nondetection of an RV trend sets an upper limit of $M \sin i < 0.8 M_{\rm J}$ for a companion at 5 au ($M \sin i < 3.2 M_{\rm J}$ at 10 au).

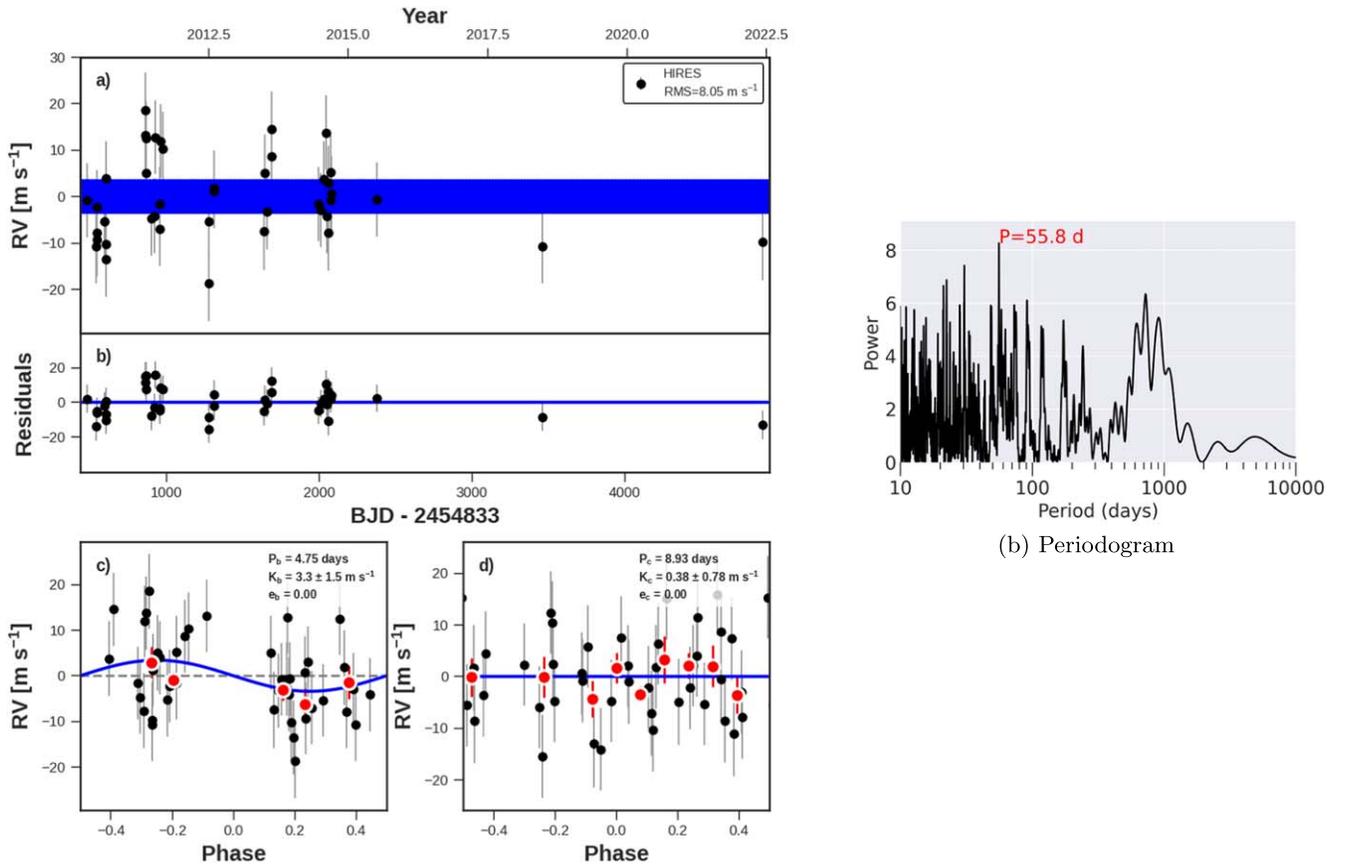

(b) Periodogram

(a) RVs

**Figure 26.** Same as Figure 8, but for KOI-153 (Kepler-113). No nontransiting companions are detected.

### A.25. KOI-261 (Kepler-96)

Kepler-96 (KOI-261) is a $V$ = 10.3 Sunlike star with one transiting planet ($P_b$ = 16.2 days, $R_b$ = 2.6 $R_\oplus$). The planet was confirmed with 44 Keck-HIRES RVs, which yielded a mass of $8.4 \pm 3.4 M_\oplus$, consistent with a volatile-enveloped planet (Marcy et al. 2014).

We have collected 11 new RVs since 2014 (Figure 32). A single RV outlier of 20 m s$^{-1}$ in 2017 might be consistent with the orbit of an eccentric nontransiting planet, although the RVs are also consistent with no giant planet if the 2017 data point is ignored. If the planet is real, it has a period of ∼4000 days and a mass of ∼$1 M_{\rm J}$. We consider this a planet candidate, although more RVs are needed to confirm its existence and characterize its orbit and mass.

### A.26. KOI-262 (Kepler-50)

Kepler-50 (KOI-262) is a $V$ = 10.7 F-type star ($T_{\rm eff}$ = 6200 K) with two transiting planets ($R_b$ = 1.7 $R_\oplus$, $R_c$ = 2.2 $R_\oplus$) confirmed in Chaplin et al. (2013). The planets' compact architecture ($P_b$ = 7.81 days, $P_c$ = 9.37 days) produces significant TTVs, which yield mass upper limits of $M_b < 8.9 M_\oplus$ and $M_c < 8.2 M_\oplus$ (Steffen et al. 2012). The star has solar-like oscillations that enabled an asteroseismic analysis, including mode splitting that revealed that the stellar rotation axis is closely aligned with the angular momentum vector of the planet orbits (Chaplin et al. 2013).

We observed this target for a KOI follow-up program that characterized planets in multiplanet systems with period ratios near mean motion resonance. We collected 39 RVs between 2012 and 2022 (10 yr baseline; Figure 33). The residual RVs to our best two-planet fit have a high rms (14 m s$^{-1}$). The star's temperature is near the Kraft break (Kraft 1967), and the star has $v \sin i = 10$ km s$^{-1}$, both of which could contribute to the high rms of the residual RVs, although we only expected a jitter of 9 m s$^{-1}$ for this target. There is no significant trend to the RV residuals, yielding a 3σ upper limit of $M \sin i < 0.70 M_{\rm J}$ at 5 au ($M \sin i < 2.8 M_{\rm J}$ at 10 au).

### A.27. KOI-265 (Kepler-507)

Kepler-507 (KOI-265) is an F-type star ($T_{\rm eff}$ = 6000 K) with one transiting planet ($P_b$ = 3.56 days, $R_b$ = 1.3 $R_\oplus$) that was statistically validated in Morton et al. (2016). The stellar and planetary properties were updated based on the Gaia parallax (Berger et al. 2018). We have collected 49 Keck-HIRES RVs between 2011 and 2022, which yield a mass of $M_b = 5.5 \pm 1.7 M_\oplus$ (Figure 34). This star was not part of the Marcy et al. (2014) results owing to stellar properties at the time falling outside of the optimal range for precise RVs.

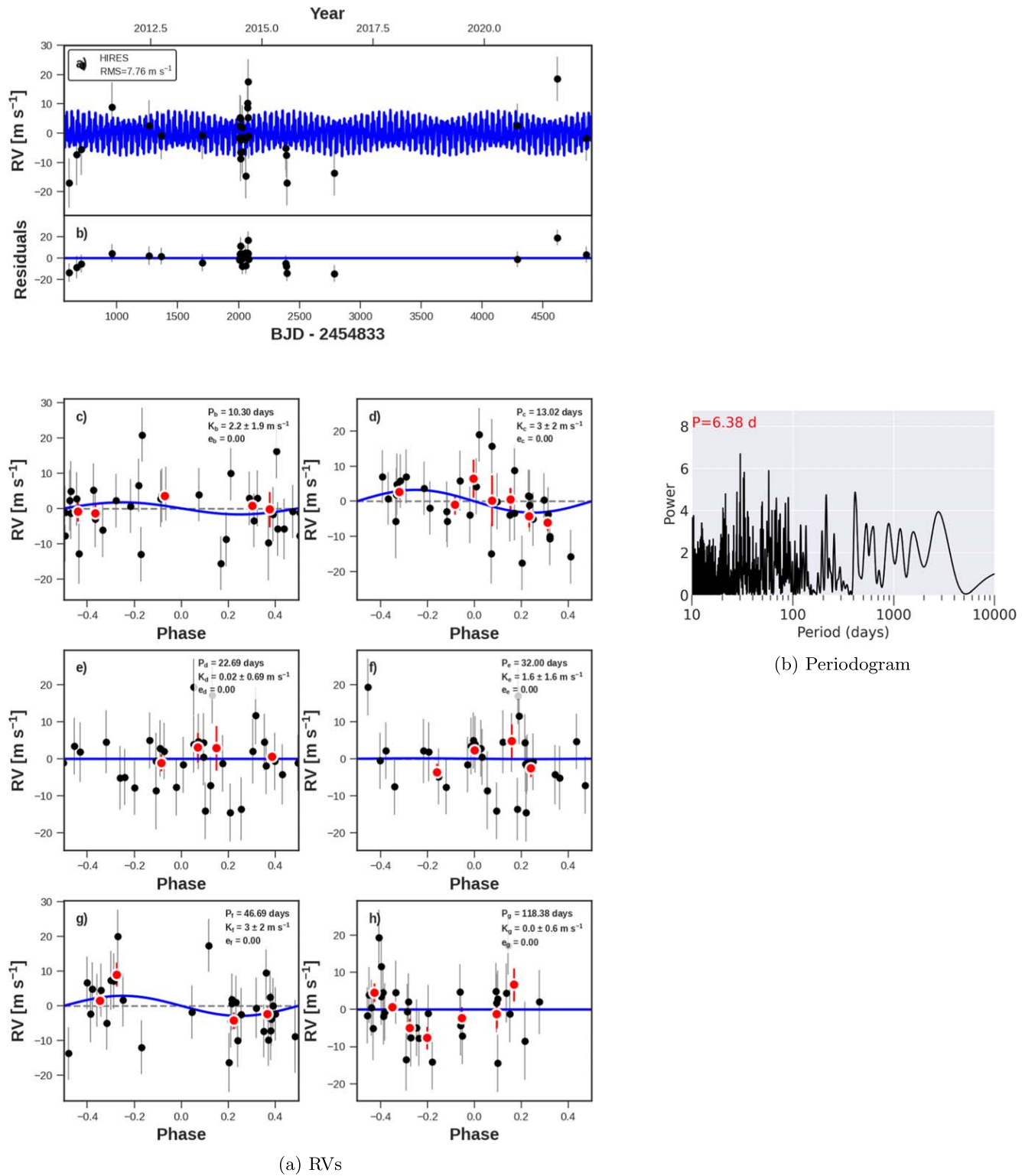

(a) RVs

Figure 27. Same as Figure 8, but for KOI-157 (Kepler-11). No nontransiting companions are detected.

The residual RVs have rms = 4.6 m s$^{-1}$, and there is no significant RV trend, resulting in a 3$\sigma$ upper limit of $M \sin i < 0.3 M_J$ at 5 au ($M \sin i < 1 M_J$ at 10 au). However, there is a marginally significant peak in the fast periodogram at 94 days (FAP = 0.04) and several harmonics and aliases of that peak. The inclusion of a Keplerian fit at 94 days reduces the rms of the RVs from 4.6 to 3.3 m s$^{-1}$ and produces $\Delta \text{BIC} = -30$. The RV amplitude of the 94 day signal is 4.5 m s$^{-1}$, corresponding to a planet candidate mass of $M \sin i = 35 M_\oplus$ (which is well within the upper limits established from the RV trend). More RVs are needed to identify the correct orbital period of this candidate second planet and determine its mass.

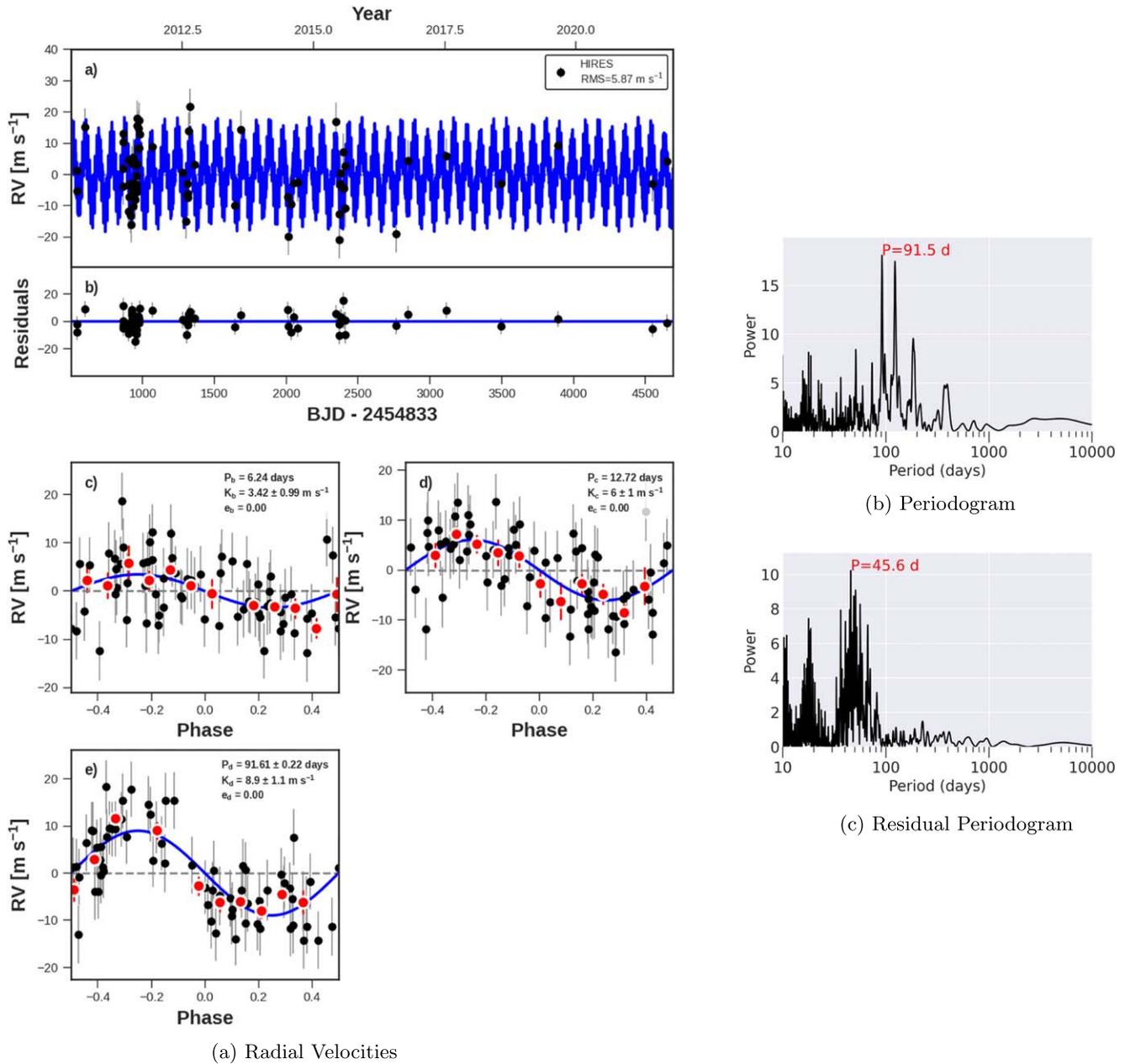

Figure 28. Same as Figure 8, but for KOI-244 (Kepler-25). The best-fit model includes a nontransiting planet, Kepler-25 d (orbital parameters inset in phase-folded panel).

### A.28. KOI-273 (Kepler-454)

KOI-273 (Kepler-454) has one transiting planet at $P = 10.57$ days. Gettel et al. (2016) used precise RVs from HIRES and HARPS-N to characterize the transiting planet (b), a non-transiting planet at $P = 524$ days (c), and a linear trend representative of a third companion. The Kepler photometry revealed solar-like oscillations of the host star, enabling a precise characterization of the stellar mass and radius. Our new RVs from Keck-HIRES are lower than expected from the linear trend, indicating the detection of curvature due to the orbit of the distant companion. Our joint fit to the HIRES and HARPS-N RVs yields planet masses of $M_b = 6 \pm 2\, M_\oplus$, $m_c \sin i = 4.6 M_J \pm 0.2 M_J$, and $m_d \sin i = 2.6 M_J \pm 0.3 M_J$ (Figure 35). Additionally, our new RVs clarify the orbits of the nontransiting planets and detect a marginally significant eccentricity for planet d: $P_c = 524 \pm 1$ days, $P_d = 4800 \pm 1700$ days, and $e_d = 0.26 \pm 0.04$. Our solution for Kepler-454 d is broadly consistent with the RV trend announced in Bonomo et al. (2023). Kepler-454 is another example of the value of long RV time baselines when detecting long-period massive planets. Because our baseline is comparable to the longest detected orbit in the system, any putative trend in the residuals to our best three-planet fit is not well constrained, and so our upper limit on additional companions is only $M \sin i < 1.7 M_J$ at 5 au ($M \sin i < 7.0 M_J$ at 10 au). Future RVs will improve the characterization of the period, eccentricity, and minimum mass of the long-period companion and will also place better constraints on the presence of additional companions.

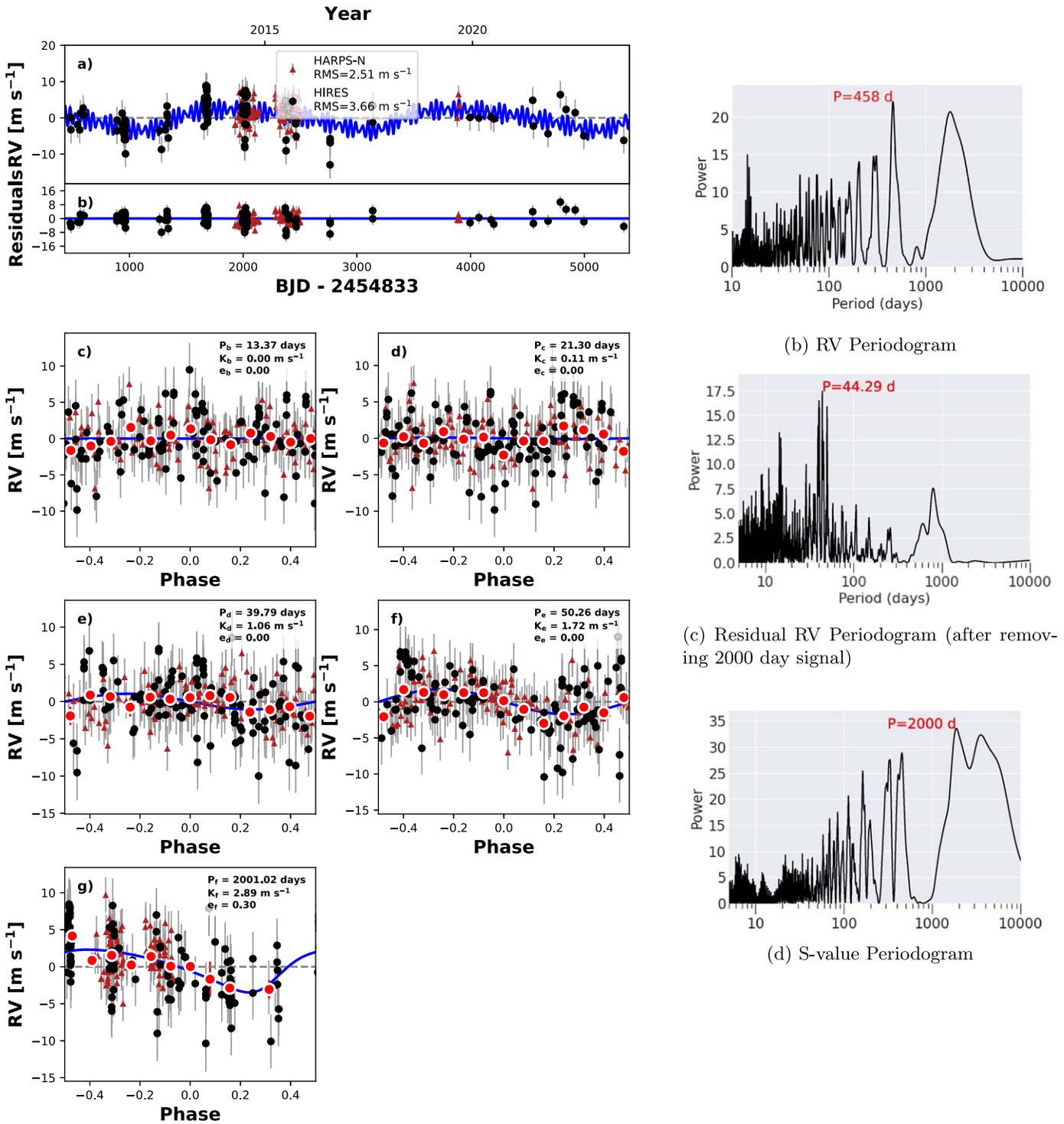

(a) Radial Velocities

(b) RV Periodogram

(c) Residual RV Periodogram (after removing 2000 day signal)

(d) S-value Periodogram

**Figure 29.** Same as Figure 8, but for KOI-245 (Kepler-37). Left: the RVs were taken at Keck-HIRES (black circles) and TNG-HARPS-N (maroon triangles). The best-fit model includes a Keplerian signal at either 460 days (pictured left) or 2000 days. Right: the periodogram of the Kepler-37 RVs after removing the signals associated with the transiting planets yields multiple peaks that are challenging to distinguish from stellar activity.

*A.29. KOI-274 (Kepler-128)*

Kepler-128 (KOI-274) is an F-type star with moderate rotation ($T_{eff} = 6000$ K, $v \sin i = 6$ km s$^{-1}$; Fulton & Petigura 2018). The star has two transiting planets with radii $R_b = 1.4 R_\oplus$ and $R_c = 1.3 R_\oplus$. The compact architecture of the transiting planets ($P_b = 15.1$ days, $P_c = 22.8$ days) produces significant TTVs (Xie 2014). An N-body dynamical characterization of the TTVs yielded planet masses of $M_b = 0.77 \pm 0.40 M_\oplus$ and $M_c = 0.90 \pm 0.45 M_\oplus$ (Hadden & Lithwick 2014). The low densities determined from the TTVs ($\rho_b = 1.5 \pm 0.8$ g cm$^{-3}$, $\rho_c = 2.1 \pm 1.1$ g cm$^{-3}$) are distinct from the density of rocky planets at this size (~7.5 g cm$^{-3}$), suggesting that the planets, though small, might contain low-density volatiles as a major constituent of their bulk compositions.

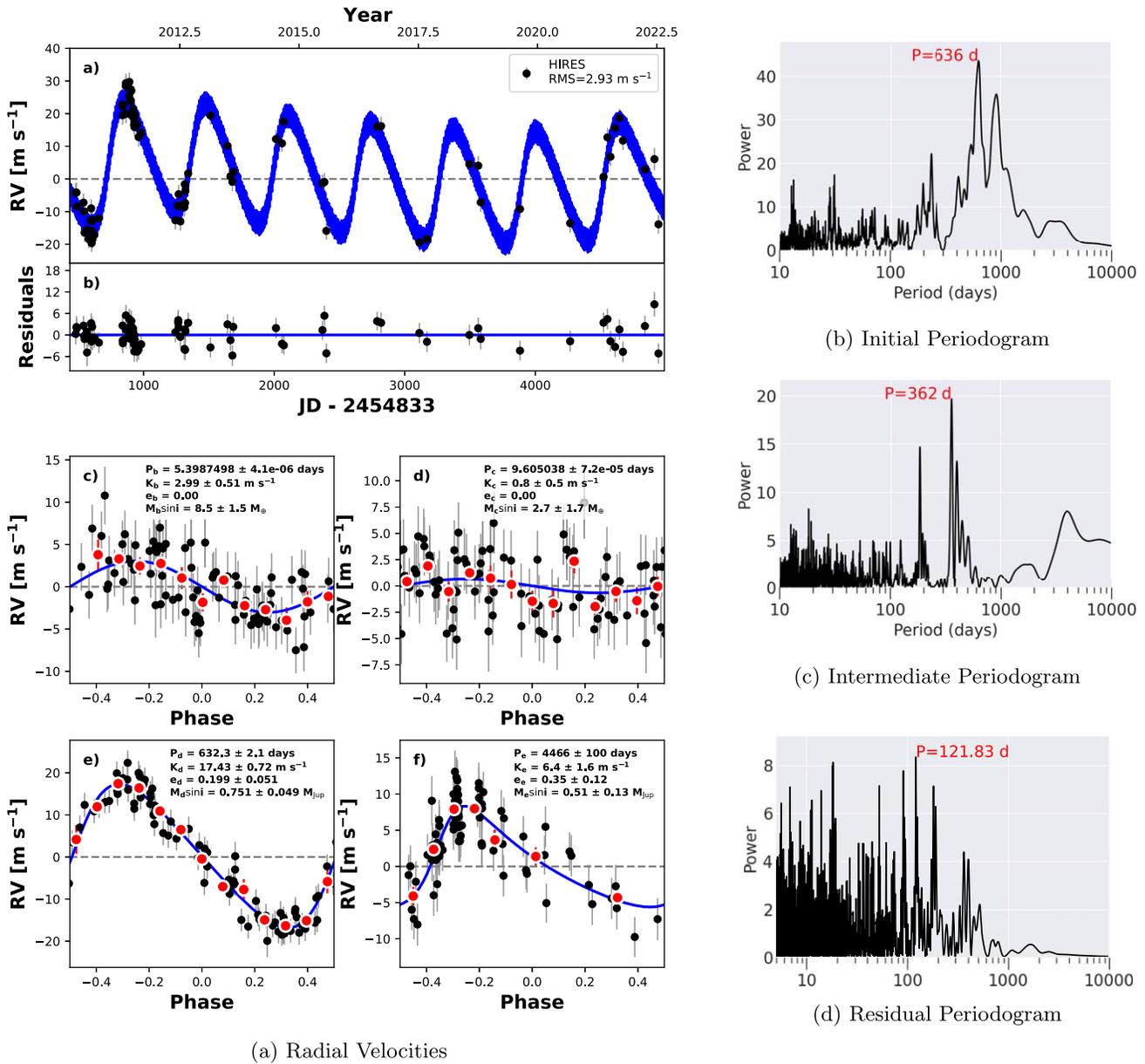

(a) Radial Velocities

**Figure 30.** Same as Figure 8, but for KOI-246. Left: the best-fit model includes two nontransiting planets, Kepler-68 d at 633 days and Kepler-68 e at 5000 days (see insets in phase-folded panels). Right: the top panel is a periodogram of the HIRES RVs after removing the transiting planets only, the middle panel is the periodogram after removing one nontransiting planet at 636 days (note the strong yearly alias at 362 days), and the bottom panel is after removing nontransiting planets at 634 and 5000 days.

We have collected 17 RVs using Keck-HIRES between 2012 and 2022 (Figure 36). This number of RVs did not meet our criterion for fitting planet masses (at least 20 dof). Assuming that the planets have typical masses for their sizes (Weiss & Marcy 2014), the residual RVs have rms = 8.1 m s$^{-1}$, which is typical for a moderately rotating F-type star. There are no strong peaks in the periodogram of the RV residuals and no apparent trends, yielding a 3$\sigma$ upper limit of $M \sin i < 0.37 M_J$ at 5 au ($M \sin i < 1.5 M_J$ at 10 au).

### A.30. KOI-275 (Kepler-129)

KOI-275 (Kepler-129) has two transiting sub-Neptune-sized planets. Zhang et al. (2021) used precise RVs from HIRES to detect and characterize a nontransiting companion at an orbital distance of a few astronomical units, with a minimum mass near the boundary between planetary- and brown-dwarf-mass objects. Solar-like oscillations of the host star revealed that the star has nonzero obliquity with respect to the transiting planets. The obliquity is consistent with a scenario in which the massive companion torques the inner planets. The inner planets have a coupling timescale much shorter than their precession time, allowing them to remain mutually transiting as their line of nodes precesses around the host star. We present nine new RVs here and their analysis, which result in an updated orbit and mass determination for the nontransiting companion (assuming no RV trend): $M \sin i_d = 6.3 M_J \pm 0.2 M_J$, $P_d = 1935$ days, $a_d = 3.26$ au, and $e_d = 0.07$. Our new RVs greatly improve our coverage of the orbit of the companion, allowing us to decisively classify it as a giant planet ($M \sin i_d < 13 M_J$) rather than a brown dwarf (Figure 37). However, the allowance of an RV trend significantly complicates our interpretation of the RVs, since the earliest RV we collected strongly influences our interpretation of the planet orbit (and any possible trend). If an RV trend is allowed in addition to our best three-planet fit, the constraints on the mass of the detected companion become $M \sin i_d =$
$6.8 M_J \pm 2.2 M_J$, which is an order of magnitude worse than if we assume no trend but still yields a planetary mass for the companion (as shown in Figure 37). Allowing an RV trend, our $3\sigma$ upper limit is $M \sin i < 13 M_J$ at 5 au ($M \sin i < 53 M_J$ at 10 au). Future RV monitoring will clarify the orbital properties of the super-Jovian companion and any additional planetary- or brown-dwarf-mass companions in the system.

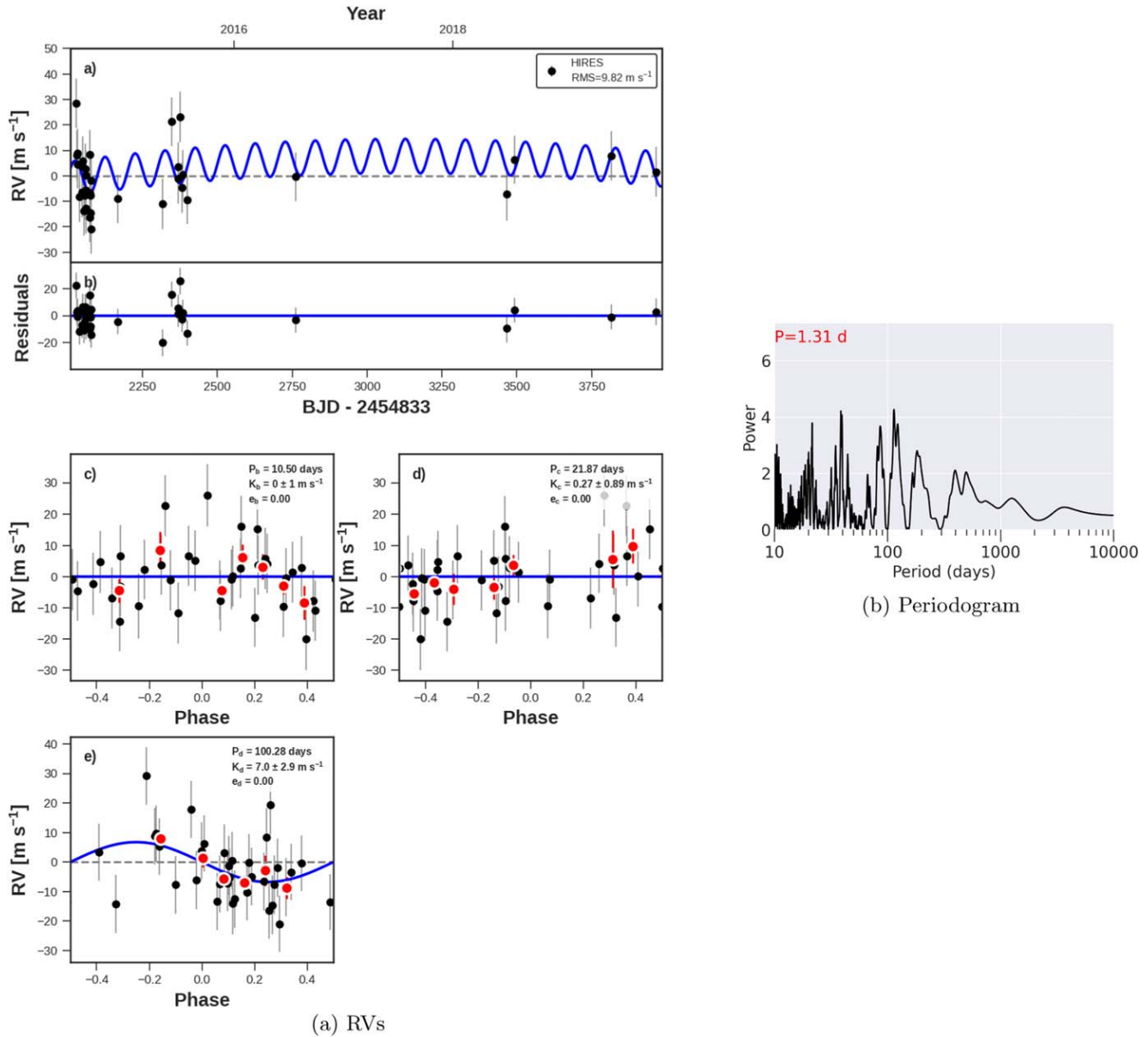

**Figure 31.** Same as Figure 8, but for KOI-260 (Kepler-126). No nontransiting companions are detected.

### A.31. KOI-277 (Kepler-36)

KOI-277 (Kepler-36) is a system with two transiting planets in orbits near a 7:6 mean motion resonance that produce significant TTVs (Carter et al. 2012). Although the planets have similar orbital periods, the inner planet is smaller, likely because its substantially smaller mass than its companion made it comparatively susceptible to photoevaporation from incident stellar X-ray and ultraviolet radiation (Lopez & Fortney 2013; Owen & Wu 2013). An additional transit of Kepler-36 c was observed from the ground, which did not significantly alter the determination of the planet masses, orbits, or sizes (Vissapragada et al. 2020).

We have collected 25 RVs of Kepler-36 between 2012 and 2022. The RVs have large errors because the host star has a modest $v \sin i$, and so they do not substantially improve the masses determined from TTVs, but they do rule out massive nontransiting planets in the vicinity of the near-resonant pair (Figure 38). There is no significant RV trend, yielding a $3\sigma$ upper limit of $M \sin i < 0.24 M_J$ at 5 au ($M \sin i < 0.95 M_J$ at 10 au). A joint analysis of the TTVs and RVs is in preparation.

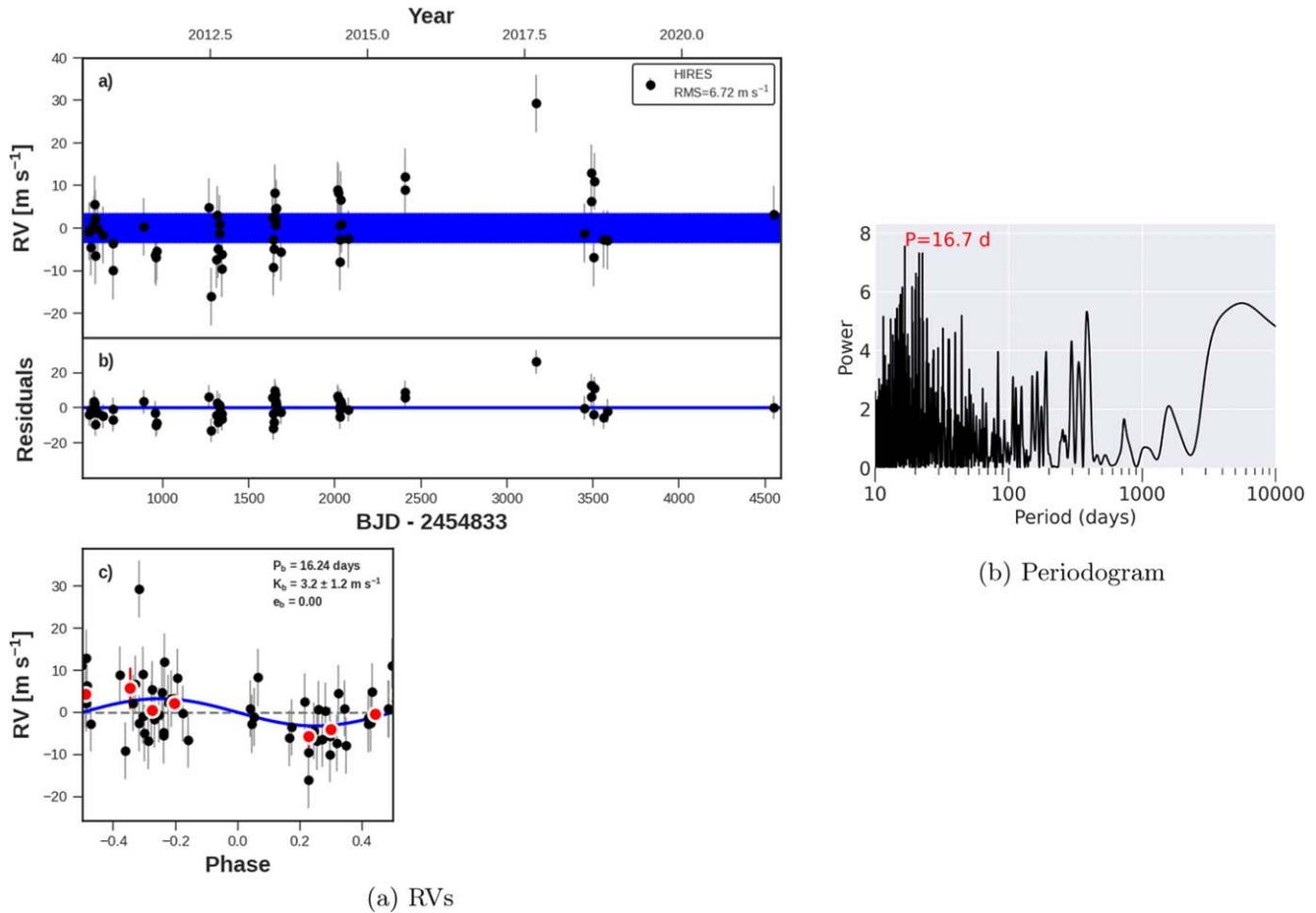

Figure 32. Same as Figure 8, but for KOI-261 (Kepler-96). No nontransiting companions are detected.

### A.32. KOI-281 (Kepler-510)

Kepler-510 is a metal-poor, G-type star ($T_{\rm eff} = 5600$ K, [Fe/H] $= -0.4$) that has one transiting planet ($P_b = 19.56$ days, $R_b = 2.5 R_\oplus$) that was statistically validated (Morton et al. 2016). We collected two RVs in 2011 and then another eight RVs between 2017 and 2021 (Figure 39). Assuming that the planet has a typical mass for its size using the Weiss & Marcy (2014) mass–radius relations, the residual RVs have rms $= 4.12$ m s$^{-1}$ and no trend, placing an upper limit of $M \sin i < 0.18 M_{\rm J}$ at 5 au ($M \sin i < 0.75 M_{\rm J}$ at 10 au).

### A.33. KOI-282 (Kepler-130)

Kepler-130 (KOI-282) is a multiplanet system that was statistically confirmed (Lissauer et al. 2014; Rowe et al. 2014). Three transiting planets orbit the primary star, which is Sunlike, although the system has a candidate wide-separation stellar companion detected in Gaia astrometry (Mugrauer 2019). The inner two planets have TTVs, and the transit durations yield upper limits on their eccentricities (Van Eylen & Albrecht 2015). This system was selected for follow-up as part of a program to survey stars with at least three transiting planets from 2015 onward. We have collected 10 Keck-HIRES RVs of Kepler-130 between 2015 and 2021 (Figure 40). Assuming typical masses for the three transiting planets (Weiss & Marcy 2014), the residual RVs have rms $= 5.9$ m s$^{-1}$ and no trend, resulting in an upper limit of $M \sin i < 0.49 M_{\rm J}$ at 5 au ($M \sin i < 2.0 M_{\rm J}$ at 10 au).

### A.34. KOI-283 (Kepler-131)

Kepler-131 (KOI-283) is a G-type star with two transiting planets at $P_b = 16.1$ days and $P_c = 25.5$ days that were statistically confirmed (Lissauer et al. 2014; Rowe et al. 2014). The planet radii are $R_b = 2.1 \pm 0.2 R_\oplus$ and $R_c = 0.82 \pm 0.06 R_\oplus$, a rare architecture in which the inner planet is larger than the outer planet (Ciardi et al. 2013; Berger et al. 2018). The planets were independently confirmed with 20 Keck-HIRES RVs, which yielded masses of $M_b = 16 \pm 4 M_\oplus$ and $M_c = 8 \pm 6 M_\oplus$ (Marcy et al. 2014). We have since collected 26 RVs on Keck-HIRES (46 RVs total), which yield a marginally improved planet mass of $M_b = 4.4 \pm 2.4 M_\oplus$ (Figure 41). We do not fit the mass of planet c because the planet is smaller than our minimum size for mass fitting; we instead assume a mass–radius relation (Weiss & Marcy 2014). The residual RVs have rms $=4.8$ m s$^{-1}$ and no trend, resulting in an upper limit of $M \sin i < 0.29 M_{\rm J}$ at 5 au ($M \sin i < 1.2 M_{\rm J}$ at 10 au). The fast periodogram of the RVs reveals a peak near 11.3 days that is not significant (FAP $= 0.3$).

### A.35. KOI-285 (Kepler-92)

Kepler-92 (KOI-285) is a bright ($V = 11.6$) Sunlike star with three transiting planets confirmed in Batalha et al. (2013):

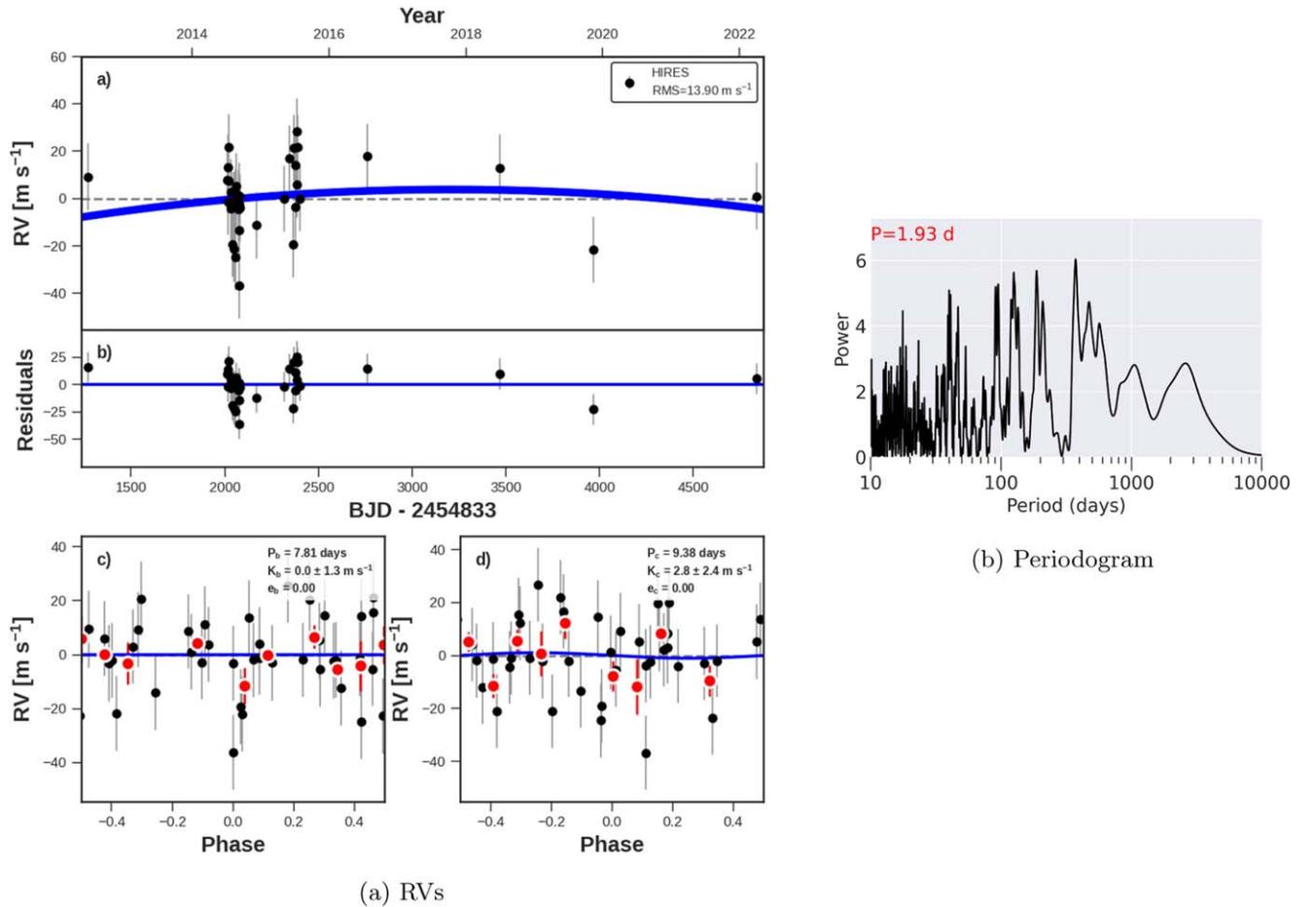

Figure 33. Same as Figure 8, but for KOI-262 (Kepler-50). No nontransiting companions are detected.

$R_b = 3.6\,R_\oplus$, $R_c = 2.5\,R_\oplus$, and $R_d = 2.1\,R_\oplus$. Solar-like oscillations were detected for this star, allowing a precise age estimate of $5.5 \pm 0.5$ Gyr (Silva Aguirre et al. 2015). The star is slightly evolved with slow rotation ($T_{\rm eff} = 5900$ K, $\log g = 4.0$ dex, $v \sin i = 3.5$ km s$^{-1}$; Fulton & Petigura 2018).

The system has three transiting planets near (but not in) a Laplace 4:2:1 mean motion resonant chain: $P_b = 13.7$ days, $P_c = 26.7$ days, and $P_d = 49.4$ days, a configuration that produces TTVs for the inner two planets (Xie 2014). The TTVs are too weak to yield precise determinations of the planet masses and eccentricities, and the orbits are consistent with circular (Van Eylen & Albrecht 2015).

Between 2011 and 2021, we collected 23 RVs on Keck-HIRES (Figure 42). The RVs yield upper limits on the planet masses: $M_b = 6.5 \pm 3.2\,M_\oplus$, $M_c = 3.9 \pm 3.1\,M_\oplus$, and $M_d = 4.4 \pm 3.6\,M_\oplus$. The residual RVs have rms = 6.0 m s$^{-1}$ and no trend, yielding an upper limit of $M \sin i < 0.29 M_J$ at 5 au ($M \sin i < 1.2 M_J$ at 10 au). The rms is higher than typical for a bright G-type star and might be a consequence of modal oscillations of this slightly evolved star.

### A.36. KOI-292 (Kepler-97)

Kepler-97 (KOI-292) is a $V = 12.9$ Sunlike star ($T_{\rm eff} = 5800$ K) with one transiting planet that was confirmed with Keck-HIRES RVs (Marcy et al. 2014). The transiting planet has $P_b = 2.59$ days and $R_b = 1.47 \pm 0.07\,R_\oplus$ (Berger et al. 2018). The RVs also indicated a trend consistent with a long-period companion ($P_c > 789$ days), which was announced as Kepler-97 c (Marcy et al. 2014).

We have collected 10 RVs since 2013, which demonstrate that the apparent trend in the earlier RVs was a statistical fluke (Figure 43). The full times series of 31 RVs, which extends from 2010 to 2019, has no trend, although the residual RVs have a somewhat high rms of 5.9 m s$^{-1}$. The RVs are consistent with a mass upper limit for the transiting planet of $M_b = 3.6 \pm 2.4\,M_\oplus$. The nondetection of an RV trend yields an upper limit of $M \sin i < 0.3 M_J$ at 5 au ($M \sin i < 1.2 M_J$ at 10 au) for any additional companions, which is inconsistent with the previously announced planet. In addition, there are no significant peaks in the fast periodogram (FAP = 0.99), suggesting that we are unlikely to have missed a Jupiter-mass planet with $P < \sim 2000$ days.

### A.37. KOI-295 (Kepler-134)

Kepler-134 (KOI-295) is a $V = 12.3$ slowly rotating, quiet Sunlike star ($T_{\rm eff} = 5900$ K, $v \sin i < 2$ km s$^{-1}$, $\log R'_{\rm HK} = -5.0$). The system has two transiting planets that were statistically confirmed (Lissauer et al. 2014; Rowe et al. 2014). The planets are interior to the 2:1 mean motion resonance ($P_b = 5.32$ days, $P_c = 10.1$ days) and are in the rare configuration in which the inner planet is larger than the outer one ($R_b = 1.8\,R_\oplus$, $R_c = 1.2\,R_\oplus$; Berger et al. 2018).

We collected 14 Keck-HIRES RVs on 13 nights between 2011 and 2022 (Figure 44). Assuming that the planets have typical masses for their sizes (Weiss & Marcy 2014), the RVs appear to be consistent with a weak long-term trend ($\Delta \mathrm{BIC} = -54$), which corresponds to a companion of $M \sin i < 3.7 M_J$ at 10 au. Even with the inclusion of a long-term trend, the residual RV scatter is uncharacteristically high for a quiet Sunlike star (11.8 m s$^{-1}$). We performed a jitter test by collecting two RVs several hours apart in a single night; the RV difference was only 5 m s$^{-1}$, which is indeed consistent with low jitter, as expected. Further observations are needed to clarify the source of the RV scatter.

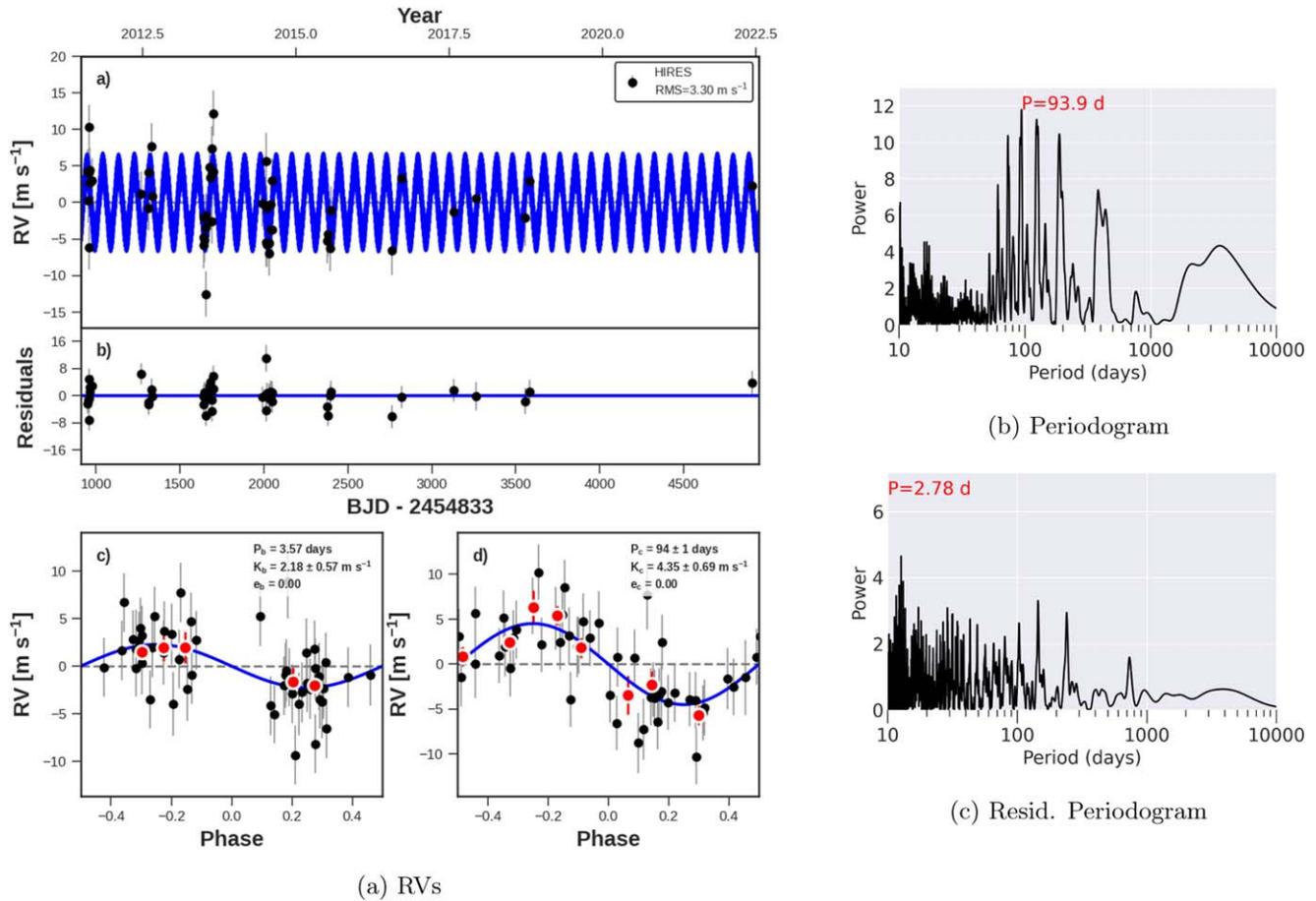

**Figure 34.** Same as Figure 8, but for KOI-265 (Kepler-507). Left: the best-fit model includes a nontransiting planet, Kepler-507 c (orbital parameters inset in phase-folded panel). Right: periodograms of the RVs after subtracting the models from the transiting planets only (top) and the best-fit model that includes the nontransiting planet (bottom).

### A.38. KOI-299 (Kepler-98)

Kepler-98 (KOI-299) is a $V = 12.9$ Sunlike star with one transiting planet ($P_b = 1.54$ days) that was confirmed with Keck-HIRES RVs, yielding a mass of $3.6 \pm 1.4 M_\oplus$ (Marcy et al. 2014). The planet radius was reported as $2.0 R_\oplus$ based on stellar spectroscopy, $1.5 R_\oplus$ based on its parallax, and $1.87 R_\oplus$ based on the combined spectroscopy and Gaia parallax (Marcy et al. 2014; Fulton & Petigura 2018; Berger et al. 2018).

We have collected 42 RVs between 2011 and 2018, including 20 since the publication of Marcy et al. (2014; see Figure 45). The new best-fit planet mass is $1.0 \pm 1.0 M_\oplus$, with residual 7.1 m s$^{-1}$. There is no apparent trend in the RVs, yielding an upper limit of $M \sin i < 0.3 M_J$ at 5 au ($M \sin i < 1.2 M_J$ at 10 au). There are no significant peaks in the fast periodogram of the residual RVs (FAP = 0.90).

### A.39. KOI-305 (Kepler-99)

Kepler-99 (KOI-305) is a $V = 13.0$ K-type star with moderate activity ($T_\mathrm{eff} = 4600$ K, $\log R'_\mathrm{HK} = -4.6$). It has one transiting planet ($P_b = 4.6$ days, $R_b = 1.5 R_\oplus$) that was confirmed with Keck-HIRES RVs. The RVs yielded a mass of $M_b = 6.2 \pm 1.3 M_\oplus$, which corresponds to a density of $10.9 \pm 2.0$ g cm$^{-3}$, consistent with a rocky composition.

We have collected 45 RVs of Kepler-99 between 2010 and 2018, including 24 since the publication of Marcy et al. (2014; see Figure 46). The new RVs produce a marginal change in the best-fit planet mass. The total time series of the RVs yields a semi-amplitude of $K = 1.74 \pm 0.88$ m s$^{-1}$, which corresponds to a mass of $2.5 \pm 1.3 M_\oplus$. The resulting density is $2.3 \pm 1.3$ g cm$^{-3}$, indicating that the planet likely has a volatile envelope. The residual RVs have rms = 5.3 ms and no apparent trend, yielding an upper limit of $M \sin i < 0.27 M_J$ at 5 au ($M \sin i < 1.1 M_J$ at 10 au). There are no significant peaks in the fast periodogram of the residual RVs (FAP = 0.99).

### A.40. KOI-316 (Kepler-139)

Kepler-139 (KOI-316) is a $V = 12.7$, quiet G-type star ($T_\mathrm{eff} = 5500$ K, $v \sin i < 2$ km s$^{-1}$, $\log R'_\mathrm{HK} = -5.1$). It has two statistically confirmed transiting planets ($P_b = 15.77$ days, $P_c = 157$ days) and a third transiting candidate at $P_d = 7.31$ days (Lissauer et al. 2014; Rowe et al. 2014; Holczer et al. 2016), which we confirm below. There is an equal-magnitude visual companion 10″ to the south of the planet-hosting star. The centroid photometry from Kepler clearly demonstrates that all three transits are around the target (northern) star (Jason Rowe, private communication). The planets have $R_b = 2.4\,R_\oplus$, $R_c = 2.4\,R_\oplus$, and $R_d = 1.7\,R_\oplus$ (Fulton & Petigura 2018).

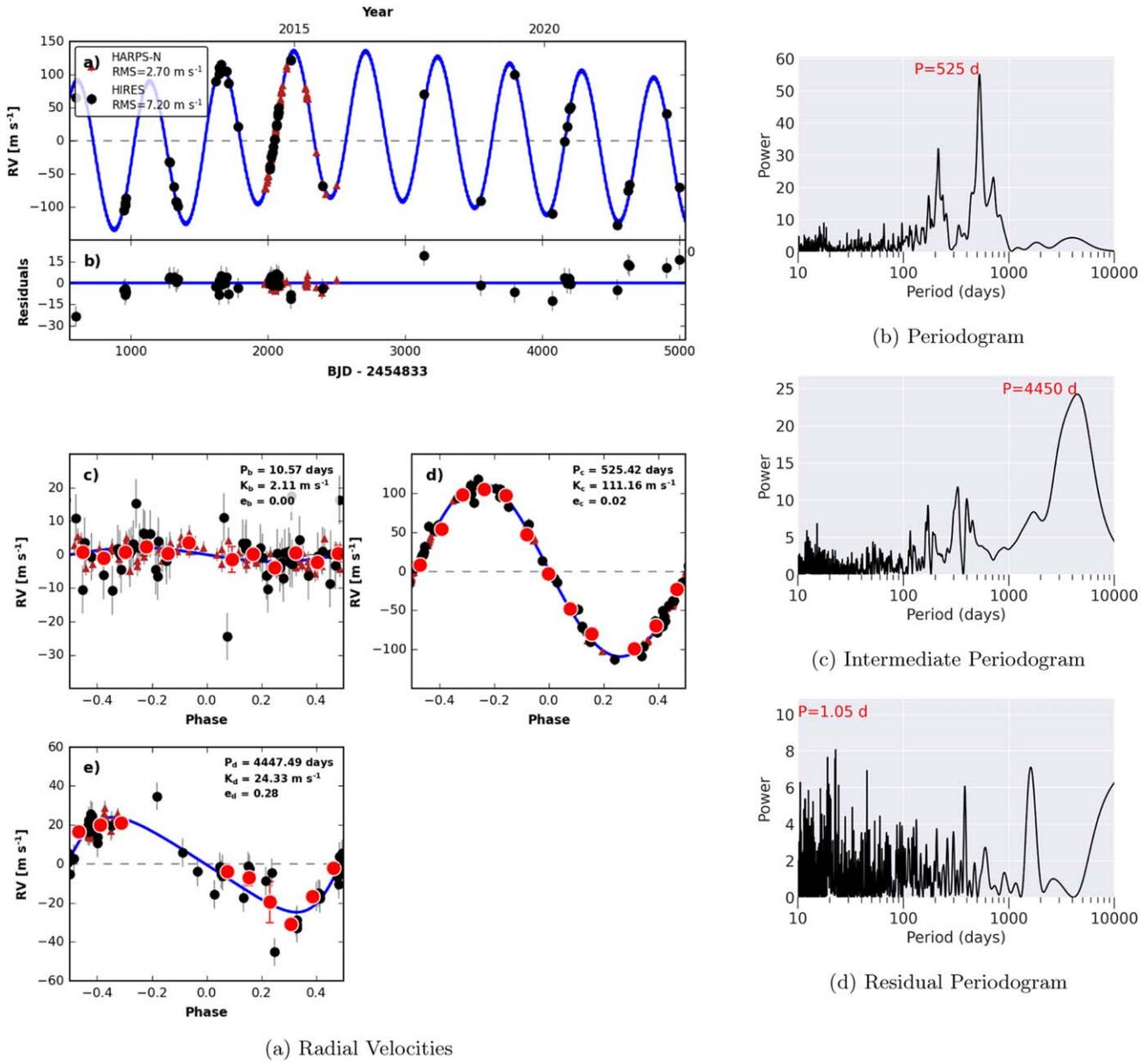

Figure 35. Same as Figure 8, but for KOI-273 (Kepler-454). Left: the RVs were taken at Keck-HIRES (black circles) and TNG-HARPS-N (maroon triangles). The best-fit model includes two nontransiting planets, Kepler-454 c at 525 days and Kepler-454 d at 4500 days (see insets in phase-folded panels). Right: the top panel is a periodogram of the RVs after removing the transiting planets only, the middle panel is the periodogram after removing one nontransiting planet at 525 days, and the bottom panel is the periodogram after removing nontransiting planets at 525 and 4500 days.

This target was observed as part of a program to search for giant planet companions in systems with three or more transiting planets. We have collected 38 Keck-HIRES RVs of Kepler-139 between 2010 and 2022, yielding a 12 yr baseline (Figure 47). The 10″ separation of the two bright stars is much larger than the typical seeing at Keck (<2″), and so the visual companion does not contaminate the HIRES C2 decker (width 1″.0, which was always rotated to avoid the companion). By coincidence, the first three RVs we collected (from 2010 to 2015) were all near the "up" quadrature of an RV signal that became apparent after our fourth RV observation. Moderate-cadence follow-up was necessary to ascertain the orbit and mass of the companion: $K_e = 19.1 \pm 1.5\,\mathrm{m\,s^{-1}}$, $P_e = 1920 \pm 33$ days, $M \sin i_e = 1.2\,M_J \pm 0.1\,M_J$, $e_e = 0.06 \pm 0.04$. Furthermore, the Keck-HIRES RVs yield masses for the transiting planets: $M_b = 4.8 \pm 2.1\,M_\oplus$, $M_c = 22.3 \pm 4.7\,M_\oplus$, and $M_d = 5.2 \pm 1.9\,M_\oplus$. The mass of planet c (at $P_c = 157$ days) is quite high compared to the other two similar-sized but closer-in planets. Perhaps planet c has a grazing transit with a radius in need of reevaluation, or perhaps the planet is rich in water-ice and high mean molecular weight volatiles, or perhaps the high mass ($3\sigma$ confidence) is spurious. The residual RVs have rms = $4\,\mathrm{m\,s^{-1}}$, consistent with the low RV scatter expected of quiet Sunlike stars. More RVs will clarify the masses and compositions of the three transiting planets, as well as the orbit

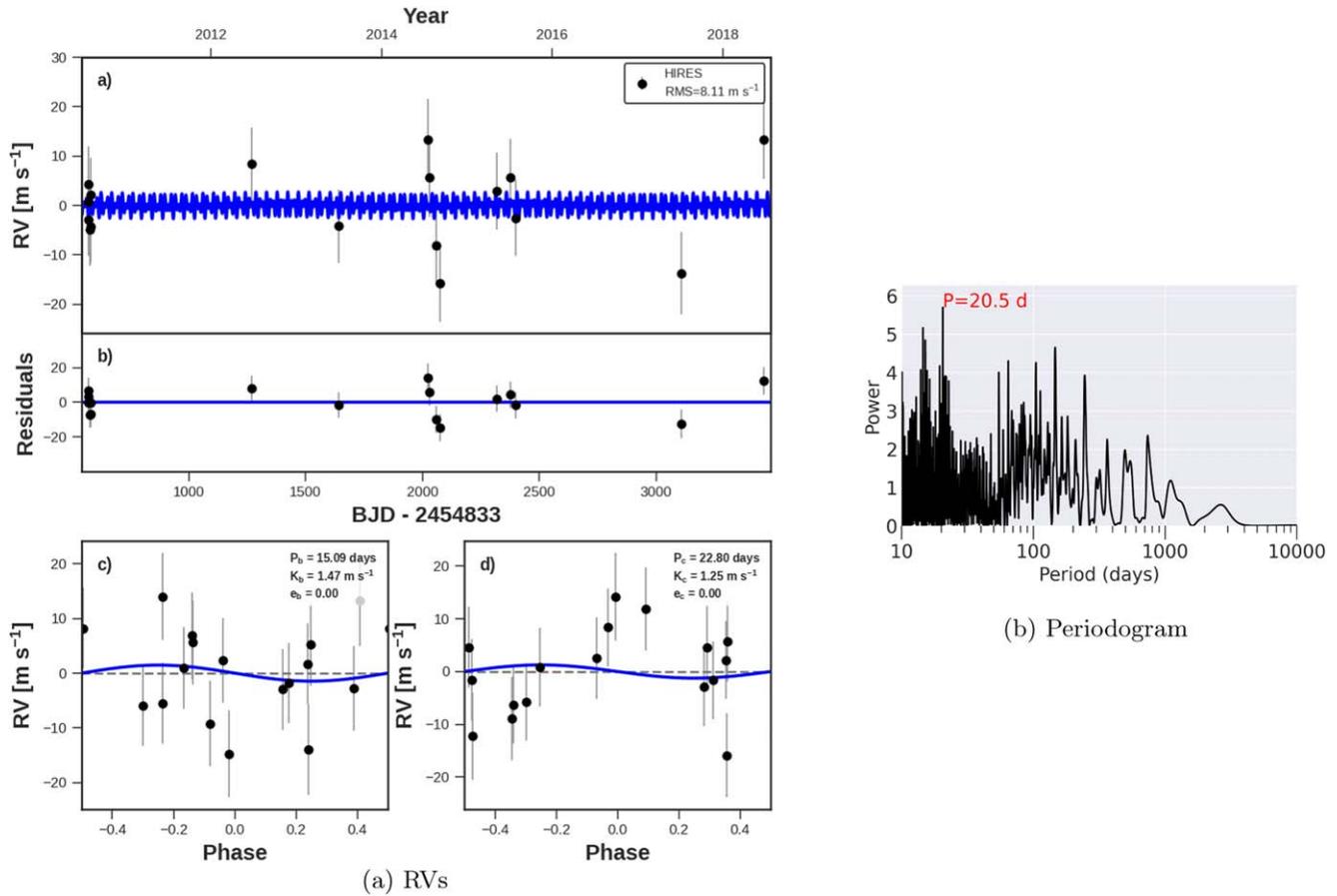

**Figure 36.** Same as Figure 8, but for KOI-274 (Kepler-128). No nontransiting companions are detected.

of the Jovian companion and the possible presence of additional companions.

### A.41. KOI-321 (Kepler-406)

Kepler-406 (KOI-321) is a $V = 12.5$ G-type star of moderate metallicity ($T_{\rm eff} = 5600$ K, [Fe/H] $= 0.3 \pm 0.1$; Fulton & Petigura 2018). It hosts two transiting planets: $P_b = 2.4$ days, $P_c = 4.6$ days). The planets are small ($R_b = 1.4 \pm 0.03$, $R_c = 0.85 \pm 0.03$) and were confirmed with Keck-HIRES RVs, which yielded $M_b = 6.4 \pm 1.4 M_\oplus$ and $M_c = 2.7 \pm 1.8 M_\oplus$ (Marcy et al. 2014).

We have collected 56 Keck-HIRES RVs between 2010 and 2022 (Figure 48). These include 14 new RVs since the publication of Marcy et al. (2014), which do not substantially change the mass determinations of the planets: $M_b = 5.4 \pm 1.9 M_\oplus$, $M_c = 2.1 \pm 1.8 M_\oplus$. The 12 yr baseline has low scatter (residual rms = 4.6 m s$^{-1}$) and does not reveal any RV trend, which corresponds to an upper limit of $M \sin i < 0.18 M_{\rm J}$ at 5 au ($M \sin i < 0.72 M_{\rm J}$ at 10 au). A periodogram of the RV residuals yields a low-significance peak at 15.9 days (FAP = 0.2). This period is not known to be associated with stellar rotation and might represent the orbit of a bona fide planet.

### A.42. KOI-351 (Kepler-90)

Kepler-90 (KOI-351) is a faint ($V = 13.8$) Sunlike star with eight confirmed transiting planets. Seven of the transiting planets (Kepler-90 b–h) were announced nearly simultaneously (Cabrera et al. 2014; Rowe et al. 2014; Schmitt et al. 2014). The compact architecture of the planets produces significant TTVs, which confirm their planetary status (Lissauer et al. 2014). The eighth planet (Kepler-90 i) was detected with candidate status at the time owing to its small size (Aviv Ofir, private communication) and was later confirmed with an independent photometric search (Shallue & Vanderburg 2018). Liang et al. (2021) found that Kepler-90 g is much lower in mass than Kepler-90 h ($15 \pm 1 M_\oplus$ for planet g vs. $203 \pm 5 M_\oplus$ for planet h) by fitting their transit timing and depth variations determined from the Kepler photometry.

We have collected 33 RVs of Kepler-90 between 2011 and 2022 (Figure 49). This 11 yr baseline was necessary to sample the orbit of Kepler-90 h, which is nearly 1 yr ($P_h = 331$ days), resulting in significant aliasing during the seasonal Kepler field RV follow-up. The RVs place upper limits on all of the small planets (Kepler-90 b, c, d, e, f, i) and yield masses for the two Saturn-sized planets: $M_g = 49 \pm 29 M_\oplus$ and $M_h = 0.63 M_{\rm J} \pm 0.15 M_{\rm J}$, consistent with the results of Liang et al. (2021). The residual RVs have significant scatter (rms = 13.7 m s$^{-1}$), which is atypical for a slowly rotating, magnetically quiet Sunlike star ($T_{\rm eff} = 6000$ K, $v \sin i = 3.6$ km s$^{-1}$, $\log R'_{\rm HK} = -5.2$, expected jitter = 9.3 m s$^{-1}$). However, there is no significant trend in the decade of RV baseline ($M \sin i < 0.3 M_{\rm J}$ at 5 au, $M \sin i < 1.4 M_{\rm J}$ at 10 au), and the fast periodogram reveals no significant peaks (FAP = 0.4). The high RV scatter might result from additional (yet-undetected) planets. A joint analysis of the TTVs and RVs of this system is currently underway (D. Shaw et al. 2024, in preparation)

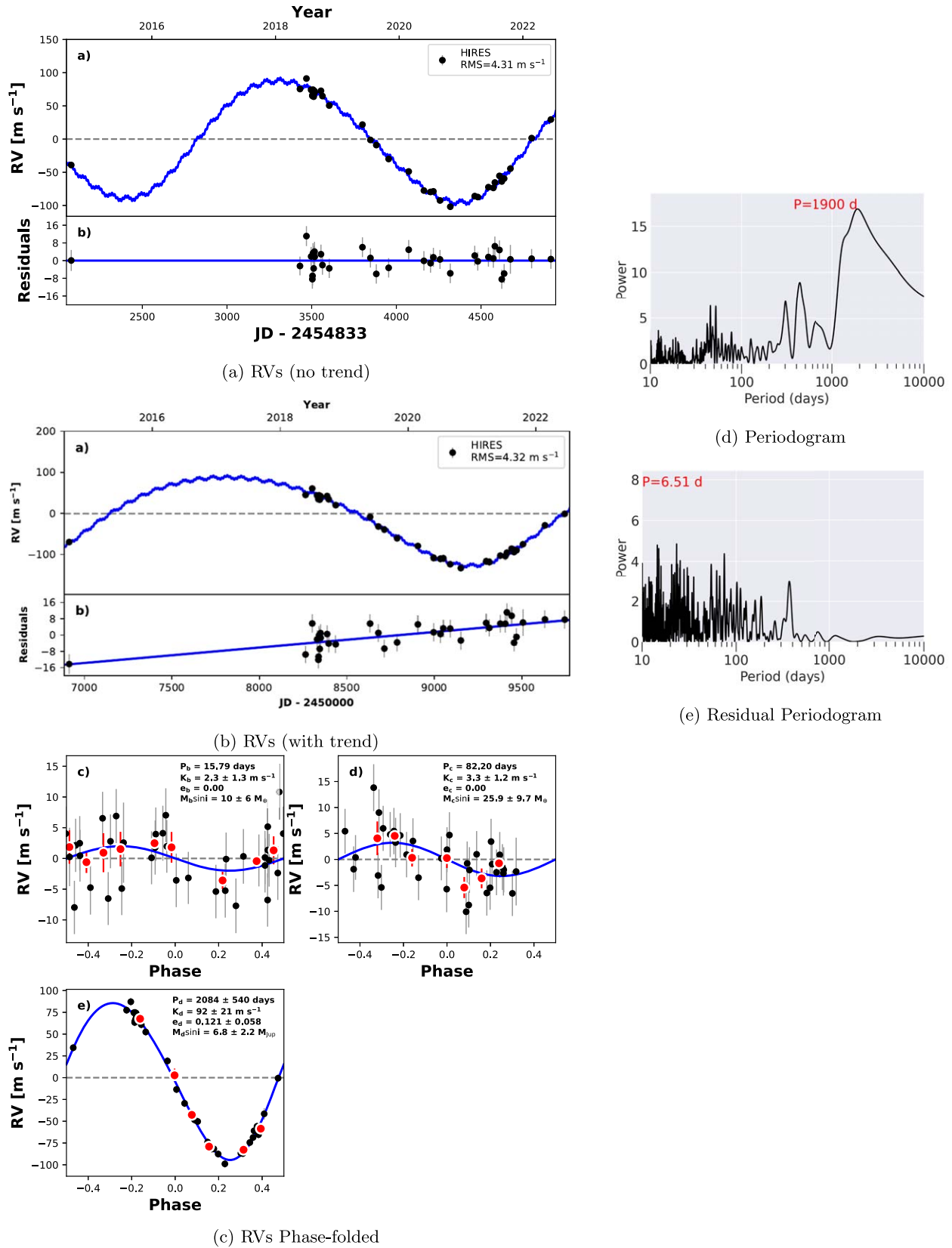

Figure 37. Same as Figure 8, but for KOI-275 (Kepler-129). Left: the best-fit models include one nontransiting planet, although the unusual time sampling is consistent with two different classes of model: either a near-circular, nontransiting giant planet (top) or a slightly longer period, eccentric giant planet with a substantial residual trend (middle). Right: periodograms after removing models of just the transiting planets (top) and a three-planet model in which the shorter, more circular period for the companion is favored (bottom).

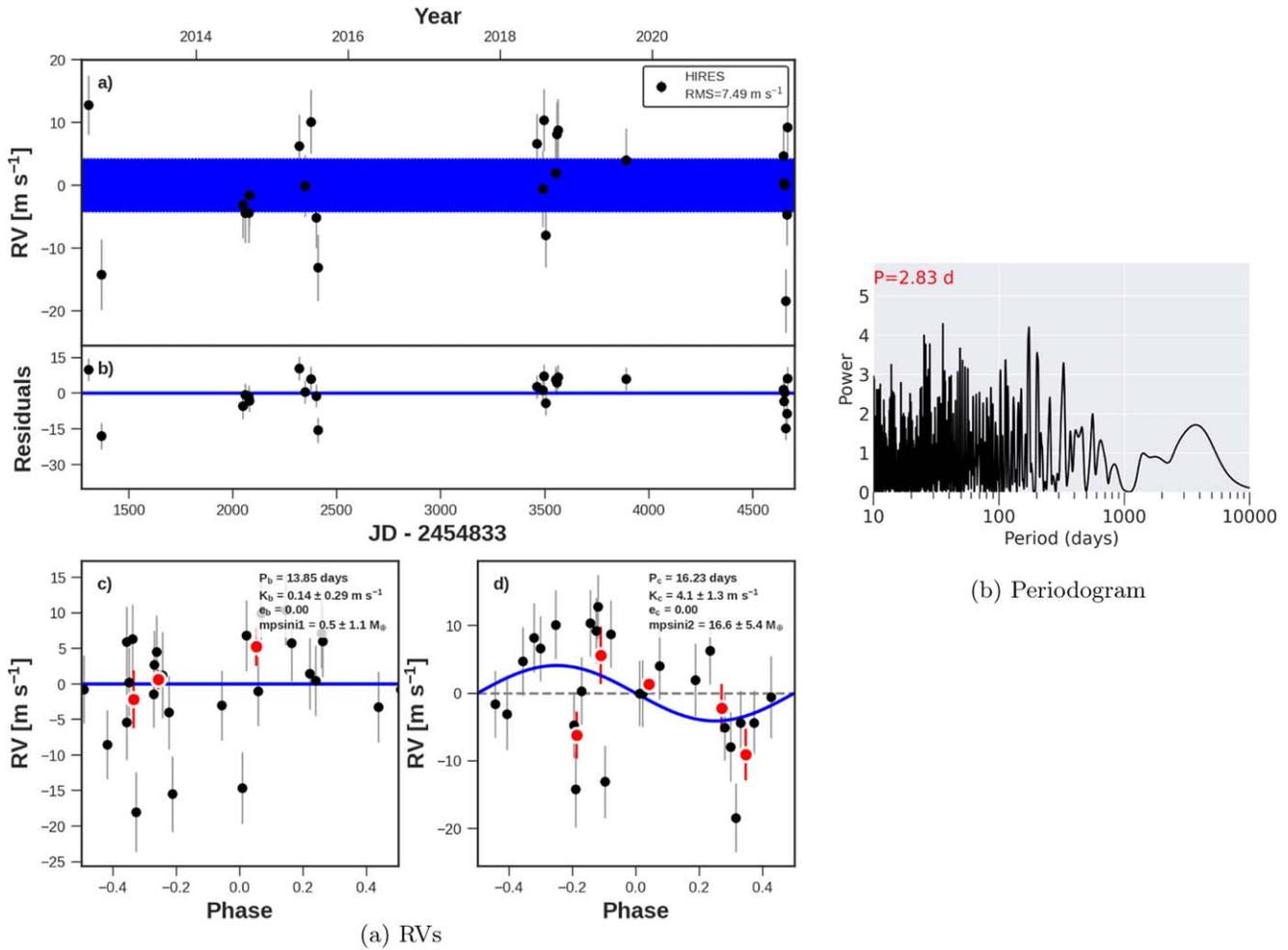

Figure 38. Same as Figure 8, but for KOI-277 (Kepler-36). No nontransiting companions are detected.

### A.43. KOI-365 (Kepler-538)

Kepler-538 (KOI-365) is a $V = 11.2$ Sunlike star ($T_{\rm eff} = 5400$ K). The star has one transiting planet ($P_b = 81.7$ days, $R_b = 2.2\,R_\oplus$) that was statistically validated (Morton et al. 2016). The system has a candidate M dwarf stellar companion, Kepler-538 B, which is 8 mag fainter than the primary at a separation of $10''$ (Mugrauer 2019). We have collected 28 HIRES RVs between 2010 and 2017 (Figure 50), yielding a mass for the transiting planet of $M_b = 6.7 \pm 3.6\,M_\oplus$ (see Mayo et al. 2019 for a detailed analysis). The residual RVs have rms $= 3.1$ m s$^{-1}$ and no apparent trend over the 7 yr baseline, consistent with $M \sin i < 0.2 M_J$ at 5 au ($M \sin i < 0.8 M_J$ at 10 au). There are no significant peaks in the fast periodogram (FAP $= 0.98$).

### A.44. KOI-370 (Kepler-145)

Kepler-145 (KOI-370) is a slightly evolved, $V = 12.0$ G-type star with moderate rotation ($T_{\rm eff} = 6000$ K, $\log g = 4.1$ dex, $v \sin i = 6.5$ km s$^{-1}$; Fulton & Petigura 2018). It has two transiting planets in a compact configuration ($P_b = 23.0$ days, $P_c = 42.9$ days) that produces TTVs (Xie 2014). The planet radii are $R_b = 2.6\,R_\oplus$ and $R_c = 4.3\,R_\oplus$. Based on their TTVs, the planets have masses of $M_b = 37 \pm 11\,M_\oplus$ and $M_c = 79 \pm 16\,M_\oplus$, which are unusually high masses for planets of these sizes (Weiss & Marcy 2014).

We have collected 10 Keck-HIRES RVs between 2014 and 2021 (Figure 51). These are too few data to constrain the planet masses effectively, although the RVs collected so far are consistent with the high TTV masses. However, the scatter of the residual RVs is high (rms $= 13$ m s$^{-1}$), which could be caused by a yet-undetected planet that might be influencing both the RV and TTV solutions. However, the star also has moderate rotation ($v \sin i = 6.5$ km s$^{-1}$), which could contribute somewhat to the high RV scatter (total expected RV jitter $= 9.3$ m s$^{-1}$). More RVs are needed to test the consistency of the RV and TTV solutions. The 7 yr RV baseline has no significant trend, although the high RV from 2014 is consistent with a variety of moderate trends, yielding a $3\sigma$ upper limit of $M \sin i < 0.6 M_J$ at 5 au ($M \sin i < 2.7 M_J$ at 10 au). There are no significant peaks in the fast periodogram (FAP $= 0.99$).

### A.45. KOI-377 (Kepler-9)

Kepler-9 (KOI-377) has three transiting planets, including two Saturn-sized planets near a 2:1 orbital resonance that were the first planets confirmed via TTVs ($P_b = 19.2$ days, $P_c = 39.0$ days; Holman et al. 2010). The third planet is a short-period super-Earth that was discovered based on additional photometry that is dynamically decoupled from the sub-Saturns and does not contribute appreciably to the TTVs ($P_d = 1.59$ days, $R_d = 1.6\,R_\oplus$; Torres et al. 2011).

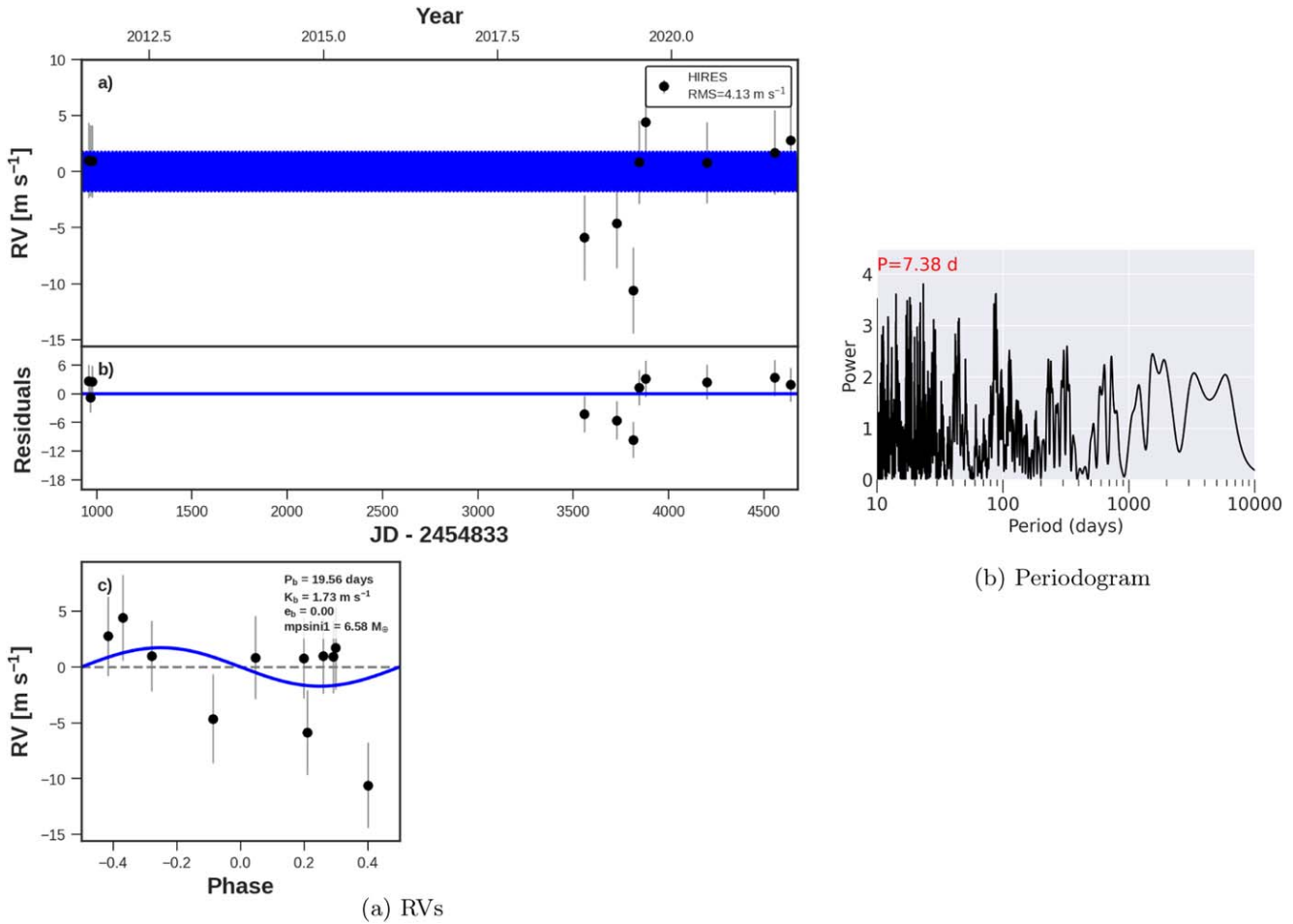

**Figure 39.** Same as Figure 8, but for KOI-281 (Kepler-510). No nontransiting companions are detected.

Holman et al. (2010) included 7 months of Kepler photometry (Q1–Q3) and six RVs from Keck-HIRES, and this initial data set was dramatically expanded over the course of the Kepler mission and a decade of ground-based follow-up. Dreizler & Ofir (2014) found masses for Kepler-9 b and c that were ∼60% of the originally reported values, based on $N$ quarters of Kepler photometry that sampled a much longer timescale of planet–planet interaction than the discovery paper. Additional transits of planets b and c were detected with ground-based photometry, leading to an improved dynamical fit of the TTVs and six HIRES RVs (Freudenthal et al. 2018). Borsato et al. (2019) collected 30 RVs with HARPS-N in 2014 and performed a joint fit to the RVs and TTVs. Their analysis yielded $R_b = 8.3\,R_\oplus$, $R_c = 8.1\,R_\oplus$, $M_b = 43 \pm 2\,M_\oplus$, and $M_c = 30 \pm 1\,M_\oplus$ (Borsato et al. 2019), although a variety of TTV solutions in the literature have produced discrepant masses (Holman et al. 2010; Hadden & Lithwick 2014, 2017). At the time Borsato et al. (2019) was published, only six HIRES RVs were publicly available, and these RVs were not included in their analysis.

We have collected 21 RVs on Keck-HIRES between 2010 and 2022 (Figure 52). Two spectra were taken with a short decker (B5 instead of C2, on 2010 June 19 and 2010 May 26), which precludes accurate sky subtraction for those spectra. Because those two observations might suffer from contamination from moonlight or other background light, we removed them from our analysis. We jointly fit the remaining 19 HIRES RVs and 30 HARPS-N RVs, fitting 49 RVs in total. The 12 yr RV baseline has rms ∼ 11 m s$^{-1}$. The most recent RV (from 2022) is consistent with either a moderate upward curvature or no trend, yielding an upper limit of $M \sin i < 0.33 M_J$ at 5 au ($M \sin i < 1.3 M_J$ at 10 au) with $3\sigma$ confidence. An analysis that combines the HIRES RVS, HARPS-N RVs, Kepler TTVs, and ground-based TTVs is beyond the scope of this paper but would likely lead to further dynamical insights about the Kepler-9 system.

### A.46. KOI-623 (Kepler-197)

Kepler-197 (KOI-623) is a $V = 11.8$ a slowly rotating, metal-poor Sunlike star ($T_{\rm eff} = 5980$ K, [Fe/H] $= -0.6 \pm 0.1$; Fulton & Petigura 2018). It has four transiting planets that range in size from 0.9 to 1.2 $R_\oplus$. The planets were statistically validated based on their multiplicity in Lissauer et al. (2014) and Rowe et al. (2014). Their compact architecture ($P_b = 5.60$ days, $P_c = 10.3$ days, $P_d = 15.7$ days, $P_e = 25.2$ days) ought to generate TTVs, although Van Eylen & Albrecht (2015) did not detect any, perhaps due to the small planet sizes (which produce small single-transit S/Ns). Hadden & Lithwick (2014) were able to constrain the mass of planet c: $M_c = 5.3 \pm 3.1\,M_\oplus$. Updated long-cadence transit times were published in Holczer et al. (2016) and Gajdoš et al. (2019).

This system was selected for follow-up as part of a program to survey stars with at least three transiting planets from 2015 onward. We have collected 11 Keck-HIRES RVs that extend from 2014 to 2021 (Figure 53). Assuming typical masses for the planet sizes (Weiss & Marcy 2014), the residual RVs have rms = 9.0 m s$^{-1}$, which is high for a Sunlike star, although the low metallicity might impact RV precision. The high rms is driven by two outliers in 2019. There is no significant peak in the periodogram (FAP = 0.99) and no apparent RV trend, yielding a mass upper limit of $M \sin i < 0.7 M_J$ at 5 au ($M \sin i < 2.7 M_J$ at 10 au).

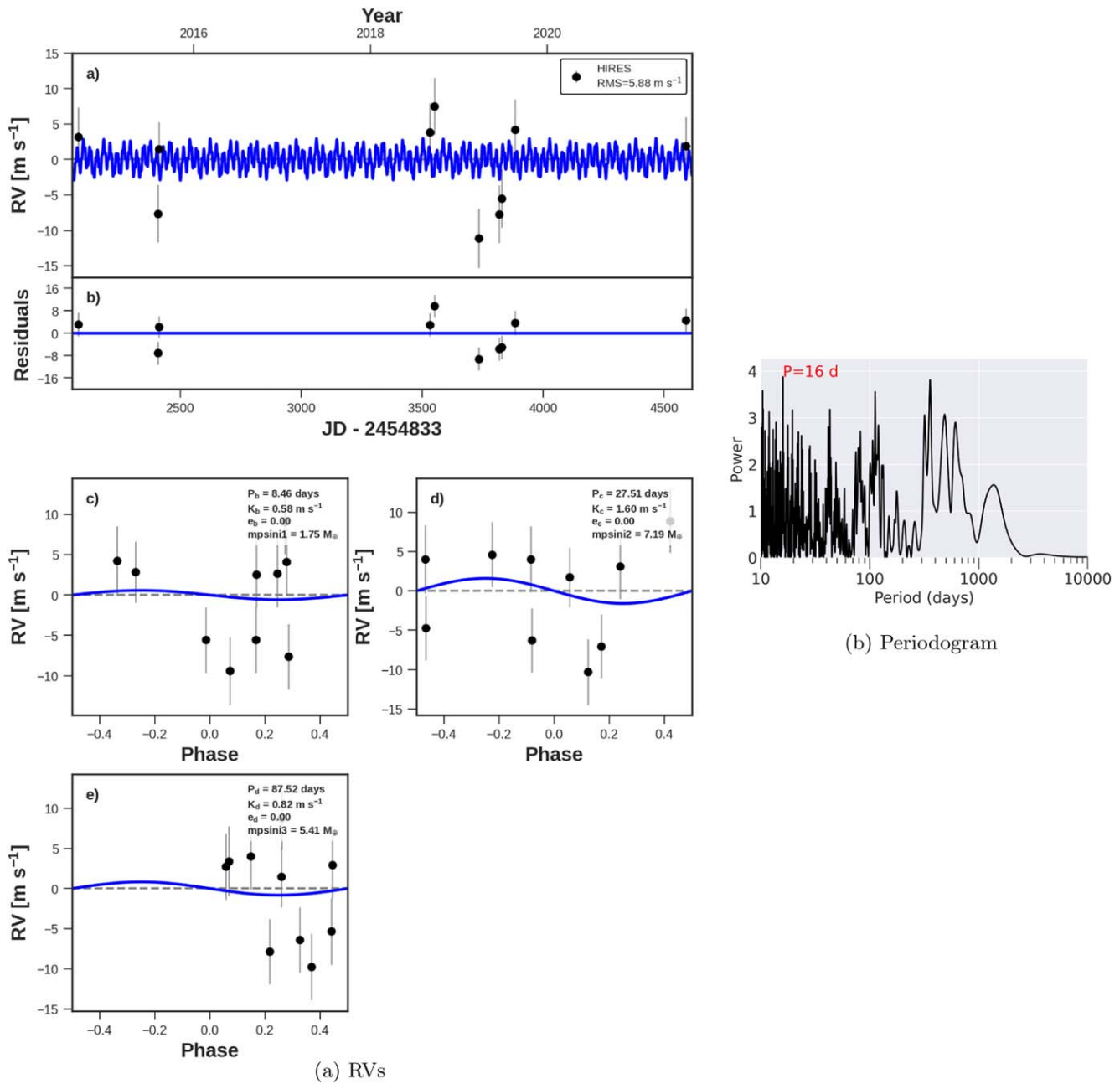

Figure 40. Same as Figure 8, but for KOI-282 (Kepler-130). No nontransiting companions are detected.

### A.47. KOI-701 (Kepler-62)

Kepler-62 (KOI-701) is a K-type star with five transiting planets, including an Earth-sized planet in the habitable zone. This is one of the only Earth-sized habitable planets detected around a G- or K-type star from the Kepler mission. The planets were first announced in Borucki et al. (2013), which included an analysis of 13 Keck-HIRES RVs obtained over a baseline of 128 days in 2012. The RVs were used to place upper limits on the planet masses, although the planets are so small that the mass upper limits are uninformative about the planet compositions. The system also has significant TTVs, which generally yielded less informative mass upper limits than the RVs did (Borucki et al. 2013, Table S4).

The host star is faint ($V = 13.7$), which makes RV follow-up challenging. We have collected five RVs on Keck-HIRES since the publication of Borucki et al. (2013), extending the RV baseline to 5 yr (Figure 54). The new RVs do not substantially improve the planet mass upper limits, but they do constrain the parameter space of possible giant planets in the system. Assuming that the transiting planets have typical masses for their sizes (Weiss & Marcy 2014), the RV residuals have rms ∼ 11 m s$^{-1}$ and no apparent trend, yielding an upper limit of $M \sin i < 0.6 M_J$ at 5 au ($M \sin i < 2.5 M_J$ at 10 au, 3σ conf.).

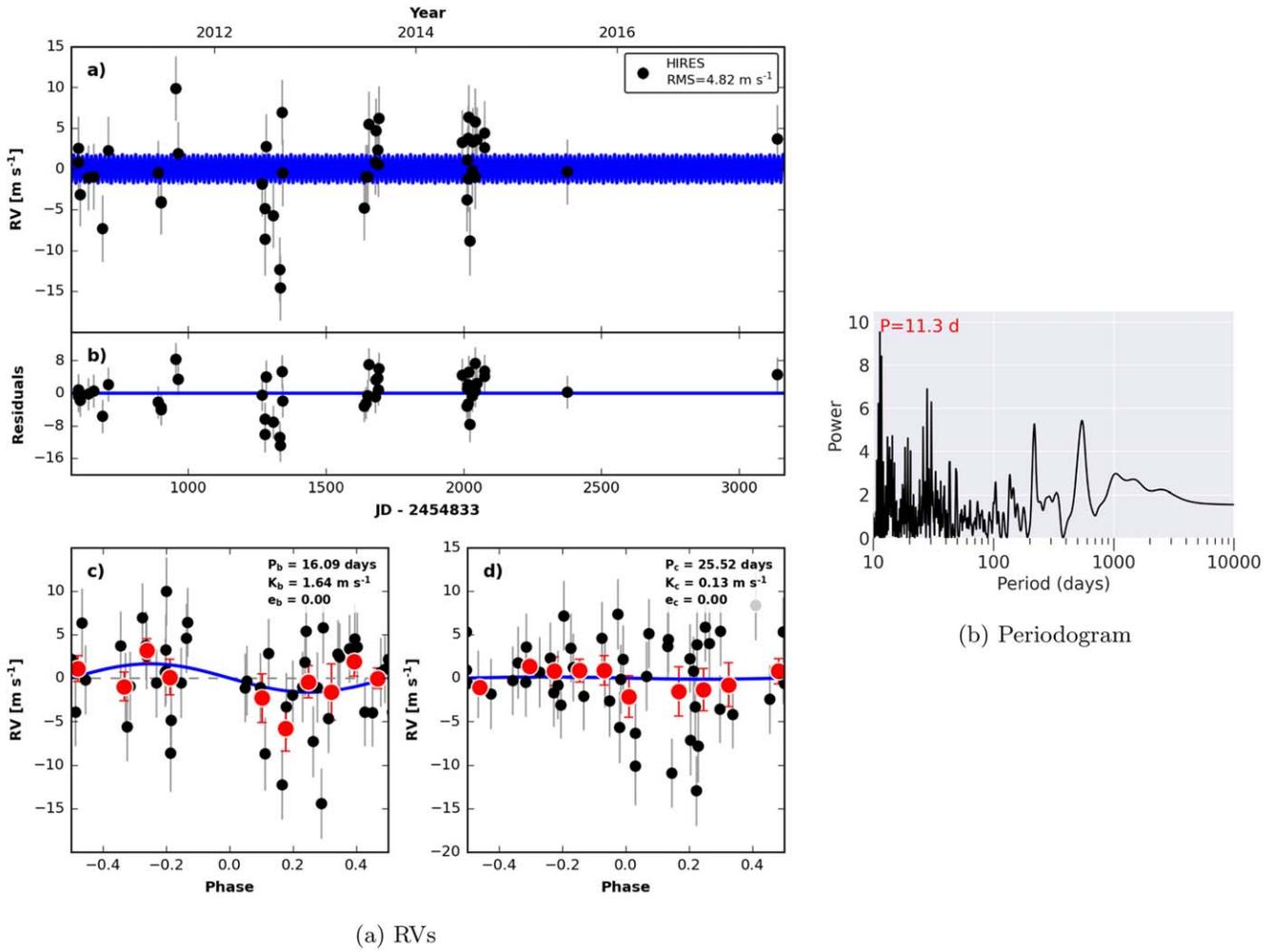

**Figure 41.** Same as Figure 8, but for KOI-283 (Kepler-131). No nontransiting companions are detected.

### A.48. KOI-719 (Kepler-220)

Kepler-220 (KOI-719) is a quiet, K-type star ($T_{\rm eff} = 4500$ K, [Fe/H] = 0.08, $v \sin i = 2$ km s$^{-1}$; Fulton & Petigura 2018). It has four transiting planets in a compact architecture, with two pairs of planets each near a 2:1 mean motion resonance, which were statistically validated (Lissauer et al. 2014; Rowe et al. 2014). The planet radii range from 0.8 to 1.6 $R_\oplus$.

We have collected 10 Keck-HIRES RVs between 2011 and 2021, with a gap in coverage between 2011 and 2019 (Figure 55). Assuming typical masses for the planet radii (Weiss & Marcy 2014), the residual RVs have rms = 5.4 m s$^{-1}$. The lack of an apparent RV trend yields an upper limit of $M \sin i < 0.3 M_J$ at 5 au ($M \sin i < 1.3 M_J$ at 10 au).

### A.49. KOI-1241 (Kepler-56)

Kepler-56 (KOI-1241) is the first transiting multiplanet system that was discovered to be misaligned by at least 40° with respect to its host star (Huber et al. 2013). This discovery was based on significant asteroseismology of the star, especially mode splitting that revealed the projected inclination of the star relative to the sky plane. The host star is evolved ($T_{\rm eff} = 4800$ K, log $g$ = 3.3; Huber et al. 2013). This system fails two criteria of the KGPS survey: (1) the star's gravity is too low, and (2) because the obliquity of the transiting planet orbits was suspected to be driven by a massive perturber, this system fails the "giant-blind" criterion.

The two transiting planets are near the 2:1 mean motion resonance, which generates TTVs ($P_b = 10.5$ days, $P_c = 21.4$ days; Steffen et al. 2013). RVs collected at Keck-HIRES had a trend indicative of a massive perturber (Huber et al. 2013). Continued RV monitoring with TNG-HARPS-N revealed the orbit and planetary mass of the perturber ($P_d = 1002 \pm 5$ days, $M \sin i_d = 5.6 M_J \pm 0.4 M_J$, $e_d = 0.2 \pm 0.01$; Otor et al. 2016). A slight misalignment of the massive perturber drives nodal precession of the transiting planets.

We have collected 29 Keck-HIRES RVs, including three since the publication of Huber et al. (2013). Our most recent RV extends the HIRES baseline from 2012 to 2022 (Figure 56). The combined HIRES and HARPS-N RVs yield $M \sin i_d = 5.9 M_J \pm 0.7 M_J$, $P_d = 1000.1 \pm 3.5$ days, $e_d = 0.21 \pm 0.03$. The residual RVs have low rms (5.1 m s$^{-1}$ on HARPS-N, 5.5 ms on HIRES). There is a significant trend in the residual RVs ($dv/dt = -0.007 \pm 0.002$ m s$^{-1}$ day$^{-1}$), corresponding to $M \sin i = 11 M_J$ at 20 au.

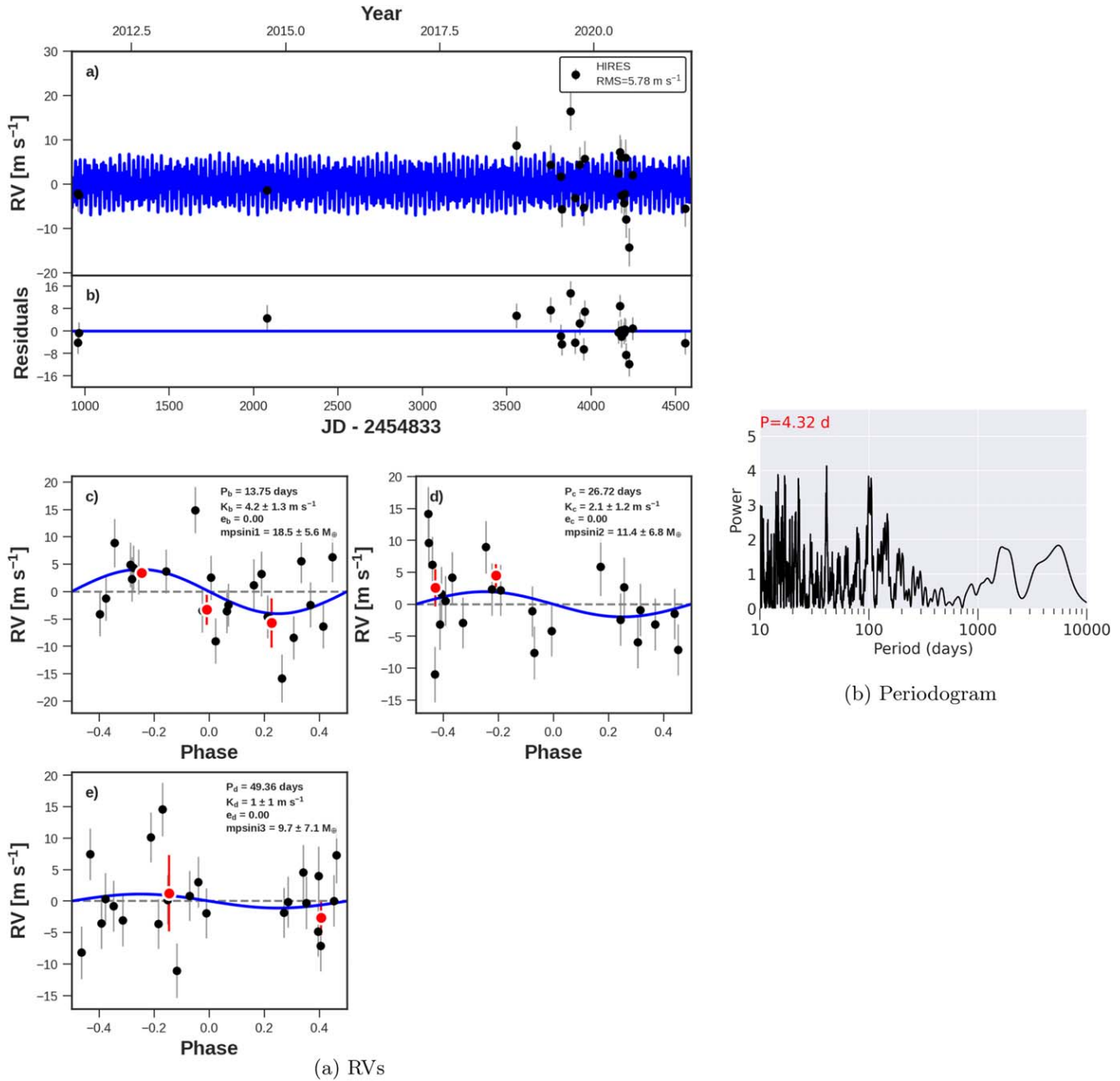

**Figure 42.** Same as Figure 8, but for KOI-285 (Kepler-92). No nontransiting companions are detected.

*A.50. KOI-1442 (Kepler-407)*

Kepler-407 (KOI-1442) is a $V = 12.5$, metal-rich, G-type star ($T_{\rm eff} = 5400$ K, $[{\rm Fe/H}] = 0.4 \pm 0.1$; Fulton & Petigura 2018). It has one transiting, ultra–short-period planet ($P_b = 0.669$ days, $R_b = 1.1\,R_\oplus$) that was confirmed with 17 Keck-HIRES RVs (Marcy et al. 2014). The RVs yielded an upper limit on the mass of $M_b = 0.6 \pm 1.2\,M_\oplus$. The RVs also showed a significant trend with curvature, which yielded preliminary parameters for a massive nontransiting planet or brown dwarf: $P_c = 3000 \pm 500$ days, $M \sin i_c = 4000 \pm 2000\,M_\oplus$.

We have collected 70 Keck-HIRES RVs between 2011 and 2022 (Figure 57). The 11 yr baseline has sampled nearly two full orbits of the companion, yielding $P_c = 2098 \pm 7$ days, $M \sin i_c = 11.2\,M_{\rm J} \pm 0.5\,M_{\rm J}$, and $e_c = 0.22 \pm 0.01$. This corresponds to an object near the boundary between brown-dwarf- and planetary-mass objects (Schlaufman & Winn 2016; Bowler et al. 2019). The additional improved characterization of the long-period companion somewhat tightens the mass constraint of the transiting planet: $M_b = 1.5 \pm 0.9\,M_\oplus$, although more RVs are needed to precisely determine its mass and composition. The residual RVs have rms $= 5.6$ ms and no apparent trend, yielding an upper limit of $M \sin i < 0.2\,M_{\rm J}$ at 5 au ($M \sin i < 0.7\,M_{\rm J}$ at 10 au). A run of high-cadence RVs collected in fall 2022 as part of a student project are not presented here.

*A.51. KOI-1612 (Kepler-408)*

Kepler-408 (KOI-1612, HD 176693) is a $V = 8.8$ F-type star ($T_{\rm eff} = 6100$ K; Fulton & Petigura 2018). It has one transiting planet ($P = 2.47$ days, $R = 0.7$, $R_\oplus$) that was statistically validated (Morton et al. 2016). Keck-HIRES RVs were collected that yielded a mass upper limit of $0.48 \pm 3.2\,M_\oplus$ (Marcy et al. 2014).

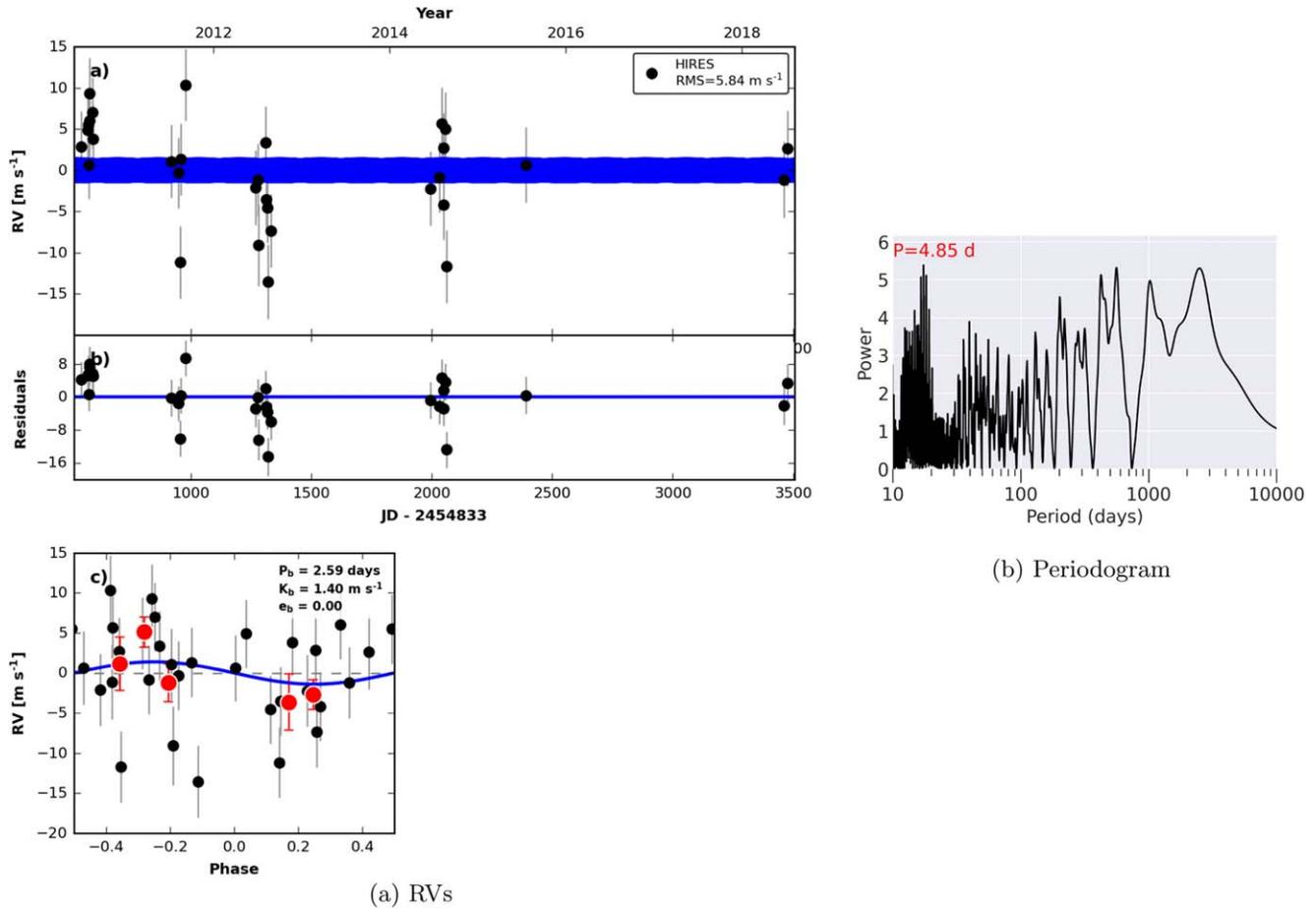

**Figure 43.** Same as Figure 8, but for KOI-292 (Kepler-97). No nontransiting companions are detected.

We have collected 55 total Keck-HIRES RV, including 17 since the publication of Marcy et al. (2014; see Figure 58). The mass upper limit for the transiting planet has not improved substantially. There is a moderate RV trend (FAP = 0.01) over the decade-long baseline. If bona fide, this trend corresponds to a companion of $M \sin i > 11 M_J$ at >20 au. The long baseline and shallow trend also yield upper limits of $M \sin i < 1.3 M_J$ at 5 au ($M \sin i < 5.2 M_J$ at 10 au, 3σ conf.).

### A.52. KOI-1692 (Kepler-314)

Kepler-314 (KOI-1692) is a $V = 12.6$, metal-rich G-type star ($T_{\rm eff} = 5400$ K, [Fe/H] = 0.37; Fulton & Petigura 2018). It has two transiting planets that were statistically validated at $P_b = 2.46$ days and $P_c = 5.96$ days (Lissauer et al. 2014; Rowe et al. 2014). The planet sizes span the radius valley, with $R_b = 0.84\,R_\oplus$ and $R_c = 2.9\,R_\oplus$ (Berger et al. 2018).

We have collected 11 Keck-HIRES RVs between 2011 and 2022, with a gap between 2011 and 2018 (Figure 59). There is an apparent downward trend in the RVs from 2018 to 2022, although the observation in 2011 is not consistent with this trend. It is possible that we are sampling the RV variations from a giant planet near 5 au, or that either the 2011 point or two points from 2021 are outliers. Including an RV trend and curvature term (two free parameters for the giant planet) fits the RVs with rms = 13.1 m s$^{-1}$, whereas a fit with no RV trend or curvature has rms = 13.5 m s$^{-1}$. Based on the stellar properties, we expect this star to have low jitter (5.2 m s$^{-1}$). To test the stellar jitter, we obtained three RVs in a single night; the maximum difference was 2.2 m s$^{-1}$, which is too low to explain the RV variations. More RVs are needed to ascertain the origin of the RV variability.

### A.53. KOI-1781 (Kepler-411)

Kepler-411 (KOI-1781) is a young, active star with four transiting planets summarized in Sun et al. (2019). Because of its high activity, KOI-1781 is not considered part of the KGPS sample. There are three transiting planets and one nontransiting planet in the system, with orbital periods of 3, 7.8, 58, and 31.5 days, respectively. All four planets have mass measurements from TTVs, with the 31.5 day planet having its presence known only through the gravitational interaction with the 58 day planet. The planets labels moving outward in orbital period are Kepler-411 b, c, e, and d.

This system was selected for follow-up as part of a program to survey stars with at least three transiting planets from 2015 onward. The star has high activity (log $R'_{\rm HK}$ value of $-4.3$), and so we expect stellar jitter to dominate the error budget, which is consistent with the observed rms of the RVs of 24 m s$^{-1}$ (Figure 60). We do not detect any of the four transiting planets in the RVs, but our RVs are consistent with the masses determined from TTVs. The relationship between RV jitter and stellar activity by Hillenbrand et al. (2015) shows that all 24 m s$^{-1}$ of RV jitter can be attributed to the activity of the star (rather than planets) and the age of the system is between 60 and 300 Myr, consistent with the age of $212 \pm 31$ Myr determined by Sun et al. (2019) based on the rotation period and the color index. The 8 yr RV baseline does not have a significant trend, yielding a $3\sigma$ upper limit of $M \sin i < 4.5 M_J$ at 10 au. Because the rms of the RVs is quite high, the nondetection of a trend is not particularly informative about lower-mass planets closer to the star—for instance, a Jupiter-mass planet at 5 au would produce an amplitude of roughly 10 m s$^{-1}$, which could easily be hidden in the 24 m s$^{-1}$ rms of the RVs.

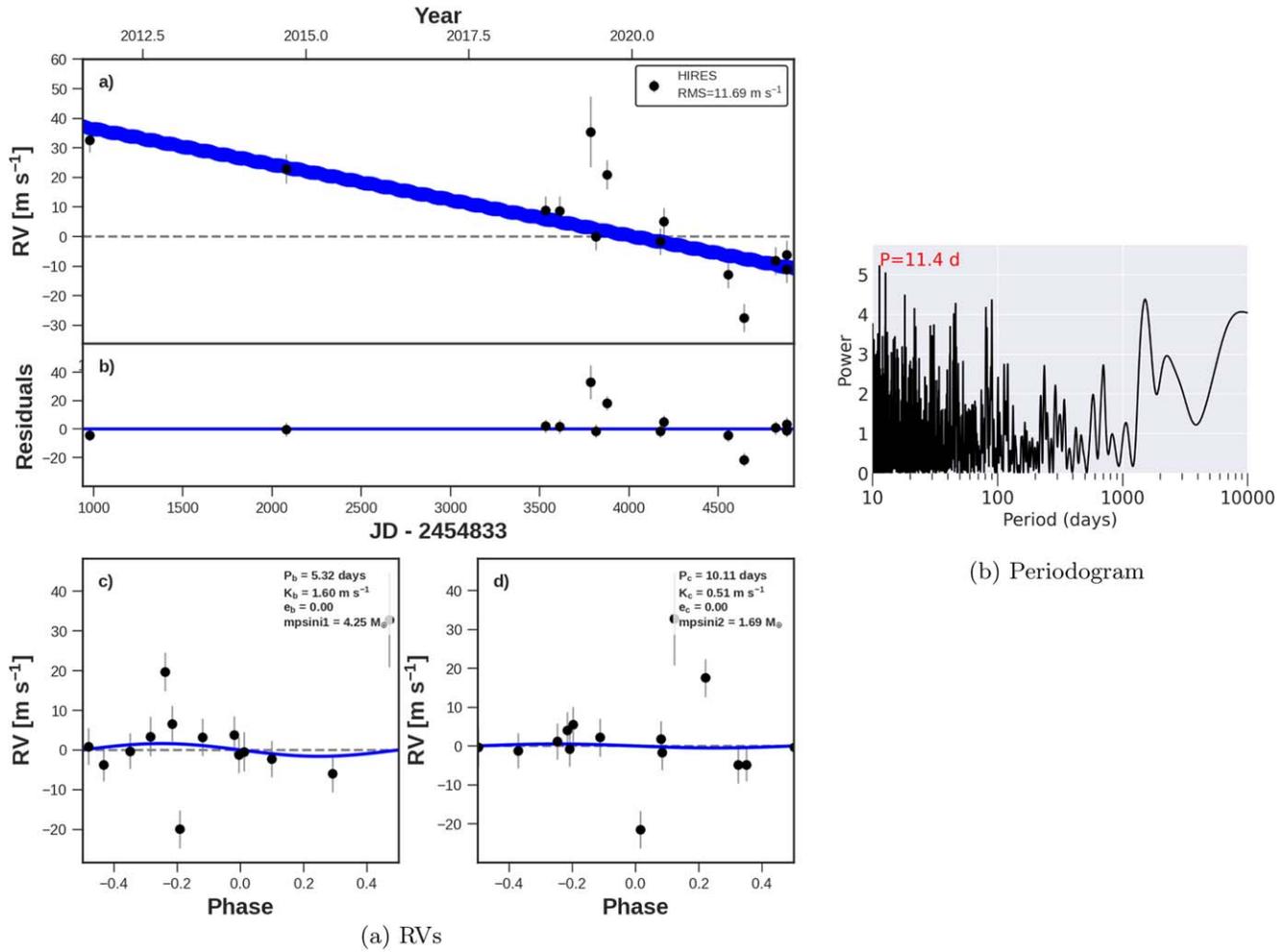

**Figure 44.** Same as Figure 8, but for KOI-295 (Kepler-134). Left: there is a marginally detected trend in the RVs ($\Delta$BIC = –54).

### A.54. KOI-1909 (Kepler-334)

Kepler-334 (KOI-1909) is a quiet, Sunlike star ($T_{\rm eff} = 5800$ K; Fulton & Petigura 2018). It has three super-Earth-sized ($R_p < 1.5 R_\oplus$) transiting planets that were statistically validated at periods of $P_b = 5.47$ days, $P_c = 12.7$ days, and $P_d = 25.1$ days (Lissauer et al. 2014; Rowe et al. 2014). The outer two planets are near the 2:1 mean motion resonance, and their transit midpoints are reported in Holczer et al. (2016). This system was selected for follow-up as part of a program to survey stars with at least three transiting planets from 2015 onward. We have collected 11 Keck-HIRES RVs between 2015 and 2021 (Figure 61). Assuming that the planets have typical masses for their sizes (Weiss & Marcy 2014), the residual RVs have rms $= 8.9$ m s$^{-1}$ and no apparent RV trend over the 6 yr baseline, yielding a $3\sigma$ upper limit of $M \sin i < 0.7 M_J$ at 5 au ($M \sin i < 2.7 M_J$ at 10 au).

### A.55. KOI-1925 (Kepler-409)

Kepler-409 (KOI-1925) is a $V = 9.4$ quiet, Sunlike star ($T_{\rm eff} = 5400$ K; Fulton & Petigura 2018). It has one transiting planet ($P_b = 70.0$ days, $R_b = 1.2 R_\oplus$) that was confirmed with Keck-HIRES RVs (Marcy et al. 2014). We have collected 97 RVs between 2012 and 2022, including 57 since 2013 (Figure 62). There is significant structure in the fast periodogram, with peaks near 340 and 3000 days (FAP = 0.001), although this structure is primarily driven by high-cadence observations in 2012 and 2014, which is not optimal sampling for detecting long-period signals. Including a Keplerian orbit at 340 days yields $\Delta$BIC $= -37$ and would correspond to a companion of $M \sin i = 40 M_\oplus$ at this orbital period. However, the 340 day signal has high eccentricity ($e = 0.4$) and is likely an alias of a signal at 3000 days, which would correspond to a companion of $M \sin i = 230 M_\oplus$. More RVs are needed to validate the planet candidate and determine its orbit and mass. Because the NEA default transit ephemeris was from Marcy et al. (2014), whereas a more updated ephemeris that accounts for TTVs is available, we tested whether using the Holczer et al. (2016) ephemeris would affect the candidate giant planet. The change in the residual RVs was negligible, and the peaks at 340 and 3000 days persisted. More RVs are needed to clarify whether the long-period signal is from an additional planet and, if so, what its orbital period is.

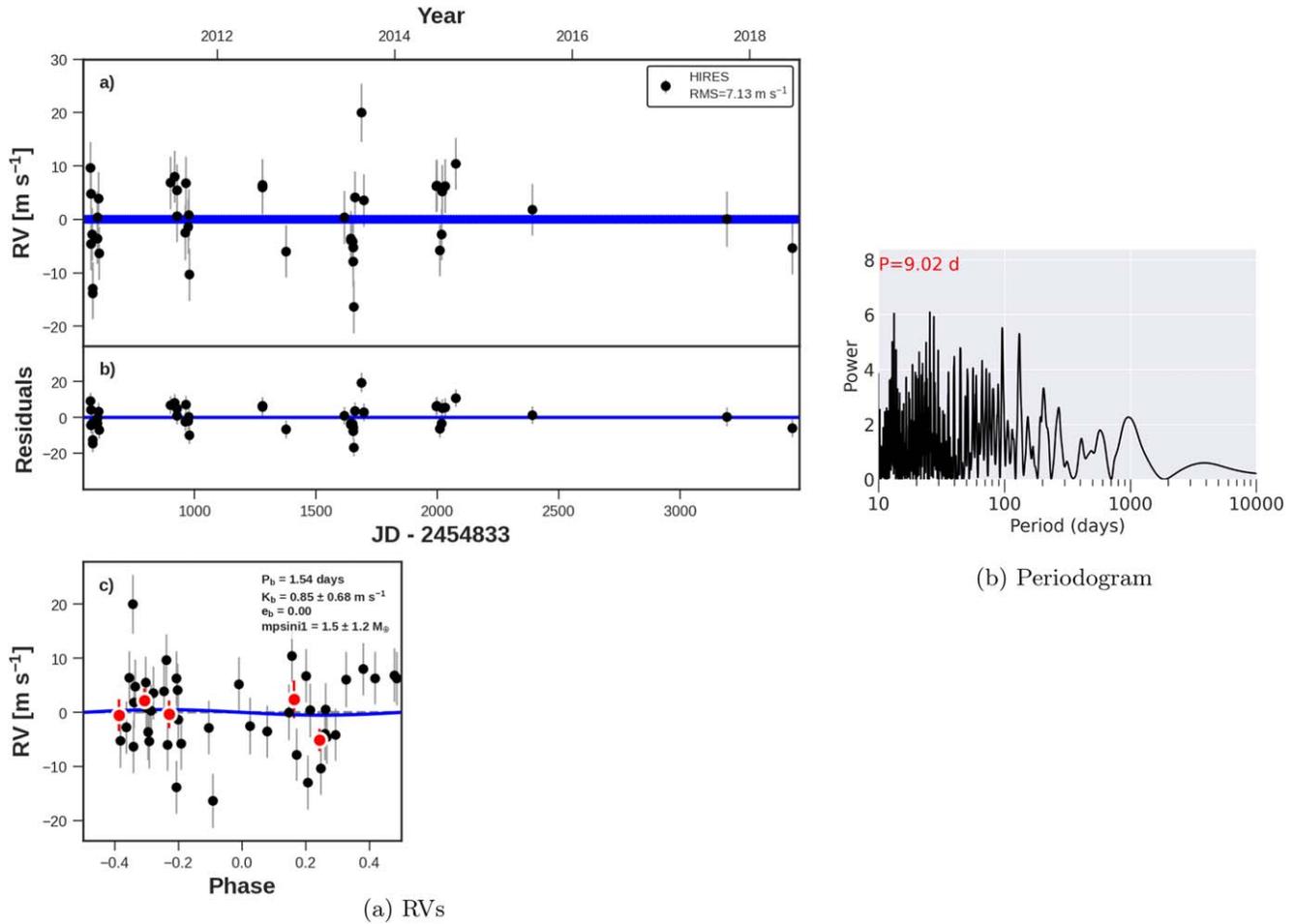

Figure 45. Same as Figure 8, but for KOI-299 (Kepler-98). No nontransiting companions are detected.

### A.56. KOI-1930 (Kepler-338)

Kepler-338 (KOI-1930) is a $V = 12.2$, slightly evolved G-type star with moderate rotation ($T_{\rm eff} = 5900$ K, $\log g = 4.0$, $v \sin i = 4.0$ km s$^{-1}$; Fulton & Petigura 2018). It has three sub-Neptune-sized transiting planets that were statistically validated (Lissauer et al. 2014; Rowe et al. 2014). The planets have TTVs, which were used to characterize the orbit of a fourth planet that also transits and is smaller than its neighbors ($P_e = 9.3$ days, $R_e = 1.6\,R_\oplus$; Hadden & Lithwick 2014).

We have collected 11 Keck-HIRES RVs between 2014 and 2021 (Figure 63). Assuming that the planets have masses that are typical for their radii, we find a residual RV scatter of 8.8 m s$^{-1}$. This RV scatter is consistent with the jitter we expect for a moderately evolved, moderately rotating star. We find no significant RV trend, yielding an upper limit of $M \sin i < 1\,M_{\rm J}$ at 5 au ($M \sin i < 3.6\,M_{\rm J}$ at 10 au). There is no significant structure in the fast periodogram.

### A.57. KOI-2169 (Kepler-1130)

Kepler-1130 (KOI-2169) is a $V = 12.4$, quiet G-type star ($T_{\rm eff} = 5400$ K; Fulton & Petigura 2018). It has four transiting planet candidates that range from 0.3 to 0.8 $R_\oplus$ (Morton et al. 2016; Valizadegan et al. 2022). Although only three of the transiting planet candidates are confirmed, including the fourth planet candidate in the RV model would not significantly influence the overall RV fit, since its small size implies a small mass. The host star also has a stellar companion at $\rho = 3\rlap{.}''5$ with a Gaia $G$-band magnitude of 17.0 that is comoving based on Gaia DR2 (Mugrauer 2019). At a distance of 250 pc, the physical separation from the primary is approximately 875 au. The Gaia data suggest that the two stars are associated, but the $2\rlap{.}''5$ companion is not the cause of the 4 km s$^{-1}$ RV signal, which has a period near 43 yr.

This system was selected for follow-up as part of a program to survey stars with at least three transiting planets from 2015 onward. We have collected 34 Keck-HIRES RVs between 2012 and 2022 (Figure 64). This decade-long baseline has a change in RV of 4000 m s$^{-1}$, consistent with the perturbation of a stellar-mass companion, which we call Kepler-1130 B. The RV time series captures both a concave-up and concave-down inflection, which likely correspond to the two quadrature times of the companion. The KGPS algorithm did not find a good fit to the orbital period of this long-period, eccentric companion. We used rvsearch to identify a period, eccentricity, and mass of a companion to use as an initial guess that we then optimized with radvel. Our best fit to the RVs yields an rms of 6.5 m s$^{-1}$ based on the inclusion of a companion at $P_B = 15{,}700 \pm 800$ days (∼12 au), $M \sin i_B = 218 M_J \pm 2 M_J$, $e_B = 0.67 \pm 0.01$. For completeness, we explored adding a linear trend to the RVs in addition to a modeled companion. The inclusion of an RV trend increases the uncertainty in the orbital period of the companion but does not significantly affect the mass or mass uncertainty of the companion.

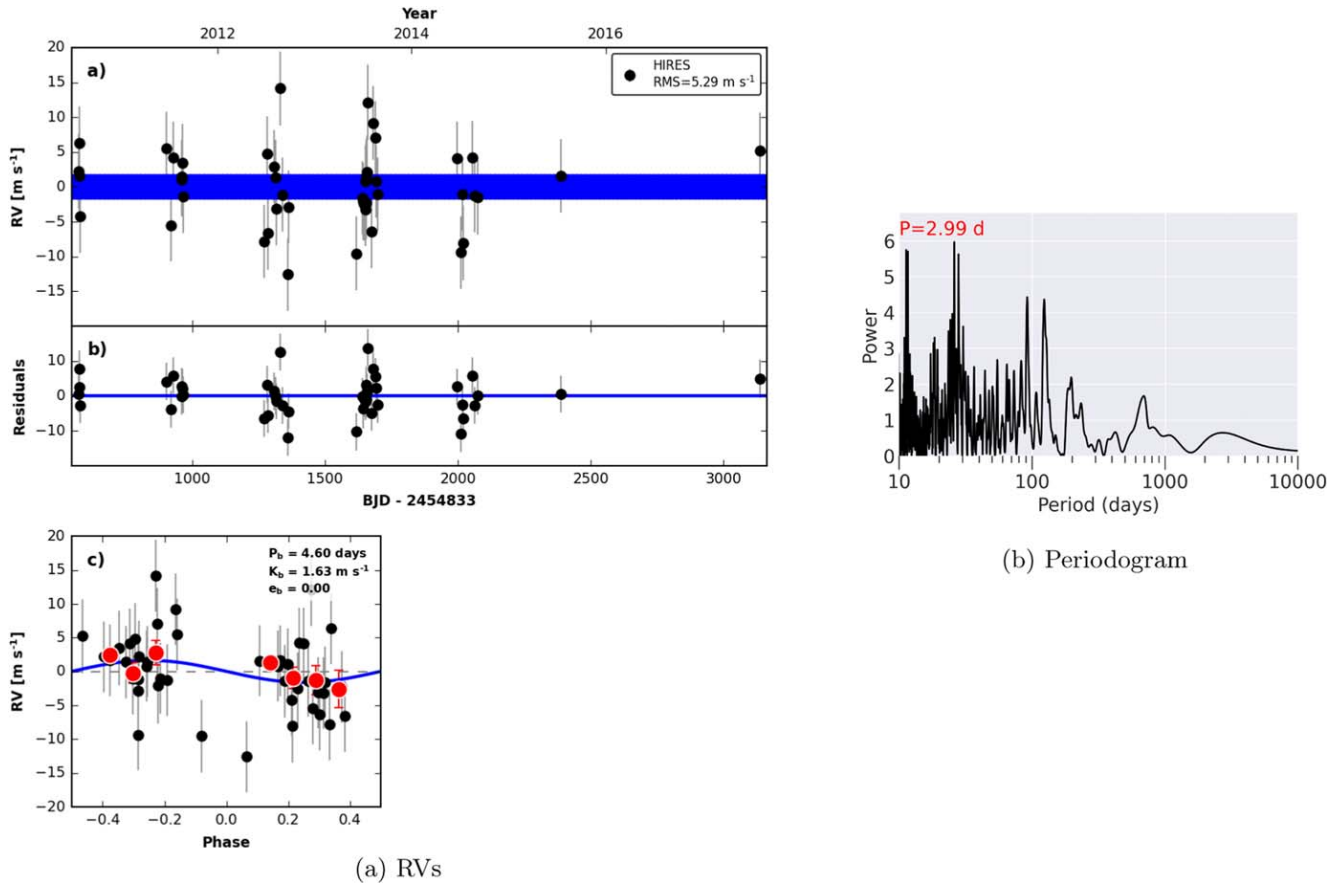

**Figure 46.** Same as Figure 8, but for KOI-305 (Kepler-99). No nontransiting companions are detected.

The periastron passage of Kepler-1130 B is ∼4 au from Kepler-1130 A, and so the transiting planets likely formed from a substantially truncated disk. Because the observational baseline is small compared to the orbital period of Kepler-1130 B, there is substantial uncertainty in any residual RV trend, allowing additional companions of $M \sin i < 5 M_J$ at 5 au ($M \sin i < 20 M_J$ at 10 au) when including the uncertainty in the orbital properties of star B. Future RV monitoring of this target is essential to map the orbit of the stellar-mass companion and accurately determine its periastron passage distance.

### A.58. KOI-2687 (Kepler-1869)

Kepler-1869 (KOI-2687) is a $V = 10.3$ Sunlike star ($T_{\rm eff} = 5800$ K; Fulton & Petigura 2018). It has two Earth-sized transiting planet candidates (Burke et al. 2014), one of which was statistically validated ($P_b = 8.17$ days; Valizadegan et al. 2022), and the other of which still has candidate status ($P_{.01} = 1.72$ days). We have collected 26 Keck-HIRES RVs between 2012 and 2021 and do not make a significant detection of the masses (Figure 65). Assuming typical masses for the planet radii (Weiss & Marcy 2014), the residual RVs have rms = 7.3 ms and no apparent trend, yielding an upper limit of $M \sin i < 0.5 M_J$ at 5 au ($M \sin i < 1.9 M_J$ at 10 au).

### A.59. KOI-2720 (KIC 8176564)

KOI-2720 (KIC 8176564) is a $V = 10.3$ Sunlike star ($T_{\rm eff} = 5900$ K; Fulton & Petigura 2018). It has one candidate transiting Earth-sized planet ($R_p = 1.2 R_\oplus$) at $P = 6.57$ days (Batalha et al. 2013). We have collected 22 Keck-HIRES RVs between 2012 and 2018, resulting in a nondetection of the planet mass (Figure 66). Assuming a typical mass for KOI-2720.01 (Weiss & Marcy 2014), the residual RVs have rms = 4.6 m s$^{-1}$ and no apparent trend, yielding an upper limit of $M \sin i < 0.4 M_J$ at 5 au ($M \sin i < 1.5 M_J$ at 10 au).

### A.60. KOI-2732 (Kepler-403)

Kepler-403 (KOI-2732) is a $V = 12.8$ F-type star with moderate rotation rate ($T_{\rm eff} = 6100$ K, $v \sin i = 4$ km s$^{-1}$; Fulton & Petigura 2018). It has three super-Earth-sized transiting planets that were statistically validated (Morton et al. 2016) and one planet candidate (Rowe et al. 2015). This system was selected for follow-up as part of a program to survey stars with at least three transiting planets from 2015 onward. We have collected 11 Keck-HIRES RVs between 2015 and 2021 (Figure 67). Assuming typical masses for the transiting planets, the residual RVs have rms = 9.6 m s$^{-1}$. This moderate RV scatter is common for a star near the Kraft break. There is no apparent RV trend over the 6 yr

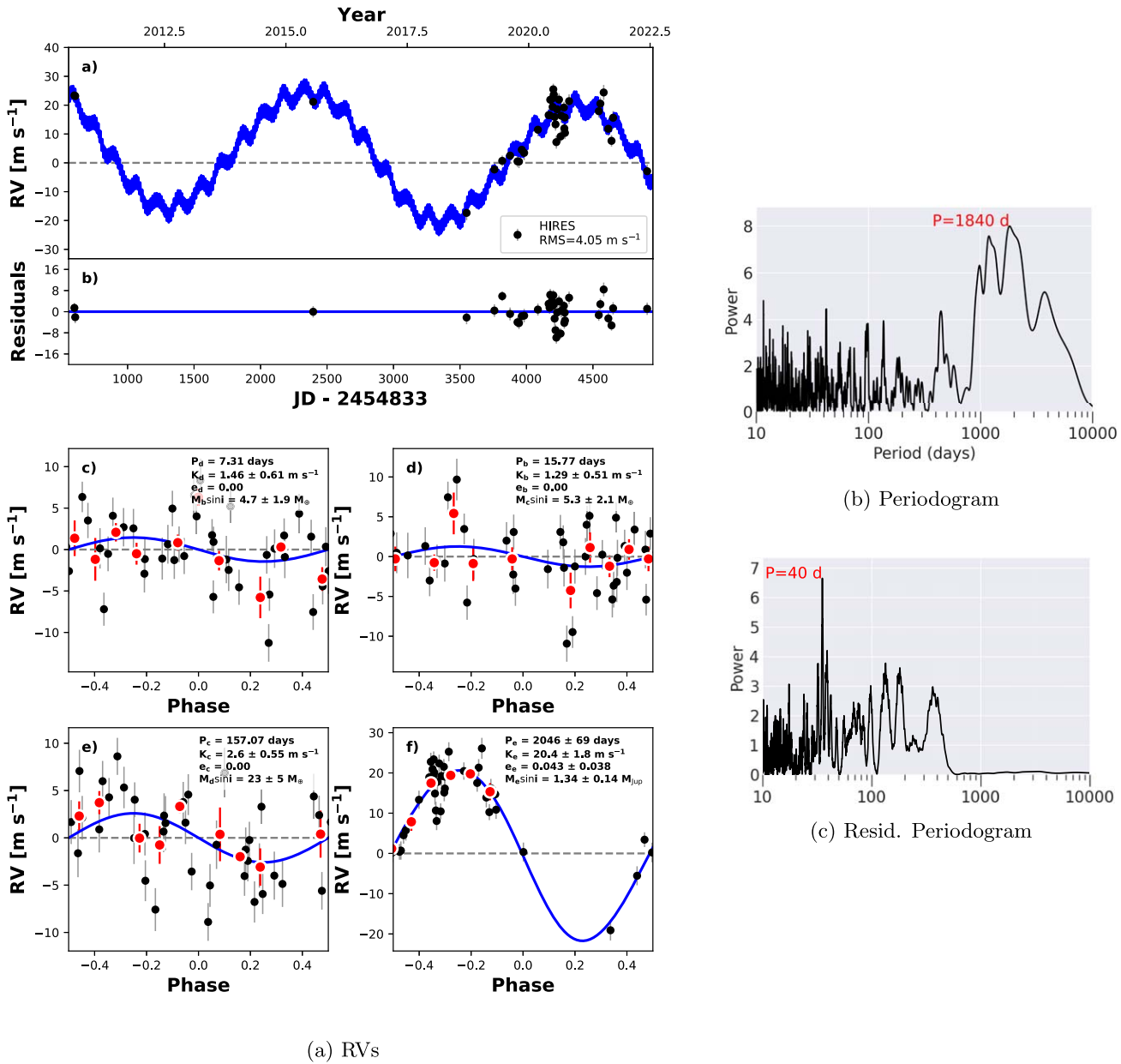

(a) RVs

**Figure 47.** Same as Figure 8, but for KOI-316 (Kepler-139). The best-fit model includes a nontransiting planet at 1800 days, although the automated KGPS algorithm did not detect this planet owing to the unusual time sampling. Right: periodograms of the RVs after removing a model of the transiting planets only (top) and our best-fit model including a giant planet at 1800 days (bottom).

baseline, yielding an upper limit of $M \sin i < 0.7 M_J$ at 5 au ($M \sin i < 3 M_J$ at 10 au).

### A.61. KOI-3083

KOI-3083 (KIC 7106173) is a $V = 12.9$ Sunlike star that is slightly metal-rich ($T_{\rm eff} = 5800$ K, [Fe/H] = 0.3 dex; Fulton & Petigura 2018). It has three sub-Earth-sized transiting planet candidates (Batalha et al. 2013). This system was selected for follow-up as part of a program to survey stars with at least three transiting planets from 2015 onward. We have collected 11 Keck-HIRES RVs between 2015 and 2021 (Figure 68). Assuming typical masses for the transiting planet candidates (Weiss & Marcy 2014), the residual RVs have rms = 6.4 m s$^{-1}$ and no apparent RV trend, yielding an upper limit of $M \sin i < 0.5 M_J$ at 5 au ($M \sin i < 2 M_J$ at 10 au).

### A.62. KOI-3179 (Kepler-1911)

KOI-3179 (Kepler-1911, KIC 6153407) is a quiet, $V = 12.3$ Sunlike star ($T_{\rm eff} = 5700$ K, $v \sin i < 2$ km s$^{-1}$, $\log R'_{\rm HK} = -4.8$; Fulton & Petigura 2018). It has one Earth-sized transiting planet candidate at $P = 6.00$ days that was statistically validated (Valizadegan et al. 2022). We have collected 11 Keck-HIRES RVs between 2013 and 2021, with a gap between 2013 and 2019. The RV have rms = 10.0 m s$^{-1}$, which is high for a quiet, Sunlike star. There is no apparent RV trend, yielding an upper limit of $M \sin i < 0.6 M_J$ at 5 au ($M \sin i < 2.4 M_J$ at 10 au), although this nondetection is primarily driven by a single RV in 2013 (Figure 69). There are no significant peaks in the fast periodogram.

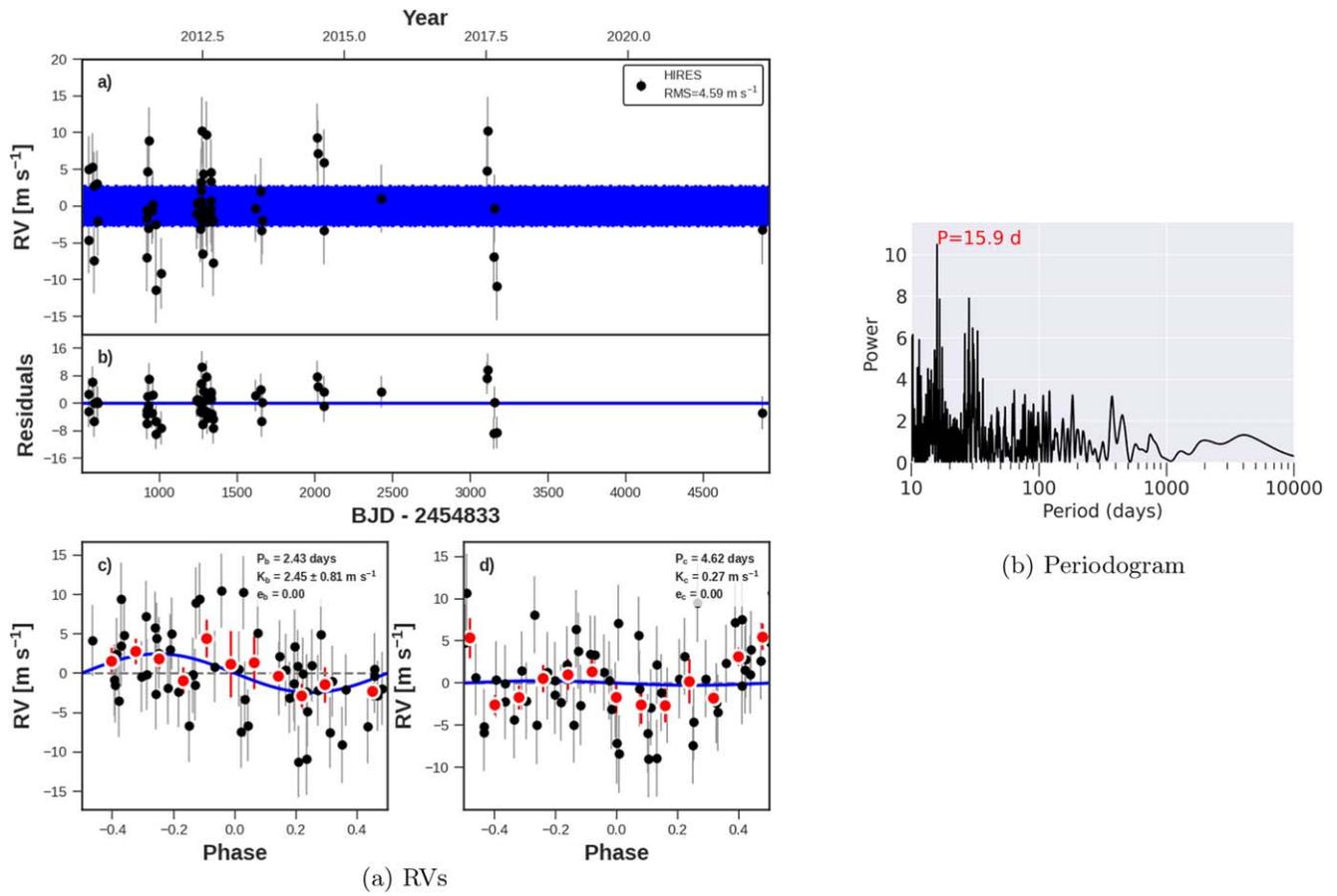

Figure 48. Same as Figure 8, but for KOI-321 (Kepler-406). No nontransiting companions are detected.

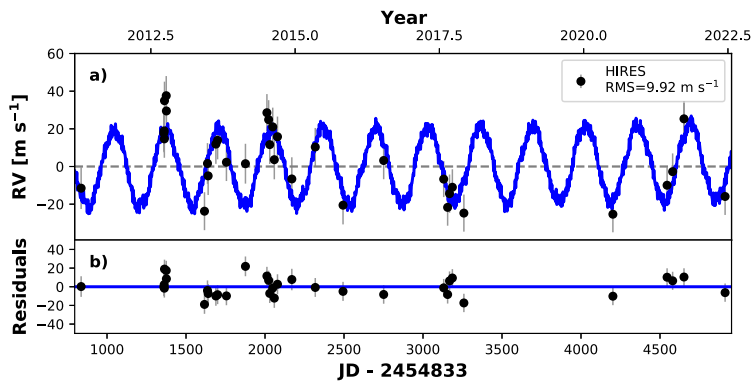
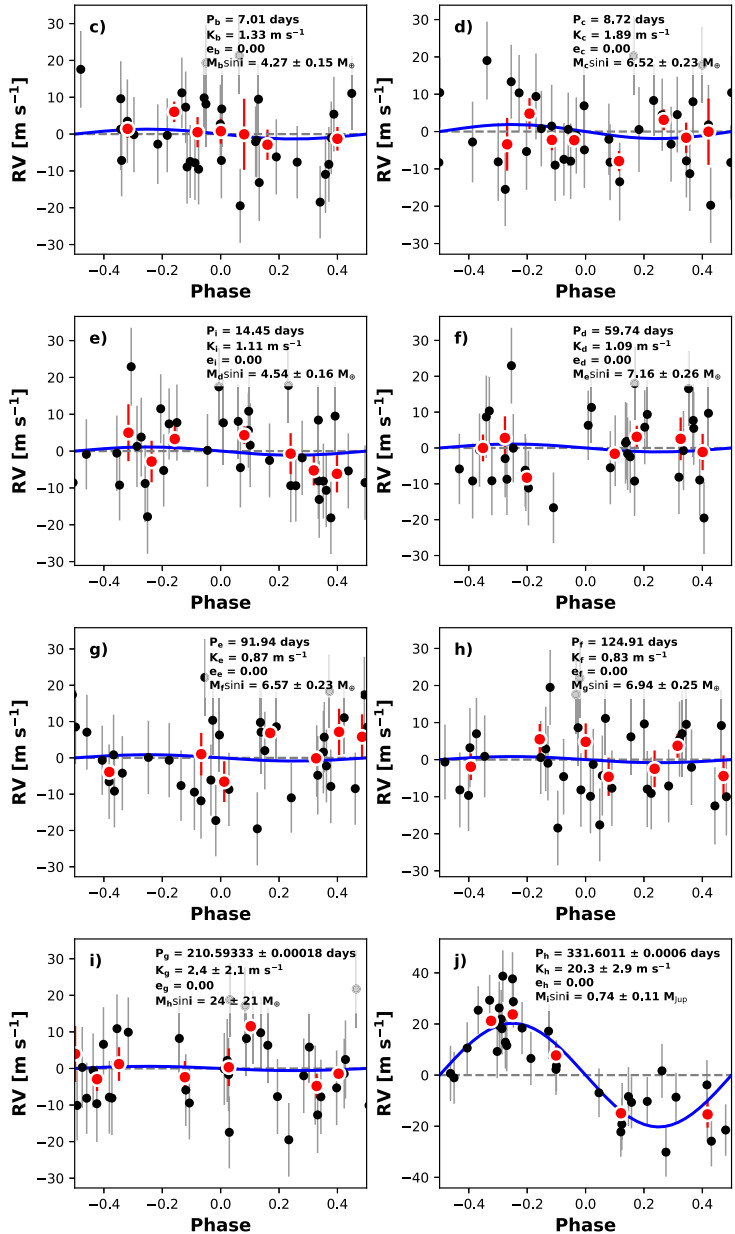
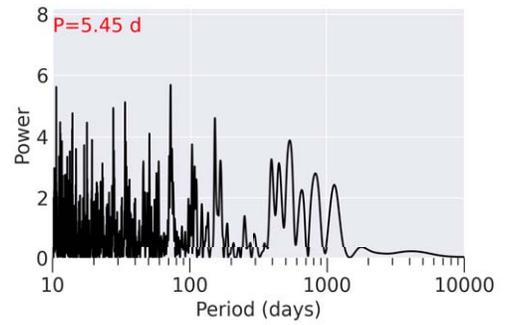

(b) Periodogram

(a) RVs

**Figure 49.** Same as Figure 8, but for KOI-351 (Kepler-90). Left: this system has eight transiting planets, of which only the most massive is detected in the RVs. Right: periodogram of the RVs after removing the best model with all transiting planets. No nontransiting companions are detected.

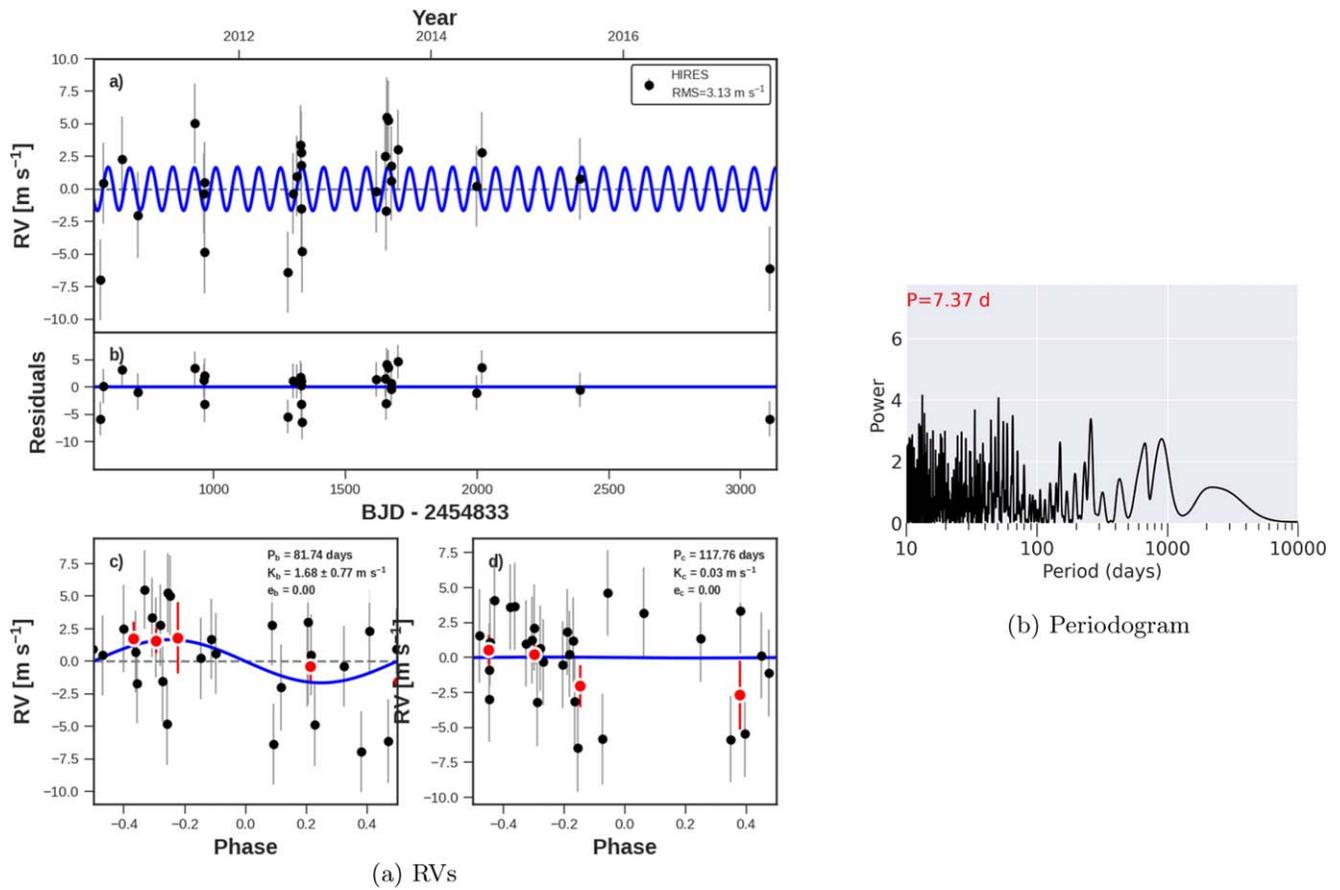

**Figure 50.** Same as Figure 8, but for KOI-365 (Kepler-538). No nontransiting companions are detected.

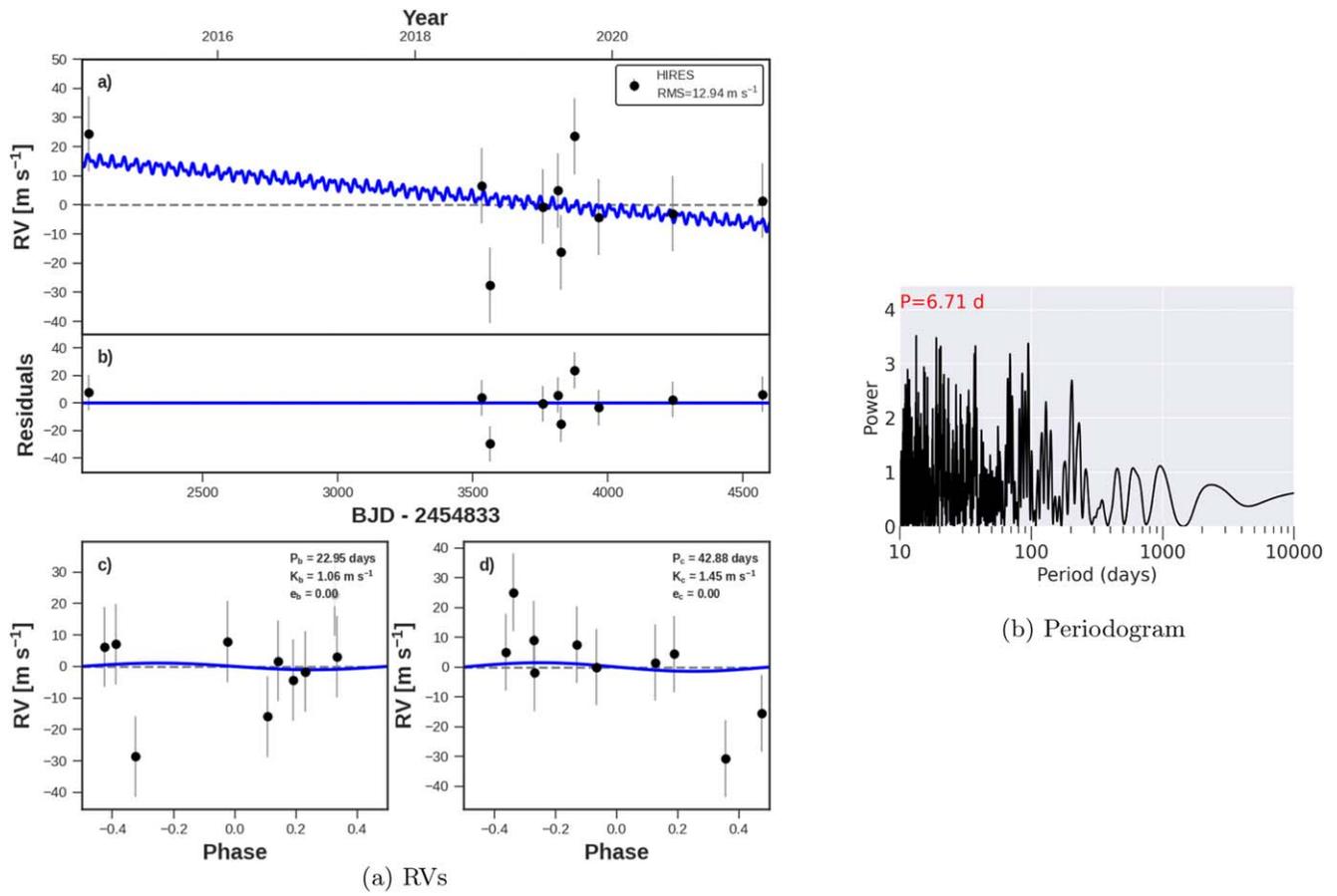

**Figure 51.** Same as Figure 8, but for KOI-370 (Kepler-145). No nontransiting companions are detected.

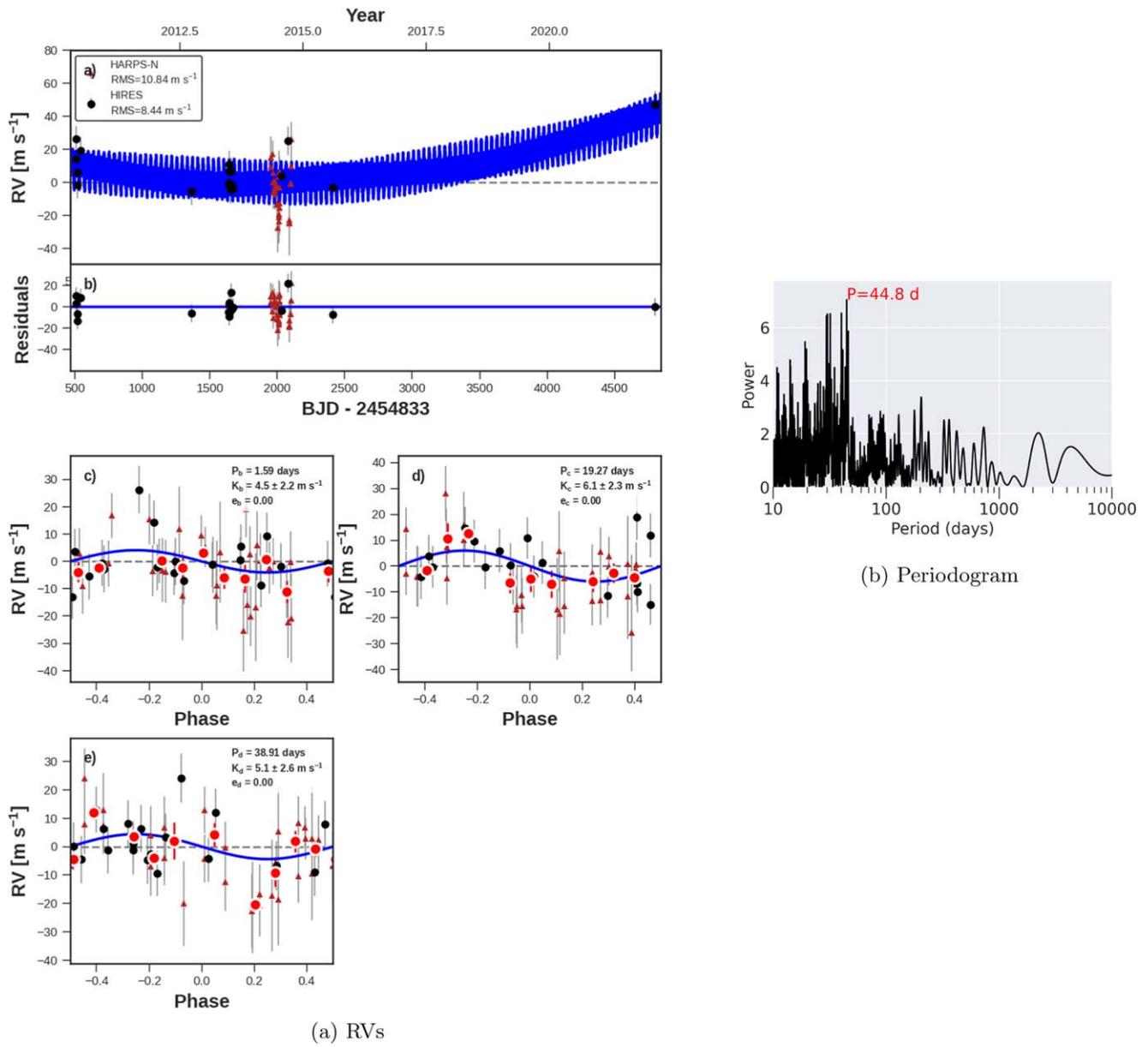

**Figure 52.** Same as Figure 8, but for KOI-377 (Kepler-9). Left: RVs are from HIRES (black dots) and HARPS-N (maroon triangles). The most recent RV is consistent with a moderate upward trend and/or curvature.

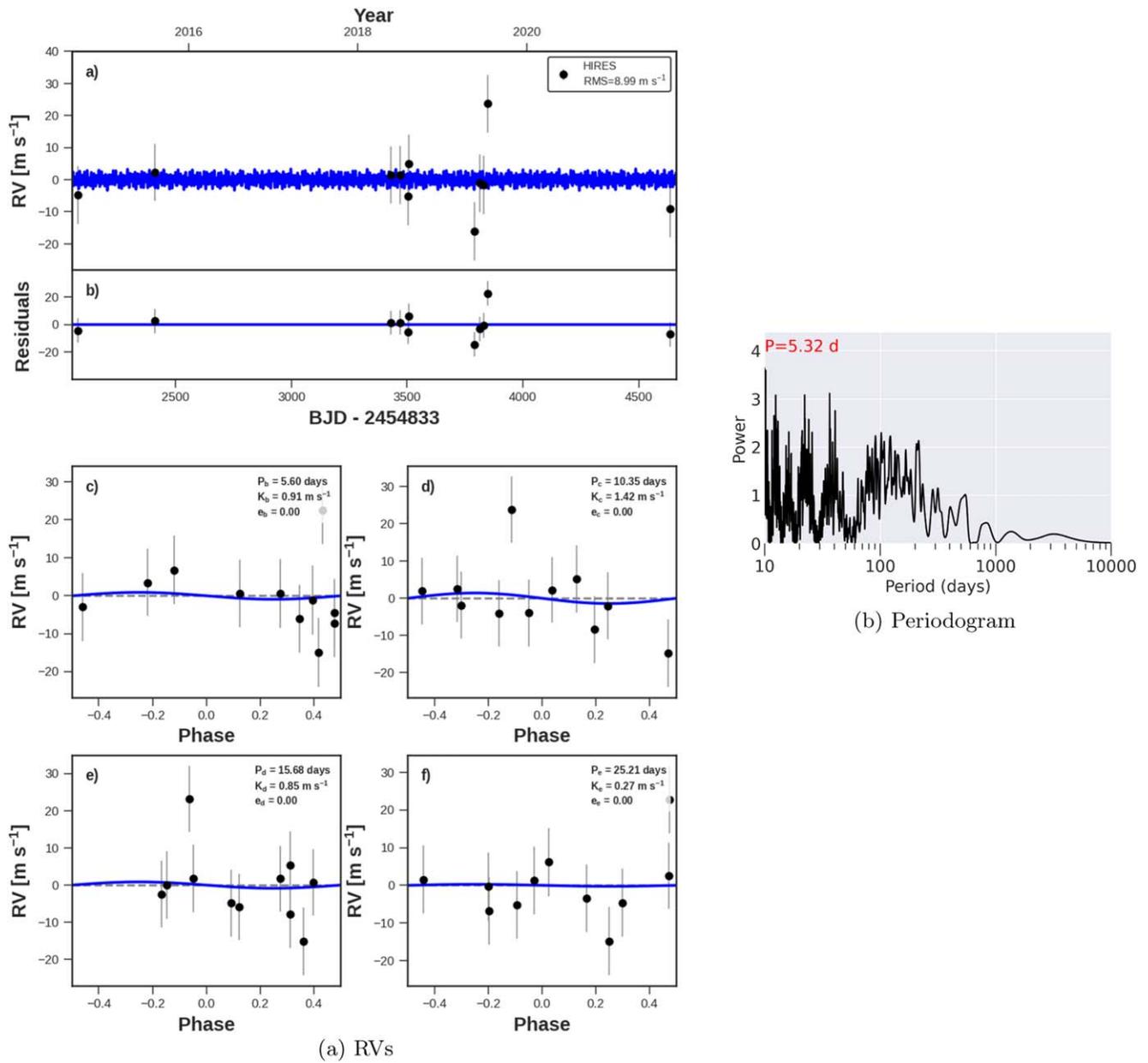

**Figure 53.** Same as Figure 8, but for KOI-623 (Kepler-197). No nontransiting companions are detected.

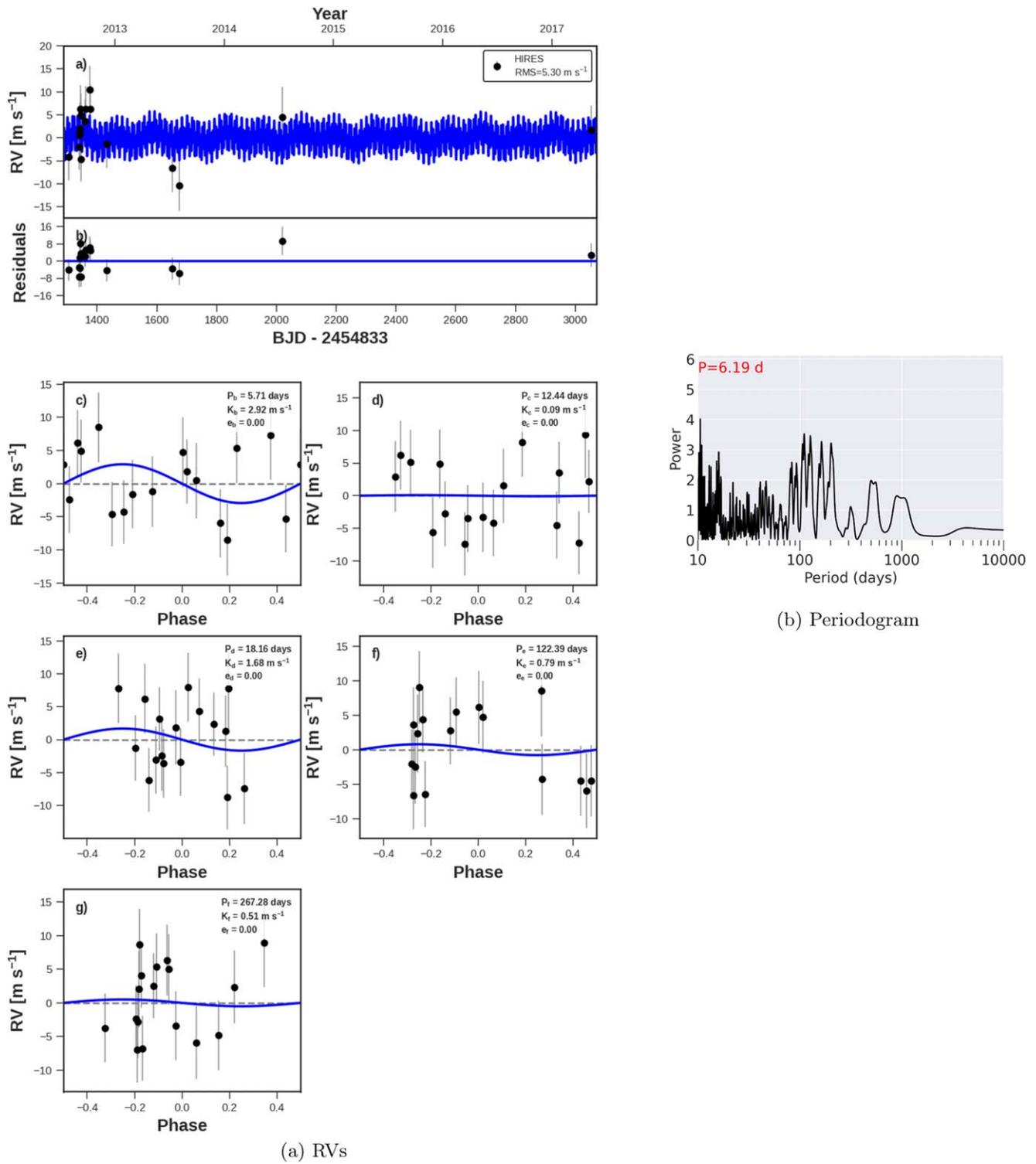

**Figure 54.** Same as Figure 8, but for KOI-701 (Kepler-62). Left: this system has five transiting companions, for which the RVs provide only upper limits (not detections). No nontransiting companions are detected.

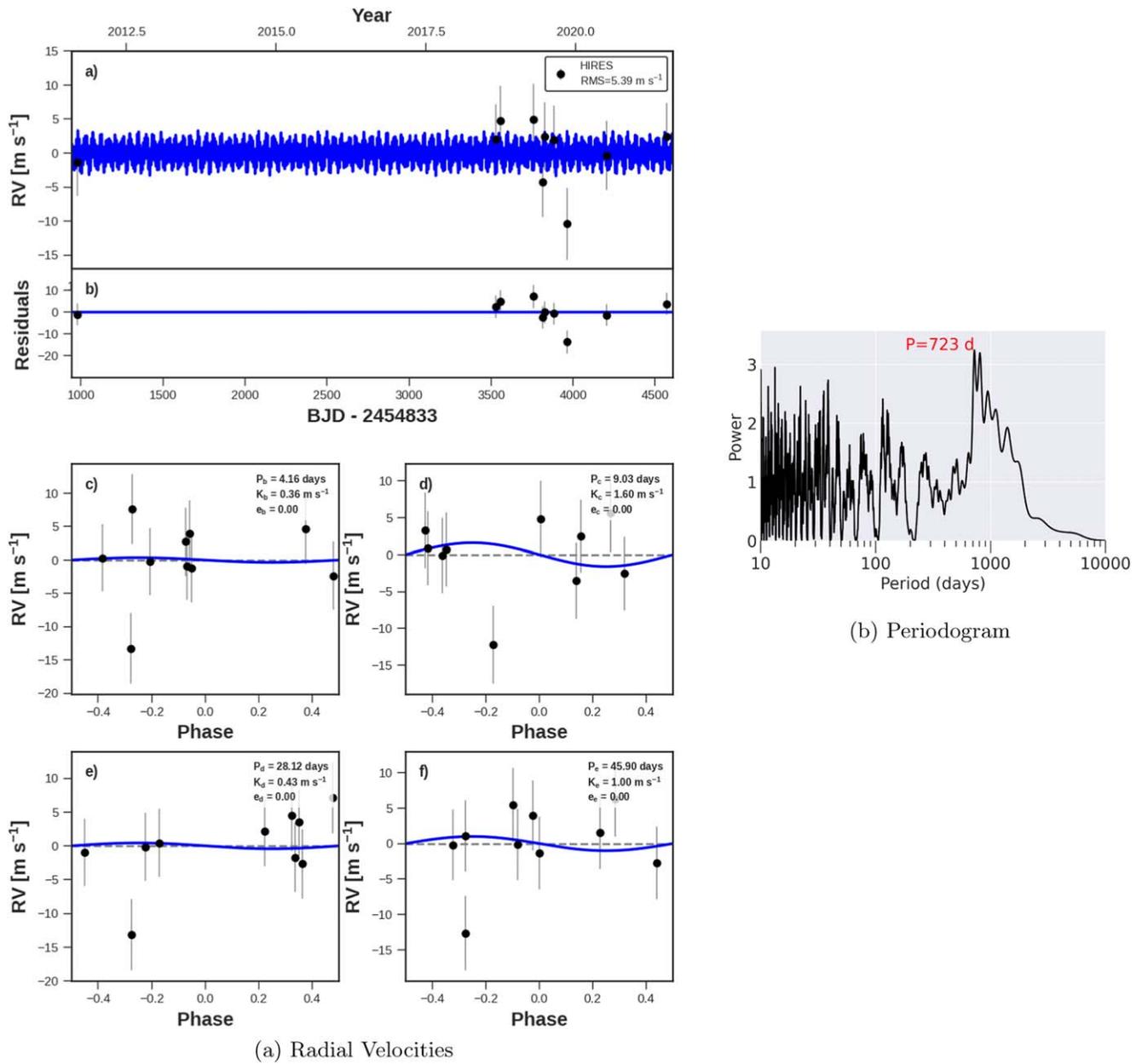

Figure 55. Same as Figure 8, but for KOI-719 (Kepler-220). No nontransiting companions are detected.

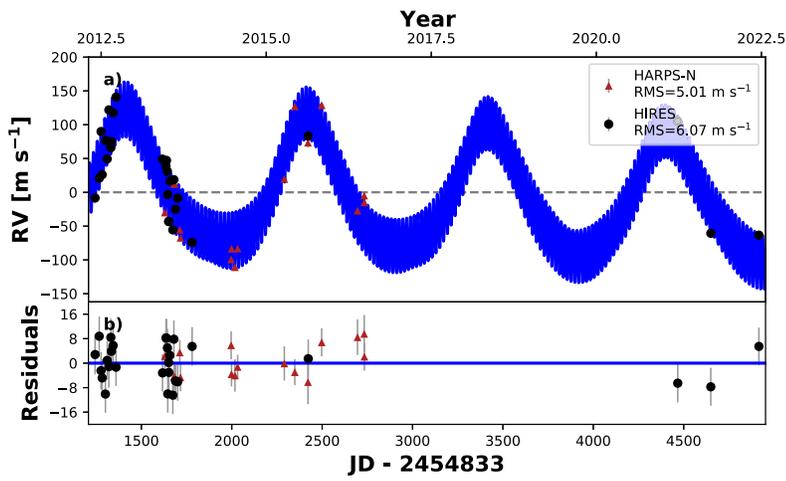
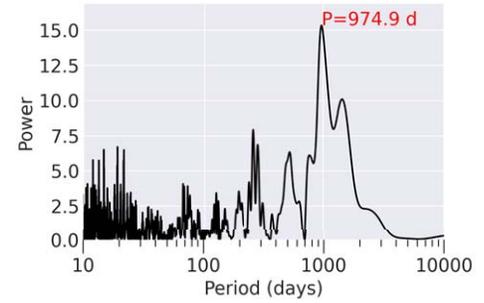

(b) Periodogram

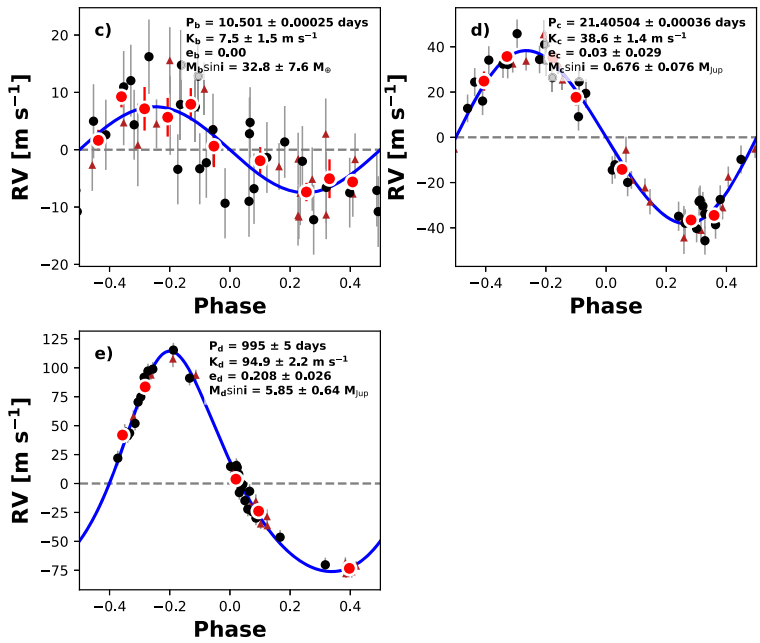
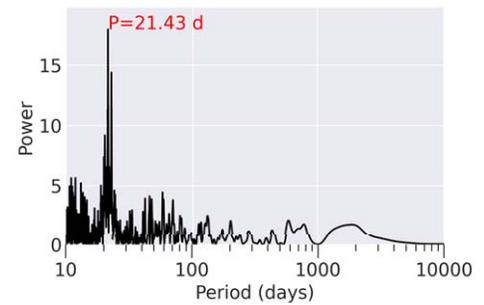

(c) Residual Periodogram

(a) RVs

**Figure 56.** Same as Figure 8, but for KOI-1241 (Kepler-56). Left: RVs are from HIRES (black circles) and HARPS-N (maroon triangles). The best-fit model has two nontransiting giant planets (details inset in RV panels, left).

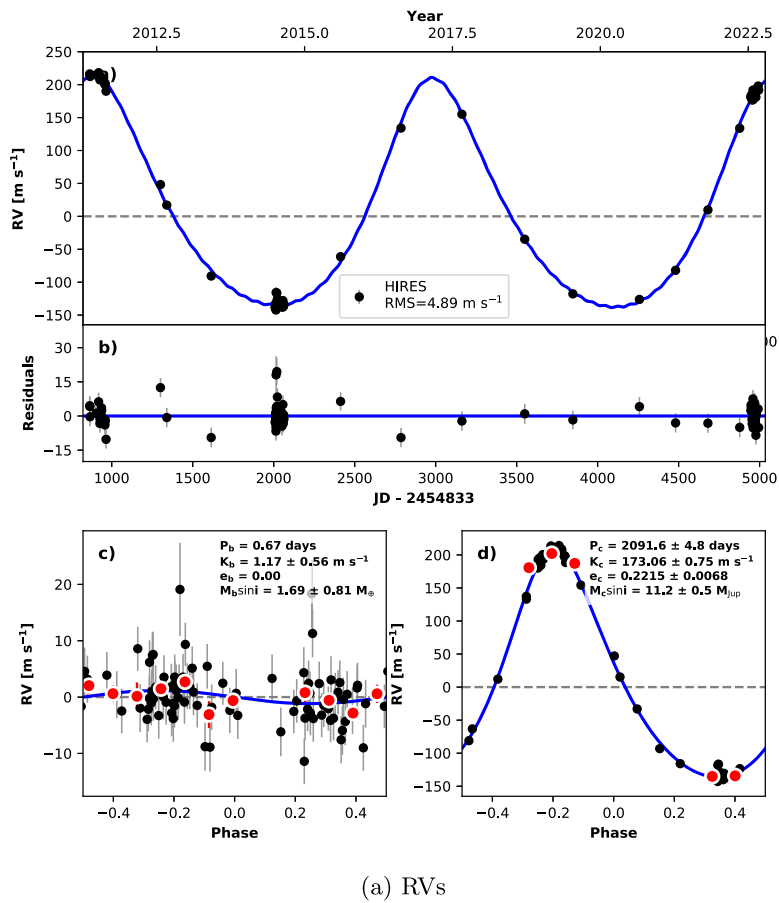

Figure 57. Same as Figure 8, but for KOI-1442 (Kepler-407). Left: the best-fit model includes a nontransiting planet, Kepler-407 c (orbital parameters inset in phase-folded panel). Right: the periodogram of the RVs after removing the signals from the transiting planet (top) and both planets (bottom).

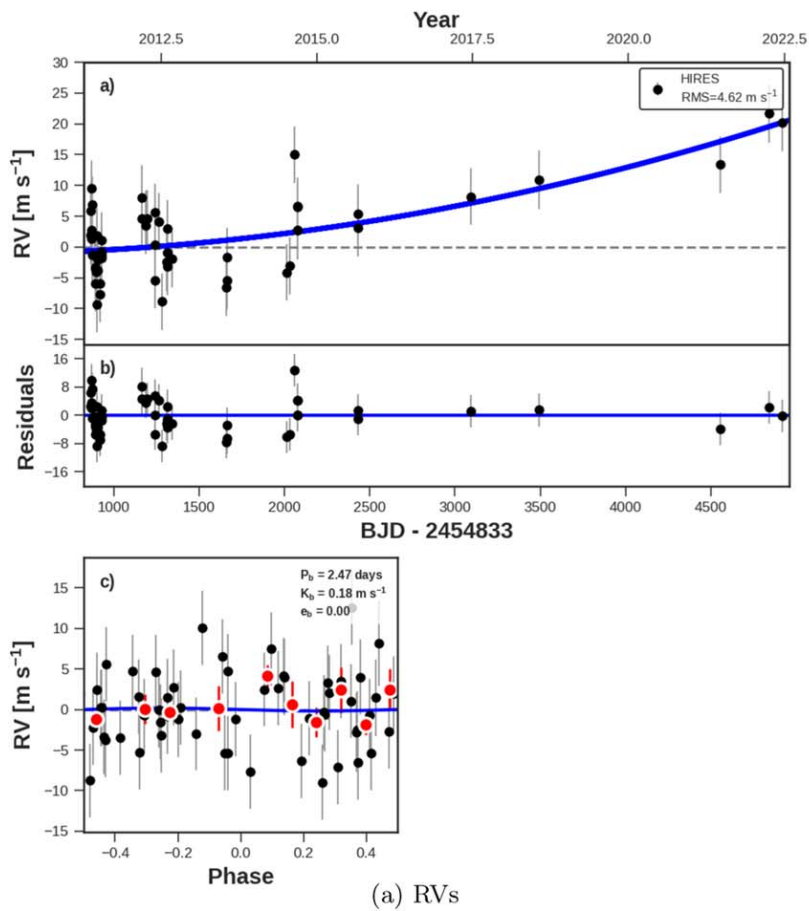
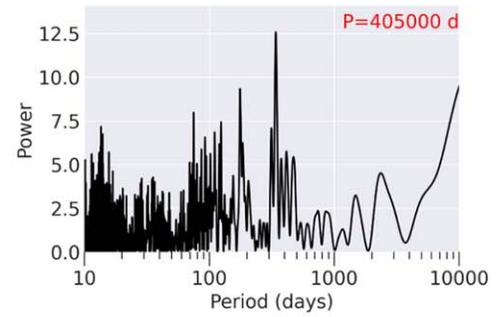

**Figure 58.** Same as Figure 8, but for KOI-1612 (Kepler-408). Left: there is a marginally significant long-term trend to the RVs consistent with a stellar-mass companion.

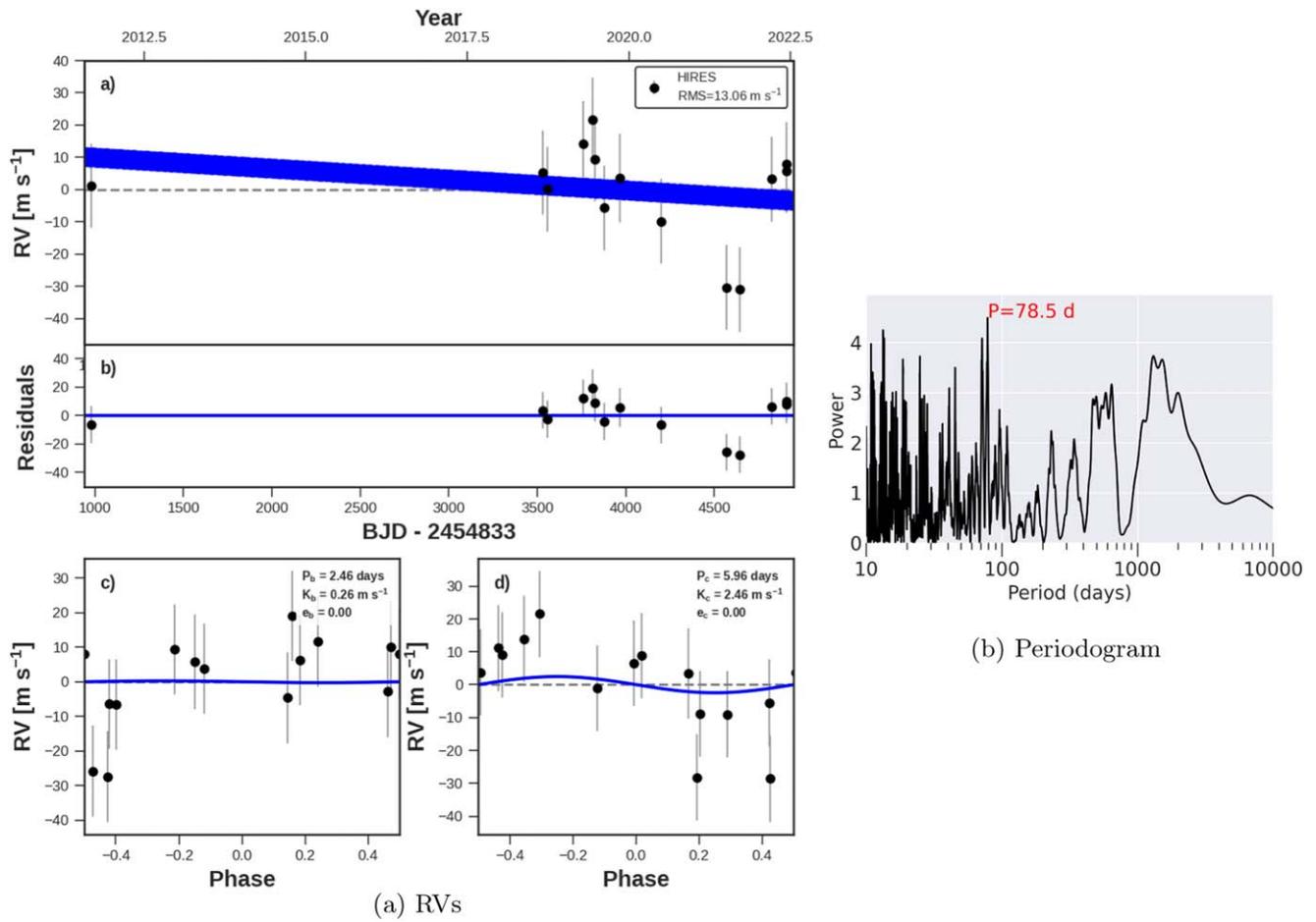

**Figure 59.** Same as Figure 8, but for KOI-1692 (Kepler-314). There is a marginal trend in the RVs.

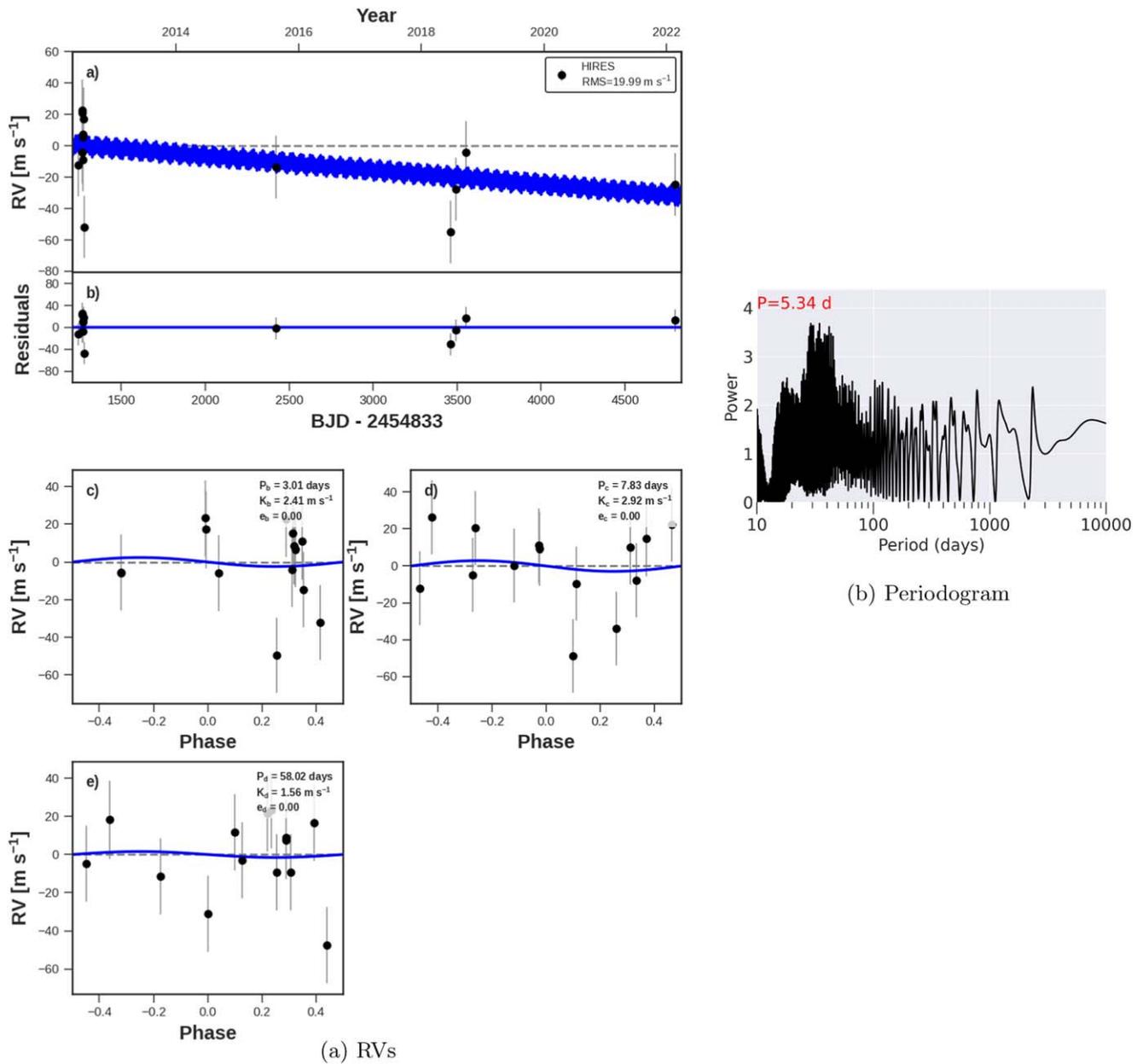

**Figure 60.** Same as Figure 8, but for KOI-1781 (Kepler-411). No nontransiting companions are detected, although the RV variability is higher than typical for our sample, likely because of high stellar activity.

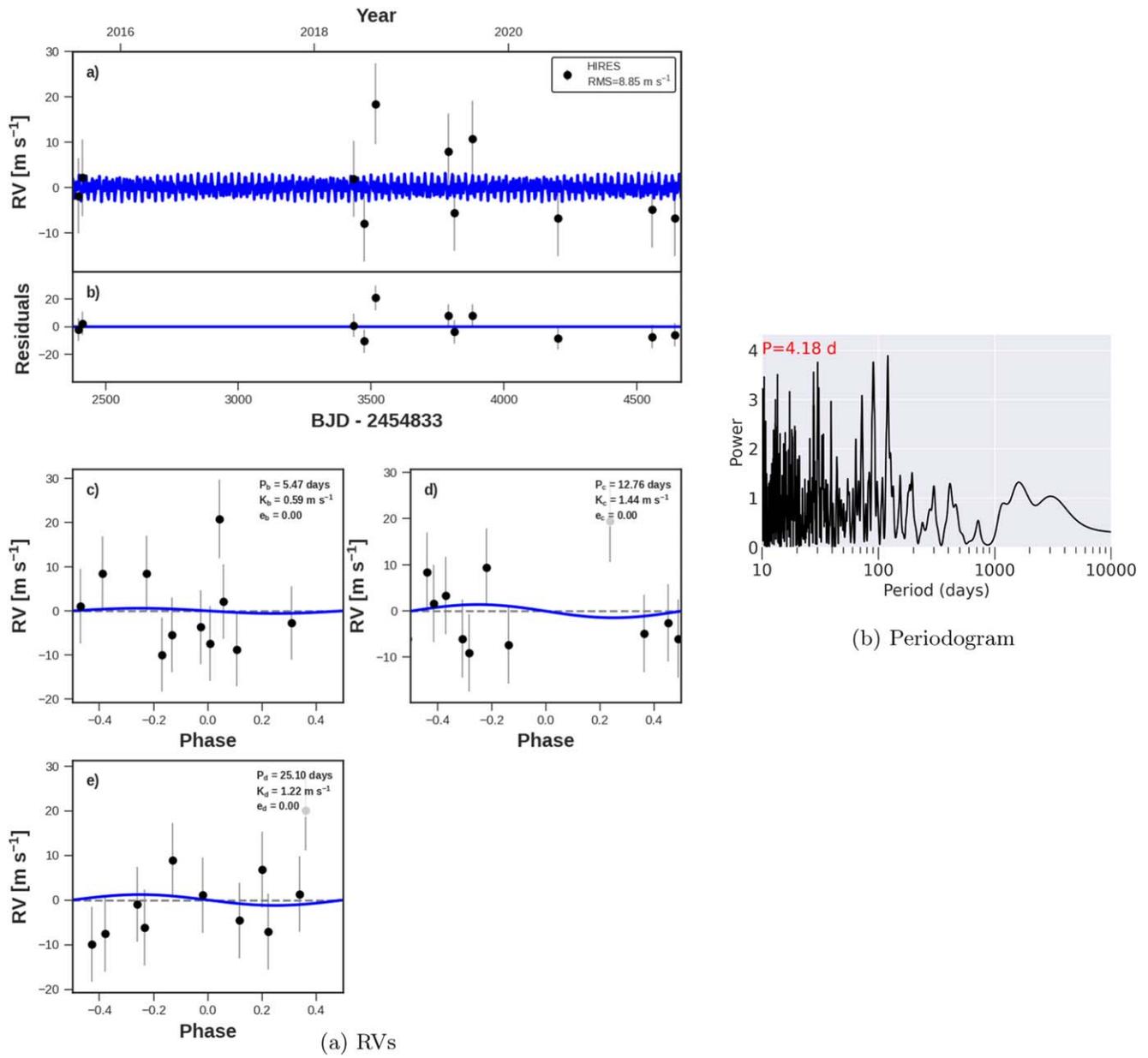

**Figure 61.** Same as Figure 8, but for KOI-1909 (Kepler-334). No nontransiting companions are detected.

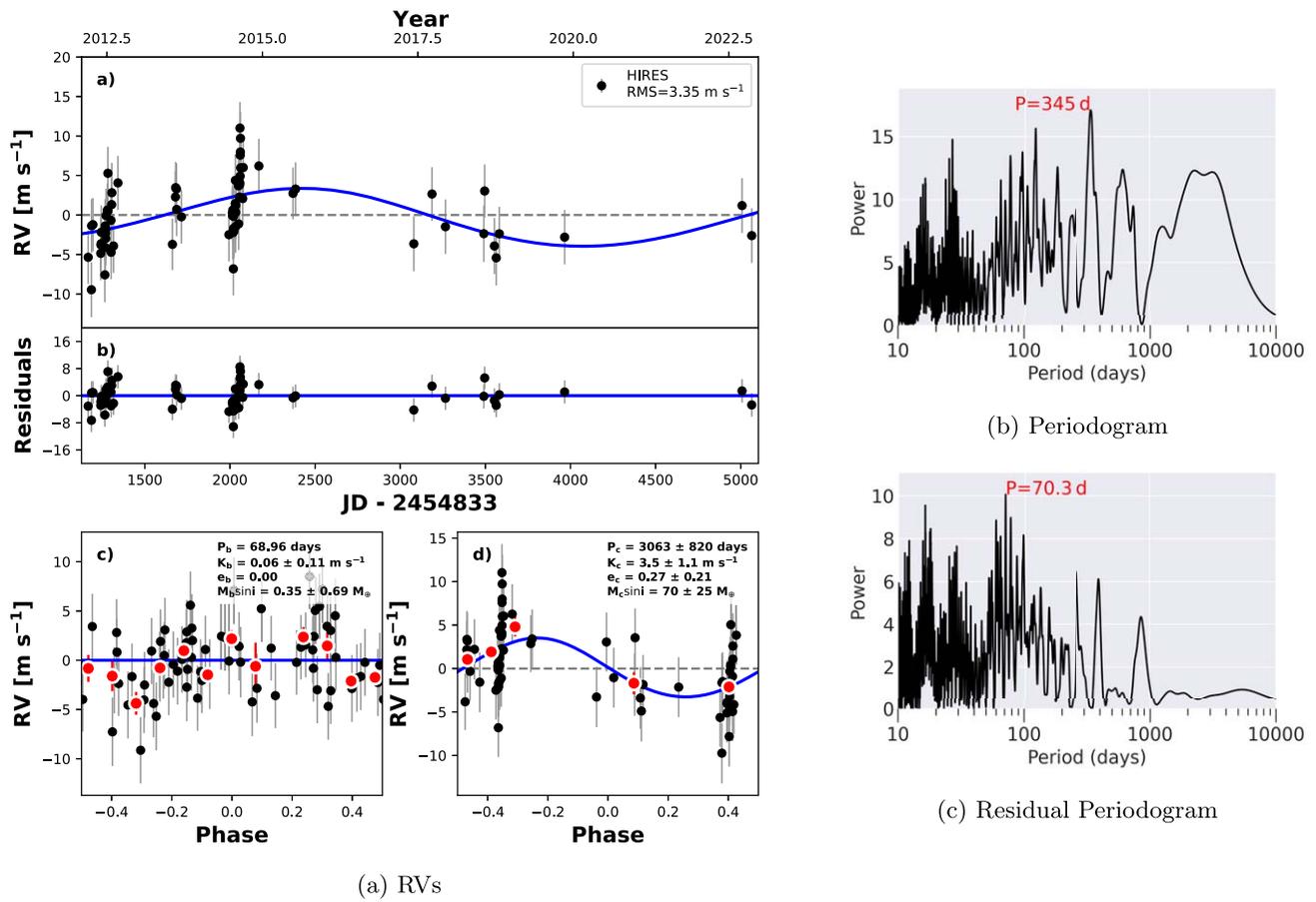

(a) RVs

(b) Periodogram

(c) Residual Periodogram

**Figure 62.** Same as Figure 8, but for KOI-1925 (Kepler-409). Left: a nontransiting planet is detected with low significance (orbital parameters inset). Right: periodograms of the RVs after removing a model of the transiting planet only (top) and the best-fit two-planet model (bottom).

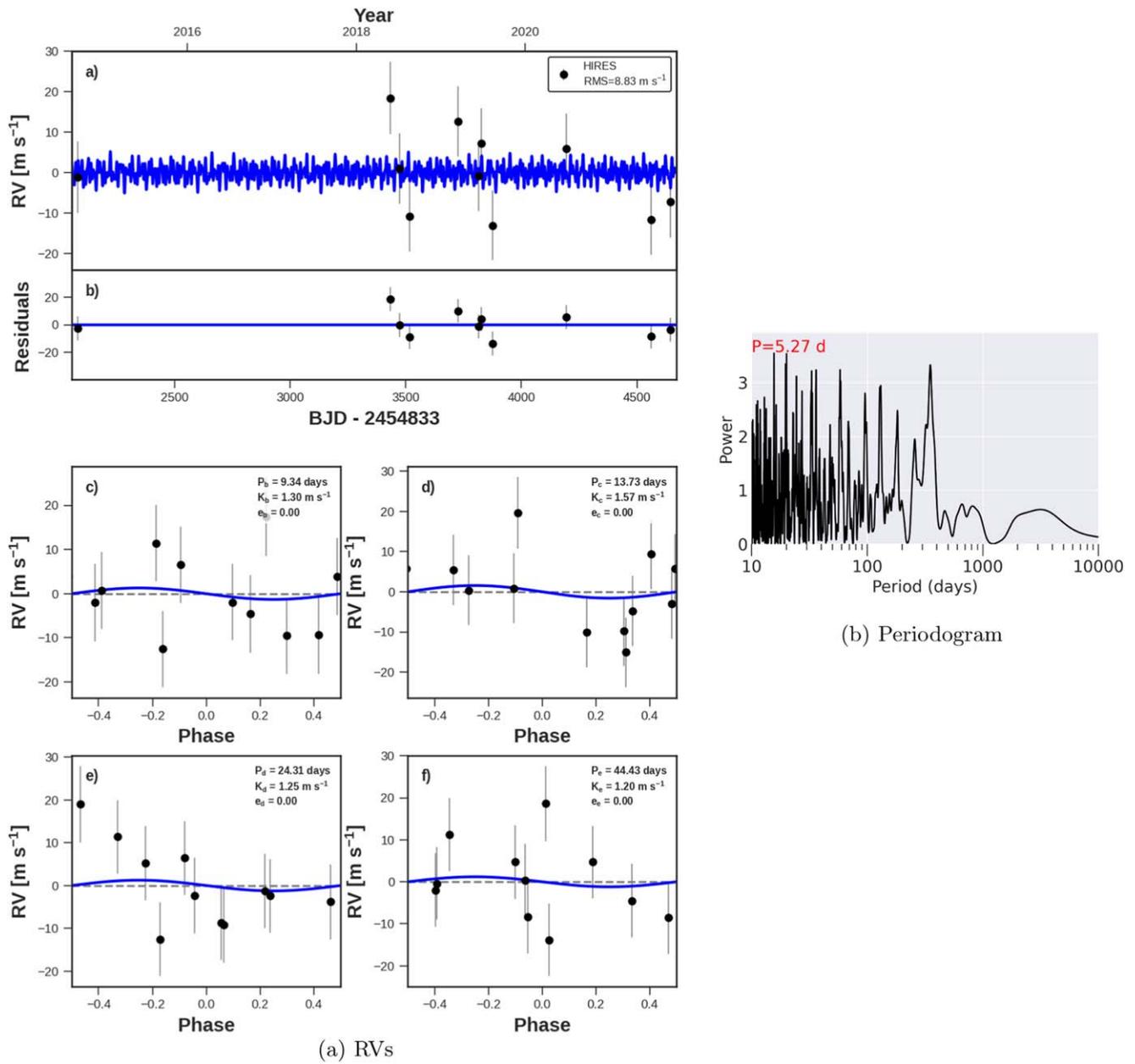

**Figure 63.** Same as Figure 8, but for KOI-1930 (Kepler-338). No nontransiting companions are detected.

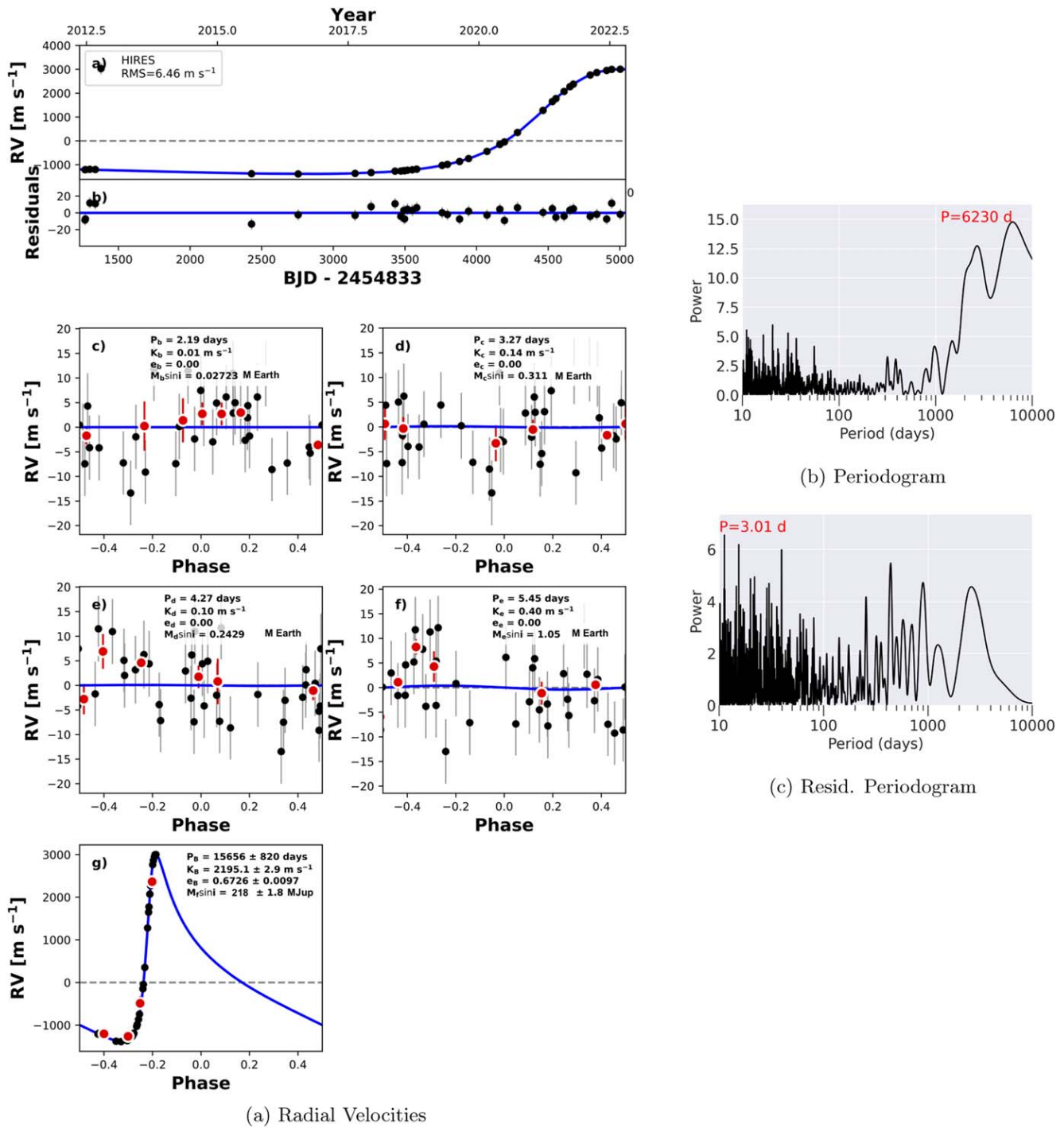

**Figure 64.** Same as Figure 8, but for KOI-2169 (Kepler-1130). Left: the best-fit model includes a long-period, highly eccentric stellar-mass companion that moved through periastron in 2022. The automated KGPS algorithm detects this companion but does not find the optimal orbital period because the full phase curve of the eccentric orbit has not been sampled yet. Right: periodograms of the RVs after removing a model of the transiting planets only (top) and our best-fit model of the stellar companion (bottom).

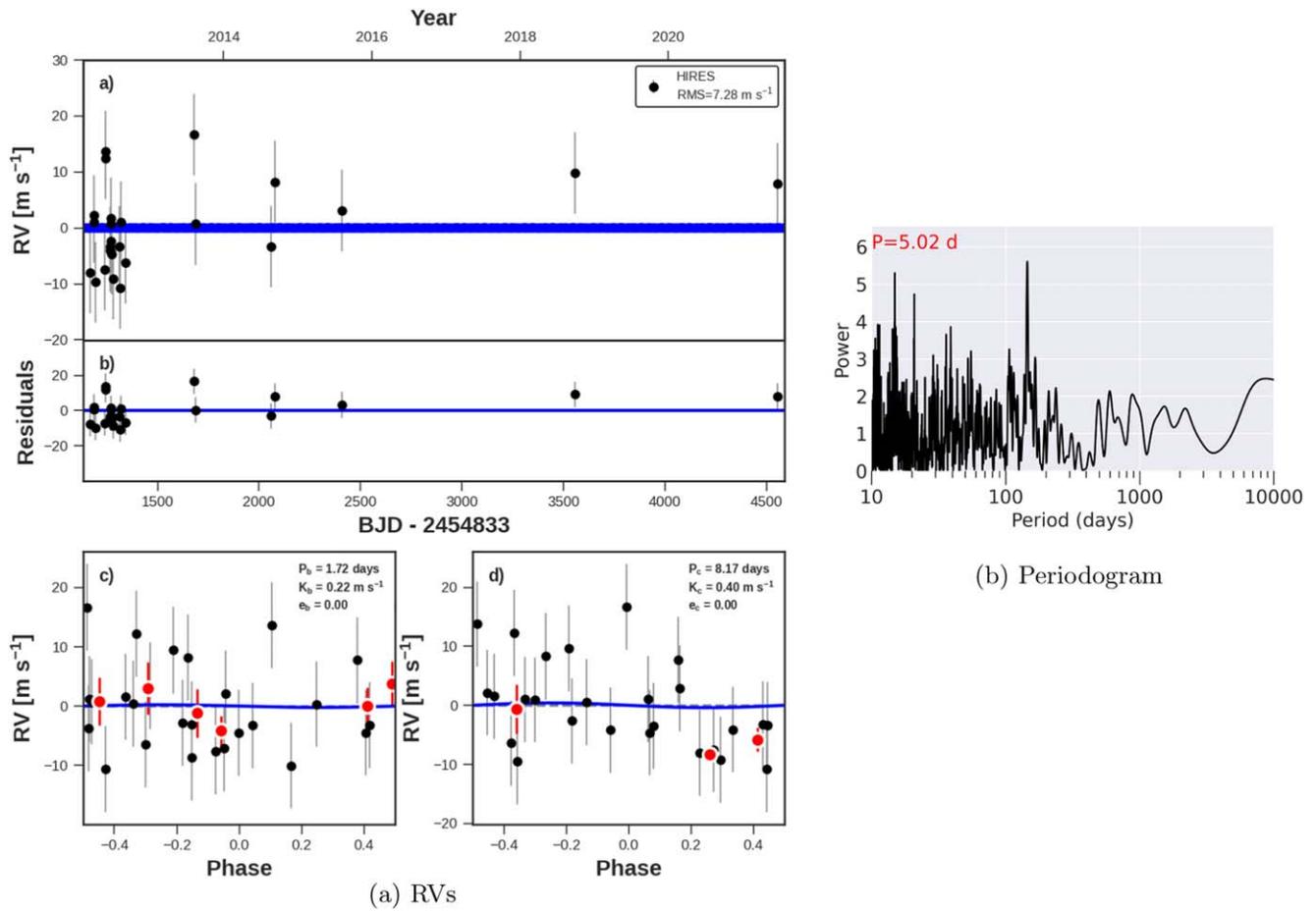

**Figure 65.** Same as Figure 8, but for KOI-2687 (Kepler-1869). No nontransiting companions are detected.

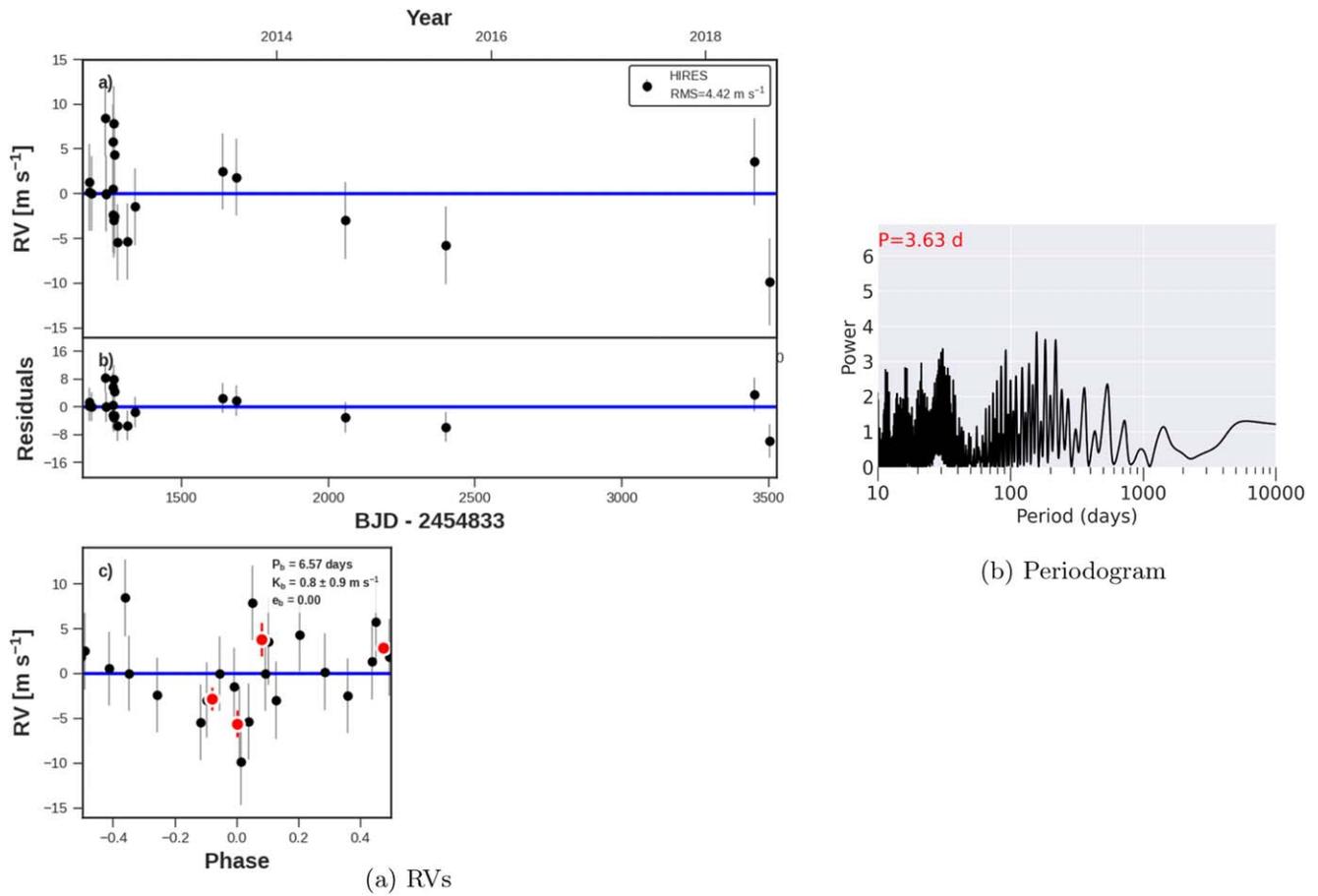

**Figure 66.** Same as Figure 8, but for KOI-2720 (KIC 8176564). No nontransiting companions are detected.

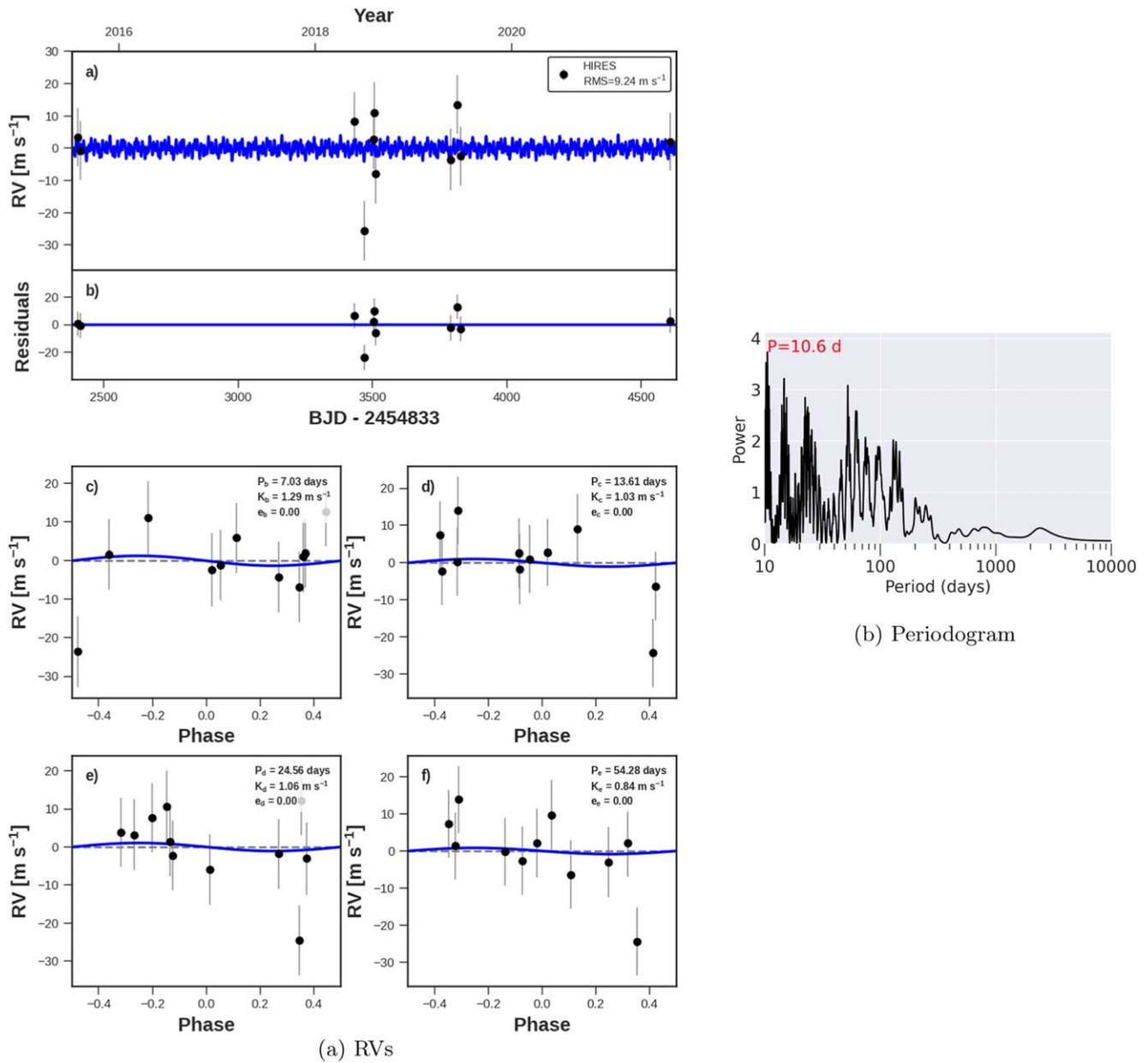

Figure 67. Same as Figure 8, but for KOI-2732 (Kepler-403). No nontransiting companions are detected.

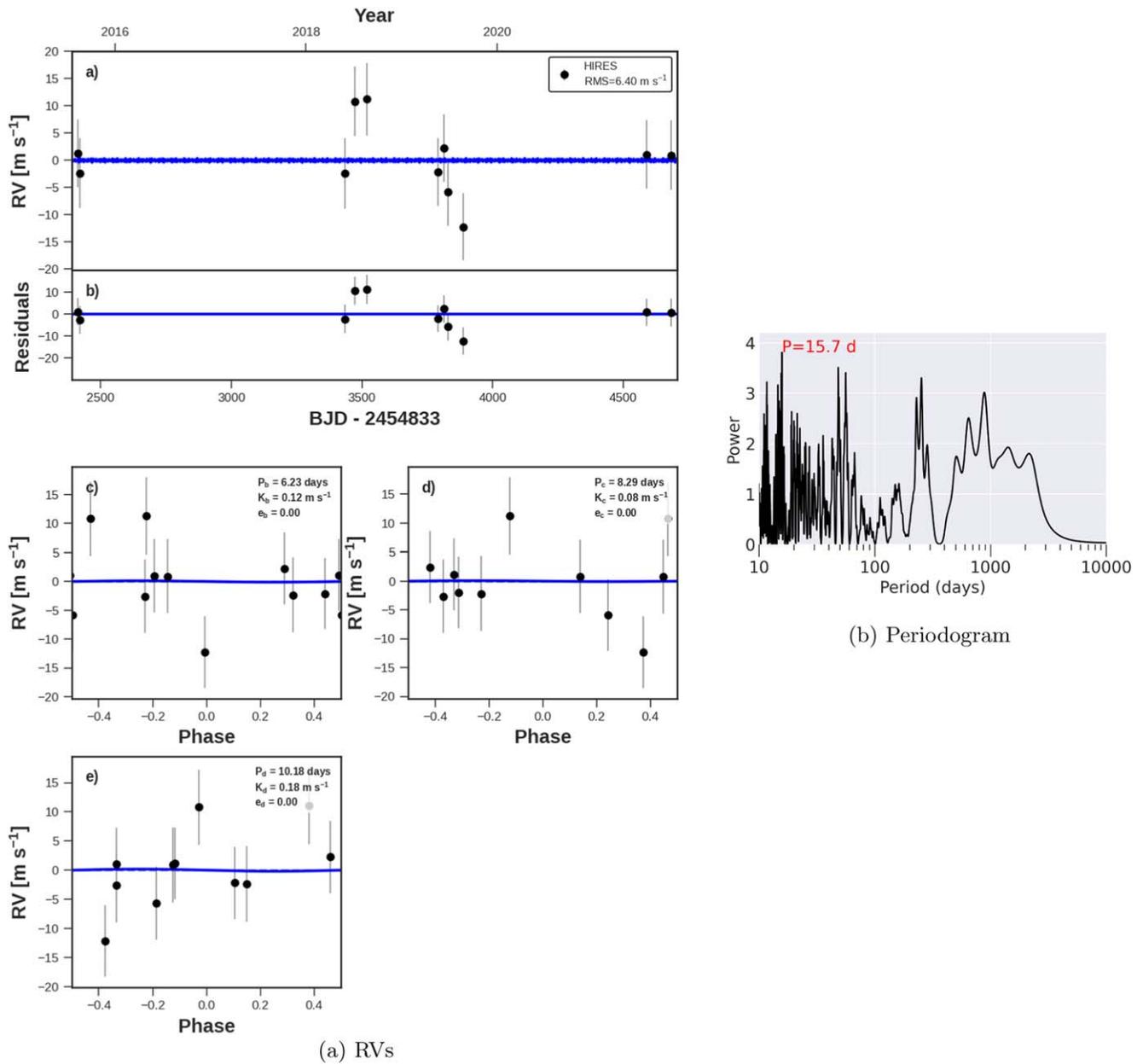

**Figure 68.** Same as Figure 8, but for KOI-3083. No nontransiting companions are detected.

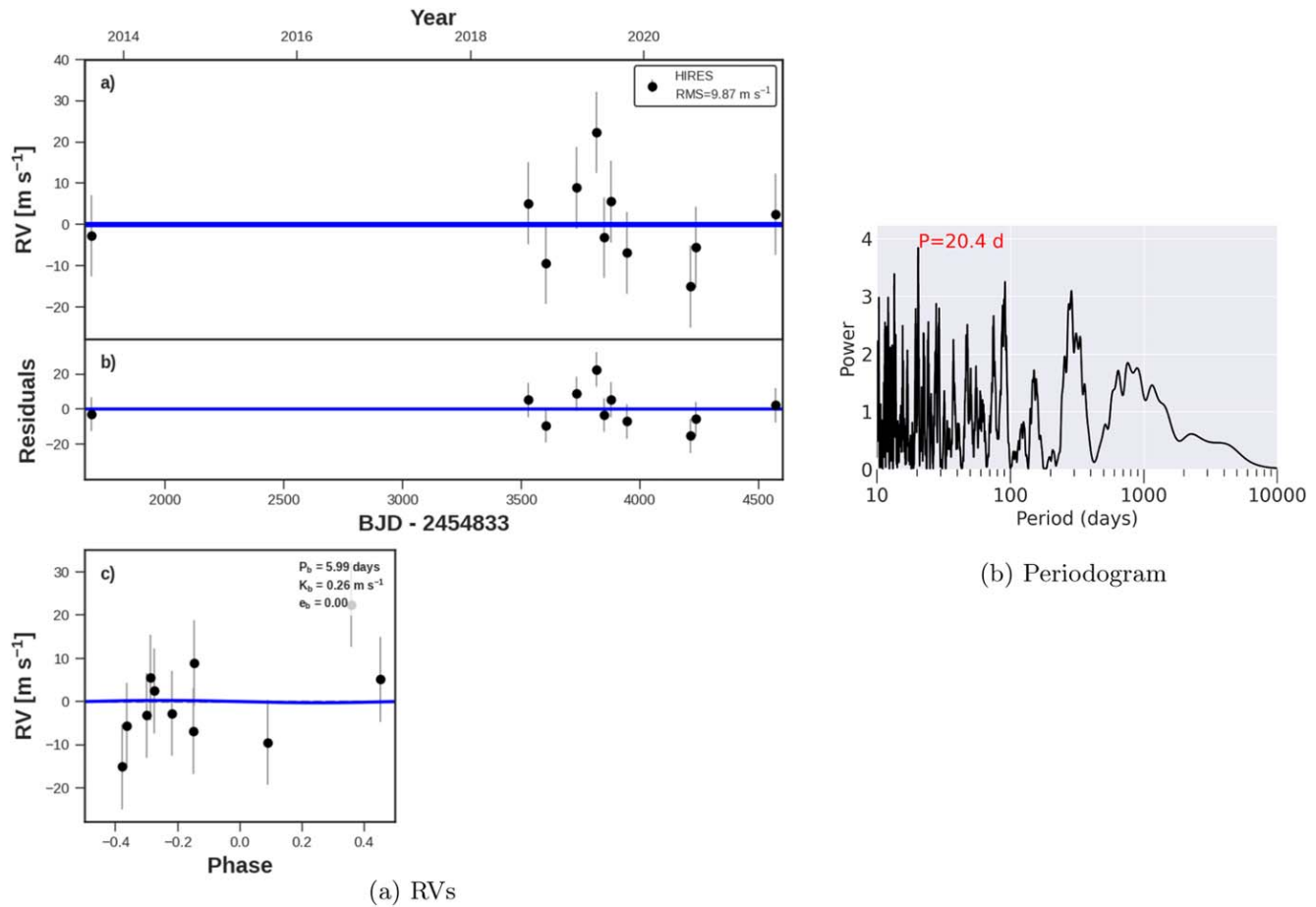

**Figure 69.** Same as Figure 8, but for KOI-3179 (Kepler-1911). No nontransiting companions are detected.

**ORCID iDs**

Lauren M. Weiss ● https://orcid.org/0000-0002-3725-3058
Howard Isaacson ● https://orcid.org/0000-0002-0531-1073
Andrew W. Howard ● https://orcid.org/0000-0001-8638-0320
Benjamin J. Fulton ● https://orcid.org/0000-0003-3504-5316
Erik A. Petigura ● https://orcid.org/0000-0003-0967-2893
Daniel Fabrycky ● https://orcid.org/0000-0003-3750-0183
Daniel Jontof-Hutter ● https://orcid.org/0000-0002-6227-7510
Jason H. Steffen ● https://orcid.org/0000-0003-2202-3847
Hilke E. Schlichting ● https://orcid.org/0000-0002-0298-8089
Jason T. Wright ● https://orcid.org/0000-0001-6160-5888
Corey Beard ● https://orcid.org/0000-0001-7708-2364
Casey L. Brinkman ● https://orcid.org/0000-0002-4480-310X
Ashley Chontos ● https://orcid.org/0000-0003-1125-2564
Steven Giacalone ● https://orcid.org/0000-0002-8965-3969
Michelle L. Hill ● https://orcid.org/0000-0002-0139-4756
Molly R. Kosiarek ● https://orcid.org/0000-0002-6115-4359
Mason G. MacDougall ● https://orcid.org/0000-0003-2562-9043
Teo Močnik ● https://orcid.org/0000-0003-4603-556X
Alex S. Polanski ● https://orcid.org/0000-0001-7047-8681
Emma V. Turtelboom ● https://orcid.org/0000-0002-1845-2617
Dakotah Tyler ● https://orcid.org/0000-0003-0298-4667
Judah Van Zandt ● https://orcid.org/0000-0002-4290-6826